\makeatletter
\disable@package@load{newtxmath}{}
\makeatother
\documentclass[acmsmall,screen,nonacm]{acmart}

\microtypesetup{expansion=false}
\usepackage{amssymb}
\usepackage{libertinust1math}
\DeclareMathAlphabet{\mathrm}{T1}{\rmdefault}{m}{n}
\DeclareMathAlphabet{\mathit}{T1}{\rmdefault}{m}{it}
\DeclareMathAlphabet{\mathbf}{T1}{\rmdefault}{b}{n}
\DeclareMathAlphabet{\mathsf}{T1}{\sfdefault}{m}{n}
\DeclareMathAlphabet{\mathtt}{T1}{zi4}{m}{n}
\DeclareMathAlphabet{\mathcal}{OMS}{lmsy}{m}{n}
\usepackage{listings}
\usepackage{lstlinebgrd}
\usepackage{needspace}
\usepackage{flafter}
\usepackage{placeins}
\usepackage{tikz}
\usetikzlibrary{decorations.pathmorphing,cd}

\pdfmapline{+inconsolata-semibold Inconsolata-SemiBold "ECEncoding ReEncodeFont" <ec.enc <./fonts/Inconsolata-SemiBold.ttf}
\pdfmapline{+inconsolata-bold Inconsolata-Bold "ECEncoding ReEncodeFont" <ec.enc <./fonts/Inconsolata-Bold.ttf}
\DeclareFontFamily{T1}{originalcode}{}
\DeclareFontShape{T1}{originalcode}{m}{n}{<-> ./fonts/inconsolata-semibold}{}
\DeclareFontShape{T1}{originalcode}{b}{n}{<-> ./fonts/inconsolata-bold}{}
\DeclareFontShape{T1}{originalcode}{bx}{n}{<-> ssub * originalcode/b/n}{}
\newcommand{\codefont}{\def\ttdefault{originalcode}\ttfamily\chardef\textasciigrave=0\relax}
\newcommand{\codesize}{\fontsize{9}{10.8}\selectfont}
\newcommand{\listingsize}{\fontsize{8.5}{10.2}\selectfont}

\definecolor{CodeKeyword}{HTML}{0000FF}
\definecolor{CodeType}{HTML}{00705B}
\definecolor{CodeView}{HTML}{9B2463}
\definecolor{CodeString}{HTML}{B02020}
\definecolor{CodeChange}{HTML}{FFF0CC}
\definecolor{CodeErrorBackground}{HTML}{FFE5E5}
\definecolor{CodeDiffRemoved}{HTML}{FDE8E7}
\definecolor{CodeDiffAdded}{HTML}{E6F4EA}

\newcommand{\CodeSquiggle}[1]{%
  \tikz[baseline=(code.base)]{%
    \node[inner sep=0pt,outer sep=0pt] (code)
      {\codefont\listingsize\mdseries\color{black}#1};
    \draw[CodeString,line width=0.8pt,decorate,
      decoration={snake,amplitude=0.55pt,segment length=2.5pt}]
      ([yshift=-1.6pt]code.base west) -- ([yshift=-1.6pt]code.base east);
  }%
}

\ExplSyntaxOn
\NewDocumentCommand{\CodeChangedLines}{mm}{
  \clist_set:Nn \l_tmpa_clist {#1}
  \exp_args:NNx \clist_if_in:NnT \l_tmpa_clist
    {\int_eval:n {\value{lstnumber}}}
    {\color{CodeChange}#2}
}
\NewDocumentCommand{\CodeErrorLines}{mm}{
  \clist_set:Nn \l_tmpa_clist {#1}
  \exp_args:NNx \clist_if_in:NnT \l_tmpa_clist
    {\int_eval:n {\value{lstnumber}}}
    {\color{CodeErrorBackground}#2}
}
\NewDocumentCommand{\CodeDiffLines}{mmm}{
  \clist_set:Nn \l_tmpa_clist {#1}
  \exp_args:NNx \clist_if_in:NnT \l_tmpa_clist
    {\int_eval:n {\value{lstnumber}}}
    {\color{CodeDiffRemoved}#3}
  \clist_set:Nn \l_tmpa_clist {#2}
  \exp_args:NNx \clist_if_in:NnT \l_tmpa_clist
    {\int_eval:n {\value{lstnumber}}}
    {\color{CodeDiffAdded}#3}
}
\ExplSyntaxOff
\newcommand{\CodeChangeBackground}[3]{%
  \smash{\rlap{\rule[-#3]{#1}{\dimexpr#2+#3\relax}}}%
}

\lstdefinelanguage{Rust}{
  keywords={struct,enum,impl,fn,let,mut,self,Self,pub,for,in,if,else,match,
    return,trait,type,where,use,as,const,static,move,loop,while,break,continue,
    async,await,dyn,ref,unsafe,extern,mod,crate,super,true,false,
    bool,char,str,u8,u16,u32,u64,u128,usize,i8,i16,i32,i64,i128,isize,f32,f64},
  sensitive=true,
  morecomment=[l]{//},
  morecomment=[l][\color{CodeString}]{//\ ERROR},
  morecomment=[s]{/*}{*/},
  morestring=[b]"
}
\lstdefinestyle{diagnostic}{
  language={},basicstyle=\codefont\listingsize\mdseries\color{black},
  keywordstyle={},identifierstyle={},commentstyle={},stringstyle={},emph={},
  moredelim=[s][\bfseries\color{CodeString}]{error[}{]}
}
\lstdefinestyle{inlinecode}{basicstyle=\upshape\codefont\codesize\mdseries}
\lstdefinestyle{diff}{
  language={},emph={},frame=single,
  framerule=0.4pt,rulecolor=\color{black!25}
}
\newcommand{\code}[1]{\mbox{\lstinline[style=inlinecode]{#1}}}
\newcommand{\codetext}[1]{{\codesize\texttt{#1}}}

\makeatletter           %
\@ifclassloaded{acmart}{}{\usepackage{amssymb}}
\makeatother
\usepackage{stmaryrd}     %
\usepackage{mathpartir}   %
\usepackage{mathtools}    %
\usepackage{float}        %
\usepackage{xspace}
\usepackage{xcolor}
\usepackage{pifont}       %
\usepackage{booktabs}
\usepackage{array}
\usepackage{longtable}   %
\usepackage{cleveref}     %
\AtBeginEnvironment{definition}{\crefalias{theorem}{definition}}

\newcommand{\kw}[1]{\mathtt{#1}}              %
\newcommand{\metafn}[1]{\mathsf{#1}}          %
\newcommand{\tyname}[1]{\mathsf{#1}}          %
\newcommand{\rulename}[1]{\textsc{#1}}        %

\newcommand{\vect}[1]{\overrightarrow{#1}}    %
\newcommand{\defeq}{\triangleq}              %
\newcommand{\grameq}{\mathrel{\vcentcolon\mkern-1.2mu\vcentcolon\mkern-1.2mu=}}  %

\newcommand{\aeneas}{\textsc{Aeneas}\xspace}

\newcommand{\invar}{\ensuremath{x_{\mathit{in}}}}
\newcommand{\retvar}{\ensuremath{x_{\mathit{ret}}}}  %
\newcommand{\locvar}{\ensuremath{x_{\mathit{loc}}}}  %

\DeclareRobustCommand{\door}{%
\hbox{\kern1.5pt\vbox to10pt{}%
\tikz[overlay,x=.7bp,y=.7bp]{%
  \draw[line width=.7bp] (2,0)--(2,12)--(8,12)--(8,0)
    (2,12)--(6,9.5)--(6,-3)--(2,0);
  \fill (4.4,2.5) rectangle +(1,1);}%
\kern 8pt}
}

\newcommand{\LLBC}{\textsf{LLBC}\xspace}
\newcommand{\LLBCsharp}{\textsf{LLBC\textsuperscript{\#}}\xspace}

\newcommand{\Mut}{\mathsf{mut}}
\newcommand{\Shr}{\mathsf{shr}}  %
\newcommand{\Imm}{\mathsf{imm}}  %
\newcommand{\MMove}{\mathsf{mov}}  %

\newcommand{\VBot}{\bot}
\newcommand{\VUnit}{()}  %
\newcommand{\VPair}[2]{(#1, #2)}  %
\newcommand{\VLeft}[1]{\mathsf{left}\; #1}  %
\newcommand{\VRight}[1]{\mathsf{right}\; #1}  %
\newcommand{\VMutLoan}[1]{\mathsf{loan}^{m}\, #1}  %
\newcommand{\VShrLoan}[2]{\mathsf{loan}^{s}\, #1\; #2}  %
\newcommand{\VMutBor}[2]{\mathsf{bor}^{m}\, #1\; #2}  %
\newcommand{\VShrBor}[1]{\mathsf{bor}^{s}\, #1}  %

\newcommand{\VSymbolicV}[1]{\sigma : #1}
\newcommand{\VIgnored}[1]{\_ : #1}

\newcommand{\TUnit}{()}
\newcommand{\TRef}[2]{\&#1\, #2}  %
\newcommand{\TShrRef}[1]{\TRef{}{#1}}    %
\newcommand{\TMutRef}[1]{\TRef{\Mut}{#1}}  %
\newcommand{\TPair}[2]{(#1, #2)}  %

\newcommand{\fndecl}[5]{\kw{fn}\ #1(#2,\, #3) \to #4\ \{\, #5\, \}}

\newcommand{\val}{v}           %
\newcommand{\ty}{\tau}         %
\newcommand{\nat}{n}           %
\newcommand{\mode}{m}          %
\newcommand{\place}{p}         %
\newcommand{\op}{\mathit{op}}  %
\newcommand{\rv}{\mathit{rv}}  %
\newcommand{\vp}{\mathit{vp}}  %
\newcommand{\view}{V}          %
\newcommand{\fn}{f}            %
\newcommand{\mfndecl}{\mathit{fndecl}}  %
\newcommand{\seg}{d}           %
\newcommand{\res}{r}           %
\newcommand{\entry}{b}         %
\newcommand{\entries}{E}       %
\newcommand{\var}{x}           %
\newcommand{\stmt}{s}          %
\newcommand{\abs}{A}           %
\newcommand{\prog}{\mathcal{P}}  %

\newcommand{\proj}{P}  %
\newcommand{\PDeref}{\mathord{*}}  %

\newcommand{\fpath}{\pi}
\newcommand{\valaddr}{\tyname{val\_addr}}  %
\newcommand{\dpath}{\delta}  %
\newcommand{\dpnil}{\varepsilon}  %
\newcommand{\dpunloan}{\mathsf{unloan}}  %
\newcommand{\dpunbor}{\mathsf{unbor}}  %
\newcommand{\dpproj}[1]{\mathsf{proj}_{#1}}  %
\newcommand{\feaddr}{\hat{\fpath}}  %

\newcommand{\st}{\Omega}

\newcommand{\frameval}[3]{\langle \locvar \mapsto #1,\ \invar \mapsto #2,\ \retvar \mapsto #3 \rangle}

\newcommand{\loc}{\ell}

\newcommand{\reads}[5][\vdash]{#1 #2(#3) \xRightarrow{#4} #5}
\newcommand{\sharpreads}{\reads[\vdash_\#]}  %
\newcommand{\plusreads}{\reads[\vdash^{+}]}  %
\newcommand{\readwrite}[7][\vdash]{#1 #2[#3 \mapsto #5 \leftarrow #6] \xRightarrow{#4} #7}
\newcommand{\sharpreadwrite}{\readwrite[\vdash_\#]}  %
\newcommand{\plusreadwrite}{\readwrite[\vdash^{+}]}  %
\newcommand{\stateupd}[5][\vdash]{#1 #2(#3) \leftarrow #4 \Rightarrow #5}
\newcommand{\sharpstateupd}{\stateupd[\vdash_\#]}  %
\newcommand{\plusstateupd}{\stateupd[\vdash^{+}]}  %
\newcommand{\fpapplies}[4][\vdash]{#1 #2(#3) \Rightarrow #4}
\newcommand{\vaapplies}[4][\vdash]{\fpapplies[#1]{#2}{#3}{#4}}
\newcommand{\fpreplace}[6][\vdash]{#1 #2(#3) \mapsto #4 \leftarrow #5 \Rightarrow #6}
\newcommand{\sharpfpreplace}{\fpreplace[\vdash_\#]}
\newcommand{\plusfpreplace}{\fpreplace[\vdash^{+}]}
\newcommand{\sharpvaapplies}{\vaapplies[\vdash_\#]}
\newcommand{\plusvaapplies}{\vaapplies[\vdash^{+}]}
\newcommand{\rvevals}[5][\vdash]{#2 #1 #3 \Downarrow (#4,\, #5)}
\newcommand{\sharprvevals}{\rvevals[\vdash_\#]}  %
\newcommand{\opevals}[5][\vdash]{#2 #1 #3 \Downarrow (#4,\, #5)}
\newcommand{\sharpopevals}{\opevals[\vdash_\#]}  %
\newcommand{\plusopevals}{\opevals[\vdash^{+}]}  %
\newcommand{\assignevals}[5][\vdash]{#2 #1 \SAssign{#3}{#4} \Downarrow #5}
\newcommand{\sharpassignevals}{\assignevals[\vdash_\#]}  %
\newcommand{\plusassignevals}{\assignevals[\vdash^{+}]}  %

\newcommand{\valty}[3][\vdash]{#1 #2 : #3}
\newcommand{\sharpvalty}{\valty[\vdash_\#]}  %
\newcommand{\plusvalty}{\valty[\vdash^{+}]}  %

\newcommand{\borrowat}[6][\vdash]{#2 #1 #4\ #3 \Downarrow (#5,\, #6)}
\newcommand{\sharpborrowat}{\borrowat[\vdash_\#]}  %
\newcommand{\borrowevals}[6][\vdash]{#2 #1 \RvBorrow{#4}{#3} \Downarrow (#5,\, #6)}

\newcommand{\dpapplies}[3]{\vdash #1(#2) \Rightarrow #3}  %

\newcommand{\vpapplies}[3]{\dpapplies{#1}{#2}{#3}}

\newcommand{\valupd}[4]{\vdash #1(#2) \leftarrow #3 \Rightarrow #4}
\newcommand{\borrowupd}[4]{\valupd{#1}{#2}{#3}{#4}}

\newcommand{\LLBCView}{\textsf{LLBC-View}\xspace}
\newcommand{\LLBCSharpView}{\textsf{LLBC\textsuperscript{\#}-View}\xspace}
\newcommand{\LLBCPlusView}{\textsf{LLBC\textsuperscript{+}-View}\xspace}
\newcommand{\LLBCPlus}{\textsf{LLBC}\textsuperscript{+}\xspace}

\newcommand{\NoFuel}{\ensuremath{\infty}}
\newcommand{\StmtFin}[2]{\ensuremath{(#1,\, #2)}}  %
\newcommand{\RRet}{\ensuremath{\kw{return}}}  %
\newcommand{\RPanic}{\ensuremath{\kw{panic}}}  %
\newcommand{\stmtres}{R}  %
\newcommand{\results}{\vect{R}}  %
\newcommand{\RUnit}{\ensuremath{()}}  %

\newcommand{\bceval}[6][\vdash]{\ensuremath{#2, #3 #1_{#5} #4 \Downarrow #6}}  %
\newcommand{\sharpeval}[4]{\ensuremath{#1, #2 \vdash_{\#} #3 \Downarrow #4}}  %
\newcommand{\pluseval}{\bceval[\vdash^{+}]}  %

\newenvironment{grammar}{\begin{array}{@{}r@{\;}l@{}l@{}}}{\end{array}}
\newcommand{\gcomment}[1]{\if\relax\detokenize{#1}\relax\else\qquad\text{\small #1}\fi}
\newcommand{\gprod}[3]{#1 &\grameq #2 &\gcomment{#3}}
\newcommand{\galt}[2]{&\mid #1 &\gcomment{#2}}
\newcommand{\gabbr}[3]{#1 &\defeq #2 &\gcomment{#3}}
\newenvironment{defs}{%
  \begin{mathpar}%
  \def\[{\begingroup\let\defsarray\array\def\array{\defsarray[t]}}%
  \def\]{\endgroup\and}%
}{\end{mathpar}}

\makeatletter
\newcommand{\releaseheading}{\if@noskipsec\leavevmode\fi}
\makeatother
\NewDocumentEnvironment{deffig}{O{H} m m}
  {\releaseheading\begin{figure}[#1]\begin{defs}}
  {\end{defs}\caption{#3}\label{#2}\end{figure}}
\NewDocumentEnvironment{rulefig}{O{H} m m}
  {\releaseheading\begin{figure}[#1]\begin{mathpar}}
  {\end{mathpar}\caption{#3}\label{#2}\end{figure}}

\newcommand{\popframe}[3]{\mathsf{pop\_frame}\ #1 \Rsh #2\ #3}  %
\newcommand{\exitfn}[3]{\mathsf{exit\_fn}\ #1\ #2\ \door\ #3}

\newcommand{\callevals}[7][\vdash]{#2, #3 #1 \SCall{#4}{#5}{#6} \Downarrow #7}
\newcommand{\sharpcallevals}{\callevals[\vdash_\#]}  %

\newcommand{\pushabs}[2]{#1,\ #2}
\newcommand{\absliteral}[1]{\abs[#1]}
\newcommand{\envreorg}[3][\hookrightarrow]{\ensuremath{#2 #1 #3}}  %
\newcommand{\sharpenvreorg}{\envreorg[\hookrightarrow_\#]}  %
\newcommand{\plusenvreorg}{\envreorg[\hookrightarrow^{+}]}  %

\newcommand{\initextern}[2]{\ensuremath{\mathsf{init\_extern}\ #1\ #2}}
\newcommand{\projinput}[2]{\ensuremath{\mathsf{proj\_input}\ #1\ #2}}  %
\newcommand{\projoutput}[3]{\ensuremath{\mathsf{proj\_output}\ #1\ #2\ #3}}  %

\newcommand{\VBorV}[1]{\ensuremath{\mathsf{bor}\, #1}}  %
\newcommand{\singleborrow}[2]{\{#1 \mapsto #2\}}
\newcommand{\doubleborrow}[2]{\{0 \mapsto #1,\ 1 \mapsto #2\}}  %
\newcommand{\singleview}[2]{\{#2\ #1\}}  %
\newcommand{\doubleview}[2]{\{#1\ 0,\ #2\ 1\}}  %
\newcommand{\fullref}[3]{\VBorV{\singleborrow{\varepsilon}{#2}} : \TRef{\singleview{\varepsilon}{#1}}{#3}}  %
\newcommand{\FullMutBor}[2]{\VMutBor{#1}{#2}}  %
\newcommand{\FullShrBor}[2]{\VShrBorTy{#1}{#2}}  %

\newcommand{\hasouterloan}[1]{#1\ \text{has an outer loan}}
\newcommand{\nohasouterloan}[1]{#1\ \text{has no outer loan}}  %
\newcommand{\noloans}[1]{\mathsf{loan} \notin #1}  %
\newcommand{\nobotsloans}[1]{\bot, \mathsf{loan} \notin #1}
\newcommand{\nobotsmutloans}[1]{\bot, \mathsf{loan}^{m} \notin #1}
\newcommand{\lfresh}[2]{#1 \notin #2}  %
\newcommand{\nolocs}[1]{#1\ \text{is loc-free}}  %

\newcommand{\abscontainsloc}[2]{#1 \in #2}
\newcommand{\statecontains}[2]{#1 \in #2}
\newcommand{\containsmutloan}[2]{\mathsf{loan}^{m}\, #1 \in #2}
\newcommand{\containsshrloan}[2]{\mathsf{loan}^{s}\, #1 \in #2}  %
\newcommand{\shrloaninside}[2]{\containsshrloan{#1}{#2}}  %
\newcommand{\containsmutbor}[2]{\mathsf{bor}^{m}\, #1 \in #2}  %
\newcommand{\notcontainsshrbor}[2]{\VShrBor{#1} \notin #2}  %

\newcommand{\notinsideborrowed}[1]{\dpunloan, \dpunbor \notin #1}  %
\newcommand{\relabel}[3]{#1[#2 \leftrightarrows #3]}
\newcommand{\absdisjoint}[2]{#1 \mathrel{\#} #2}  %
\newcommand{\valabsdisjoint}[4]{(#1, #2) \mathrel{\#} (#3, #4)}  %
\newcommand{\alldead}[1]{#1\ \text{is dead}}

\newcommand{\reffree}[1]{#1\ \text{is ref-free}}
\newcommand{\notreffree}[1]{#1\ \text{is not ref-free}}  %
\newcommand{\unloanin}[1]{\dpunloan \in #1}

\newcommand{\toabs}[2]{#1 \prec^{\mathsf{to\text{-}abs}} #2}
\newcommand{\mergeabs}[4][\vdash]{#1 #2 \bowtie #3 \Rightarrow #4}  %
\newcommand{\sharpmergeabs}{\mergeabs[\vdash_\#]}  %
\newcommand{\plusmergeabs}{\mergeabs[\vdash^{+}]}  %

\newcommand{\SSeq}[2]{#1;\ #2}
\newcommand{\SAssign}[2]{#1 := #2}  %
\newcommand{\SMatch}[3]{\kw{match}\ #1\ \{#2 \mid #3\}}  %
\newcommand{\SCall}[3]{#1 := #2(#3)}  %
\newcommand{\SRet}{\kw{return}}  %
\newcommand{\SPanic}{\kw{panic}}  %
\newcommand{\SNil}{\kw{skip}}  %
\newcommand{\OpMove}[1]{\mathsf{move}\ #1}  %
\newcommand{\OpUnit}{()}  %
\newcommand{\OpPair}[2]{(#1, #2)}  %
\newcommand{\OpLeft}[2]{\VLeftTy{#1}{#2}}  %
\newcommand{\OpRight}[2]{\VRightTy{#1}{#2}}  %
\newcommand{\RvRefMut}[1]{\&\Mut\, #1}  %
\newcommand{\RvRefShr}[1]{\&\, #1}  %
\newcommand{\RvBorrow}[2]{\&#1\; #2}  %
\newcommand{\TSum}[2]{#1 + #2}
\newcommand{\VBotTy}[1]{\VBot : #1}  %
\newcommand{\VLeftTy}[2]{\VLeft{#1} : \TSum{\_}{#2}}   %
\newcommand{\VRightTy}[2]{\VRight{#2} : \TSum{#1}{\_}}  %
\newcommand{\VMutLoanTy}[2]{\VMutLoan{#1} : #2}  %
\newcommand{\VShrBorTy}[2]{\VShrBor{#1} : \TShrRef{#2}}  %
\newcommand{\VBorrow}[2]{\VBorV{#2} : \TRef{\viewof{#2}}{#1}}  %
\newcommand{\BMut}[2]{\Mut\, #1\; #2}  %
\newcommand{\BShared}[2]{\Shr\, #1 : \TShrRef{#2}}  %
\newcommand{\BDead}[1]{\dagger^{#1}}  %
\newcommand{\InFrames}[2]{#2^{#1}}  %
\newcommand{\Anon}[1]{\_{}_{#1}}  %
\newcommand{\InAbs}[2]{\abs_{#1}.#2}  %
\newcommand{\pathcomp}[2]{#1 \circ #2}  %
\newcommand{\fplit}[2]{\pathcomp{#2}{#1}}  %
\newcommand{\dealit}[2]{\pathcomp{#2}{#1}}  %
\newcommand{\placelit}[2]{#1(#2)}  %
\newcommand{\frameget}[2]{#2[#1]}  %
\newcommand{\progfn}[2]{#1[#2]}
\newcommand{\frameset}[3]{#3[#1 \mapsto #2]}  %
\newcommand{\pushanon}[2]{#2,\ \_ \mapsto #1}  %
\newcommand{\placeaddr}[5][\vdash]{#2 #1 #3 \overset{#4}{\leadsto} #5}  %
\newcommand{\sharpplaceaddr}{\placeaddr[\vdash_\#]}  %
\newcommand{\readat}[6][\vdash]{\placeaddr[#1]{#2}{#3}{#4}{#5} \Rightarrow #6}  %
\newcommand{\sharpreadat}{\readat[\vdash_\#]}  %
\newcommand{\plusreadat}{\readat[\vdash^{+}]}  %
\newcommand{\placeaddraux}[7][\vdash]{#2 #1 #5(#3 \mapsto #4) \overset{#6}{\leadsto} #7}  %
\newcommand{\sharpplaceaddraux}{\placeaddraux[\vdash_\#]}  %
\newcommand{\fnbody}[1]{#1.\mathsf{body}}  %
\newcommand{\feapath}[1]{\mathsf{path}(#1)}  %
\newcommand{\entrycontent}[1]{\mathsf{content}(#1)}  %
\newcommand{\viewof}[1]{\mathsf{view}(#1)}  %
\newcommand{\refines}[4][]{#3 \leq_{#2}^{#1} #4}  %
\newcommand{\lesharp}{\refines{\sharp}}  %
\newcommand{\lesharpstar}{\refines[*]{\sharp}}  %
\newcommand{\leplus}{\refines{+}}  %
\newcommand{\leplusstar}{\refines[*]{+}}  %
\newcommand{\leplusstarsym}{\leq_{+}^{*}}  %
\newcommand{\nleplusstar}[2]{#1 \not\leq_{+}^{*} #2}  %

\newcommand{\Permutation}[2]{#1 \text{ is a permutation of } #2}  %
\newcommand{\borrowchecks}[2]{\metafn{borrow\_checks}(#1, #2)}  %
\newcommand{\botof}[1]{\metafn{bot\_of}(#1)}  %
\newcommand{\abstractionend}[2]{#1 \succ^{\mathsf{from\text{-}abs}} #2}  %
\newcommand{\elemend}[2]{#1 \succ^{\mathsf{from\text{-}abs}} #2}  %
\newcommand{\entrymode}[1]{\mathsf{mode}(#1)}  %
\newcommand{\entrytoabsarrow}[2]{\toabs{#1}{#2}}  %
\newcommand{\eqb}[2]{#1 = #2}  %
\newcommand{\fninitendstates}[3]{\{#2\}\ #1\ \{#3\}}  %
\newcommand{\fpinsideregionabs}[1]{\abs \in #1}  %
\newcommand{\notinsideregionabs}[1]{\abs \notin #1}  %
\newcommand{\giveback}[6][\vdash]{#1 #2(#4) \mapsto #3 \xRightarrow{\mathsf{give}\ #5} #6}  %
\newcommand{\sharpgiveback}{\giveback[\vdash_\#]}  %
\newcommand{\plusgiveback}{\giveback[\vdash^{+}]}  %
\newcommand{\initexternentry}[3]{\metafn{init\_extern\_entry}(#1, #2, #3)}  %
\newcommand{\instsig}[5]{\metafn{inst\_sig}(#1, #2, #3, #4, #5)}  %
\newcommand{\length}[1]{\metafn{length}(#1)}  %
\newcommand{\lset}[3]{#1[#2 \mapsto #3]}

\newcommand{\nonunit}[1]{\metafn{non\_unit}(#1)}  %
\newcommand{\lindex}[2]{#1[#2]}
\newcommand{\projinputentry}[2]{\metafn{proj\_input\_entry}(#1, #2)}  %
\newcommand{\projoutputentry}[4]{\metafn{proj\_output\_entry}(#1, #2, #3, #4)}  %
\newcommand{\shrborrowat}[6][\vdash]{#1 #2(#4) \mapsto #3 \xRightarrow{\mathsf{share}\ #5} #6}  %
\newcommand{\sharpshrborrowat}{\shrborrowat[\vdash_\#]}  %
\newcommand{\stripouterloans}[1]{\metafn{strip\_outer\_loans}(#1)}  %
\newcommand{\symof}[1]{\metafn{sym\_of}(#1)}  %
\newcommand{\tyof}[1]{\metafn{ty\_of}(#1)}  %
\newcommand{\unitstates}[1]{\metafn{unit\_states}(#1)}  %
\newcommand{\valcontainsloc}[2]{#1 \in #2}  %
\newcommand{\viewapplicable}[2]{\vdash \TRef{#2}{#1}\ \mathsf{wf}}  %
\newcommand{\projty}[3]{\vdash #2(#1) \Rightarrow #3}
\newcommand{\placety}[3]{#1 \vdash #2 : #3}

\newcommand{\circled}[1]{\textcircled{\scriptsize #1}}

\title{Partial Borrows in Rust}

\author{Sasha Pak}
\affiliation{\institution{Australian National University}\country{Australia}}
\email{sasha.pak@anu.edu.au}
\author{Richard Willie}
\affiliation{\institution{National University of Singapore}\country{Singapore}}
\email{richardw@u.nus.edu}
\author{Umang Mathur}
\affiliation{\institution{National University of Singapore}\country{Singapore}}
\email{umathur@nus.edu.sg}
\author{Fabian Muehlboeck}
\affiliation{\institution{Australian National University}\country{Australia}}
\email{fabian.muehlboeck@anu.edu.au}
\author{Alex Potanin}
\affiliation{\institution{Australian National University}\country{Australia}}
\email{alex.potanin@anu.edu.au}
\renewcommand{\shortauthors}{Pak et al.}

\begin{document}

\begin{abstract}
Rust ensures memory safety by requiring exclusive access to mutably borrowed
data. However, a borrow can prevent access to more data than necessary.
For example, a method with an \code{&mut self} parameter
borrows the entire struct, even when the method uses only some of its fields.
Calling the method
can therefore conflict with an existing borrow of a field the method never uses,
causing Rust to reject the program.
Resolving such conflicts can require programmers to compromise useful
abstractions, make non-local changes to their programs, or worse, resort to
unsafe code.

We introduce \emph{view types}, a backward-compatible refinement of
reference types that specifies which fields a reference may access.
This allows methods to partially borrow a struct, accessing only the fields
they need without blocking access to the rest.
We establish the first fully mechanized foundation for Aeneas in Rocq.
Along the way, we uncover and fix gaps in the original semantics and proofs.
Building on this foundation, we formalize partial borrows and prove the
extended borrow checking sound. We implement view types in rustc and extend
the Aeneas interpreter to support partial borrows.
Our evaluation shows how view types systematically resolve borrowing conflicts
across method boundaries at no runtime cost.

\end{abstract}

\maketitle

\section{Introduction}
\label{sec:introduction}

Rust is a systems programming language that combines low-level control over
memory with strong memory safety guarantees~\cite{rustbook}. These guarantees rely on a type
system based on ownership and borrowing. Each value in Rust has a unique owner.
To access a value without taking ownership,
one can borrow it through a reference. Shared references (\code{&T}) can coexist,
but a mutable reference (\code{&mut T}) cannot be used alongside other references
to the same value. Rust enforces this ownership and borrowing discipline
through a process known as borrow checking. Borrow checking is modular: each
function body is checked against its own signature, and each call is checked
using the callee's signature without inspecting its body.

For example, an \code{&mut self} parameter allows a method to read or modify
any field of the struct. The borrow checker, being conservative, therefore
treats the entire struct as mutably borrowed. This can be unnecessarily
restrictive when the method uses only some fields. Because the remaining
fields are borrowed too, an existing borrow of one of them can cause Rust to
reject the call.

Consider Helix~\cite{helix}, a text editor that supports editing multiple
documents in separate panes. Each pane shows one document, and editing commands
act on the active pane. The following method, adapted from Helix, retrieves
that pane and its document~\cite{cortier2021helix}:
\begin{lstlisting}
fn current(&mut self) -> (&mut Pane, &mut Document) {
    let pane = &mut self.tree.panes[self.tree.focus];
    let doc = &mut self.documents[pane.doc];
    (pane, doc)
}
\end{lstlisting}

Helix lets users cut text from a document and paste it elsewhere. The cut text is stored in a register, which works like a clipboard and can be selected by name. A straightforward implementation would retrieve the active pane and document through current, then look up the chosen register. A helper shared by Helix commands would then save the selected text to the register and delete it from the document:

\begin{lstlisting}
fn delete_selection(editor: &mut Editor, name: char) {
    let (pane, doc) = editor.current();
    let register = editor.registers.get_or_insert(name);
    delete_selection_impl(register, doc, pane.id);
}
\end{lstlisting}

Yet Rust rejects this implementation:
\begin{lstlisting}[style=diagnostic]
error[E0499]: cannot borrow `editor.registers` as mutable more than once at a time
43 |     let (pane, doc) = editor.current();
   |                       ------ first mutable borrow occurs here
44 |     let register = editor.registers.get_or_insert(name);
   |                    ^^^^^^^^^^^^^^^^ second mutable borrow occurs here
45 |     delete_selection_impl(register, doc, pane.id);
   |                                          -------
   |                                          first borrow later used here
\end{lstlisting}

\begingroup\emergencystretch=2em
Calling \code{current} borrows the entire editor through \code{&mut self}.
Because the returned references \code{pane} and \code{doc} are used later in
the call to \code{delete_selection_impl}, this borrow remains active after
\code{current} returns.
Looking up the register requires another mutable borrow before
\code{delete_selection_impl} is called. This second borrow overlaps the first, even though \code{current}
never accesses the registers.\par
\endgroup

But this program is nonetheless safe! Rust rejects it because it does not have enough information to establish that the borrows are disjoint. Helix's developers tried to work around these restrictions, but continued
borrowing complications ultimately led them to replace the methods with
macros~\cite{cortier2021helix}. As you will see in our case studies (\S\ref{sec:eval-cases}), Helix is not an isolated case. Other common workarounds, such as inlining methods, passing fields explicitly,
or splitting structs, compromise useful abstractions or make code harder to
maintain. We argue in \S\ref{sec:motivation} that the only satisfactory solution is for Rust
to let methods declare which fields they borrow. Ultimately, we believe developers should be able to write code that reflects their intent without the borrow checker getting in their way.

To this end, we introduce view types, a backward-compatible refinement of reference
types that specifies which fields a reference may access.
Returning to our Helix example, \code{current}
needs access only to \code{tree} and \code{documents}. We express this by annotating its
\code{self} parameter:
\begin{lstlisting}[emph={[2]tree,documents}]
fn current(&mut {mut tree, mut documents} self) -> (&mut Pane, &mut Document)
\end{lstlisting}

The view permits mutable access to \code{tree} and \code{documents}, but not
\code{registers}. The references returned by \code{current} now keep
only the tree and documents borrowed. Looking up the chosen register through
\code{editor.registers} no longer conflicts with these borrows, so we can pass
the register alongside \code{pane} and \code{doc} to \code{delete_selection_impl}.
The compiler now accepts \code{delete_selection} as written. Unlike the macro, this fix is local, meaning that only the signature of \code{current} changes, while its body and its callers remain the same.

View types preserve the modularity of borrow checking. Much like how lifetime
annotations relate the lifetimes of input and output references, view types
specify a contract that restricts the set of fields a reference may access and,
crucially, determines which fields remain borrowed while output references are
active. The compiler checks that each method body respects this contract.
As in ordinary Rust, each call can be checked using the signature alone.
View types allow this checking to operate at a finer granularity.

We implement view types in rustc and establish their soundness by extending
the semantics underlying Aeneas~\cite{ho2022aeneas,ho2024symbolic}, a Rust
verification framework. Aeneas provides a concrete semantics for program
execution and a symbolic semantics for borrow checking. We prove that
programs accepted by the extended symbolic borrow checker respect Rust's
borrowing discipline. Mechanizing these semantics in Rocq also uncovered
gaps in the original semantics and soundness argument, which we repair
(\S\ref{sec:aeneas}).

In summary, we make the following contributions:
\nopagebreak
\begin{enumerate}
  \item We present view types, a refinement of Rust's reference types that
  specifies the set of fields a reference may access (\S\ref{sec:design}). Methods declare these
  fields in their signatures, so borrow checking remains modular. Our rustc
  implementation (\S\ref{sec:implementation}) is backward compatible and introduces no runtime cost
  (\S\ref{sec:evaluation}).
  \item We establish the first fully mechanized foundation for Aeneas in Rocq.
  Along the way, we uncover and fix gaps in the original semantics and proofs
  (\S\ref{sec:aeneas}). Building on this foundation, we formalize partial
  borrows and prove that programs accepted by the extended borrow checking
  respect the borrowing discipline of Rust (\S\ref{sec:semantics}).
  We also extend the Aeneas interpreter to support partial borrows.
  \item We show how view types systematically resolve borrowing conflicts
  without compromising useful abstractions (\S\ref{sec:motivation}), and evaluate these benefits
  through case studies drawn from real-world Rust projects (\S\ref{sec:evaluation}).
\end{enumerate}

\section{The Case for View Types}
\label{sec:motivation}

We propose view types to resolve borrowing conflicts across method boundaries.
But why extend Rust when we can already work around these conflicts?
We could inline methods or split structs, for instance. Or we could even
abandon methods altogether and use free functions that take the fields as
separate arguments. These workarounds are far from the only options, but none
is without drawbacks, as we will see. These problems reflect an underlying limitation in the
expressiveness of Rust. View types address that limitation directly and allow
us to resolve these conflicts without those drawbacks.

\subsection{Replacing Methods}
\label{sec:workarounds-methods}

One family of workarounds sidesteps the problem by giving up methods.
To borrow the words of Matsakis, co-lead of Rust's language design
team~\cite{matsakis2024borrow}, these workarounds are
\begin{quote}
\centering
\begin{tikzpicture}
\def\quotationwords{non-obvious, annoying, and non-local.}
\node[inner sep=0pt,text=black,
      font=\normalsize\bfseries] (quotation)
  {\quotationwords};
\node[anchor=base east,inner sep=0pt,text=CodeView,
      font=\fontsize{27.14}{27.14}\selectfont] (openingmark)
  at ([xshift=-3.57pt,yshift=-10.71pt]quotation.base west) {\textquotedblleft};
\node[anchor=base west,inner sep=0pt,text=CodeView,
      font=\fontsize{27.14}{27.14}\selectfont] (closingmark)
  at ([xshift=3.57pt,yshift=-15.71pt]quotation.base east) {\textquotedblright};
\pgfresetboundingbox
\path[use as bounding box] (openingmark.north west)
  rectangle ([yshift=-8.57pt]closingmark.north east);
\end{tikzpicture}\par
\end{quote}
\vspace{-10pt}

\paragraph{Passing fields separately.}
We can convert a method into a free function that takes the
required fields as separate arguments.
For example, a terminal's \code{handle_key} method might need its cells,
dimensions, cursor position, and scroll offset:
\begin{lstlisting}[language=Rust, numbers=none]
fn handle_key(cells: &mut Vec<Cell>, cols: &usize, rows: &usize,
              cursor_x: &mut usize, cursor_y: &mut usize,
              scroll_offset: &mut usize, key: Key) { ... }
\end{lstlisting}
This workaround sacrifices three benefits of method-based APIs.
First, it duplicates every field's type in the function's signature.
Changing a field's type now requires updating that signature as well.
Second, callers must supply each field explicitly, so adding one field to
\code{handle_key} requires updating every call site.
Third, and potentially the worst, is that nothing requires these arguments
to come from the same instance. A caller can accidentally pass \code{other.cells} together with
the dimensions and cursor position from \code{self}:
\begin{lstlisting}[language=Rust, numbers=none,escapeinside={(*@}{@*)}]
handle_key(&mut (*@\CodeSquiggle{other.cells}@*), &self.cols, &self.rows, &mut self.cursor_x,
           &mut self.cursor_y, &mut self.scroll_offset, key);
\end{lstlisting}
\paragraph{Inlining method calls.}
Inlining replaces a method call with the method's body.
The workaround carries two costs.
First, it gives up the abstraction provided by the method. The caller now
contains implementation details that the method was meant to encapsulate.
Second, changes to the method must be repeated in every inlined copy,
and a forgotten update can go unnoticed.

\paragraph{Macros.}
A macro invocation expands into code at the call site.
Our Helix case study (\S\ref{sec:eval-cases}) illustrates this approach.
Macro definitions are more cumbersome to write and maintain than ordinary
methods. They also cannot themselves be passed as function arguments or
used as trait methods. Unlike methods, there is nowhere in a macro to
specify a borrowing contract. So changing the macro's body can introduce
borrowing conflicts in callers without any change to its invocation syntax
to indicate the new requirements.
Macro libraries~\cite{borrow2,partialref} can support some form of partial
borrowing, but require programmers to learn a separate vocabulary.
And worse, they make debugging harder.

\subsection{Reorganizing Data}
\label{sec:workarounds-data}

Another family of workarounds keeps methods but reorganizes the data they
borrow. We highlight one such workaround.

\paragraph{Struct decomposition.}
We can split a struct into smaller structs and move methods onto them.
Each method then borrows only the smaller struct it operates on.
Our moonfire-nvr case study (\S\ref{sec:eval-cases}) illustrates this.
This workaround has two primary drawbacks.
First, the resulting structs need not be useful abstractions on their own.
For example, Matsakis~\cite{matsakis2021view} recounts his experience working
on rustc, where borrowing conflicts led him to move CFG data from the MIR
\code{Builder} into a \code{cfg} sub-struct.
Although this data was conceptually distinct, the new type was not independently useful.
Second, this refactoring is non-local. Fields and the methods that operate
on them must move to the new structs. Existing constructors, destructuring
patterns, and accesses to the relocated fields must be updated.

\subsection{Other Workarounds}
\label{sec:workarounds-other}

There are many other workarounds, from changing the types of fields to
bypassing borrow checking altogether. We highlight a few below.

\paragraph{Copying or moving data.}
Cloning a field's data lets us work on an independent copy without
borrowing from the struct. For collections such as \code{Vec} or
\code{HashMap}, however, cloning can require a runtime allocation
proportional to their size. We can avoid this cost by temporarily
moving the data out, for example by wrapping the field in
\lstinline[style=inlinecode]{Option<T>} and extracting its contents with
\code{Option::take}. This leaves \code{None} behind, which existing
accesses to the field must now handle. If the field's type implements
\code{Default}, \code{mem::take} instead leaves a default value without
changing its type.
Whichever approach we use, we must eventually propagate any changes back
to the original struct. If we forget this step, or an early return or panic
skips it, the type system will not catch the mistake.

\paragraph{Interior mutability.}
A field wrapped in \lstinline[style=inlinecode]{Cell<T>} or
\lstinline[style=inlinecode]{RefCell<T>} can be mutated through a shared reference,
allowing a method that writes to the field to take \code{&self} instead of
\code{&mut self}.
This workaround is quite limited. It allows calls alongside shared borrows
of other fields, but an active mutable borrow of any field still prevents
the call because \code{&self} borrows the entire struct.
Even when it applies, we must rewrite existing field accesses to use
the wrapper's methods. Each wrapper also has its own complications.
\code{Cell} requires \code{T} to implement \code{Copy}
to read its contents without moving them out.
\code{RefCell} allows borrowing non-\code{Copy} data, but checks those borrows
at runtime and panics on a conflict.
Both also inherit memory-layout caveats
from \code{UnsafeCell}~\cite{rustunsafecell}.

\paragraph{\textup{\codetext{unsafe}} code.}
When no safe workaround is available, the borrow checker can be bypassed
entirely by using \lstinline[style=inlinecode]{unsafe} code to access fields
through raw pointers. But doing so forfeits the protection
of the borrow checker and leaves us fully responsible for the safety
of our code.

\section{Design}
\label{sec:design}

View types extend Rust's reference-type syntax~\cite{rustreference} with an
optional view before the referent type. A view lists one or more
comma-separated field paths in braces. Each path consists of field names
separated by dots and may be prefixed with \code{mut}.
Omitting the view gives an ordinary reference. The type \code{&S}
corresponds to a full view with shared permission for every field, and
\code{&mut S} to one with mutable permission.
For a view to be well-formed, it must satisfy the following requirements:
\begin{enumerate}
\item The referent type must be a struct.
\item Every path must identify an existing field. In a nested path, every
  field except the last must refer to a struct.
\item Paths in the same view must not overlap. That is, no path may be a
  prefix of another.
\end{enumerate}

Consider a reference of type
\lstinline[style=inlinecode]|&mut {f, mut g.i} S|.
The entry \code{f} grants shared permission for \code{f} and all its
subfields. The entry \code{mut g.i} grants mutable permission for
\code{g.i} and its subfields, but not for other fields of \code{g}.
We apply Rust's borrowing discipline at this finer granularity.
Two active borrows of the same struct conflict when their views overlap
on a field for which at least one grants mutable permission.
In this example, other shared borrows of \code{f} can coexist with the
reference, and fields outside the view, such as \code{g.j} or \code{h},
are available for borrowing.

This finer control over borrowing is what we mean by view types being a
refinement of reference types. The referent type remains unchanged.
View types thus provide a zero-cost abstraction because they make borrow
checking more precise without changing runtime representation
(\S\ref{sec:impl-rustc}).
In earlier online discussions,
Matsakis~\cite{matsakis2021view,matsakis2025redux} proposes attaching views
to struct types, with the broader aim of supporting owned partial values
as well. Realizing this alternative design would require resolving
complications in how partial values are represented and how they interact
with Rust's existing features.
In this work, we focus on our goal of resolving borrowing conflicts across method
boundaries, for which our design suffices.

\subsection{Narrowing Is Reborrowing}
\label{sec:design-composition}

A method may need fewer permissions than its caller has.
For instance, a method whose receiver has type
\lstinline[style=inlinecode]|&{g.i} S| needs only shared access to \code{g.i}.
The reference of type \lstinline[style=inlinecode]|&mut {f, mut g.i} S|
introduced above should be usable for the call.
We use reborrowing to supply the required view.
For a reference \code{r} of type
\lstinline[style=inlinecode]|&mut {f, mut g.i} S|, we write
\lstinline[style=inlinecode]|&{g.i} *r| to reborrow \code{r}.
This reborrow omits \code{f} and weakens the permission for \code{g.i}
from mutable to shared.
We call this narrowing the view. One view can be narrowed to another
if every path in the target view has a prefix in the source view with
at least the same permission.
As in ordinary Rust, the reborrow cannot outlive the borrow from which it
is obtained. While the shared reborrow of \code{g.i} is active, the
original reference cannot be used to modify \code{g.i}.

These reborrows need not be written explicitly. At method calls, our
compiler inserts them automatically, so callers can pass ordinary
references or references with wider views without changing the call
sites (\S\ref{sec:impl-rustc}).

\subsection{What Remains Borrowed After a Call?}
\label{sec:design-lifetimes}

If a method returns a reference whose lifetime is tied to its receiver,
the receiver's view determines which fields remain borrowed while the
output reference is active. Recall the Helix example from
\S\ref{sec:introduction}. The references returned by \code{current}
keep \code{tree} and \code{documents} borrowed while those references
remain active, but not the other fields, such as \code{registers}.

\section{Aeneas, Mechanized}
\label{sec:aeneas}
\label{sec:background}

We want to show that view types are a sound extension of Rust.
We therefore build on \aeneas{}~\cite{ho2022aeneas,ho2024symbolic},
a Rust formalization covering the features that view types interact with.
\aeneas{} defines alias and ownership safety by a runtime semantics, \LLBC{},
and borrow checking by a modular symbolic semantics, \LLBCsharp{}, proved sound
against \LLBC{} on paper~\cite{ho2024thesis}.
Attempting to reproduce this proof in Rocq uncovered many gaps.
We address these gaps by adjusting the semantics, reformulating lemmas or
reworking individual proof steps,
and obtain a fully mechanized proof of soundness of \aeneas{} borrow checking.
This is the foundation on which \S\ref{sec:semantics} adds view types.

\subsection{The model}
\label{sec:aeneas-models}

As a tool, \aeneas{} is a modular symbolic interpreter of Rust programs.
It translates programs into pure functional code for verification and also acts as a borrow checker.
In this work we use \aeneas{} only as a borrow checker.
We discuss potential applications of view types to functional translation and
verification as future work (\S\ref{sec:conclusion}).

\paragraph{Syntax}
\aeneas{} works on the Low-Level Borrow Calculus (\LLBC{}), an imperative
language resembling Rust, with moves and borrows of places, pairs and sums, and function calls.
\[
\begin{array}{@{}r@{\;}c@{\;}l@{\qquad}r@{\;}c@{\;}l@{}}
  \place & \grameq & \var \mid {\PDeref}\place \mid \dpproj{\nat}\;\place\text{\footnotemark} &
  \rv & \grameq & \op \mid \RvRefMut{\place} \mid \RvRefShr{\place} \\
  \op & \grameq & \multicolumn{4}{@{}l}{\OpMove{\place} \mid \OpUnit \mid \OpPair{\op_{0}}{\op_{1}} \mid \OpLeft{\op}{\ty} \mid \OpRight{\ty}{\op}} \\
  \stmt & \grameq & \multicolumn{4}{@{}l}{\SNil \mid \SSeq{\stmt_{0}}{\stmt_{1}} \mid \SAssign{\place}{\rv} \mid \SMatch{\place}{\stmt_{0}}{\stmt_{1}} \mid \SRet \mid \SPanic \mid \SCall{\place}{\fn}{\op}}
\end{array}
\]
\footnotetext{We simplify the definition of places $\place$; it is equivalent to the full grammar in the appendix, where a place is a variable with a projection.}

\paragraph{LLBC}
\label{sec:formal-llbc}
The runtime semantics of \LLBC{} has no pointers.
A reference is a \emph{borrow} value, and the part of a value it borrows is
replaced in its owner by a \emph{loan} with the same label $\loc$:
\[
\val \grameq \VUnit \mid \VPair{\val_{0}}{\val_{1}} \mid \VLeftTy{\val}{\ty} \mid \VRightTy{\ty}{\val} \mid \VBot
\mid \VMutLoan{\loc} \mid \VShrLoan{\loc}{\val} \mid \VMutBor{\loc}{\val} \mid \VShrBor{\loc}
\]
$\VBot$ is a moved-out or uninitialized value.
A state is a stack of frames and a list of anonymous values:
\[
\st \grameq (\vect{\frameval{\val_{\mathit{loc}}}{\val_{\mathit{in}}}{\val_{\mathit{ret}}}},\ \vect{\val_{\mathit{anon}}})
\]
A frame has three variables, local, argument and return; more variables are
nested pairs. A call pushes a frame. Anonymous values hold values no variable
owns, such as an overwritten value.

The semantics is big-step and indexed by fuel $n$~\cite{appel2001indexed}:
\[
\bceval{\prog\text{\footnotemark}}{\st}{\stmt}{n}{\stmtres}
\qquad
\stmtres \grameq \NoFuel \mid \StmtFin{\res}{\st'}
\qquad
\res \grameq \RUnit \mid \RRet \mid \RPanic
\]
\footnotetext{$\prog$ is the program: a set of function declarations, each a function name with its argument, local and return types and its body.}
where $\NoFuel$ indicates that evaluation ran out of fuel.
Statements evaluate rvalues, $\rvevals{\st}{\rv}{\val}{\st'}$, which evaluate
operands, $\opevals{\st}{\op}{\val}{\st'}$. Operands build values and move
them out of places. Rvalues also create references. For example (omitting empty lists of anonymous values), let $\val = \VPair{\VUnit}{\VUnit}$ and
$\st = \frameval{\val}{\VBot}{\VBot}$. Then
\begin{align*}
  &\opevals{\st}{\OpMove{\locvar}}{\val}{\frameval{\VBot}{\VBot}{\VBot}} \\
  &\rvevals{\st}{\RvRefMut{\locvar}}{\VMutBor{\loc}{\val}}{\frameval{\VMutLoan{\loc}}{\VBot}{\VBot}}
\end{align*}

\LLBC{} enforces Rust's aliasing discipline by getting stuck: no read, write,
move or borrow passes through $\VMutLoan{\loc}$, so $\locvar$ is unusable while
its borrow lives. Borrows end by \emph{reorganisation} steps
$\envreorg{\st}{\st'}$, administrative steps allowed before any statement.
Ending a mutable borrow puts its value back in place of the loan; ending a
shared borrow leaves the loan, which ends once no borrow of its label remains ($\retvar$, holding $\VBot$, omitted):
\begin{align*}
  \langle \locvar \mapsto \VMutLoan{\loc},\ \invar \mapsto \VMutBor{\loc}{\val} \rangle
  &\hookrightarrow \langle \locvar \mapsto \val,\ \invar \mapsto \VBot \rangle \\
  \langle \locvar \mapsto \VShrLoan{\loc}{\val},\ \invar \mapsto \VShrBor{\loc} \rangle
  &\hookrightarrow \langle \locvar \mapsto \VShrLoan{\loc}{\val},\ \invar \mapsto \VBot \rangle
  \hookrightarrow \langle \locvar \mapsto \val,\ \invar \mapsto \VBot \rangle
\end{align*}

\paragraph{LLBC\textsuperscript{\#}}
\label{sec:formal-llbcsharp}
\LLBCsharp{} is a modular symbolic execution of \LLBC{} that acts as a borrow
checker. It differs from \LLBC{} in three ways.
First, values gain a symbolic value $\VSymbolicV{\ty}$, an unknown value of type
$\ty$.
Second, a call does not enter the callee's body. The call is modelled by a
\emph{region abstraction} $\abs$, a list of values: the borrows the callee
takes from the caller and the loans of the references it returns. A state
holds one frame, the anonymous values and the abstractions,
$\st = (\mathit{frame},\ \vect{\val_{\mathit{anon}}},\ \vect{\abs})$.
Third, a match on a symbolic value runs both branches and collects their
results: $\st \vdash_{\#} \stmt \Downarrow \results$, each a pair $\StmtFin{\res}{\st'}$.

For example, take a callee $f : \TMutRef{\ty} \to \TMutRef{\ty}$ that returns a
reference into its argument, and the call $\SCall{\invar}{f}{\RvRefMut{\locvar}}$
from a state where $\locvar$ owns a value $\val$:
\begin{align*}
&\langle \locvar \mapsto \val \rangle \vdash_{\#} \SCall{\invar}{f}{\RvRefMut{\locvar}} \Downarrow (\RUnit,\ \st_{1}) \\
&\st_{1} = \langle \locvar \mapsto \VMutLoan{\loc},\ \invar \mapsto \VMutBor{\loc'}{\sigma} \rangle,\ \absliteral{\VMutBor{\loc}{\VIgnored{\ty}},\ \VMutLoan{\loc'}}
\end{align*}
The argument's borrow moves into the abstraction with ignored content, and
the abstraction holds the loan of the returned reference. By design, an
abstraction holding a loan cannot end, so $\locvar$ is unavailable while
$\invar$ is in use. Once the program no longer uses the returned reference,
its borrow can end, which removes the loan from the abstraction. The
abstraction can then end too, and its borrow ends as in \LLBC{}, giving
$\locvar$ back:
\[
\hookrightarrow \langle \locvar \mapsto \VMutLoan{\loc} \rangle,\ \absliteral{\VMutBor{\loc}{\VIgnored{\ty}}}
\hookrightarrow \langle \locvar \mapsto \VMutLoan{\loc} \rangle,\ \VMutBor{\loc}{\sigma'}
\hookrightarrow \langle \locvar \mapsto \sigma' \rangle
\]

A function $f$ borrow-checks when its body $\stmt$, run from an entry state
$\mathit{entry}(f)$ built from its signature, ends every run in a panic or in a
return whose state refines an exit state $\mathit{exit}(f)$, also built from the
signature:
\[
\mathit{entry}(f) \vdash_{\#} \stmt \Downarrow \vect{(\res_{i}, \st_{i})}
\quad\text{and}\quad
\forall i.\ \res_{i} = \RPanic \ \vee\ (\res_{i} = \RRet \wedge \lesharpstar{\st_{i}}{\mathit{exit}(f)})
\]
Checking the body of $f$ is the mirror image of the call above. At the call,
an abstraction simulates the callee for the caller; in the body, an
abstraction simulates the caller for the callee. It is a barrier that shows
only what crosses the boundary. At entry it holds the borrow $\loc_{0}$ of the
caller's loan and the loan of the argument; at exit, the same borrow and the
loan of the returned reference:
\begin{align*}
  \mathit{entry}(f) &= \langle \invar \mapsto \VMutBor{\loc}{\sigma} \rangle,\ \absliteral{\VMutBor{\loc_{0}}{\VIgnored{\ty}},\ \VMutLoan{\loc}} \\
  \mathit{exit}(f) &= \langle \retvar \mapsto \VMutBor{\loc'}{\sigma'} \rangle,\ \absliteral{\VMutBor{\loc_{0}}{\VIgnored{\ty}},\ \VMutLoan{\loc'}}
\end{align*}

Central to borrow checking is \emph{refinement} ($\lesharpstar{\st}{\st'}$), the transitive closure of local rules that loosen the aliasing graph by
forgetting information. For example, \rulename{Le-ToSymbolic} replaces a value
with a symbolic value, and \rulename{Le-MergeAbs} merges two
abstractions, cancelling a borrow in one against its loan in the other:
\begin{gather*}
  \lesharp{\langle \locvar \mapsto \VLeftTy{\VUnit}{\TUnit} \rangle}{\langle \locvar \mapsto \sigma \rangle} \\
  \lesharp{\absliteral{\VMutLoan{\loc_{0}},\ \VMutBor{\loc}{\sigma}},\ \absliteral{\VMutLoan{\loc},\ \VMutBor{\loc_{1}}{\sigma'}}}{\absliteral{\VMutLoan{\loc_{0}},\ \VMutBor{\loc_{1}}{\sigma'}}}
\end{gather*}

\subsection{The theorem}
\label{sec:soundness}

We now have two semantics: \LLBC{}, which runs programs, and \LLBCsharp{},
which borrow-checks them. Soundness links the two: a program whose functions
all borrow-check never gets stuck.

\begin{theorem}[Soundness]
Let $\prog$ be a program whose functions all borrow-check, and whose argument-free function
$\mathit{main}$\footnote{Formally, $\mathit{main}$ takes a unit argument, $\invar \mapsto \VUnit$, so that every function has exactly one argument.} has body $\stmt_{\mathit{main}}$. Then for every
fuel $n$ there is an $\stmtres$ with
\[
\bceval{\prog}{\langle \locvar \mapsto \VBot,\ \retvar \mapsto \VBot \rangle}{\stmt_{\mathit{main}}}{n}{\stmtres}
\quad\text{and}\quad
\stmtres = \NoFuel \ \vee\ \stmtres = \StmtFin{\RRet}{\st'} \ \vee\ \stmtres = \StmtFin{\RPanic}{\st'}.
\]
\end{theorem}

\noindent With any fuel, $\mathit{main}$ either runs out of fuel, returns or
panics; its evaluation never gets stuck, so it has a valid execution and never breaks the aliasing
discipline that \LLBC{} enforces.

For \LLBC{}, which models safe Rust, this is the only theorem we need.
Whoever runs a program starts from $\mathit{main}$, and before that has borrow-checked the program
and every library it depends on. So all we care about is the execution of
$\mathit{main}$.\footnote{A guarantee starting from arbitrary states would
matter for safe code entered from outside safe Rust, such as a function
called through a foreign function interface or after unsafe code, whose
state the borrow checker never saw. This is out of our scope; RustBelt~\cite{lambda_rust}
addresses it with semantic types.}

The proof goes through a more general lemma: any statement, run from any
state that refines a symbolic one, never gets stuck, and its results refine
the symbolic results. Stating it needs a semantics where concrete and
symbolic structures live side by side, which we explain next.

\subsection{The general lemma and its repair}
\label{sec:proof-structure}

To compare a concrete state with a symbolic one, the proof uses a third
semantics, \LLBCPlus{}. Its states have a stack of frames, anonymous values
and abstractions, so both \LLBC{} and \LLBCsharp{} states are valid \LLBCPlus{}
states, and refinement extends to them, $\leplusstar{\st}{\st'}$.
Ho's thesis~\cite{ho2024thesis} proves this general lemma (its Theorem~7,
step-indexed as Lemma~21) for any concrete state $\st$ that refines a symbolic
state $\st^{\#}$: if every function borrow-checks,
$\leplusstar{\st}{\st^{\#}}$ and $\st^{\#} \vdash_{\#} \stmt \Downarrow \results$, then
for every fuel $n$, $\bceval{\prog}{\st}{\stmt}{n}{\stmtres}$ for some $\stmtres$ that is
$\NoFuel$, a panic, or $\StmtFin{\res}{\st_{1}}$ with
$\leplusstar{\st_{1}}{\st_{1}^{\#}}$ for some $\StmtFin{\res}{\st_{1}^{\#}} \in \results$.

\paragraph{The statement does not hold}
This is the biggest step of the thesis's proof, and unfortunately it fails.
Consider the identity function $\mathit{id} : \TShrRef{\ty} \to \TShrRef{\ty}$,
and a caller that borrows a field through its result:
\[
y := \mathit{id}(\RvRefShr{x});\quad z := \RvRefShr{(\PDeref y).0}
\]
After the call, with $x$ holding $\VPair{\VUnit}{\VUnit}$, the concrete evaluation
has $y$ borrow $x$. The symbolic evaluation reaches the same situation, but
through the abstraction generated by the call: $y$ borrows a copy
$\sigma$ held by the abstraction, which itself borrows $x$ with $\loc_{0}$:
\begin{align*}
  \st &= \langle x \mapsto \VShrLoan{\loc_{0}}{\VPair{\VUnit}{\VUnit}},\ y \mapsto \VShrBor{\loc_{0}} \rangle \\
  \st^{\#} &= \langle x \mapsto \VShrLoan{\loc_{0}}{\VPair{\VUnit}{\VUnit}},\ y \mapsto \VShrBor{\loc_{1}} \rangle,\ \absliteral{\VShrBor{\loc_{0}},\ \VShrLoan{\loc_{1}}{\sigma}}
\end{align*}
and $\leplusstar{\st}{\st^{\#}}$: refinement may retarget $y$ to a fresh loan
of a copy of the borrowed value, held by an abstraction that borrows the
original.
The symbolic evaluation of $z := \RvRefShr{(\PDeref y).0}$
expands $\sigma$ into $\VPair{\sigma_{0}}{\sigma_{1}}$, follows $y$ to the
loan $\loc_{1}$ in the abstraction, and puts the new shared loan there:
\[
\st^{\#}_{1} = \langle x \mapsto \VShrLoan{\loc_{0}}{\VPair{\VUnit}{\VUnit}},\ y \mapsto \VShrBor{\loc_{1}},\ z \mapsto \VShrBor{\loc_{2}} \rangle,\
\absliteral{\VShrBor{\loc_{0}},\ \VShrLoan{\loc_{1}}{\VPair{\VShrLoan{\loc_{2}}{\sigma_{0}}}{\sigma_{1}}}}
\]
The concrete evaluation follows $y$ to the loan $\loc_{0}$ in $x$, and puts the new
loan there:
\[
\st_{1} = \langle x \mapsto \VShrLoan{\loc_{0}}{\VPair{\VShrLoan{\loc}{\VUnit}}{\VUnit}},\ y \mapsto \VShrBor{\loc_{0}},\ z \mapsto \VShrBor{\loc} \rangle
\]
It has no other choice: the only reorganisation available beforehand ends
$y$, after which $\PDeref y$ cannot be read.
But $\nleplusstar{\st_{1}}{\st^{\#}_{1}}$. In $\st_{1}$ the shared loan in $x$
holds a loan, and every refinement rule keeps a loan or a symbolic value
there, while in $\st^{\#}_{1}$ it holds $\VPair{\VUnit}{\VUnit}$. The two runs
put the new loan in different places, and refinement, which only forgets
information, cannot move it.
The concrete evaluation is not stuck, though: what fails is the refinement
the proof maintains, not the execution. So the counterexample does not make
the borrow checker of \aeneas{} unsound; it leaves its soundness unproven.

\paragraph{The repair}
We change the rule that creates a shared borrow only where the borrowed value
lies inside a region abstraction (\rulename{Shared-Borrow-Abs}). There,
$z := \RvRefShr{(\PDeref y).0}$ no longer writes into the copy: it adds an
abstraction of its own, which borrows the copy and lends $z$ a fresh symbolic
value of the borrowed part's type:
\begin{align*}
\st^{\#}_{1} = {} & \langle x \mapsto \VShrLoan{\loc_{0}}{\VPair{\VUnit}{\VUnit}},\ y \mapsto \VShrBor{\loc_{1}},\ z \mapsto \VShrBor{\loc_{2}} \rangle, \\
& \absliteral{\VShrBor{\loc_{0}},\ \VShrLoan{\loc_{1}}{\sigma}},\
\absliteral{\VShrBor{\loc_{1}},\ \VShrLoan{\loc_{2}}{\sigma'}}
\end{align*}
The concrete result now refines the symbolic one:
$\leplusstar{\st_{1}}{\st^{\#}_{1}}$.

\paragraph{The lemma we prove}
With this rule the counterexample is gone. Other cases of the proof need
properties of reachable states that the thesis uses without stating them,
such as that no place has two active mutable borrows. We collect them as
$\mathit{wf}(\st)$ (\S\ref{sec:fixes}), as an extra hypothesis.

\begin{lemma}[General soundness]
If every function of $\prog$ borrow-checks, $\mathit{wf}(\st)$,
$\leplusstar{\st}{\st^{\#}}$ and $\st^{\#} \vdash_{\#} \stmt \Downarrow \results$, then
for every fuel $n$ there is an $\stmtres$ with $\bceval{\prog}{\st}{\stmt}{n}{\stmtres}$, and
$\stmtres$ is $\NoFuel$, a panic, or $\StmtFin{\res}{\st_{1}}$ with
$\leplusstar{\st_{1}}{\st_{1}^{\#}}$ for some $\StmtFin{\res}{\st_{1}^{\#}} \in \results$.
\end{lemma}

\noindent The soundness theorem of \S\ref{sec:soundness} is its instance for
$\mathit{main}$'s initial state, which is well formed and related to itself.

\paragraph{The proof}
The proof starts from the symbolic evaluation
$\st^{\#} \vdash_{\#} \stmt \Downarrow \results$, and \emph{embeds} it into
\LLBCPlus{} as $\st^{\#} \vdash_{+} \stmt \Downarrow \results$.
It then \emph{simulates} this evaluation down to $\st$: since
$\leplusstar{\st}{\st^{\#}}$, $\st$ evaluates $\stmt$ too, and its results refine
those in $\results$.
Finally, $\st$ is concrete, and an \LLBCPlus{} evaluation from a
concrete state is an \LLBC{} evaluation (\emph{stability}), with a single
result: $\bceval{\prog}{\st}{\stmt}{n}{\stmtres}$.

\paragraph{Calls}
Calls need special care: \LLBCsharp{} replaces the callee's body by
the abstraction its signature instantiates, while \LLBCPlus{} runs the body.
For a call $\SCall{\place}{\fn}{\op}$ we have three facts. First, $f$
borrow-checks: its body runs symbolically from $\mathit{entry}(f)$ to results
that refine $\mathit{exit}(f)$, both built from its signature. Second, the call
runs symbolically from $\st^{\#}$ to $\st^{\#}_{1}$: it evaluates $\op$ to
$\val$, instantiates the signature as an abstraction $\abs$ and assigns the
returned value $v_{\mathit{out}}$ to $\place$. Third, by induction on fuel, the
general lemma holds for the body. From them we must build an \LLBCPlus{}
evaluation of the call from $\st^{\#}$ that runs $f$'s body, and show that its
results refine $\st^{\#}_{1}$.

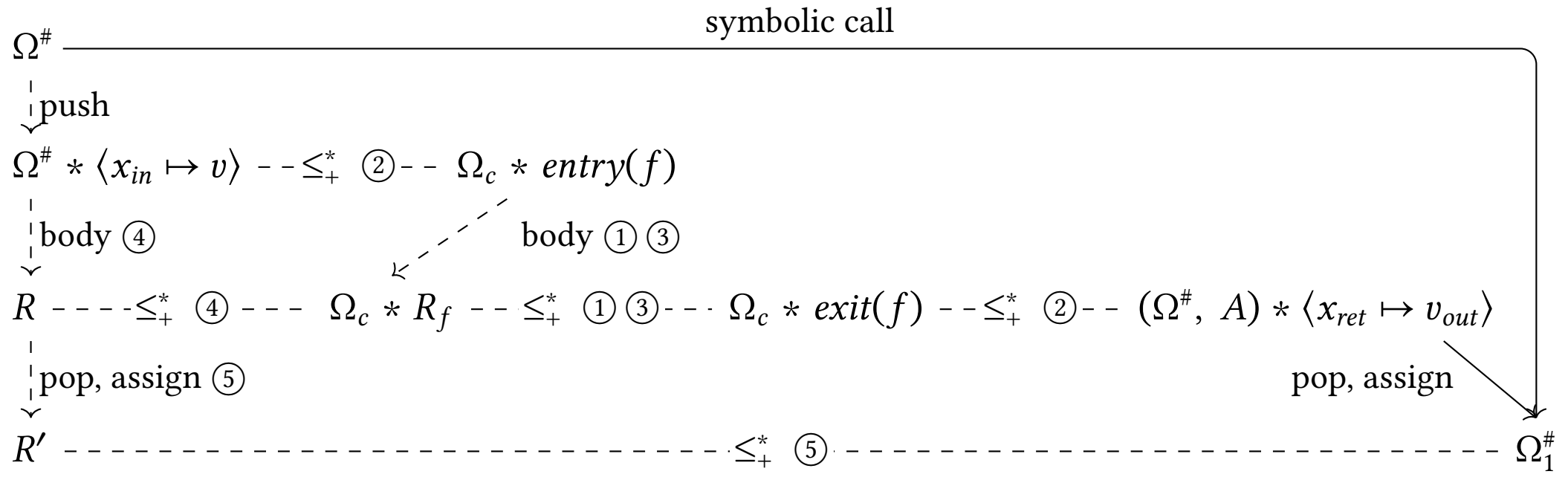
\begin{figure}[!ht]
\raggedright
\begin{tikzpicture}[>={cm to}, baseline,
  every node/.style={inner sep=2pt},
  built/.style={dashed},
  conn/.style={dashed, -, shorten <=2pt, shorten >=2pt},
  lab/.style={fill=white, inner sep=1pt, font=\small},
  side/.style={font=\small}]
\def\rowA{0}\def\rowB{-3em}\def\rowC{-6.6em}\def\rowD{-10.2em}
\node[anchor=base west] (s)  at (0,\rowA) {$\st^{\#}$};
\node[anchor=base west] (in) at (0,\rowB) {$\st^{\#} \ast \langle \invar \mapsto \val \rangle$};
\node[anchor=base west] (en) at ([xshift=5em]in.base east) {$\st_{c} \ast \mathit{entry}(f)$};
\node[anchor=base west] (R)  at (0,\rowC) {$R$};
\node[anchor=base west] (Rp) at (0,\rowD) {$R'$};
\node[anchor=base east] (s1) at (\linewidth,\rowD) {$\st^{\#}_{1}$};
\node[anchor=base east] (out) at ([xshift=-1.6em]\linewidth,\rowC) {$(\pushabs{\st^{\#}}{\abs}) \ast \langle \retvar \mapsto v_{\mathit{out}} \rangle$};
\node[anchor=base east] (ex) at ([xshift=-5em]out.base west) {$\st_{c} \ast \mathit{exit}(f)$};
\node[anchor=base east] (Rf) at ([xshift=-6.5em]ex.base west) {$\st_{c} \ast R_{f}$};
\def\ax{0.6ex}\def\dn{0.55em}\def\up{1.1em}\def\vx{0.7em}
\draw[->, rounded corners] ([yshift=\ax]s.base east) -| ([yshift=\up]s1.base) node[pos=0.25, above, side] {symbolic call};
\draw[->, built] (\vx,\rowA-\dn) -- (\vx,\rowB+\up) node[midway, anchor=mid west, side] {push};
\draw[->, built] (\vx,\rowB-\dn) -- (\vx,\rowC+\up) node[midway, anchor=mid west, side] {body\ \circled{4}};
\draw[->, built] (\vx,\rowC-\dn) -- (\vx,\rowD+\up) node[midway, anchor=mid west, side] {pop, assign\ \circled{5}};
\draw[->, built] ([xshift=1.5em, yshift=-\dn]en.base west) -- ([yshift=\up]Rf.base) node[midway, anchor=mid west, xshift=1.6em, side] {body\ \circled{1}\,\circled{3}};
\draw[->] ([xshift=-1.5em, yshift=-\dn]out.base east) -- ([yshift=\up]s1.base) node[midway, anchor=mid east, xshift=-0.7em, side] {pop, assign};
\draw[conn] ([yshift=\ax]in.base east) -- ([yshift=\ax]en.base west) node[midway, lab] {$\leplusstarsym\ \circled{2}$};
\draw[conn] ([yshift=\ax]R.base east) -- ([yshift=\ax]Rf.base west) node[midway, lab] {$\leplusstarsym\ \circled{4}$};
\draw[conn] ([yshift=\ax]Rf.base east) -- ([yshift=\ax]ex.base west) node[midway, lab] {$\leplusstarsym\ \circled{1}\,\circled{3}$};
\draw[conn] ([yshift=\ax]ex.base east) -- ([yshift=\ax]out.base west) node[midway, lab] {$\leplusstarsym\ \circled{2}$};
\draw[conn] ([yshift=\ax]Rp.base east) -- ([yshift=\ax]s1.base west) node[midway, lab] {$\leplusstarsym\ \circled{5}$};
\end{tikzpicture}
\caption{A symbolic call implies an \LLBCPlus{} call. Solid arrows are given:
the symbolic call, and its last step, assigning $v_{\mathit{out}}$ to the
place, here from a callee frame holding it. Everything dashed is built, by the
result with that number.}
\label{fig:proof-call}
\end{figure}

To get there, we follow the construction of Figure~\ref{fig:proof-call}. It
combines states: $\st \ast \st'$ stacks the frames, anonymous values and
abstractions of $\st'$ on top of those of $\st$. As in the separating
conjunction of separation logic~\cite{reynolds2002separation}, the two parts
are disjoint, with one exception: a label may occur in both, but then only as
borrows in $\st'$ of a loan in $\st$. The
construction rests on five results.
\begin{itemize}
\item[\circled{1}] \emph{Translation of a borrow check}: by the induction
  hypothesis, $f$'s check holds in \LLBCPlus{}, its body evaluating
  from $\mathit{entry}(f)$ to results $R_{f}$ that refine $\mathit{exit}(f)$,
  with labels fresh with respect to the caller's.
\item[\circled{2}] \emph{Call reborrows}: the check's entry and exit states are built by the
  same projections as the call's abstraction $\abs$. Reborrowing the argument's
  borrows moves them into abstractions added to $\st^{\#}$, giving $\st_{c}$ and
  $\leplusstar{\st^{\#} \ast \langle \invar \mapsto \val \rangle}{\st_{c} \ast \mathit{entry}(f)}$.
  Merging gives
  $\leplusstar{\st_{c} \ast \mathit{exit}(f)}{(\pushabs{\st^{\#}}{\abs}) \ast \langle \retvar \mapsto v_{\mathit{out}} \rangle}$,
  which holds the call's output in a callee frame.
\item[\circled{3}] \emph{Framing}: an evaluation on top of $\st_{c}$ leaves $\st_{c}$ alone,
  so the body evaluates from $\st_{c} \ast \mathit{entry}(f)$ to
  $\st_{c} \ast R_{f}$.
\item[\circled{4}] \emph{Simulation}: the body evaluates from
  $\st_{c} \ast \mathit{entry}(f)$ to $\st_{c} \ast R_{f}$ (\circled{3}), and
  $\st^{\#} \ast \langle \invar \mapsto \val \rangle$ refines
  $\st_{c} \ast \mathit{entry}(f)$ (\circled{2}); so the body evaluates from
  $\st^{\#} \ast \langle \invar \mapsto \val \rangle$ too, to results $R$
  refining $\st_{c} \ast R_{f}$.
\item[\circled{5}] \emph{Exit}: the more concrete state pops whenever the more abstract one
  does, and the abstract one pops to exactly $\st^{\#}_{1}$, so $R$ pops to results $R'$
  refining $\st^{\#}_{1}$.
\end{itemize}
With calls handled, every symbolic evaluation embeds into \LLBCPlus{}, and
embedding, simulation and stability together give the lemma.

\subsection{Other fixes}
\label{sec:fixes}

Our goal was to reproduce the proof of \aeneas{} in Rocq, so that we can later
add view types and check that soundness is preserved. We followed the proof
strategy of \S\ref{sec:proof-structure}, which is that of the
thesis~\cite{ho2024thesis}, and hit various obstacles along the way. Sometimes
restating a lemma was enough; other times we had to change the semantics or
the approach. Besides the repair of \S\ref{sec:proof-structure}, this
subsection describes the main decisions we made;
Appendix~\ref{app:thesis} relates every rule of our semantics to its
counterpart in the thesis.

Our mechanized proof establishes soundness of the revised semantics.
Changes can still affect expressiveness:
\LLBC{} must keep modelling how Rust programs behave, and \LLBCsharp{} should
keep accepting the programs Rust accepts. Staying close to the thesis's
semantics gives us confidence in both.

\paragraph{Well-formed states}
The proof relies on properties of reachable states. We state them as
$\mathit{wf}$, the hypothesis of the general lemma, and prove that evaluation
and refinement preserve them. The rules, the thesis's and ours, rely on these
properties too, but their premises only hint at them without naming them;
collected, they are the aliasing discipline, stated explicitly as an
invariant. Some of the invariants are administrative and can be found in the mechanization; three are worth
mentioning:
\begin{itemize}
\item \emph{aliasing xor mutability}: no label is used both shared and
mutably, no label has two mutable borrows, and the content of a shared loan
holds no mutable loan;
\item \emph{borrowed, not owned}: no part of a borrowed value is ever moved
out, so no $\VBot$ appears under a mutable borrow or a shared loan;
\item \emph{abstraction elements}: each is a loan or a borrow whose content
holds no label.
\end{itemize}
The third makes abstractions flat: an abstraction is a list of loans and borrows, with
no loan or borrow nested inside them. That simplifies the rules: ending a loan
that an abstraction holds removes the element, and a shared loan joins an
abstraction only once its content holds no label.

\paragraph{States and scope}

Our states have the same parts as the thesis's: a stack of frames, anonymous
values and, in \LLBCsharp{}, region abstractions. Two differences are only in
representation: anonymous values form a list instead of being named, and a
$\bot$ or $\sigma$ of pair type is the pair of its components' $\bot$s or
$\sigma$s, which makes it easier to model several variables as one tuple. Our
calculus leaves out loops, boxes, copies and two-phase borrows, and
\LLBCsharp{} handles signatures with a single region (lifetime), where the
thesis allows several. \aeneas{} handles all of them on paper; adding them
means porting them to the mechanization and closing the gaps this exposes,
which we leave for future work.
Every other
evaluation rule of the thesis's \LLBC{} has a counterpart in ours, so \LLBC{}
gives programs the same behaviour.

\paragraph{Notation}
Some changes are purely notational. The thesis's
rules locate a value by a context with a hole: in $\st = \st'[x \mapsto
\VShrLoan{\loc}{(\square, w)}]$, the hole stands inside $x$, under the shared
loan, in the first component. We name the position by an address instead, a
\emph{full path}: the root it starts from (a variable, an anonymous value or
an element of an abstraction), followed by a direct path into that value,
composed like functions. Here the address is $\fpath =
\placelit{\pathcomp{\dpproj{0}}{\dpunloan}}{x}$, and
$\fpapplies{\fpath}{\st}{\val}$ reads the value $\val$ there. Addresses also
let symbolic values go without names. A match on a symbolic $\sigma$ of sum
type continues in two branches; the thesis's \rulename{E-Match-Symbolic}
substitutes $\VLeft{\sigma_{0}}$ (resp.\ $\VRight{\sigma_{1}}$) for $\sigma$
wherever it occurs, $\st[\VLeft{\sigma_{0}}/\sigma]$, while ours writes it at
the matched address. The two agree: no thesis rule duplicates a symbolic, so
each name occurs once.

\paragraph{Bookkeeping}
The other changes keep the rules consistent with each other or with what the
proof needs, and Appendix~\ref{app:thesis} lists them: a few rules gain a
premise that the thesis's rule likely omits by oversight; our simpler calculus
and the invariants of its states let us drop eight refinement rules of the
thesis; two rules move to the refinement of \LLBCPlus{}, where only the proof
uses them; and four rules are added: one that drops a redundant borrow from an
abstraction, and three that permute the lists that stand for the thesis's
unordered collections. Our refinement thus has 11 rules, where the thesis's
has 17.

\paragraph{Tests}
To check that these changes keep \LLBCsharp{} expressive, we encoded by hand
86 distinct functions from \aeneas{}'s test suite, without
changing their control flow or erasing a type, and all of them borrow-check,
with machine-checked proofs. Most of the other functions use features we leave
out: loops, boxes and collections, recursive types, traits. We also encode
35 more functions after cutting them down to our calculus (erasing a box or a
vector, bounding a recursive type, or merging two lifetimes); they
borrow-check too. We do not encode the negative tests. Like the original, our \LLBCsharp{} is
nondeterministic, so showing that a program does not borrow-check means
ruling out every possible derivation.
A deterministic \LLBCsharp{}, for which a failed check would be a result, is
future work.

\section{View Types Formally}
\label{sec:semantics}
\label{sec:formal}

We extend \LLBC{} and \LLBCsharp{} with views to obtain \LLBCView{} and
\LLBCSharpView{}. We present the new syntax and rules below. The full rules are in
Appendix~\ref{app:views}. The formalization covers views one field deep:
nested field paths such as \lstinline[style=inlinecode]|f.g|
(\S\ref{sec:design}) are left for future work.

\subsection{The semantics}
\label{sec:formal-syntax}

\paragraph{Views}
A \emph{viewpoint} $\vp$ selects a field of a pair: the empty viewpoint
$\varepsilon$ selects the value itself, $0$ and $1$ its two fields.
A \emph{view} $\view$ either borrows one viewpoint with a mode $k$,
$\singleview{\vp}{k}$, or both fields, each with its own mode,
$\doubleview{k_0}{k_1}$.
A reference type $\TRef{\view}{\ty}$ carries its view and the type of the
whole referent.

\[
\begin{array}{@{}c@{}}
\vp \grameq \varepsilon \mid 0 \mid 1 \qquad
k \grameq \Mut \mid \Shr \qquad
\view \grameq \singleview{\vp}{k} \mid \doubleview{k}{k} \\
\ty \grameq \dots \mid \TRef{\view}{\ty} \qquad
\rv \grameq \dots \mid \RvBorrow{\view}{\place}
\end{array}
\]

\paragraph{References}
\label{sec:formal-llbc-view}

Partial borrows operate at the level of references, leaving most of the
semantics unchanged. The key change in \LLBCView{} is the structure of
the borrow: a reference holds one entry per field its view borrows, and
carries the type of the whole referent:
\[
  \val \grameq \dots \mid \VBorV{\singleborrow{\vp}{\entry}} : \TRef{\singleview{\vp}{\entrymode{\entry}}}{\ty}
  \mid \VBorV{\doubleborrow{\entry_{0}}{\entry_{1}}} : \TRef{\doubleview{\entrymode{\entry_{0}}}{\entrymode{\entry_{1}}}}{\ty}
\]

For example, a reference can now borrow only the first component of a pair:
\[
  \VBorV{\singleborrow{0}{\BMut{\loc_0}{\VUnit}}} : \TRef{\singleview{0}{\Mut}}{\TPair{\TUnit}{\TUnit}}
\]

Each entry $\entry ::= \BMut{\loc}{\val} \mid \BShared{\loc}{\ty} \mid
\BDead{k}$ holds the previous borrow forms, plus a \emph{dead} entry
$\BDead{k}$ for a borrow that has ended. Ordinary borrows become
references with one entry at the empty viewpoint,
$\VMutBor{\loc}{\val} \triangleq
\VBorrow{\ty}{\singleborrow{\varepsilon}{\BMut{\loc}{\val}}}$, and similarly
for shared borrows.

\paragraph{Reference creation}
A borrow $\RvBorrow{\view}{\place}$ borrows each field its view names,
delegating the checks to the rule that borrows a single place,
$\borrowat{\st}{\place}{k}{\entry}{\st'}$: \LLBC{}'s mutable or shared
borrow, producing an entry.
\begin{mathpar}
  \inferrule*[lab={\rulename{E-Borrow}}]
    {\placety{\st}{\place}{\ty} \\ \viewapplicable{\ty}{\view} \\ \borrowevals{\st}{\place}{\view}{\entries}{\st'}}
    {\rvevals{\st}{\RvBorrow{\view}{\place}}{\VBorrow{\ty}{\entries}}{\st'}}

  \inferrule*[lab={\rulename{Borrow-Single}}]
    {\borrowat{\st}{(\pathcomp{\vp}{\place})}{k}{\entry}{\st'}}
    {\borrowevals{\st}{\place}{\singleview{\vp}{k}}{\singleborrow{\vp}{\entry}}{\st'}}

  \inferrule*[lab={\rulename{Borrow-Double}}]
    {\borrowat{\st_{0}}{(\pathcomp{0}{\place})}{k_{0}}{\entry_{0}}{\st_{1}} \\\\ \borrowat{\pushanon{\entrycontent{\entry_{0}}}{\st_{1}}}{(\pathcomp{1}{\place})}{k_{1}}{\entry_{1}}{\pushanon{\entrycontent{\entry_{0}}}{\st_{2}}}}
    {\borrowevals{\st_{0}}{\place}{\doubleview{k_{0}}{k_{1}}}{\doubleborrow{\entry_{0}}{\entry_{1}}}{\st_{2}}}
\end{mathpar}

\noindent \textsc{E-Borrow} asks that the view fits the type of the place,
$\viewapplicable{\ty}{\view}$, so no view enters a sum.
\textsc{Borrow-Double} borrows the two fields in order, so the state updates
for each entry are sequenced. While the second field is borrowed, the content
taken by the first entry is held as an anonymous value.
A mutable entry takes the value out and leaves a mutable loan in the owner;
a shared entry reuses the shared loan at the place or wraps the value in a new
one. Because these checks apply independently for each field, a new partial borrow
conflicts with an active one only when their viewpoints overlap at a field
where at least one access is mutable.
When a view borrows the whole value at $\varepsilon$, this recovers
\LLBC{}'s ordinary exclusivity rule.

\paragraph{State reorganisation}

A reference ends one entry at a time. Ending a mutable entry marks it dead,
$\BDead{\Mut}$, and gives its value back to the owner's loan; ending a shared
entry marks it dead, $\BDead{\Shr}$, and its loan ends as in \LLBC{}; a
reference whose entries are all dead becomes $\VBot$
(\rulename{Reorg-End-MutEntry}, \rulename{Reorg-End-SharedEntry},
\rulename{Reorg-Drop-Dead}).
Together they implement borrow termination at field granularity: each mutable
field is returned to its owner independently, and a partial borrow
terminates only when all its entries have ended.

\paragraph{Memory model}
Dereferencing a reference looks ahead at the projection that follows it: it
enters the entry that covers the field the projection names, and goes on only
if that entry permits the access, a mutable entry for a write and any live
entry for a read.

\paragraph{Calls in \LLBCSharpView{}}
\label{sec:formal-llbcsharp-view}

\LLBCSharpView{} extends \LLBCsharp{} as \LLBCView{} extends \LLBC{}. The one
new step is at calls: before a partial borrow moves into a region abstraction,
it is split into ordinary borrows, one per entry. When the abstraction ends,
they are given back one entry at a time. Region abstractions thus contain only
ordinary, single-entry borrows. Dead entries make this possible:
without them, entries could not be released independently.

\subsection{Soundness of View Types}
\label{sec:views-proof}

The soundness theorem of \S\ref{sec:soundness} holds for \LLBCView{} and
\LLBCSharpView{}, and its proof follows \S\ref{sec:proof-structure} over
\LLBCPlusView{}, the combined semantics with views. The new proof obligations
concern types. In \LLBC{}, the type of a reference is read off the reference
itself: a mutable borrow from the value it holds, a shared borrow from its
annotation. A view cannot be typed from its entries, since it may cover only
part of the referent: it carries the type of the whole referent, and each
live entry carries a type of its own, through its value or its annotation,
and these types can disagree. They matter at calls: a callee is
borrow-checked against its signature alone, independently of its callers, so
the proof must know that a caller passes values of the types the signature
names. So $\mathit{wf}$ gains two invariants, where
$\tyof{\BMut{\loc}{\val}} = \tyof{\val}$ and $\tyof{\BShared{\loc}{\ty}} = \ty$:
\begin{gather*}
\tyof{\entry} = \ty.\vp \quad \text{for every entry } \vp \mapsto \entry \text{ of a reference of type } \TRef{\view}{\ty} \\
\exists\, \Gamma : \mathit{Loc} \to \mathit{Ty}.\ \ \tyof{\mathit{lb}} = \Gamma(\loc) \quad \text{for every loan or entry } \mathit{lb} \text{ of label } \loc
\end{gather*}
In the second, $\Gamma$ is a typing context, from labels to types, in which
each loan and the entries that borrow it agree. Both invariants are
preserved by every rule. With them, a reference's type consistently tells the type
of what it borrows, as it does in \LLBC{}.

Views increase the semantics from 2\,422 to 3\,101 lines of Rocq and the
soundness proof from 23\,980 to 30\,024 lines. Most of the additional proof
work concerns the invariants, which now carry the types of the entries.

\subsection{Views conservatively extend Aeneas}
\label{sec:views-conservative}

View types are an addition to Rust, so they must not change the meaning of
programs that do not use them. To state this, we translate an \LLBC program into
\LLBCView by giving every reference the full view: a
reference type $\TRef{k}{\ty}$ becomes $\TRef{\singleview{\varepsilon}{k}}{\ty}$,
and each borrow becomes a single-entry borrow with the same mutability.
We write $\overline{\prog}$, $\overline{\st}$,
$\overline{\stmt}$ and $\overline{\stmtres}$ for the translations.
Both theorems are mechanized without additional axioms.

\begin{theorem}[Borrow checking agrees]
A function $f$ of a program $\prog$ without views borrow-checks under
\LLBCsharp{} if and only if $\overline{f}$ borrow-checks under
\LLBCSharpView{} in $\overline{\prog}$.
\end{theorem}

\begin{theorem}[Executions correspond]
Let $\st$ be a well-typed \LLBC{} state: the loan and the borrows of each label
agree on its type. Every \LLBC{} evaluation
$\bceval{\prog}{\st}{\stmt}{n}{\stmtres}$ is an \LLBCView{} evaluation
$\bceval{\overline{\prog}}{\overline{\st}}{\overline{\stmt}}{n}{\overline{\stmtres}}$.
Conversely, for every \LLBCView{} evaluation
$\bceval{\overline{\prog}}{\overline{\st}}{\overline{\stmt}}{n}{\StmtFin{\res}{\st'}}$
whose final state holds no dead entry, there exists $\st_1$ such that $\st' = \overline{\st_{1}}$ and
 $\bceval{\prog}{\st}{\stmt}{n}{\StmtFin{\res}{\st_{1}}}$.\footnote{\LLBCView{}
ends a reference one entry at a time and then drops it, where \LLBC{} ends it
in one step, so an arbitrary \LLBCView{} state may hold a dead reference where
the \LLBC{} one holds $\VBot$.}
\end{theorem}

\section{Implementation}
\label{sec:implementation}

We briefly describe our implementations of rustc and Aeneas.

\subsection{rustc}
\label{sec:impl-rustc}

We extend rustc's frontend to parse the syntax of view types
(\S\ref{sec:design}) and record the views in the internal representation of
reference types. During type checking, rustc uses the receiver's declared
view to restrict field accesses through \code{self} in the method body.
At a method call, rustc checks whether the caller's reference can be narrowed
to the method's declared view (\S\ref{sec:design-composition}).
When the call occurs inside a closure, capture analysis combines the fields
required by the method with those accessed directly by the closure.

After type checking, rustc constructs its mid-level intermediate
representation (MIR) and performs borrow checking on it. The borrow checker
first creates a record for each borrow, identifying the borrowed
location and whether the borrow is shared or mutable. We carry the view into
the MIR borrow expression and store it in the corresponding borrow record.
The view identifies which fields the partial borrow covers.
Next, rustc infers lifetimes from constraints in
MIR. A dataflow analysis uses those lifetimes to compute the set of live
borrows at each program point. With these sets computed, rustc walks the
MIR and checks each read, write, or borrow against the live borrows at that point.
We extend these checks to compare accesses with the fields covered by each
partial borrow.

During code generation, rustc uses the same pointer representation for
view-typed references as for ordinary references. This makes view types a
zero-cost abstraction that increases Rust's expressiveness through more
precise borrow checking.

\subsection{Aeneas}
\label{sec:impl-aeneas}

Aeneas uses Charon~\cite{charon} as its frontend to translate rustc's MIR into
LLBC. We extend both this frontend and Aeneas's interpreter with support for
partial borrows. The implementation largely follows our semantics in
\S\ref{sec:semantics}.

\section{Evaluation}
\label{sec:evaluation}

We evaluate how view types help resolve borrowing conflicts in existing
Rust code. Our case studies illustrate the benefits of view types
(\S\ref{sec:eval-cases}).
We then examine how often method signatures overapproximate the fields
their bodies access
(\S\ref{sec:eval-prevalence}). We also assess backward compatibility
(\S\ref{sec:eval-compat}), compilation overhead (\S\ref{sec:eval-perf}),
and runtime cost (\S\ref{sec:eval-runtime}).

\subsection{Case Studies}
\label{sec:eval-cases}

We study which borrowing conflicts view types can resolve in existing Rust
programs. We collected Rust projects that contain borrowing workarounds,
including those discussed in \S\ref{sec:motivation}. Starting from
code that compiles, we reconstruct a version without the workaround.
When Rust rejects that version, we try to resolve the conflict with views.

We first note that view types do not resolve every borrowing conflict.
For example, some conflicts require distinguishing elements at different
indices in the same vector. Indeed, we encountered such cases during our
search (Appendix~\ref{app:workaround-study}). The bottom line is that view
types resolve a specific class of borrowing conflicts, namely those caused
by the loss of field-level precision across method boundaries.

Across 12 repositories, we successfully remove borrowing workarounds at 16
sites using view types. Table~\ref{tab:workaround-characterization}
summarizes the results. View types let us reuse helpers and keep related fields
in one struct. They also let us use borrowed data directly, eliminating
copies and repeated lookups, among other things.
We believe these changes improve code quality overall and help developers
maintain their code.
We highlight some of these in two detailed case studies below.

\begin{table}[htbp]
\centering
\caption{Workarounds we remove with view types.}
\label{tab:workaround-characterization}
\small
\setlength{\tabcolsep}{4pt}
\begin{tabular*}{\linewidth}{@{\extracolsep{\fill}}lll@{}}
\toprule
\textbf{Existing workaround} & \textbf{With view types} & \textbf{Projects} \\
\midrule
Pass fields as separate arguments & Avoid extra arguments &
  Apicula~\cite{apicularepo}; RhoJVM~\cite{rhojvmrepo}; Replica~\cite{replicarepo} \\
Inline function calls & Use reusable helpers &
  diamond-types~\cite{genco2021diamond}; Wasmtime~\cite{wasmtimerepo} \\
Use macros & Avoid macros &
  Varisat~\cite{varisatrepo}; Helix~\cite{helix} \\
Combining conflicting functions & Keep functions separate &
  Wasmtime~\cite{wasmtimerepo} \\
Separate structs & Keep related fields together &
  moonfire-nvr~\cite{lamb2025moonfire}; Replica~\cite{replicarepo} \\
Copy data & Use data directly &
  grok-build~\cite{grokbuildrepo}; Binius~\cite{biniusrepo} \\
Move data out, then restore it & Keep data in place &
  fish~\cite{fishrepo}; Wasmtime~\cite{wasmtimerepo} \\
Release and reacquire references & Retain references across calls &
  Sway~\cite{swayrepo}; diamond-types~\cite{genco2021diamond} \\
\bottomrule
\end{tabular*}
\end{table}

\Needspace{7\baselineskip}
\paragraph{Case Study 1: \textup{\codetext{moonfire-nvr}}~\cite{lamb2025moonfire}.}

\codetext{moonfire-nvr} is an application for recording video from security cameras.
The implementation assembles these recordings into MP4 files with a
\code{FileBuilder} struct. \code{FileBuilder} stores segments of the input
recordings in \code{segments} and uses \code{slices} to describe the pieces that
make up the output file. Assembling the output file involves traversing
the video segments and adding their metadata to the output.

Figure~\ref{fig:eval-moonfire}(i) shows how we would like to implement
this traversal in \code{append_truns}. Each iteration calls
\code{append_slice} to update \code{slices}.
The iterator keeps \code{segments} borrowed throughout the loop, but
\code{append_slice} takes \code{&mut self} and therefore borrows the
entire \code{FileBuilder}.
Rust rejects the call, even though the helper only accesses
\code{slices}.

\FloatBarrier
\Needspace{7\baselineskip}
\noindent\begin{minipage}{\linewidth}
\captionsetup{position=top,hypcap=false}
\captionof{figure}{MP4 construction in moonfire-nvr (adapted): (i) calling a method
while iterating over segments causes a conflict; (ii) the existing workaround
moves output state into a separate struct; (iii) a view annotation permits
keeping the fields and methods together.
Red highlights mark borrowing errors; amber highlights mark changes relative to (i).}
\label{fig:eval-moonfire}
\end{minipage}
\nopagebreak[4]
\begin{lstlisting}[emph={PageNum,FileBuilder,BodyState,Segment,Slices,Slice,SliceType,Error,Vec,Result},linebackgroundcolor={\CodeErrorLines{17}}]
// (i) Desired: rejected
struct FileBuilder {
    segments: Vec<Segment>,
    slices: Slices<Slice>,
    buf: Vec<u8>,
    unflushed_buf_pos: usize,
    /* ... */
}
impl FileBuilder {
    fn flush_buf(&mut self) -> Result<(), Error> { /* ... */ }
    fn append_slice(&mut self, len: u64, t: SliceType, p: usize)
        -> Result<(), Error> { /* only accesses self.slices */ }
    fn append_truns(&mut self) -> Result<(), Error> {
        self.flush_buf()?;
        for (i, s) in self.segments.iter().enumerate() {
            // ERROR: cannot borrow `*self` as mutable: `self.segments` is borrowed
            self.append_slice(s.truns_len() as u64, SliceType::Truns, i)?;
        }
        Ok(())
    }
}
\end{lstlisting}

\begin{lstlisting}[emph={PageNum,FileBuilder,BodyState,Segment,Slices,Slice,SliceType,Error,Vec,Result},linebackgroundcolor={\CodeChangedLines{4,7,8,9,10,11,12,13,14,15,16,19,21}}]
// (ii) Workaround: group the output-state fields into a sub-struct
struct FileBuilder {
    segments: Vec<Segment>,
    body: BodyState,
    /* ... */
}
struct BodyState {
    slices: Slices<Slice>,
    buf: Vec<u8>,
    unflushed_buf_pos: usize
}
impl BodyState {
    fn flush_buf(&mut self) -> Result<(), Error> { /* ... */ }
    fn append_slice(&mut self, len: u64, t: SliceType, p: usize)
        -> Result<(), Error> { /* ... */ }
}
impl FileBuilder {
    fn append_truns(&mut self) -> Result<(), Error> {
        self.body.flush_buf()?;
        for (i, s) in self.segments.iter().enumerate() {
            self.body.append_slice(s.truns_len() as u64, SliceType::Truns, i)?;
        }
        Ok(())
    }
}
\end{lstlisting}

\begin{lstlisting}[emph={PageNum,FileBuilder,BodyState,Segment,Slices,Slice,SliceType,Error,Vec,Result},emph={[2]next_free_page,slices},linebackgroundcolor={\CodeChangedLines{5}}]
// (iii) View-typed: helpers declare which fields they touch
impl FileBuilder {
    fn flush_buf(&mut self) -> Result<(), Error> { /* ... */ }
    fn append_slice(
        &mut {mut slices} self, len: u64, t: SliceType, p: usize,
    ) -> Result<(), Error> { /* unchanged */ }
    fn append_truns(&mut self) -> Result<(), Error> { /* unchanged */ }
}
\end{lstlisting}

The workaround in Figure~\ref{fig:eval-moonfire}(ii) moves \code{slices},
\code{buf}, and \code{unflushed_buf_pos} from \code{FileBuilder} into
a new \code{BodyState} struct.
The helper methods that operate on these fields, including \code{flush_buf}
and \code{append_slice}, must move to \code{BodyState} as well.
These helpers now borrow \code{self.body}, while the iterator borrows
\code{self.segments}, so the two borrows no longer conflict.
The developer explicitly attributes the split to borrowing:
``\textit{separated out from the rest
  so that it can be borrowed in a loop over
  \code{FileBuilder::segments};
otherwise this would cause a double-self-borrow}.''

But fixing this one loop now means changing code elsewhere in the program.
Wherever the existing code constructs a \code{FileBuilder}, we have to
change that code to initialize \code{BodyState}.
We also have to rewrite every use of the moved fields and helpers to go
through the new \code{body} field.
View types let us keep these fields and methods in \code{FileBuilder}.
Adding
\lstinline[style=inlinecode]|{mut slices}|
to the receiver type of \code{append_slice} restricts its borrow to
\code{slices}, so the call no longer conflicts with the iterator's
borrow of \code{segments}.
The implementation in Figure~\ref{fig:eval-moonfire}(i) then compiles
without changing the method bodies, callers, or struct definitions.

\paragraph{Case Study 2: Helix~\cite{cortier2021helix}.}
Recall the Helix example in \S\ref{sec:introduction}.
Calling \code{current} borrows the whole editor, preventing access to the
registers while the returned pane and document references remain active.
Helix initially worked around this conflict by adding \code{current_with_registers},
which returned the registers together with the pane and document.
This allowed callers to obtain all three references in one call,
but duplicated the lookup.

But the developer encountered borrowing complications again when adding support
for copying and pasting between Helix and other applications through the system
clipboard. This required accessing the editor's
clipboard provider while retaining the pane and document references, which
\code{current_with_registers} did not allow. Frustrated by these recurring
borrowing problems, the developer ultimately gave up on the accessor methods
and replaced them with
macros:
\begin{lstlisting}
macro_rules! current {
    ($editor:ident) => {{
        let pane = &mut $editor.tree.panes[$editor.tree.focus];
        let doc = &mut $editor.documents[pane.doc];
        (pane, doc)
    }};
}
\end{lstlisting}
The call to \code{current!(editor)} now expands into accesses to
\code{tree} and \code{documents}, leaving \code{registers} available for
a separate borrow. The body of \code{delete_selection} becomes:
\begin{lstlisting}
let (pane, doc) = current!(editor);
let register = editor.registers.get_or_insert(name);
delete_selection_impl(register, doc, pane.id);
\end{lstlisting}
With view types, \code{current} can remain a method even as new features
require access to other editor fields.

\subsection{How Often Can Methods Borrow Less?}
\label{sec:eval-prevalence}

\paragraph{Methodology.}
We start from the 100 most-downloaded crates on crates.io as of 27 June 2026.
Some belong to the same repository, so we group them into 76 workspaces.
Two workspaces do not compile, so we exclude them. We also exclude four
predominantly macro-based workspaces that would require substantial effort
to annotate. The remaining 70 workspaces cover 86 of the ranked crates.
We consider methods with \code{&self} or \code{&mut self}
receivers.
We use one host configuration per workspace.
See Appendix~\ref{app:annotation-counts} for more details.
For each method, we determine which fields it reads or writes and annotate
its receiver type with a view. We retain the annotation only if it narrows the borrow
to a strict subset of the struct's fields, and the annotated code passes
\code{cargo check --lib} under our modified \code{rustc}.

\paragraph{Results.}
Table~\ref{tab:prevalence} summarizes the results.
Of the 7,767 methods considered, we successfully annotate 2,184 (28.1\%).
These annotations substantially narrow the borrows. On average, they exclude
62.2\% of the struct's fields.
We stress that these figures should be read as a measure of opportunities for more precise
borrowing, not as a recommendation to annotate every method as narrowly as
possible. View annotations are part of a method's contract, and an
unnecessarily narrow view can constrain future implementation changes.

\begin{table}[tp]
\centering
\caption{View annotations across 70 workspaces covering 86 of the top 100
  ranked crates. We show the 20 workspaces with the most methods in scope and aggregate the remaining 50.
  \emph{Methods} counts the methods within the scope of our study.
  \emph{Annotated} counts successfully annotated methods.
  \emph{Excl.} is the mean percentage of struct fields excluded per annotated
  method. Dashes denote undefined rates. For the full table, please refer
  to Appendix~\ref{app:annotation-counts}.}
\label{tab:prevalence}
\small
\setlength{\tabcolsep}{3pt}
\begin{tabular}{@{}lrrr@{\hspace{6pt}}|@{\hspace{6pt}}lrrr@{}}
\toprule
\textbf{Workspace} & \textbf{Methods} & \textbf{Annotated} & \textbf{Excl.} &
\textbf{Workspace} & \textbf{Methods} & \textbf{Annotated} & \textbf{Excl.} \\
\midrule
\codetext{regex-syntax} & 1,350 & 342 (25.3\%) & 62.9\% & \codetext{hyper} & 213 & 138 (64.8\%) & 60.5\% \\
\codetext{tokio} & 1,343 & 366 (27.3\%) & 57.5\% & \codetext{futures-core} & 205 & 44 (21.5\%) & 55.0\% \\
\codetext{h2} & 516 & 110 (21.3\%) & 62.7\% & \codetext{chrono} & 165 & 40 (24.2\%) & 80.0\% \\
\codetext{clap} & 466 & 172 (36.9\%) & 71.0\% & \codetext{memchr} & 143 & 45 (31.5\%) & 52.2\% \\
\codetext{rustls} & 385 & 182 (47.3\%) & 61.4\% & \codetext{http} & 121 & 16 (13.2\%) & 48.1\% \\
\codetext{toml} & 321 & 57 (17.8\%) & 60.7\% & \codetext{socket2} & 115 & 2 (1.7\%) & 50.0\% \\
\codetext{indexmap} & 287 & 10 (3.5\%) & 48.3\% & \codetext{syn} & 103 & 18 (17.5\%) & 54.1\% \\
\codetext{tracing-core} & 284 & 79 (27.8\%) & 56.6\% & \codetext{rustix} & 96 & 16 (16.7\%) & 56.1\% \\
\codetext{hashbrown} & 224 & 77 (34.4\%) & 57.1\% & \codetext{linux-raw-sys} & 94 & 86 (91.5\%) & 89.8\% \\
\codetext{aho-corasick} & 213 & 89 (41.8\%) & 63.7\% & \codetext{idna} & 87 & 24 (27.6\%) & 68.3\% \\
\midrule
\multicolumn{8}{c}{\textbf{Remaining 50 workspaces}: 1,036 methods; 271 (26.2\%) annotated; 59.6\% average exclusion.}\\
\multicolumn{8}{c}{\textbf{Total 70 workspaces}: 7,767 methods; 2,184 (28.1\%) annotated; 62.2\% average exclusion.}\\
\bottomrule
\end{tabular}
\end{table}

\subsection{Backward Compatibility}
\label{sec:eval-compat}

View types are backward compatible in two ways.
First, code without view annotations compiles unchanged.
Second, view annotations can be added to method signatures without
changing existing calls.

We compiled the top 500 crates by all-time download count from
crates.io using both our modified \code{rustc} (based on commit
\codetext{c871d09}, nightly 2025-11-25) and the unmodified \code{rustc} of
the same date as our baseline compiler.
Of these, 13 crates failed to compile under both compilers for reasons
unrelated to view types, so we excluded them.
All remaining 487 compiled unchanged under our modified \code{rustc},
without additional diagnostics.
In \S\ref{sec:eval-prevalence}, we added view annotations
to 2,184 methods across 70 workspaces without changing their call sites.
All 70 passed \code{cargo check --lib} under our modified \code{rustc}.

\subsection{Compile-Time Overhead}
\label{sec:eval-perf}

The first experiment measures whether supporting view types slows
compilation of existing code.
The second measures the overhead when programs make use of view types.

\paragraph{Experimental setup.}
We measure clean \codetext{cargo check} wall time using \codetext{hyperfine}.
In the first experiment, we compile the 487 unannotated crates of
\S\ref{sec:eval-compat} with both the baseline and modified \code{rustc}.
In the second experiment, we use the 70 workspaces of
\S\ref{sec:eval-prevalence}.
We compile their original sources with the baseline \code{rustc}.
We then compile the annotated sources with our modified \code{rustc}
using \codetext{cargo check --lib}.
We perform one warmup run and 12--30 timed runs for each crate, using more
repetitions for faster-compiling crates to reduce noise.
We report each crate's compile time as a 20\% trimmed mean, discarding the
fastest and slowest 10\% of runs to reduce the effect of outliers.
All measurements were taken on an Intel Xeon w9-3495X workstation
with 256\,GB memory running Ubuntu 24.04 LTS
(Linux 6.8.0). We report overall overhead as the time-weighted mean,
calculated as the increase in summed compilation time relative to the baseline.

\paragraph{Results on unannotated code.}
Across the 487 crates, our modified \code{rustc} increased total compilation
time by 0.29\%. Per-crate overhead ranged from $-3.02\%$ to $+3.05\%$.
For the full results, please refer to Appendix~\ref{app:compilation-overhead}.

\paragraph{Results on annotated code.}
Across the 70 workspaces, overall compilation overhead rounded to 0.00\%.
Per-workspace overhead ranged from $-1.09\%$ to $+1.14\%$.
For the full results, please refer to Appendix~\ref{app:compilation-overhead}.

\subsection{Runtime Cost}
\label{sec:eval-runtime}

We evaluate runtime cost by comparing the machine code generated for the
70 workspaces of \S\ref{sec:eval-prevalence}. We build the original sources
with the baseline \code{rustc} and the annotated sources with our modified version.
Both builds use the same dependencies and features, with release optimization,
one codegen unit, and no link-time optimization.
Six workspaces emit no machine code. Ten others match after
normalizing symbol names, addresses, and compiler-generated type identifiers.
For the remaining 54 workspaces, the machine code matches byte for byte.
For more details, please refer to Appendix~\ref{app:runtime-comparison}.

\section{Related Work}
\label{sec:related}

\paragraph{Rust Formalizations.}
RustBelt~\cite{lambda_rust} interprets the types of its core language
$\lambda_{\mathrm{Rust}}$ in Iris to prove type soundness and verify
libraries that encapsulate unsafe code.
Oxide~\cite{oxide} formalizes Rust's ownership and borrowing rules, including
non-lexical lifetimes, by tracking reference origins as sets of loans.
Featherweight Rust~\cite{fr} combines type checking and borrow checking in a
small calculus with a flow-sensitive type system.
Other work develops executable semantics for Rust~\cite{wang2018krust,kan2020rustsem},
models its aliasing rules~\cite{stacked,villani2025tree}, and establishes
foundations for verifying Rust programs~\cite{rusthorn,matsushita2022rusthornbelt,gaher2024refinedrust}.

We build on \aeneas{}~\cite{ho2022aeneas,ho2024symbolic} because it is
actively maintained~\cite{aeneasrepo}.
It checks each function using only the signatures of its callees, matching
our goal of preserving modular borrow checking.
Its semantics also accounts for reborrowing and functions that return
references, both central to our design.
\aeneas{}'s frontend, Charon~\cite{ho2025charon}, translates Rust programs
into LLBC. By extending this translation to preserve views, we can test
our extension directly on Rust programs without manually rewriting them
in the calculus.
\aeneas{} is also used as a verification framework~\cite{ho2022aeneas,ho2026scaling},
giving our extension practical relevance beyond having a formal model.

\paragraph{Uniqueness and Fractional Permissions.}
Marshall and Orchard~\cite{fractional} extend the functional language Granule
with Rust-like ownership and borrowing. They build on uniqueness typing, which
permits mutation when a value has a single reference. Supporting shared
borrowing requires temporarily allowing multiple readers, then recovering
exclusive access when they finish. Their approach, fractional uniqueness,
tracks this access through permissions. A whole permission permits mutation.
Splitting it into fractions allows readers to share access but suspends
mutation. The type system prevents these fractions from being duplicated,
so recovering the whole permission requires collecting every outstanding
fraction. Only then can mutation resume.

Their system also supports borrowing the components of a pair separately.
The operation \texttt{push} turns a borrow of a pair into borrows of its two
components, and \texttt{pull} recombines them. Our semantics similarly
represents a partial borrow as individual borrows of the selected fields.
Granule exposes this decomposition through \texttt{push} and \texttt{pull}
in the source language, whereas our design keeps it internal to the semantics.

Fractional uniqueness builds on uniqueness typing~\cite{smetsers1994uniqueness}
and fractional permissions~\cite{boyland2003fractional}.
Fractional permissions also support reasoning about shared mutable data in
program logics~\cite{bornat2005permissions}.

\paragraph{Types for Aliasing and Ownership.}
Mezzo~\cite{balabonski2016mezzo} is an ML-style language whose type system
controls access to mutable data through permissions. To express borrowing,
it tracks how temporarily transferred permissions can be recovered.
For example, suppose we have permission to access a list and call
\texttt{find} to select a mutable element \texttt{x}. The call returns
permission to access \texttt{x} and a restoration callback, while permission
for the whole list becomes unavailable. When we are done using \texttt{x},
invoking the callback relinquishes access to the element and restores
permission for the list.

\aeneas{} records how access can be restored through borrows and loans
(\S\ref{sec:aeneas}). Our semantics applies this mechanism to each field
selected by a view, ending the component borrows individually.

Earlier work uses ownership types to control aliasing~\cite{ownership_flex},
permits safe destructive operations in the presence of aliasing~\cite{smith2000alias},
and supports temporary exclusive access to aliased objects~\cite{fahndrich2002adoption}.

\paragraph{Extending Rust's Expressiveness.}
Research on Rust has introduced refinement types~\cite{lehmann2023flux,gaher2024refinedrust,ogawa2025thrust}
and generic refinements for trait contracts~\cite{lehmann2025generic}.
Verus adds proof-oriented constructs and linear ghost types~\cite{verus}.
GhostCell~\cite{ghostcell} separates access permissions from data through a
library interface.
Rust's \code{split_at_mut} permits simultaneous mutable borrows of disjoint
slice regions~\cite{rustsplitatmut}. View types instead describe statically
named struct fields and do not express disjointness between collection
indices or ranges.

\section{Conclusion and Future Work}
\label{sec:conclusion}

View types recover the field-level precision that the borrow checker
loses at method boundaries. They are backward compatible and provide a
zero-cost abstraction. We establish their soundness through a mechanized
semantics and demonstrate their benefits in existing Rust projects. We hope
this work lays the groundwork for the adoption of view types in Rust.

\paragraph{Traits and abstract fields.}
Views name concrete fields, so our design does not yet support trait
methods. Trait signatures account for 13,120 method exclusions in our
study (Appendix~\ref{app:annotation-counts}), making support for traits
a promising next step.
Disjoint abstract fields~\cite{matsakis2025redux} offer a direction: a trait
declares abstract fields, the views of its methods name them, and each
implementation maps them to its own concrete fields.

\paragraph{Independent lifetimes within a view.}
All fields in a view share the lifetime of the reference. In Helix, both
\code{tree} and \code{documents} therefore stay borrowed for as long as either
result of \code{current} is in use. Giving each view entry its own lifetime
would release \code{tree} as soon as the pane is no longer used, while
\code{documents} stays borrowed through the document.

\paragraph{Program verification.}
View types could reduce verification specifications. A receiver view
\lstinline[style=inlinecode]|{mut x}| could let Verus~\cite{verus} establish
that other integer fields remain unchanged without explicit postconditions.
We plan to extend the functional translation of \aeneas{} to support view types
and investigate their use in other Rust verifiers.

\paragraph{AI Use.}
We used Claude and ChatGPT primarily to assist with experimental scripts, the Rocq mechanization, and manuscript revision.

\section*{Data-Availability Statement}
We intend to release the rustc, Charon, and Aeneas implementations, the Rocq
mechanization, and the experimental scripts and data.

\bibliographystyle{ACM-Reference-Format}
\bibliography{references}


\begin{thebibliography}{54}


\ifx \showCODEN    \undefined \def \showCODEN     #1{\unskip}     \fi
\ifx \showISBNx    \undefined \def \showISBNx     #1{\unskip}     \fi
\ifx \showISBNxiii \undefined \def \showISBNxiii  #1{\unskip}     \fi
\ifx \showISSN     \undefined \def \showISSN      #1{\unskip}     \fi
\ifx \showLCCN     \undefined \def \showLCCN      #1{\unskip}     \fi
\ifx \shownote     \undefined \def \shownote      #1{#1}          \fi
\ifx \showarticletitle \undefined \def \showarticletitle #1{#1}   \fi
\ifx \showURL      \undefined \def \showURL       {\relax}        \fi
\providecommand\bibfield[2]{#2}
\providecommand\bibinfo[2]{#2}
\providecommand\natexlab[1]{#1}
\providecommand\showeprint[2][]{arXiv:#2}
\makeatletter
\@ifundefined{NAT@parse@date}{}{\let\NAT@parse@date@orig\NAT@parse@date}
\@ifundefined{NAT@parse@date}{}{\def\NAT@parse@date#1#2#3#4#5#6@@{\NAT@parse@date@orig#1#2#3#4#5#6@@\def\NAT@tempyear{0000}\def\NAT@tempexlab{{?}}\ifx\NAT@year\NAT@tempyear\ifx\NAT@exlab\NAT@tempexlab\def\NAT@date{[n.\,d.]}\else\edef\NAT@date{[n.\,d.]\NAT@exlab}\fi\fi}}
\makeatother

\bibitem[{Aeneas contributors}(nd)]%
        {aeneasrepo}
\bibfield{author}{\bibinfo{person}{{Aeneas contributors}}.}
  \bibinfo{year}{n.d.}\natexlab{}.
\newblock \bibinfo{title}{{Aeneas}}.
\newblock
\urldef\tempurl%
\url{https://github.com/AeneasVerif/aeneas}
\showURL{%
\tempurl}


\bibitem[{Apicula contributors}(nd)]%
        {apicularepo}
\bibfield{author}{\bibinfo{person}{{Apicula contributors}}.}
  \bibinfo{year}{n.d.}\natexlab{}.
\newblock \bibinfo{title}{{Apicula}}.
\newblock
\urldef\tempurl%
\url{https://github.com/scurest/apicula}
\showURL{%
\tempurl}
\newblock
\shownote{Commit 3d4e91e140}.


\bibitem[Appel and McAllester(2001)]%
        {appel2001indexed}
\bibfield{author}{\bibinfo{person}{Andrew~W. Appel} {and}
  \bibinfo{person}{David McAllester}.} \bibinfo{year}{2001}\natexlab{}.
\newblock \showarticletitle{An indexed model of recursive types for
  foundational proof-carrying code}.
\newblock \bibinfo{journal}{\emph{ACM Transactions on Programming Languages and
  Systems}} \bibinfo{volume}{23}, \bibinfo{number}{5} (\bibinfo{date}{Sept.}
  \bibinfo{year}{2001}), \bibinfo{pages}{657--683}.
\newblock
\showISSN{1558-4593}
\href{https://doi.org/10.1145/504709.504712}{doi:\nolinkurl{10.1145/504709.504712}}


\bibitem[Balabonski et~al\mbox{.}(2016)]%
        {balabonski2016mezzo}
\bibfield{author}{\bibinfo{person}{Thibaut Balabonski},
  \bibinfo{person}{Fran{\c{c}}ois Pottier}, {and} \bibinfo{person}{Jonathan
  Protzenko}.} \bibinfo{year}{2016}\natexlab{}.
\newblock \showarticletitle{The Design and Formalization of {Mezzo}, a
  Permission-Based Programming Language}.
\newblock \bibinfo{journal}{\emph{ACM Transactions on Programming Languages and
  Systems}} \bibinfo{volume}{38}, \bibinfo{number}{4}, Article
  \bibinfo{articleno}{14} (\bibinfo{year}{2016}), \bibinfo{numpages}{94}~pages.
\newblock
\href{https://doi.org/10.1145/2837022}{doi:\nolinkurl{10.1145/2837022}}


\bibitem[{Binius contributors}(nd)]%
        {biniusrepo}
\bibfield{author}{\bibinfo{person}{{Binius contributors}}.}
  \bibinfo{year}{n.d.}\natexlab{}.
\newblock \bibinfo{title}{{Binius}}.
\newblock
\urldef\tempurl%
\url{https://github.com/IrreducibleOSS/binius}
\showURL{%
\tempurl}
\newblock
\shownote{Commit 1b09443bbb}.


\bibitem[Bornat et~al\mbox{.}(2005)]%
        {bornat2005permissions}
\bibfield{author}{\bibinfo{person}{Richard Bornat}, \bibinfo{person}{Cristiano
  Calcagno}, \bibinfo{person}{Peter O'Hearn}, {and} \bibinfo{person}{Matthew
  Parkinson}.} \bibinfo{year}{2005}\natexlab{}.
\newblock \showarticletitle{Permission Accounting in Separation Logic}. In
  \bibinfo{booktitle}{\emph{Proceedings of the 32nd ACM SIGPLAN-SIGACT
  Symposium on Principles of Programming Languages}}
  \emph{(\bibinfo{series}{POPL '05})}. \bibinfo{publisher}{Association for
  Computing Machinery}, \bibinfo{address}{New York, NY, USA},
  \bibinfo{pages}{259--270}.
\newblock
\href{https://doi.org/10.1145/1040305.1040327}{doi:\nolinkurl{10.1145/1040305.1040327}}


\bibitem[{borrow contributors}(nd)]%
        {borrow2}
\bibfield{author}{\bibinfo{person}{{borrow contributors}}.}
  \bibinfo{year}{n.d.}\natexlab{}.
\newblock \bibinfo{title}{borrow: Partial Borrows for {Rust}}.
\newblock
\urldef\tempurl%
\url{https://docs.rs/borrow/2.0.0/borrow/}
\showURL{%
\tempurl}
\newblock
\shownote{Version 2.0.0 documentation}.


\bibitem[Boyland(2003)]%
        {boyland2003fractional}
\bibfield{author}{\bibinfo{person}{John Boyland}.}
  \bibinfo{year}{2003}\natexlab{}.
\newblock \showarticletitle{Checking Interference with Fractional Permissions}.
  In \bibinfo{booktitle}{\emph{Static Analysis (SAS 2003)}}
  \emph{(\bibinfo{series}{Lecture Notes in Computer Science},
  Vol.~\bibinfo{volume}{2694})}. \bibinfo{publisher}{Springer},
  \bibinfo{address}{Berlin, Heidelberg}, \bibinfo{pages}{55--72}.
\newblock
\href{https://doi.org/10.1007/3-540-44898-5_4}{doi:\nolinkurl{10.1007/3-540-44898-5_4}}


\bibitem[{Charon contributors}(nd)]%
        {charon}
\bibfield{author}{\bibinfo{person}{{Charon contributors}}.}
  \bibinfo{year}{n.d.}\natexlab{}.
\newblock \bibinfo{title}{{Charon}}.
\newblock
\urldef\tempurl%
\url{https://github.com/AeneasVerif/charon}
\showURL{%
\tempurl}


\bibitem[Clarke et~al\mbox{.}(1998)]%
        {ownership_flex}
\bibfield{author}{\bibinfo{person}{David~G. Clarke}, \bibinfo{person}{John~M.
  Potter}, {and} \bibinfo{person}{James Noble}.}
  \bibinfo{year}{1998}\natexlab{}.
\newblock \showarticletitle{Ownership types for flexible alias protection}. In
  \bibinfo{booktitle}{\emph{Proceedings of the 13th ACM SIGPLAN Conference on
  Object-Oriented Programming, Systems, Languages, and Applications}}
  (Vancouver, British Columbia, Canada) \emph{(\bibinfo{series}{OOPSLA '98})}.
  \bibinfo{publisher}{Association for Computing Machinery},
  \bibinfo{address}{New York, NY, USA}, \bibinfo{pages}{48–64}.
\newblock
\showISBNx{1581130058}
\href{https://doi.org/10.1145/286936.286947}{doi:\nolinkurl{10.1145/286936.286947}}


\bibitem[Cortier(2021)]%
        {cortier2021helix}
\bibfield{author}{\bibinfo{person}{Beno{\^\i}t Cortier}.}
  \bibinfo{year}{2021}\natexlab{}.
\newblock \bibinfo{title}{Replace {Editor::current} by a Macro}.
\newblock \bibinfo{howpublished}{Helix pull request \#291}.
\newblock
\urldef\tempurl%
\url{https://github.com/helix-editor/helix/pull/291}
\showURL{%
\tempurl}


\bibitem[F{\"a}hndrich and DeLine(2002)]%
        {fahndrich2002adoption}
\bibfield{author}{\bibinfo{person}{Manuel F{\"a}hndrich} {and}
  \bibinfo{person}{Robert DeLine}.} \bibinfo{year}{2002}\natexlab{}.
\newblock \showarticletitle{Adoption and Focus: Practical Linear Types for
  Imperative Programming}. In \bibinfo{booktitle}{\emph{Proceedings of the ACM
  SIGPLAN 2002 Conference on Programming Language Design and Implementation}}
  \emph{(\bibinfo{series}{PLDI '02})}. \bibinfo{publisher}{Association for
  Computing Machinery}, \bibinfo{address}{New York, NY, USA},
  \bibinfo{pages}{13--24}.
\newblock
\href{https://doi.org/10.1145/512529.512532}{doi:\nolinkurl{10.1145/512529.512532}}


\bibitem[{fish contributors}(nd)]%
        {fishrepo}
\bibfield{author}{\bibinfo{person}{{fish contributors}}.}
  \bibinfo{year}{n.d.}\natexlab{}.
\newblock \bibinfo{title}{{fish}}.
\newblock
\urldef\tempurl%
\url{https://github.com/fish-shell/fish-shell}
\showURL{%
\tempurl}
\newblock
\shownote{Commit 76a3afcab9}.


\bibitem[G{\"a}her et~al\mbox{.}(2024)]%
        {gaher2024refinedrust}
\bibfield{author}{\bibinfo{person}{Lennard G{\"a}her}, \bibinfo{person}{Michael
  Sammler}, \bibinfo{person}{Ralf Jung}, \bibinfo{person}{Robbert Krebbers},
  {and} \bibinfo{person}{Derek Dreyer}.} \bibinfo{year}{2024}\natexlab{}.
\newblock \showarticletitle{{RefinedRust}: A Type System for High-Assurance
  Verification of {Rust} Programs}.
\newblock \bibinfo{journal}{\emph{Proc. ACM Program. Lang.}}
  \bibinfo{volume}{8}, \bibinfo{number}{PLDI}, Article \bibinfo{articleno}{192}
  (\bibinfo{date}{June} \bibinfo{year}{2024}), \bibinfo{numpages}{25}~pages.
\newblock
\href{https://doi.org/10.1145/3656422}{doi:\nolinkurl{10.1145/3656422}}


\bibitem[Gentle(2024)]%
        {genco2021diamond}
\bibfield{author}{\bibinfo{person}{Seph Gentle}.}
  \bibinfo{year}{2024}\natexlab{}.
\newblock \bibinfo{title}{diamond-types}.
\newblock
\urldef\tempurl%
\url{https://github.com/josephg/diamond-types}
\showURL{%
\tempurl}
\newblock
\shownote{Commit 00f722d}.


\bibitem[{grok-build contributors}(nd)]%
        {grokbuildrepo}
\bibfield{author}{\bibinfo{person}{{grok-build contributors}}.}
  \bibinfo{year}{n.d.}\natexlab{}.
\newblock \bibinfo{title}{{grok-build}}.
\newblock
\urldef\tempurl%
\url{https://github.com/xai-org/grok-build}
\showURL{%
\tempurl}
\newblock
\shownote{Commit 2bdd1d6a63}.


\bibitem[{Helix contributors}(nd)]%
        {helix}
\bibfield{author}{\bibinfo{person}{{Helix contributors}}.}
  \bibinfo{year}{n.d.}\natexlab{}.
\newblock \bibinfo{title}{Helix}.
\newblock
\urldef\tempurl%
\url{https://helix-editor.com/}
\showURL{%
\tempurl}


\bibitem[Ho(2024)]%
        {ho2024thesis}
\bibfield{author}{\bibinfo{person}{Son Ho}.} \bibinfo{year}{2024}\natexlab{}.
\newblock \emph{\bibinfo{title}{Formal Verification of Rust Programs by
  Functional Translation}}.
\newblock \bibinfo{thesistype}{Ph.\,D. Dissertation}.
  \bibinfo{school}{Universit{\'e} PSL}.
\newblock
\urldef\tempurl%
\url{https://theses.hal.science/tel-05004605}
\showURL{%
\tempurl}


\bibitem[Ho et~al\mbox{.}(2025)]%
        {ho2025charon}
\bibfield{author}{\bibinfo{person}{Son Ho}, \bibinfo{person}{Guillaume
  Boisseau}, \bibinfo{person}{Lucas Franceschino}, \bibinfo{person}{Yoann
  Prak}, \bibinfo{person}{Aymeric Fromherz}, {and} \bibinfo{person}{Jonathan
  Protzenko}.} \bibinfo{year}{2025}\natexlab{}.
\newblock \showarticletitle{{Charon}: An Analysis Framework for {Rust}}. In
  \bibinfo{booktitle}{\emph{Computer Aided Verification}}
  \emph{(\bibinfo{series}{Lecture Notes in Computer Science},
  Vol.~\bibinfo{volume}{15934})}. \bibinfo{publisher}{Springer Nature
  Switzerland}, \bibinfo{address}{Cham}, \bibinfo{pages}{377--391}.
\newblock
\href{https://doi.org/10.1007/978-3-031-98685-7_18}{doi:\nolinkurl{10.1007/978-3-031-98685-7_18}}


\bibitem[Ho et~al\mbox{.}(2026)]%
        {ho2026scaling}
\bibfield{author}{\bibinfo{person}{Son Ho}, \bibinfo{person}{C{\'e}dric
  Fournet}, \bibinfo{person}{Jonathan Protzenko}, \bibinfo{person}{Michael
  Naehrig}, \bibinfo{person}{Joshua Clune}, \bibinfo{person}{Patrick Longa},
  \bibinfo{person}{Guillaume Boisseau}, \bibinfo{person}{Fernando
  Leal~S{\'a}nchez}, \bibinfo{person}{Aymeric Fromherz}, {and}
  \bibinfo{person}{Antoine Delignat-Lavaud}.} \bibinfo{year}{2026}\natexlab{}.
\newblock \bibinfo{title}{Scaling Verification of Cryptographic Software with
  {Aeneas}, {Rust}, and {Lean}}.
\newblock
\showeprint[arxiv]{2609.15648}~[cs.CR]
\urldef\tempurl%
\url{https://arxiv.org/abs/2609.15648}
\showURL{%
\tempurl}


\bibitem[Ho et~al\mbox{.}(2024)]%
        {ho2024symbolic}
\bibfield{author}{\bibinfo{person}{Son Ho}, \bibinfo{person}{Aymeric Fromherz},
  {and} \bibinfo{person}{Jonathan Protzenko}.} \bibinfo{year}{2024}\natexlab{}.
\newblock \showarticletitle{Sound Borrow-Checking for {Rust} via Symbolic
  Semantics}.
\newblock \bibinfo{journal}{\emph{Proceedings of the ACM on Programming
  Languages}} \bibinfo{volume}{8}, \bibinfo{number}{ICFP}, Article
  \bibinfo{articleno}{251} (\bibinfo{year}{2024}),
  \bibinfo{numpages}{29}~pages.
\newblock
\href{https://doi.org/10.1145/3674640}{doi:\nolinkurl{10.1145/3674640}}


\bibitem[Ho and Protzenko(2022)]%
        {ho2022aeneas}
\bibfield{author}{\bibinfo{person}{Son Ho} {and} \bibinfo{person}{Jonathan
  Protzenko}.} \bibinfo{year}{2022}\natexlab{}.
\newblock \showarticletitle{Aeneas: {Rust} Verification by Functional
  Translation}.
\newblock \bibinfo{journal}{\emph{Proceedings of the ACM on Programming
  Languages}} \bibinfo{volume}{6}, \bibinfo{number}{ICFP}, Article
  \bibinfo{articleno}{116} (\bibinfo{year}{2022}),
  \bibinfo{numpages}{31}~pages.
\newblock
\href{https://doi.org/10.1145/3547647}{doi:\nolinkurl{10.1145/3547647}}


\bibitem[Jung et~al\mbox{.}(2019)]%
        {stacked}
\bibfield{author}{\bibinfo{person}{Ralf Jung}, \bibinfo{person}{Hoang-Hai
  Dang}, \bibinfo{person}{Jeehoon Kang}, {and} \bibinfo{person}{Derek Dreyer}.}
  \bibinfo{year}{2019}\natexlab{}.
\newblock \showarticletitle{Stacked borrows: an aliasing model for Rust}.
\newblock \bibinfo{journal}{\emph{Proc. ACM Program. Lang.}}
  \bibinfo{volume}{4}, \bibinfo{number}{POPL}, Article \bibinfo{articleno}{41}
  (\bibinfo{date}{Dec.} \bibinfo{year}{2019}), \bibinfo{numpages}{32}~pages.
\newblock
\href{https://doi.org/10.1145/3371109}{doi:\nolinkurl{10.1145/3371109}}


\bibitem[Jung et~al\mbox{.}(2017)]%
        {lambda_rust}
\bibfield{author}{\bibinfo{person}{Ralf Jung}, \bibinfo{person}{Jacques-Henri
  Jourdan}, \bibinfo{person}{Robbert Krebbers}, {and} \bibinfo{person}{Derek
  Dreyer}.} \bibinfo{year}{2017}\natexlab{}.
\newblock \showarticletitle{RustBelt: securing the foundations of the Rust
  programming language}.
\newblock \bibinfo{journal}{\emph{Proc. ACM Program. Lang.}}
  \bibinfo{volume}{2}, \bibinfo{number}{POPL}, Article \bibinfo{articleno}{66}
  (\bibinfo{date}{Dec.} \bibinfo{year}{2017}), \bibinfo{numpages}{34}~pages.
\newblock
\href{https://doi.org/10.1145/3158154}{doi:\nolinkurl{10.1145/3158154}}


\bibitem[Kan et~al\mbox{.}(2020)]%
        {kan2020rustsem}
\bibfield{author}{\bibinfo{person}{Shuanglong Kan}, \bibinfo{person}{Zhe Chen},
  \bibinfo{person}{David San{\'a}n}, \bibinfo{person}{Shang-Wei Lin}, {and}
  \bibinfo{person}{Yang Liu}.} \bibinfo{year}{2020}\natexlab{}.
\newblock \bibinfo{title}{An Executable Operational Semantics for {Rust} with
  the Formalization of Ownership and Borrowing}.
\newblock
\showeprint[arxiv]{1804.07608}~[cs.PL]
\urldef\tempurl%
\url{https://arxiv.org/abs/1804.07608v2}
\showURL{%
\tempurl}


\bibitem[Klabnik et~al\mbox{.}(nd)]%
        {rustbook}
\bibfield{author}{\bibinfo{person}{Steve Klabnik}, \bibinfo{person}{Carol
  Nichols}, {and} \bibinfo{person}{Chris Krycho}.}
  \bibinfo{year}{n.d.}\natexlab{}.
\newblock \bibinfo{title}{The {Rust} Programming Language}.
\newblock
\urldef\tempurl%
\url{https://doc.rust-lang.org/book/}
\showURL{%
\tempurl}


\bibitem[Lamb(2025)]%
        {lamb2025moonfire}
\bibfield{author}{\bibinfo{person}{Scott Lamb}.}
  \bibinfo{year}{2025}\natexlab{}.
\newblock \bibinfo{title}{moonfire-nvr}.
\newblock
\urldef\tempurl%
\url{https://github.com/scottlamb/moonfire-nvr}
\showURL{%
\tempurl}
\newblock
\shownote{Commit 2985214}.


\bibitem[Lattuada et~al\mbox{.}(2023)]%
        {verus}
\bibfield{author}{\bibinfo{person}{Andrea Lattuada}, \bibinfo{person}{Travis
  Hance}, \bibinfo{person}{Chanhee Cho}, \bibinfo{person}{Matthias Brun},
  \bibinfo{person}{Isitha Subasinghe}, \bibinfo{person}{Yi Zhou},
  \bibinfo{person}{Jon Howell}, \bibinfo{person}{Bryan Parno}, {and}
  \bibinfo{person}{Chris Hawblitzel}.} \bibinfo{year}{2023}\natexlab{}.
\newblock \showarticletitle{Verus: Verifying Rust Programs using Linear Ghost
  Types}.
\newblock \bibinfo{journal}{\emph{Proc. ACM Program. Lang.}}
  \bibinfo{volume}{7}, \bibinfo{number}{OOPSLA1}, Article
  \bibinfo{articleno}{85} (\bibinfo{date}{April} \bibinfo{year}{2023}),
  \bibinfo{numpages}{30}~pages.
\newblock
\href{https://doi.org/10.1145/3586037}{doi:\nolinkurl{10.1145/3586037}}


\bibitem[Lehmann et~al\mbox{.}(2023)]%
        {lehmann2023flux}
\bibfield{author}{\bibinfo{person}{Nico Lehmann}, \bibinfo{person}{Adam~T.
  Geller}, \bibinfo{person}{Niki Vazou}, {and} \bibinfo{person}{Ranjit Jhala}.}
  \bibinfo{year}{2023}\natexlab{}.
\newblock \showarticletitle{Flux: Liquid Types for {Rust}}.
\newblock \bibinfo{journal}{\emph{Proc. ACM Program. Lang.}}
  \bibinfo{volume}{7}, \bibinfo{number}{PLDI}, Article \bibinfo{articleno}{169}
  (\bibinfo{date}{June} \bibinfo{year}{2023}), \bibinfo{numpages}{25}~pages.
\newblock
\href{https://doi.org/10.1145/3591283}{doi:\nolinkurl{10.1145/3591283}}


\bibitem[Lehmann et~al\mbox{.}(2025)]%
        {lehmann2025generic}
\bibfield{author}{\bibinfo{person}{Nico Lehmann}, \bibinfo{person}{Cole
  Kurashige}, \bibinfo{person}{Nikhil Akiti}, \bibinfo{person}{Niroop
  Krishnakumar}, {and} \bibinfo{person}{Ranjit Jhala}.}
  \bibinfo{year}{2025}\natexlab{}.
\newblock \showarticletitle{Generic Refinement Types}.
\newblock \bibinfo{journal}{\emph{Proc. ACM Program. Lang.}}
  \bibinfo{volume}{9}, \bibinfo{number}{POPL}, Article \bibinfo{articleno}{49}
  (\bibinfo{year}{2025}), \bibinfo{numpages}{29}~pages.
\newblock
\href{https://doi.org/10.1145/3704885}{doi:\nolinkurl{10.1145/3704885}}


\bibitem[Marshall and Orchard(2024)]%
        {fractional}
\bibfield{author}{\bibinfo{person}{Danielle Marshall} {and}
  \bibinfo{person}{Dominic Orchard}.} \bibinfo{year}{2024}\natexlab{}.
\newblock \showarticletitle{Functional Ownership through Fractional
  Uniqueness}.
\newblock \bibinfo{journal}{\emph{Proc. ACM Program. Lang.}}
  \bibinfo{volume}{8}, \bibinfo{number}{OOPSLA1}, Article
  \bibinfo{articleno}{131} (\bibinfo{date}{April} \bibinfo{year}{2024}),
  \bibinfo{numpages}{31}~pages.
\newblock
\href{https://doi.org/10.1145/3649848}{doi:\nolinkurl{10.1145/3649848}}


\bibitem[Matsakis(2021)]%
        {matsakis2021view}
\bibfield{author}{\bibinfo{person}{Nicholas~D. Matsakis}.}
  \bibinfo{year}{2021}\natexlab{}.
\newblock \bibinfo{title}{View Types for {Rust}}.
\newblock
\urldef\tempurl%
\url{https://smallcultfollowing.com/babysteps/blog/2021/11/05/view-types/}
\showURL{%
\tempurl}
\newblock
\shownote{Blog post, 5 November}.


\bibitem[Matsakis(2024)]%
        {matsakis2024borrow}
\bibfield{author}{\bibinfo{person}{Nicholas~D. Matsakis}.}
  \bibinfo{year}{2024}\natexlab{}.
\newblock \bibinfo{title}{The Borrow Checker Within}.
\newblock
\urldef\tempurl%
\url{https://smallcultfollowing.com/babysteps/blog/2024/06/02/the-borrow-checker-within/}
\showURL{%
\tempurl}
\newblock
\shownote{Blog post, 2 June}.


\bibitem[Matsakis(2025)]%
        {matsakis2025redux}
\bibfield{author}{\bibinfo{person}{Nicholas~D. Matsakis}.}
  \bibinfo{year}{2025}\natexlab{}.
\newblock \bibinfo{title}{View Types Redux and Abstract Fields}.
\newblock
\urldef\tempurl%
\url{https://smallcultfollowing.com/babysteps/blog/2025/02/25/view-types-redux/}
\showURL{%
\tempurl}
\newblock
\shownote{Blog post, 25 February}.


\bibitem[Matsushita et~al\mbox{.}(2022)]%
        {matsushita2022rusthornbelt}
\bibfield{author}{\bibinfo{person}{Yusuke Matsushita}, \bibinfo{person}{Xavier
  Denis}, \bibinfo{person}{Jacques-Henri Jourdan}, {and} \bibinfo{person}{Derek
  Dreyer}.} \bibinfo{year}{2022}\natexlab{}.
\newblock \showarticletitle{{RustHornBelt}: A Semantic Foundation for
  Functional Verification of {Rust} Programs with Unsafe Code}. In
  \bibinfo{booktitle}{\emph{Proceedings of the 43rd ACM SIGPLAN International
  Conference on Programming Language Design and Implementation}}
  \emph{(\bibinfo{series}{PLDI '22})}. \bibinfo{publisher}{Association for
  Computing Machinery}, \bibinfo{address}{New York, NY, USA},
  \bibinfo{pages}{841--856}.
\newblock
\href{https://doi.org/10.1145/3519939.3523704}{doi:\nolinkurl{10.1145/3519939.3523704}}


\bibitem[Matsushita et~al\mbox{.}(2021)]%
        {rusthorn}
\bibfield{author}{\bibinfo{person}{Yusuke Matsushita}, \bibinfo{person}{Takeshi
  Tsukada}, {and} \bibinfo{person}{Naoki Kobayashi}.}
  \bibinfo{year}{2021}\natexlab{}.
\newblock \showarticletitle{RustHorn: CHC-based Verification for Rust
  Programs}.
\newblock \bibinfo{journal}{\emph{ACM Trans. Program. Lang. Syst.}}
  \bibinfo{volume}{43}, \bibinfo{number}{4}, Article \bibinfo{articleno}{15}
  (\bibinfo{date}{Oct.} \bibinfo{year}{2021}), \bibinfo{numpages}{54}~pages.
\newblock
\showISSN{0164-0925}
\href{https://doi.org/10.1145/3462205}{doi:\nolinkurl{10.1145/3462205}}


\bibitem[Ogawa et~al\mbox{.}(2025)]%
        {ogawa2025thrust}
\bibfield{author}{\bibinfo{person}{Hiromi Ogawa}, \bibinfo{person}{Taro
  Sekiyama}, {and} \bibinfo{person}{Hiroshi Unno}.}
  \bibinfo{year}{2025}\natexlab{}.
\newblock \showarticletitle{{Thrust}: A Prophecy-Based Refinement Type System
  for {Rust}}.
\newblock \bibinfo{journal}{\emph{Proc. ACM Program. Lang.}}
  \bibinfo{volume}{9}, \bibinfo{number}{PLDI}, Article \bibinfo{articleno}{230}
  (\bibinfo{year}{2025}), \bibinfo{numpages}{25}~pages.
\newblock
\href{https://doi.org/10.1145/3729333}{doi:\nolinkurl{10.1145/3729333}}


\bibitem[{partial\_ref contributors}(nd)]%
        {partialref}
\bibfield{author}{\bibinfo{person}{{partial\_ref contributors}}.}
  \bibinfo{year}{n.d.}\natexlab{}.
\newblock \bibinfo{title}{partial\_ref: Type Checked Partial References}.
\newblock
\urldef\tempurl%
\url{https://docs.rs/partial_ref/0.3.4/partial_ref/}
\showURL{%
\tempurl}
\newblock
\shownote{Version 0.3.4 documentation}.


\bibitem[Pearce(2021)]%
        {fr}
\bibfield{author}{\bibinfo{person}{David~J. Pearce}.}
  \bibinfo{year}{2021}\natexlab{}.
\newblock \showarticletitle{A Lightweight Formalism for Reference Lifetimes and
  Borrowing in Rust}.
\newblock \bibinfo{journal}{\emph{ACM Trans. Program. Lang. Syst.}}
  \bibinfo{volume}{43}, \bibinfo{number}{1}, Article \bibinfo{articleno}{3}
  (\bibinfo{date}{April} \bibinfo{year}{2021}), \bibinfo{numpages}{73}~pages.
\newblock
\showISSN{0164-0925}
\href{https://doi.org/10.1145/3443420}{doi:\nolinkurl{10.1145/3443420}}


\bibitem[{Replica contributors}(nd)]%
        {replicarepo}
\bibfield{author}{\bibinfo{person}{{Replica contributors}}.}
  \bibinfo{year}{n.d.}\natexlab{}.
\newblock \bibinfo{title}{{Replica}}.
\newblock
\urldef\tempurl%
\url{https://github.com/josephg/replica}
\showURL{%
\tempurl}
\newblock
\shownote{Commit 73ec5c52ce}.


\bibitem[Reynolds(2002)]%
        {reynolds2002separation}
\bibfield{author}{\bibinfo{person}{John~C. Reynolds}.}
  \bibinfo{year}{2002}\natexlab{}.
\newblock \showarticletitle{Separation logic: a logic for shared mutable data
  structures}. In \bibinfo{booktitle}{\emph{Proceedings 17th Annual IEEE
  Symposium on Logic in Computer Science}} \emph{(\bibinfo{series}{LICS-02})}.
  \bibinfo{publisher}{IEEE Comput. Soc}, \bibinfo{pages}{55--74}.
\newblock
\href{https://doi.org/10.1109/lics.2002.1029817}{doi:\nolinkurl{10.1109/lics.2002.1029817}}


\bibitem[{RhoJVM contributors}(nd)]%
        {rhojvmrepo}
\bibfield{author}{\bibinfo{person}{{RhoJVM contributors}}.}
  \bibinfo{year}{n.d.}\natexlab{}.
\newblock \bibinfo{title}{{RhoJVM}}.
\newblock
\urldef\tempurl%
\url{https://github.com/MinusGix/RhoJVM-old}
\showURL{%
\tempurl}
\newblock
\shownote{Commit de5ae7103c}.


\bibitem[Smetsers et~al\mbox{.}(1994)]%
        {smetsers1994uniqueness}
\bibfield{author}{\bibinfo{person}{Sjaak Smetsers}, \bibinfo{person}{Erik
  Barendsen}, \bibinfo{person}{Marko van Eekelen}, {and} \bibinfo{person}{Rinus
  Plasmeijer}.} \bibinfo{year}{1994}\natexlab{}.
\newblock \showarticletitle{Guaranteeing Safe Destructive Updates Through a
  Type System with Uniqueness Information for Graphs}. In
  \bibinfo{booktitle}{\emph{Graph Transformations in Computer Science}}
  \emph{(\bibinfo{series}{Lecture Notes in Computer Science},
  Vol.~\bibinfo{volume}{776})}. \bibinfo{publisher}{Springer},
  \bibinfo{address}{Berlin, Heidelberg}, \bibinfo{pages}{358--379}.
\newblock
\href{https://doi.org/10.1007/3-540-57787-4_23}{doi:\nolinkurl{10.1007/3-540-57787-4_23}}


\bibitem[Smith et~al\mbox{.}(2000)]%
        {smith2000alias}
\bibfield{author}{\bibinfo{person}{Frederick Smith}, \bibinfo{person}{David
  Walker}, {and} \bibinfo{person}{Greg Morrisett}.}
  \bibinfo{year}{2000}\natexlab{}.
\newblock \showarticletitle{Alias Types}. In
  \bibinfo{booktitle}{\emph{Programming Languages and Systems (ESOP 2000)}}
  \emph{(\bibinfo{series}{Lecture Notes in Computer Science},
  Vol.~\bibinfo{volume}{1782})}. \bibinfo{publisher}{Springer},
  \bibinfo{address}{Berlin, Heidelberg}, \bibinfo{pages}{366--381}.
\newblock
\href{https://doi.org/10.1007/3-540-46425-5_24}{doi:\nolinkurl{10.1007/3-540-46425-5_24}}


\bibitem[{Sway contributors}(nd)]%
        {swayrepo}
\bibfield{author}{\bibinfo{person}{{Sway contributors}}.}
  \bibinfo{year}{n.d.}\natexlab{}.
\newblock \bibinfo{title}{{Sway}}.
\newblock
\urldef\tempurl%
\url{https://github.com/FuelLabs/sway}
\showURL{%
\tempurl}
\newblock
\shownote{Commit 7dafb1c7ae}.


\bibitem[{The Rust Project Developers}(nda)]%
        {rustreference}
\bibfield{author}{\bibinfo{person}{{The Rust Project Developers}}.}
  \bibinfo{year}{n.d.}\natexlab{a}.
\newblock \bibinfo{title}{The {Rust} Reference: Pointer Types}.
\newblock
\urldef\tempurl%
\url{https://doc.rust-lang.org/reference/types/pointer.html}
\showURL{%
\tempurl}


\bibitem[{The Rust Project Developers}(ndb)]%
        {rustsplitatmut}
\bibfield{author}{\bibinfo{person}{{The Rust Project Developers}}.}
  \bibinfo{year}{n.d.}\natexlab{b}.
\newblock \bibinfo{title}{{Rust} Standard Library: {slice::split\_at\_mut}}.
\newblock
\urldef\tempurl%
\url{https://doc.rust-lang.org/std/primitive.slice.html#method.split_at_mut}
\showURL{%
\tempurl}


\bibitem[{The Rust Project Developers}(ndc)]%
        {rustunsafecell}
\bibfield{author}{\bibinfo{person}{{The Rust Project Developers}}.}
  \bibinfo{year}{n.d.}\natexlab{c}.
\newblock \bibinfo{title}{{Rust} Standard Library: {std::cell::UnsafeCell}}.
\newblock
\urldef\tempurl%
\url{https://doc.rust-lang.org/std/cell/struct.UnsafeCell.html}
\showURL{%
\tempurl}


\bibitem[{Varisat contributors}(nd)]%
        {varisatrepo}
\bibfield{author}{\bibinfo{person}{{Varisat contributors}}.}
  \bibinfo{year}{n.d.}\natexlab{}.
\newblock \bibinfo{title}{{Varisat}}.
\newblock
\urldef\tempurl%
\url{https://github.com/jix/varisat}
\showURL{%
\tempurl}
\newblock
\shownote{Commit 33e876937c}.


\bibitem[Villani et~al\mbox{.}(2025)]%
        {villani2025tree}
\bibfield{author}{\bibinfo{person}{Neven Villani}, \bibinfo{person}{Johannes
  Hostert}, \bibinfo{person}{Derek Dreyer}, {and} \bibinfo{person}{Ralf Jung}.}
  \bibinfo{year}{2025}\natexlab{}.
\newblock \showarticletitle{Tree Borrows}.
\newblock \bibinfo{journal}{\emph{Proc. ACM Program. Lang.}}
  \bibinfo{volume}{9}, \bibinfo{number}{PLDI}, Article \bibinfo{articleno}{188}
  (\bibinfo{date}{June} \bibinfo{year}{2025}), \bibinfo{numpages}{24}~pages.
\newblock
\href{https://doi.org/10.1145/3735592}{doi:\nolinkurl{10.1145/3735592}}


\bibitem[Wang et~al\mbox{.}(2018)]%
        {wang2018krust}
\bibfield{author}{\bibinfo{person}{Feng Wang}, \bibinfo{person}{Fu Song},
  \bibinfo{person}{Min Zhang}, \bibinfo{person}{Xiaoran Zhu}, {and}
  \bibinfo{person}{Jun Zhang}.} \bibinfo{year}{2018}\natexlab{}.
\newblock \showarticletitle{{KRust}: A Formal Executable Semantics of {Rust}}.
  In \bibinfo{booktitle}{\emph{2018 International Symposium on Theoretical
  Aspects of Software Engineering (TASE)}}. \bibinfo{publisher}{IEEE},
  \bibinfo{pages}{44--51}.
\newblock
\href{https://doi.org/10.1109/TASE.2018.00014}{doi:\nolinkurl{10.1109/TASE.2018.00014}}


\bibitem[{Wasmtime contributors}(nd)]%
        {wasmtimerepo}
\bibfield{author}{\bibinfo{person}{{Wasmtime contributors}}.}
  \bibinfo{year}{n.d.}\natexlab{}.
\newblock \bibinfo{title}{{Wasmtime}}.
\newblock
\urldef\tempurl%
\url{https://github.com/bytecodealliance/wasmtime}
\showURL{%
\tempurl}
\newblock
\shownote{Commit e2f9ca6be1}.


\bibitem[Weiss et~al\mbox{.}(2021)]%
        {oxide}
\bibfield{author}{\bibinfo{person}{Aaron Weiss}, \bibinfo{person}{Olek
  Gierczak}, \bibinfo{person}{Daniel Patterson}, {and} \bibinfo{person}{Amal
  Ahmed}.} \bibinfo{year}{2021}\natexlab{}.
\newblock \bibinfo{title}{Oxide: The Essence of Rust}.
\newblock
\showeprint[arxiv]{1903.00982}~[cs.PL]
\urldef\tempurl%
\url{https://arxiv.org/abs/1903.00982v4}
\showURL{%
\tempurl}


\bibitem[Yanovski et~al\mbox{.}(2021)]%
        {ghostcell}
\bibfield{author}{\bibinfo{person}{Joshua Yanovski}, \bibinfo{person}{Hoang-Hai
  Dang}, \bibinfo{person}{Ralf Jung}, {and} \bibinfo{person}{Derek Dreyer}.}
  \bibinfo{year}{2021}\natexlab{}.
\newblock \showarticletitle{GhostCell: separating permissions from data in
  Rust}.
\newblock \bibinfo{journal}{\emph{Proc. ACM Program. Lang.}}
  \bibinfo{volume}{5}, \bibinfo{number}{ICFP}, Article \bibinfo{articleno}{92}
  (\bibinfo{date}{Aug.} \bibinfo{year}{2021}), \bibinfo{numpages}{30}~pages.
\newblock
\href{https://doi.org/10.1145/3473597}{doi:\nolinkurl{10.1145/3473597}}


\end{thebibliography}

\clearpage
\appendix

\section{Common syntax definitions}
\label{app:common-syntax}

The following definitions are shared by \LLBC and \LLBCView.

\begin{defs}
\[\begin{grammar}
  \gabbr{\fn \in \tyname{fn}}{\tyname{string}}{}
\end{grammar}\]

\[\begin{grammar}
  \gprod{\var \in \tyname{var}}{\locvar}{} \\
  \galt{\invar}{} \\
  \galt{\retvar}{}
\end{grammar}\]

\[\begin{grammar}
  \gprod{\proj \in \tyname{projection}}{\dpnil}{} \\
  \galt{\pathcomp{\proj}{\PDeref}}{} \\
  \galt{\pathcomp{\proj}{\dpproj{\nat}}}{}
\end{grammar}\]

\[\begin{grammar}
  \gabbr{\place \in \tyname{place}}{\placelit{\proj}{\var}}{}
\end{grammar}\]
\end{defs}

\section{\LLBC Syntax and Semantics}
\label{app:llbc}

\subsection{\LLBC Syntax}\label{app:llbc:syntax}

\begin{defs}

\[\begin{grammar}
  \gprod{\ty}{\TShrRef{\ty}}{} \\
  \galt{\TMutRef{\ty}}{} \\
  \galt{\TUnit}{} \\
  \galt{\TPair{\ty_{0}}{\ty_{1}}}{} \\
  \galt{\TSum{\ty_{0}}{\ty_{1}}}{}
\end{grammar}\]

\[\begin{grammar}
  \gprod{\op}{\OpMove{\place}}{} \\
  \galt{\OpUnit}{} \\
  \galt{\OpPair{\op_{0}}{\op_{1}}}{} \\
  \galt{\OpLeft{\op}{\ty}}{} \\
  \galt{\OpRight{\ty}{\op}}{}
\end{grammar}\]

\[\begin{grammar}
  \gprod{\rv}{\op}{} \\
  \galt{\RvRefMut{\place}}{} \\
  \galt{\RvRefShr{\place}}{}
\end{grammar}\]

\[\begin{grammar}
  \gprod{\stmt}{\SNil}{} \\
  \galt{\SSeq{\stmt_{0}}{\stmt_{1}}}{} \\
  \galt{\SAssign{\place}{\rv}}{} \\
  \galt{\SMatch{\place}{\stmt_{0}}{\stmt_{1}}}{} \\
  \galt{\SRet}{} \\
  \galt{\SPanic}{} \\
  \galt{\SCall{\place}{\fn}{\op}}{}
\end{grammar}\]

\[\begin{grammar}
  \gprod{\mfndecl}{\fndecl{f}{\tau_{\mathit{in}}}{\tau_{\mathit{loc}}}{\tau_{\mathit{ret}}}{s}}{}
\end{grammar}\]

\[\begin{grammar}
  \gabbr{\prog}{\vect{\fn \times \mfndecl}}{}
\end{grammar}\]

\end{defs}

A program $\prog$ is a list of functions, each with its declaration. We write $\progfn{\prog}{\fn} = \mfndecl$ when $(\fn, \mfndecl) \in \prog$.

\subsection{Concrete Semantics (\LLBC)}
\label{app:llbc:concrete}

\subsubsection{Values and states}\label{app:llbc:concrete:values}

\begin{deffig}{fig:llbc:concrete:values}{\LLBC values and states.}
\[\begin{grammar}
  \gabbr{\loc}{\tyname{nat}}{Location: the label connecting a loan and its borrows}
\end{grammar}\]

\[\begin{grammar}
  \gprod{\val}{\VUnit}{} \\
  \galt{\VPair{\val_{0}}{\val_{1}}}{} \\
  \galt{\VBotTy{\ty}}{} \\
  \galt{\VMutLoanTy{\loc}{\ty}}{} \\
  \galt{\VShrLoan{\loc}{\val}}{} \\
  \galt{\VLeftTy{\val}{\ty}}{} \\
  \galt{\VRightTy{\ty}{\val}}{} \\
  \galt{\VMutBor{\loc}{\val}}{} \\
  \galt{\VShrBorTy{\loc}{\ty}}{}
\end{grammar}\]

\[\begin{grammar}
  \gprod{\tyname{frame}}{\frameval{v_{\mathit{loc}}}{v_{\mathit{in}}}{v_{\mathit{ret}}}}{}
\end{grammar}\]

\[\begin{grammar}
  \gprod{\st}{(\vect{\tyname{frame}},\ \vect{\val_{\mathit{anon}}})}{}
\end{grammar}\]
\end{deffig}

\subsubsection{Auxiliary semantic data types}\label{app:llbc:concrete:aux-data}

\begin{deffig}{fig:llbc:modes-results}{Access modes and execution results.}
\[\begin{grammar}
  \gprod{\mode}{\Imm}{} \\
  \galt{\Mut}{} \\
  \galt{\MMove}{}
\end{grammar}\]

\[\begin{grammar}
  \gprod{\res}{\RRet}{} \\
  \galt{\RPanic}{} \\
  \galt{\RUnit}{}
\end{grammar}\]

\[\begin{grammar}
  \gprod{\tyname{stmt{\_}res}}{\NoFuel}{Fuel exhausted} \\
  \galt{\StmtFin{\res}{\st}}{Converging result}
\end{grammar}\]
\end{deffig}

\begin{deffig}{fig:llbc:direct-path}{Addresses of sub-values in a value.}
\[\begin{grammar}
  \gprod{\seg}{\dpunloan}{Direct path segment} \\
  \galt{\dpproj{\nat}}{} \\
  \galt{\dpunbor}{}
\end{grammar}\]

\[\begin{grammar}
  \gabbr{\dpath}{\vect{\seg}}{Direct path}
\end{grammar}\]
\end{deffig}

\begin{deffig}{fig:llbc:state-addr}{Addresses of values in a state.}
\[\begin{grammar}
  \gprod{\valaddr}{\InFrames{\nat}{\var}}{Variable $x$ in $n$-th frame} \\
  \galt{\Anon{\nat}}{$n$-th anonymous variable}
\end{grammar}\]

\[\begin{grammar}
  \gabbr{\fpath}{\fplit{\valaddr}{\dpath}}{Full path}
\end{grammar}\]
\end{deffig}

\subsubsection{Operations on values}\label{app:llbc:concrete:value-ops}

\begin{definition}[Relabelling]\label{def:llbc:relabel}
\[\begin{array}{@{}r@{\;}l@{}}
  \relabel{\loc}{\loc_{0}}{\loc_{1}} &\defeq \mathsf{if}\ \eqb{\loc}{\loc_{0}}\ \mathsf{then}\ \loc_{1}\ \mathsf{else}\ \mathsf{if}\ \eqb{\loc}{\loc_{1}}\ \mathsf{then}\ \loc_{0}\ \mathsf{else}\ \loc
\end{array}\]
\end{definition}
\noindent $\relabel{\loc}{\loc_{0}}{\loc_{1}}$ swaps the labels $\loc_{0}$ and $\loc_{1}$, in both directions, and leaves every other label unchanged.

\begin{definition}[Bottom of a type]\label{def:llbc:botof}
$\mathtt{bot\_of} : \ty \to \val$
\[\begin{array}{@{}r@{\;}l@{}}
  \botof{\TPair{\ty_{a}}{\ty_{b}}} &= \VPair{\botof{\ty_{a}}}{\botof{\ty_{b}}} \\
  \botof{\ty} &= \VBotTy{\ty}
\end{array}\]
\end{definition}
\noindent $\botof{\ty}$ builds a tree of bottoms that recurses on pairs.

\begin{definition}[Type of a value]\label{def:llbc:tyof}
$\mathtt{ty\_of} : \val \to \ty$
\[\begin{array}{@{}r@{\;}l@{}}
  \tyof{\VUnit} &= \TUnit \\
  \tyof{\VPair{\val_{0}}{\val_{1}}} &= \TPair{\tyof{\val_{0}}}{\tyof{\val_{1}}} \\
  \tyof{\VBotTy{\ty}} &= \ty \\
  \tyof{\VMutLoanTy{\_}{\ty}} &= \ty \\
  \tyof{\VShrLoan{\_}{\val_{0}}} &= \tyof{\val_{0}} \\
  \tyof{\VLeftTy{\val_{0}}{\ty_{1}}} &= \TSum{\tyof{\val_{0}}}{\ty_{1}} \\
  \tyof{\VRightTy{\ty_{0}}{\val_{1}}} &= \TSum{\ty_{0}}{\tyof{\val_{1}}} \\
  \tyof{\VMutBor{\_}{\val_{0}}} &= \TMutRef{\tyof{\val_{0}}} \\
  \tyof{\VShrBorTy{\_}{\ty}} &= \TShrRef{\ty}
\end{array}\]
\end{definition}

\begin{rulefig}{fig:llbc:dp-applies}{Reading a value along a direct path, $\dpapplies{\dpath}{\val}{\val'}$.}
\inferrule*[lab={\rulename{DP-Nil}}]
    {}
    {\dpapplies{\dpnil}{\val}{\val}}
\and
\inferrule*[lab={\rulename{DP-Pair-0}}]
    {\dpapplies{\dpath}{\val_{0}}{\val}}
    {\dpapplies{\pathcomp{\dpath}{\dpproj{0}}}{\VPair{\val_{0}}{\val_{1}}}{\val}}
\and
\inferrule*[lab={\rulename{DP-Pair-1}}]
    {\dpapplies{\dpath}{\val_{1}}{\val}}
    {\dpapplies{\pathcomp{\dpath}{\dpproj{1}}}{\VPair{\val_{0}}{\val_{1}}}{\val}}
\and
\inferrule*[lab={\rulename{DP-Left}}]
    {\dpapplies{\dpath}{\val_{0}}{\val}}
    {\dpapplies{\pathcomp{\dpath}{\dpproj{0}}}{\VLeft{\val_{0}}}{\val}}
\and
\inferrule*[lab={\rulename{DP-Right}}]
    {\dpapplies{\dpath}{\val_{0}}{\val}}
    {\dpapplies{\pathcomp{\dpath}{\dpproj{1}}}{\VRight{\val_{0}}}{\val}}
\and
\inferrule*[lab={\rulename{DP-Unloan}}]
    {\dpapplies{\dpath}{\val_{0}}{\val}}
    {\dpapplies{\pathcomp{\dpath}{\dpunloan}}{\VShrLoan{\loc}{\val_{0}}}{\val}}
\and
\inferrule*[lab={\rulename{DP-Unbor}}]
    {\dpapplies{\dpath}{\val_{0}}{\val}}
    {\dpapplies{\pathcomp{\dpath}{\dpunbor}}{\VMutBor{\loc}{\val_{0}}}{\val}}
\end{rulefig}

\begin{rulefig}{fig:llbc:val-updates}{Writing a value along a direct path, $\valupd{\dpath}{\val}{\val'}{\val''}$.}
\inferrule*[lab={\rulename{V-Base}}]
    {\tyof{w} = \tyof{v}}
    {\valupd{\dpnil}{v}{w}{w}}
\and
\inferrule*[lab={\rulename{V-Sum-Left}}]
    {\valupd{\dpath}{v}{w}{\val'}}
    {\valupd{\pathcomp{\dpath}{\dpproj{0}}}{\VLeftTy{v}{\ty}}{w}{\VLeftTy{\val'}{\ty}}}
\and
\inferrule*[lab={\rulename{V-Sum-Right}}]
    {\valupd{\dpath}{v}{w}{\val'}}
    {\valupd{\pathcomp{\dpath}{\dpproj{1}}}{\VRightTy{\ty}{v}}{w}{\VRightTy{\ty}{\val'}}}
\and
\inferrule*[lab={\rulename{V-Pair-Left}}]
    {\valupd{\dpath}{\val_{0}}{\val}{\val_{0}'}}
    {\valupd{\pathcomp{\dpath}{\dpproj{0}}}{\VPair{\val_{0}}{\val_{1}}}{\val}{\VPair{\val_{0}'}{\val_{1}}}}
\and
\inferrule*[lab={\rulename{V-Pair-Right}}]
    {\valupd{\dpath}{\val_{1}}{\val}{\val_{1}'}}
    {\valupd{\pathcomp{\dpath}{\dpproj{1}}}{\VPair{\val_{0}}{\val_{1}}}{\val}{\VPair{\val_{0}}{\val_{1}'}}}
\and
\inferrule*[lab={\rulename{V-SharedUnloan}}]
    {\valupd{\dpath}{v}{w}{\val'}}
    {\valupd{\pathcomp{\dpath}{\dpunloan}}{\VShrLoan{\loc}{v}}{w}{\VShrLoan{\loc}{\val'}}}
\and
\inferrule*[lab={\rulename{V-Unbor}}]
    {\valupd{\dpath}{v}{w}{\val'}}
    {\valupd{\pathcomp{\dpath}{\dpunbor}}{\VMutBor{\loc}{v}}{w}{\VMutBor{\loc}{\val'}}}
\end{rulefig}

\begin{definition}[Outer loan]\label{def:llbc:outer-loan}
\begin{mathpar}
\inferrule*[lab={\rulename{Outer-Loan-Mut}}]
    {\dpapplies{\dpath}{\val}{\VMutLoan{\loc}} \\ \notinsideborrowed{\dpath}}
    {\hasouterloan{\val}}
\and
\inferrule*[lab={\rulename{Outer-Loan-Shared}}]
    {\dpapplies{\dpath}{\val}{\VShrLoan{\loc}{\val'}} \\ \notinsideborrowed{\dpath}}
    {\hasouterloan{\val}}
\end{mathpar}
\end{definition}

\subsubsection{Operations on states}\label{app:llbc:concrete:state-ops}

Square brackets look up and replace entries of a list: $\lindex{l}{n}$ is the
$n$-th entry of the list $l$, so a premise $\lindex{l}{n} = x$ also says that $l$
has an $n$-th entry, and $\lset{l}{n}{x}$ is $l$ with its $n$-th entry replaced
by $x$. On a frame, $\frameget{\var}{\mathit{frame}}$ is the value of the variable
$\var$, and $\frameset{\var}{\val}{\mathit{frame}}$ replaces it in the same way.

\begin{definition}[Adding an anonymous value]\label{def:llbc:push-anon}
\[\pushanon{\val}{(\mathit{frames},\ \mathit{anons})} \defeq (\mathit{frames},\ \val :: \mathit{anons})\]
\end{definition}

\begin{definition}[Reading the value at an address]\label{def:llbc:state-get}
\begin{mathpar}
\inferrule*[lab={\rulename{State-Get-Frames}}]
    {\lindex{\mathit{frames}}{\nat} = \mathit{frame}}
    {\vaapplies{\InFrames{\nat}{\var}}{(\mathit{frames},\ \mathit{anons})}{\frameget{\var}{\mathit{frame}}}}
\and
\inferrule*[lab={\rulename{State-Get-Anon}}]
    {\lindex{\mathit{anons}}{\nat} = \val}
    {\vaapplies{\Anon{\nat}}{(\mathit{frames},\ \mathit{anons})}{\val}}
\end{mathpar}
\end{definition}

\begin{rulefig}{fig:llbc:state-set}{Writing a value at an address of a state, $\stateupd{\valaddr}{\st}{\val}{\st'}$.}
\inferrule*[lab={\rulename{State-Set-Anon}}]
    {}
    {\stateupd{\Anon{\nat}}{(\mathit{frames},\ \mathit{anons})}{\val}{(\mathit{frames},\ \lset{\mathit{anons}}{\nat}{\val})}}
\and
\inferrule*[lab={\rulename{State-Set-Frames}}]
    {\lindex{\mathit{frames}}{\nat} = \mathit{frame}}
    {\stateupd{\InFrames{\nat}{\var}}{(\mathit{frames},\ \mathit{anons})}{\val}{(\lset{\mathit{frames}}{\nat}{\frameset{\var}{\val}{\mathit{frame}}},\ \mathit{anons})}}
\end{rulefig}

\begin{definition}[Reading along a full path]\label{def:llbc:fp-applies}
\begin{mathpar}
\inferrule*[lab={\rulename{FP-Applies}}]
    {\vaapplies{\valaddr}{\st}{\val'} \\ \dpapplies{\dpath}{\val'}{\val}}
    {\fpapplies{\fplit{\valaddr}{\dpath}}{\st}{\val}}
\end{mathpar}
\end{definition}

\begin{definition}[Writing along a full path]\label{def:llbc:state-updates}
\begin{mathpar}
\inferrule*[lab={\rulename{State-Updates}}]
    {\vaapplies{\valaddr}{\st}{\val} \\ \valupd{\dpath}{\val}{w}{\val'} \\ \stateupd{\valaddr}{\st}{\val'}{\st'}}
    {\stateupd{\fplit{\valaddr}{\dpath}}{\st}{w}{\st'}}
\end{mathpar}
\end{definition}
\noindent We write $\fpreplace{\fpath}{\st}{w}{w'}{\st'}$ for $\fpapplies{\fpath}{\st}{w}$ and $\stateupd{\fpath}{\st}{w'}{\st'}$: the value $w$ at $\fpath$ is replaced by $w'$. The same shorthand reads and writes along a direct path in a value, $\fpreplace{\dpath}{\val}{w}{w'}{\val'}$ for $\dpapplies{\dpath}{\val}{w}$ and $\valupd{\dpath}{\val}{w'}{\val'}$.

\begin{rulefig}{fig:llbc:place-address-aux}{Following a projection: from the value $\val$ at the full path $\fpath$, the projection $\proj$ in mode $\mode$ leads to the full path $\fpath'$, $\placeaddraux{\st}{\fpath}{\val}{\proj}{\mode}{\fpath'}$.}
\inferrule*[lab={\rulename{PA-Base}}]
    {}
    {\placeaddraux{\st}{\fpath_{\mathit{cur}}}{\val}{\dpnil}{\mode}{\fpath_{\mathit{cur}}}}
\and
\inferrule*[lab={\rulename{PA-Proj-Sum-Left}}]
    {\placeaddraux{\st}{\pathcomp{\dpproj{0}}{\fpath_{\mathit{cur}}}}{\val}{\proj}{\mode}{\fpath_{\mathit{final}}}}
    {\placeaddraux{\st}{\fpath_{\mathit{cur}}}{\VLeft{\val}}{\pathcomp{\proj}{\dpproj{0}}}{\mode}{\fpath_{\mathit{final}}}}
\and
\inferrule*[lab={\rulename{PA-Proj-Sum-Right}}]
    {\placeaddraux{\st}{\pathcomp{\dpproj{1}}{\fpath_{\mathit{cur}}}}{\val}{\proj}{\mode}{\fpath_{\mathit{final}}}}
    {\placeaddraux{\st}{\fpath_{\mathit{cur}}}{\VRight{\val}}{\pathcomp{\proj}{\dpproj{1}}}{\mode}{\fpath_{\mathit{final}}}}
\and
\inferrule*[lab={\rulename{PA-Proj-Pair-Left}}]
    {\placeaddraux{\st}{\pathcomp{\dpproj{0}}{\fpath_{\mathit{cur}}}}{\val_{0}}{\proj}{\mode}{\fpath_{\mathit{final}}}}
    {\placeaddraux{\st}{\fpath_{\mathit{cur}}}{\VPair{\val_{0}}{\val_{1}}}{\pathcomp{\proj}{\dpproj{0}}}{\mode}{\fpath_{\mathit{final}}}}
\and
\inferrule*[lab={\rulename{PA-Proj-Pair-Right}}]
    {\placeaddraux{\st}{\pathcomp{\dpproj{1}}{\fpath_{\mathit{cur}}}}{\val_{1}}{\proj}{\mode}{\fpath_{\mathit{final}}}}
    {\placeaddraux{\st}{\fpath_{\mathit{cur}}}{\VPair{\val_{0}}{\val_{1}}}{\pathcomp{\proj}{\dpproj{1}}}{\mode}{\fpath_{\mathit{final}}}}
\and
\inferrule*[lab={\rulename{PA-SharedLoan}}]
    {\proj \neq \dpnil \\ \placeaddraux{\st}{\pathcomp{\dpunloan}{\fpath_{\mathit{cur}}}}{\val}{\proj}{\Imm}{\fpath_{\mathit{final}}}}
    {\placeaddraux{\st}{\fpath_{\mathit{cur}}}{\VShrLoan{\loc}{\val}}{\proj}{\Imm}{\fpath_{\mathit{final}}}}
\and
\inferrule*[lab={\rulename{PA-Deref-MutBorrow}}]
    {\mode \neq \MMove \\ \placeaddraux{\st}{\pathcomp{\dpunbor}{\fpath_{\mathit{cur}}}}{\val}{\proj}{\mode}{\fpath_{\mathit{final}}}}
    {\placeaddraux{\st}{\fpath_{\mathit{cur}}}{\VMutBor{\loc}{\val}}{\pathcomp{\proj}{\PDeref}}{\mode}{\fpath_{\mathit{final}}}}
\and
\inferrule*[lab={\rulename{PA-Deref-SharedBorrow}}]
    {\fpapplies{\fpath_{\mathit{loan}}}{\st}{\VShrLoan{\loc}{\val}} \\ \placeaddraux{\st}{\fpath_{\mathit{loan}}}{\VShrLoan{\loc}{\val}}{\proj}{\Imm}{\fpath_{\mathit{final}}}}
    {\placeaddraux{\st}{\fpath_{\mathit{cur}}}{\VShrBor{\loc}}{\pathcomp{\proj}{\PDeref}}{\Imm}{\fpath_{\mathit{final}}}}
\end{rulefig}

\begin{definition}[Address of a place]\label{def:llbc:place-address}
\begin{mathpar}
\inferrule*[lab={\rulename{Place-Address}}]
    {\vaapplies{\InFrames{0}{\var}}{\st}{\val} \\ \placeaddraux{\st}{\InFrames{0}{\var}}{\val}{\proj}{\mode}{\fpath}}
    {\placeaddr{\st}{\placelit{\proj}{\var}}{\mode}{\fpath}}
\end{mathpar}
\end{definition}

\begin{definition}[Reading a place along its path]\label{def:llbc:read-at}
\begin{mathpar}
\inferrule*[lab={\rulename{Read-At}}]
    {\placeaddr{\st}{\place}{\mode}{\fpath} \\ \fpapplies{\fpath}{\st}{\val}}
    {\readat{\st}{\place}{\mode}{\fpath}{\val}}
\end{mathpar}
\end{definition}

\begin{definition}[Reading a place]\label{def:llbc:read}
\begin{mathpar}
\inferrule*[lab={\rulename{Read}}]
    {\readat{\st}{\place}{\mode}{\fpath}{\val}}
    {\reads{\st}{\place}{\mode}{\val}}
\end{mathpar}
\end{definition}

\begin{definition}[Reading and writing a place]\label{def:llbc:read-write}
\begin{mathpar}
\inferrule*[lab={\rulename{Read-Write}}]
    {\readat{\st}{\place}{\mode}{\fpath}{\val} \\ \stateupd{\fpath}{\st}{w}{\st'}}
    {\readwrite{\st}{\place}{\mode}{\val}{w}{\st'}}
\end{mathpar}
\end{definition}

\subsubsection{Evaluation}\label{app:llbc:concrete:eval}

\begin{rulefig}{fig:llbc:op-evals}{Evaluating an operand, $\opevals{\st}{\op}{\val}{\st'}$.}
\inferrule*[lab={\rulename{E-Move}}]
    {\readwrite{\st}{\place}{\MMove}{\val}{\botof{\tyof{\val}}}{\st'} \\ \nobotsloans{\val}}
    {\opevals{\st}{\OpMove{\place}}{\val}{\st'}}
\and
\inferrule*[lab={\rulename{E-Pair}}]
    {\opevals{\st_{0}}{\op_{0}}{\val_{0}}{\st_{1}} \\ \opevals{\st_{1}}{\op_{1}}{\val_{1}}{\st_{2}}}
    {\opevals{\st_{0}}{\OpPair{\op_{0}}{\op_{1}}}{\VPair{\val_{0}}{\val_{1}}}{\st_{2}}}
\and
\inferrule*[lab={\rulename{E-Sum-Left}}]
    {\opevals{\st}{\op}{\val}{\st'}}
    {\opevals{\st}{\OpLeft{\op}{\ty}}{\VLeftTy{\val}{\ty}}{\st'}}
\and
\inferrule*[lab={\rulename{E-Sum-Right}}]
    {\opevals{\st}{\op}{\val}{\st'}}
    {\opevals{\st}{\OpRight{\ty}{\op}}{\VRightTy{\ty}{\val}}{\st'}}
\and
\inferrule*[lab={\rulename{E-Unit}}]
    {}
    {\opevals{\st}{\OpUnit}{\VUnit}{\st}}
\end{rulefig}

\begin{rulefig}{fig:llbc:rv-evals}{Evaluating an rvalue, $\rvevals{\st}{\rv}{\val}{\st'}$.}
\inferrule*[lab={\rulename{E-Op}}]
    {\opevals{\st}{\op}{\val}{\st'}}
    {\rvevals{\st}{\op}{\val}{\st'}}
\and
\inferrule*[lab={\rulename{E-MutBorrow}}]
    {\readwrite{\st}{\place}{\Mut}{\val}{\VMutLoanTy{\loc}{\tyof{\val}}}{\st'} \\ \nobotsloans{\val} \\ \lfresh{\loc}{\st}}
    {\rvevals{\st}{\RvRefMut{\place}}{\VMutBor{\loc}{\val}}{\st'}}
\and
\inferrule*[lab={\rulename{E-SharedBorrow-Reuse}}]
    {\reads{\st}{\place}{\Imm}{\VShrLoan{\loc}{\val}} \\ \nobotsmutloans{\val}}
    {\rvevals{\st}{\RvRefShr{\place}}{\VShrBorTy{\loc}{\tyof{\VShrLoan{\loc}{\val}}}}{\st}}
\and
\inferrule*[lab={\rulename{E-SharedBorrow-Wrap}}]
    {\readwrite{\st}{\place}{\Imm}{\val}{\VShrLoan{\loc}{\val}}{\st'} \\ \val \neq \VShrLoan{\_}{\_} \\ \nobotsmutloans{\val} \\ \lfresh{\loc}{\st}}
    {\rvevals{\st}{\RvRefShr{\place}}{\VShrBorTy{\loc}{\tyof{\val}}}{\st'}}
\end{rulefig}

\begin{rulefig}{fig:llbc:env-reorg}{State reorganisation, $\envreorg{\st}{\st'}$.}
\inferrule*[lab={\rulename{Reorg-End-SharedLoan}}]
    {\notcontainsshrbor{\loc}{\st} \\ \fpreplace{\fpath}{\st}{\VShrLoan{\loc}{\val}}{\val}{\st'}}
    {\envreorg{\st}{\st'}}
\and
\inferrule*[lab={\rulename{Reorg-End-SharedBorrow}}]
    {\notinsideborrowed{\fpath} \\ \fpreplace{\fpath}{\st}{\VShrBorTy{\loc}{\ty}}{\VBotTy{\TShrRef{\ty}}}{\st'}}
    {\envreorg{\st}{\st'}}
\and
\inferrule*[lab={\rulename{Reorg-End-MutBorrow}}]
    {\notinsideborrowed{\fpath} \\ \fpreplace{\fpath}{\st}{\VMutBor{\loc}{\val}}{\VBotTy{\TMutRef{\tyof{\val}}}}{\st_{1}} \\ \noloans{\val} \\\\ \fpreplace{\fpath_{\mathsf{own}}}{\st_{1}}{\VMutLoan{\loc}}{\val}{\st_{2}}}
    {\envreorg{\st}{\st_{2}}}
\end{rulefig}

\begin{rulefig}{fig:llbc:assign}{Assigning to a place: $\assignevals{\st}{\place}{\rv}{\st'}$}
\inferrule*[lab={\rulename{Assign-Val-Evals}}]
    {\readwrite{\st}{\place}{\Mut}{\val_{p}}{\val}{\st'} \\ \nohasouterloan{\val_{p}}}
    {\assignevals{\st}{\place}{\val}{\pushanon{\val_{p}}{\st'}}}
\and
\inferrule*[lab={\rulename{Assign-Evals}}]
    {\rvevals{\st}{\rv}{\val}{\st'} \\ \assignevals{\st'}{\place}{\val}{\st''}}
    {\assignevals{\st}{\place}{\rv}{\st''}}
\end{rulefig}

\begin{definition}[Popping a frame]\label{def:llbc:pop-frame}
\begin{mathpar}
\inferrule*[lab={\rulename{Pop-Frame}}]
    {\nohasouterloan{\val_{l}} \\ \nohasouterloan{\val_{i}} \\ \nobotsloans{\val}}
    {\popframe{(\frameval{\val_{l}}{\val_{i}}{\val} :: \mathit{frames},\ \mathit{anons})}{\val}{(\mathit{frames},\ \val_{i} :: \val_{l} :: \mathit{anons})}}
\end{mathpar}
\end{definition}

\begin{rulefig}{fig:llbc:exit-fn}{Leaving a called function, $\exitfn{\place}{r}{r'}$.}
\inferrule*[lab={\rulename{Exit-No-Fuel}}]
    {}
    {\exitfn{\place}{\NoFuel}{\NoFuel}}
\and
\inferrule*[lab={\rulename{Exit-Panic}}]
    {}
    {\exitfn{\place}{\StmtFin{\RPanic}{\st_{\mathit{end}}}}{\StmtFin{\RPanic}{\st_{\mathit{end}}}}}
\and
\inferrule*[lab={\rulename{Exit-Ret}}]
    {\popframe{\st_{\mathit{end}}}{\val}{\st'} \\ \assignevals{\st'}{\place}{\val}{\st''}}
    {\exitfn{\place}{\StmtFin{\RRet}{\st_{\mathit{end}}}}{\StmtFin{\RUnit}{\st''}}}
\end{rulefig}

\begin{rulefig}{fig:llbc:stmt-evals}{Evaluating a statement with fuel $\nat$, $\bceval{\prog}{\st}{\stmt}{\nat}{r}$.}
\inferrule*[lab={\rulename{E-Step-Zero}}]
    {}
    {\bceval{\prog}{\st}{\stmt}{0}{\NoFuel}}
\and
\inferrule*[lab={\rulename{E-Return}}]
    {}
    {\bceval{\prog}{\st}{\SRet}{\nat + 1}{\StmtFin{\RRet}{\st}}}
\and
\inferrule*[lab={\rulename{E-Panic}}]
    {}
    {\bceval{\prog}{\st}{\SPanic}{\nat + 1}{\StmtFin{\RPanic}{\st}}}
\and
\inferrule*[lab={\rulename{E-Reorg}}]
    {\envreorg{\st_{0}}{\st_{1}} \\ \bceval{\prog}{\st_{1}}{\stmt}{\nat + 1}{\stmtres}}
    {\bceval{\prog}{\st_{0}}{\stmt}{\nat + 1}{\stmtres}}
\and
\inferrule*[lab={\rulename{E-Seq-Unit}}]
    {\bceval{\prog}{\st_{0}}{\stmt_{0}}{\nat + 1}{\StmtFin{\RUnit}{\st_{1}}} \\ \bceval{\prog}{\st_{1}}{\stmt_{1}}{\nat + 1}{\stmtres}}
    {\bceval{\prog}{\st_{0}}{\SSeq{\stmt_{0}}{\stmt_{1}}}{\nat + 1}{\stmtres}}
\and
\inferrule*[lab={\rulename{E-Seq-Propagate}}]
    {\forall \st'.\ \stmtres \neq \StmtFin{\RUnit}{\st'} \\ \bceval{\prog}{\st}{\stmt_{0}}{\nat + 1}{\stmtres}}
    {\bceval{\prog}{\st}{\SSeq{\stmt_{0}}{\stmt_{1}}}{\nat + 1}{\stmtres}}
\and
\inferrule*[lab={\rulename{E-Match-Left}}]
    {\reads{\st}{\place}{\Imm}{w} \\ w = \VLeft{\val} \vee w = \VShrLoan{\loc}{\VLeft{\val}} \\ \bceval{\prog}{\st}{\stmt_{1}}{\nat + 1}{\stmtres}}
    {\bceval{\prog}{\st}{\SMatch{\place}{\stmt_{1}}{\stmt_{2}}}{\nat + 1}{\stmtres}}
\and
\inferrule*[lab={\rulename{E-Match-Right}}]
    {\reads{\st}{\place}{\Imm}{w} \\ w = \VRight{\val} \vee w = \VShrLoan{\loc}{\VRight{\val}} \\ \bceval{\prog}{\st}{\stmt_{2}}{\nat + 1}{\stmtres}}
    {\bceval{\prog}{\st}{\SMatch{\place}{\stmt_{1}}{\stmt_{2}}}{\nat + 1}{\stmtres}}
\and
\inferrule*[lab={\rulename{E-Assign}}]
    {\assignevals{\st}{\place}{\rv}{\st'}}
    {\bceval{\prog}{\st}{\SAssign{\place}{\rv}}{\nat + 1}{\StmtFin{\RUnit}{\st'}}}
\and
\inferrule*[lab={\rulename{E-Call}}]
    {\progfn{\prog}{\fn} = \fndecl{\fn}{\ty_{\mathit{in}}}{\ty_{\mathit{loc}}}{\ty_{\mathit{ret}}}{\stmt} \\ \opevals{\st}{\op}{\val}{(\mathit{frames},\ \mathit{anons})} \\\\ \bceval{\prog}{(\frameval{\botof{\ty_{\mathit{loc}}}}{\val}{\botof{\ty_{\mathit{ret}}}} :: \mathit{frames},\ \mathit{anons})}{\stmt}{\nat}{\mathit{res}_{\mathit{body}}} \\ \exitfn{\place}{\mathit{res}_{\mathit{body}}}{\mathit{res}_{\mathit{fn}}}}
    {\bceval{\prog}{\st}{\SCall{\place}{\fn}{\op}}{\nat + 1}{\mathit{res}_{\mathit{fn}}}}
\and
\inferrule*[lab={\rulename{E-Nil}}]
    {}
    {\bceval{\prog}{\st}{\SNil}{\nat + 1}{\StmtFin{\RUnit}{\st}}}
\end{rulefig}

\subsection{Symbolic Semantics (\LLBCsharp)}
\label{app:llbc:symbolic}

\subsubsection{Values and states}\label{app:llbc:symbolic:values}

Frames are as in \LLBC (\Cref{fig:llbc:concrete:values}).

\begin{deffig}{fig:llbc:symbolic:values}{\LLBCsharp values and states.}
\[\begin{grammar}
  \gprod{\val}{\VUnit}{} \\
  \galt{\VSymbolicV{\ty}}{} \\
  \galt{\VPair{\val_{0}}{\val_{1}}}{} \\
  \galt{\VLeftTy{\val}{\ty}}{} \\
  \galt{\VRightTy{\ty}{\val}}{} \\
  \galt{\VBotTy{\ty}}{} \\
  \galt{\VMutLoanTy{\loc}{\ty}}{} \\
  \galt{\VShrLoan{\loc}{\val}}{} \\
  \galt{\VMutBor{\loc}{\val}}{} \\
  \galt{\VShrBorTy{\loc}{\ty}}{} \\
  \galt{\VIgnored{\ty}}{}
\end{grammar}\]

\[\begin{grammar}
  \gabbr{\abs}{\vect{\val}}{Region abstraction}
\end{grammar}\]

\[\begin{grammar}
  \gprod{\st}{(\tyname{frame},\ \vect{\val_{\mathit{anon}}},\ \vect{\abs})}{
    State. Notice a single frame instead of a stack of frames}
\end{grammar}\]
\end{deffig}

\subsubsection{Auxiliary semantic data types}\label{app:llbc:symbolic:aux-data}

Full paths are as in \LLBC (\Cref{fig:llbc:state-addr}), over the addresses below.

\begin{deffig}{fig:llbc:symbolic:state-addr}{Addresses of values in a state.}
\[\begin{grammar}
  \gprod{\valaddr}{\var}{Variable $x$ of the frame} \\
  \galt{\Anon{\nat}}{$n$-th anonymous variable} \\
  \galt{\InAbs{\nat_{0}}{\nat_{1}}}{$n_1$-th value of the $n_0$-th region abstraction}
\end{grammar}\]

\end{deffig}

\begin{definition}[Paths into a region abstraction]\label{def:llbc:symbolic:fp-inside-region-abs}
A full path goes into a region abstraction, $\fpinsideregionabs{\fpath}$, when its address has the form $\InAbs{\nat_{0}}{\nat_{1}}$; otherwise $\notinsideregionabs{\fpath}$.
\begin{mathpar}
\inferrule*[lab={\rulename{In-Abs}}]
    { }
    {\fpinsideregionabs{\fplit{\InAbs{\nat_{0}}{\nat_{1}}}{\dpath}}}
\end{mathpar}
\end{definition}

\subsubsection{Operations on values}\label{app:llbc:symbolic:value-ops}

Bottoms of a type (\Cref{def:llbc:botof}), reading and writing along a direct path
(\Cref{fig:llbc:dp-applies,fig:llbc:val-updates}) and outer loans
(\Cref{def:llbc:outer-loan}) are as in \LLBC.

\begin{definition}[Type of a value]\label{def:llbc:symbolic:tyof}
$\mathtt{ty\_of} : \val \to \ty$
\[\begin{array}{@{}r@{\;}l@{}}
  \tyof{\VUnit} &= \TUnit \\
  \tyof{\VPair{\val_{0}}{\val_{1}}} &= \TPair{\tyof{\val_{0}}}{\tyof{\val_{1}}} \\
  \tyof{\VLeftTy{\val_{0}}{\ty_{1}}} &= \TSum{\tyof{\val_{0}}}{\ty_{1}} \\
  \tyof{\VRightTy{\ty_{0}}{\val_{1}}} &= \TSum{\ty_{0}}{\tyof{\val_{1}}} \\
  \tyof{\VShrLoan{\_}{\val_{0}}} &= \tyof{\val_{0}} \\
  \tyof{\VMutBor{\_}{\val_{0}}} &= \TMutRef{\tyof{\val_{0}}} \\
  \tyof{\VShrBorTy{\_}{\ty}} &= \TShrRef{\ty} \\
  \tyof{\VSymbolicV{\ty}} &= \ty \\
  \tyof{\VBotTy{\ty}} &= \ty \\
  \tyof{\VMutLoanTy{\_}{\ty}} &= \ty \\
  \tyof{\VIgnored{\ty}} &= \ty
\end{array}\]
\end{definition}

\begin{definition}[Value typing judgement]\label{def:llbc:symbolic:val-ty}
\begin{mathpar}
\inferrule
    {\tyof{\val} = \ty}
    {\sharpvalty{\val}{\ty}}
\end{mathpar}
\end{definition}

\begin{definition}[Symbolic value of a type]\label{def:llbc:symbolic:symof}
$\mathtt{sym\_of} : \ty \to \val$
\[\begin{array}{@{}r@{\;}l@{}}
  \symof{\TPair{\ty_{a}}{\ty_{b}}} &= \VPair{\symof{\ty_{a}}}{\symof{\ty_{b}}} \\
  \symof{\ty} &= \VSymbolicV{\ty}
\end{array}\]
\end{definition}
\noindent $\symof{\ty}$ builds a tree of symbolic values that recurses on pairs, as $\botof{\ty}$ does with bottoms.

\begin{definition}[Relabelling a value]\label{def:llbc:symbolic:relabel-val}
$\relabel{\val}{\loc_{0}}{\loc_{1}}$ swaps the labels $\loc_{0}$ and $\loc_{1}$ everywhere in $\val$ (\Cref{def:llbc:relabel}).
\end{definition}

\begin{definition}[Stripping outer shared loans]\label{def:llbc:symbolic:strip-outer-loans}
$\mathtt{strip\_outer\_loans} : \val \to \val$
\[\begin{array}{@{}r@{\;}l@{}}
  \stripouterloans{\VPair{a}{b}} &= \VPair{\stripouterloans{a}}{\stripouterloans{b}} \\
  \stripouterloans{\VLeftTy{a}{\ty}} &= \VLeftTy{\stripouterloans{a}}{\ty} \\
  \stripouterloans{\VRightTy{\ty}{a}} &= \VRightTy{\ty}{\stripouterloans{a}} \\
  \stripouterloans{\VShrLoan{\_}{u}} &= \stripouterloans{u} \\
  \stripouterloans{\val} &= \val
\end{array}\]
\end{definition}
\noindent $\stripouterloans{\val}$ removes the shared loans reached from the root of $\val$ through pairs and sums only, keeping their contents.

\subsubsection{Operations on states}\label{app:llbc:symbolic:state-ops}

Reading and writing along a full path
(\Cref{def:llbc:fp-applies,def:llbc:state-updates}) are as in \LLBC, with
$\vdash_\#$ for $\vdash$, over the \LLBCsharp addresses
(\Cref{fig:llbc:symbolic:state-addr}).

\begin{definition}[Adding an anonymous value]\label{def:llbc:symbolic:push-anon}
\[\pushanon{\val}{(\mathit{frame},\ \mathit{anons},\ \vect{\abs})} \defeq (\mathit{frame},\ \val :: \mathit{anons},\ \vect{\abs})\]
\end{definition}

\begin{definition}[Adding a region abstraction]\label{def:llbc:symbolic:push-abs}
\[\pushabs{(\mathit{frame},\ \mathit{anons},\ \vect{\abs})}{\abs} \defeq (\mathit{frame},\ \mathit{anons},\ \abs :: \vect{\abs})\]
\end{definition}

\begin{definition}[Reading the value at an address]\label{def:llbc:symbolic:state-get}
\begin{mathpar}
\inferrule*[lab={\rulename{State-Get-Var}}]
    {}
    {\sharpvaapplies{\var}{(\mathit{frame},\ \mathit{anons},\ \vect{\abs})}{\frameget{\var}{\mathit{frame}}}}
\and
\inferrule*[lab={\rulename{State-Get-Anon}}]
    {\lindex{\mathit{anons}}{\nat} = \val}
    {\sharpvaapplies{\Anon{\nat}}{(\mathit{frame},\ \mathit{anons},\ \vect{\abs})}{\val}}
\and
\inferrule*[lab={\rulename{State-Get-Abs}}]
    {\lindex{\vect{\abs}}{\nat_{0}} = \abs \\ \lindex{\abs}{\nat_{1}} = \val}
    {\sharpvaapplies{\InAbs{\nat_{0}}{\nat_{1}}}{(\mathit{frame},\ \mathit{anons},\ \vect{\abs})}{\val}}
\end{mathpar}
\end{definition}

\begin{rulefig}{fig:llbc:symbolic:state-set}{Writing a value at an address of a state, $\sharpstateupd{\valaddr}{\st}{\val}{\st'}$.}
\inferrule*[lab={\rulename{State-Set-Anon}}]
    {}
    {\sharpstateupd{\Anon{\nat}}{(\mathit{frame},\ \mathit{anons},\ \vect{\abs})}{\val}{(\mathit{frame},\ \lset{\mathit{anons}}{\nat}{\val},\ \vect{\abs})}}
\and
\inferrule*[lab={\rulename{State-Set-Var}}]
    {}
    {\sharpstateupd{\var}{(\mathit{frame},\ \mathit{anons},\ \vect{\abs})}{\val}{(\frameset{\var}{\val}{\mathit{frame}},\ \mathit{anons},\ \vect{\abs})}}
\and
\inferrule*[lab={\rulename{State-Set-Abs}}]
    {\lindex{\vect{\abs}}{\nat_{0}} = \abs}
    {\sharpstateupd{\InAbs{\nat_{0}}{\nat_{1}}}{(\mathit{frame},\ \mathit{anons},\ \vect{\abs})}{\val}{(\mathit{frame},\ \mathit{anons},\ \lset{\vect{\abs}}{\nat_{0}}{\lset{\abs}{\nat_{1}}{\val}})}}
\end{rulefig}

The following are as in \LLBC, with $\vdash_\#$ for $\vdash$:
\begin{itemize}
  \item following a projection, $\sharpplaceaddraux{\st}{\fpath}{\val}{\proj}{\mode}{\fpath'}$ (\Cref{fig:llbc:place-address-aux});
  \item the address of a place, $\sharpplaceaddr{\st}{\place}{\mode}{\fpath}$ (\Cref{def:llbc:place-address}), which starts from the variable's address $\var$ in the single frame instead of $\InFrames{0}{\var}$ in the top frame;
  \item reading a place along its path, $\sharpreadat{\st}{\place}{\mode}{\fpath}{\val}$ (\Cref{def:llbc:read-at});
  \item reading a place, $\sharpreads{\st}{\place}{\mode}{\val}$ (\Cref{def:llbc:read});
  \item reading and writing a place, $\sharpreadwrite{\st}{\place}{\mode}{\val}{w}{\st'}$ (\Cref{def:llbc:read-write}).
\end{itemize}

\subsubsection{Evaluation}\label{app:llbc:symbolic:eval}

The following are as in \LLBC, with $\vdash_\#$ for $\vdash$:
\begin{itemize}
  \item evaluating an operand, $\sharpopevals{\st}{\op}{\val}{\st'}$ (\Cref{fig:llbc:op-evals});
  \item assigning to a place, $\sharpassignevals{\st}{\place}{\rv}{\st'}$ and $\sharpassignevals{\st}{\place}{\val}{\st'}$ (\Cref{fig:llbc:assign}).
\end{itemize}

\begin{rulefig}{fig:llbc:symbolic:rv-evals}{Evaluating an rvalue, $\sharprvevals{\st}{\rv}{\val}{\st'}$.}
\inferrule*[lab={\rulename{E-Op}}]
    {\sharpopevals{\st}{\op}{\val}{\st'}}
    {\sharprvevals{\st}{\op}{\val}{\st'}}
\and
\inferrule*[lab={\rulename{E-MutBorrow}}]
    {\sharpreadwrite{\st}{\place}{\Mut}{\val}{\VMutLoanTy{\loc}{\tyof{\val}}}{\st'} \\ \nobotsloans{\val} \\ \lfresh{\loc}{\st}}
    {\sharprvevals{\st}{\RvRefMut{\place}}{\VMutBor{\loc}{\val}}{\st'}}
\and
\inferrule*[lab={\rulename{E-SharedBorrow-Reuse}}]
    {\sharpreads{\st}{\place}{\Imm}{\VShrLoan{\loc}{\val}} \\ \nobotsmutloans{\val}}
    {\sharprvevals{\st}{\RvRefShr{\place}}{\VShrBorTy{\loc}{\tyof{\VShrLoan{\loc}{\val}}}}{\st}}
\and
\inferrule*[lab={\rulename{E-SharedBorrow}}]
    {\sharpreadat{\st}{\place}{\Imm}{\fpath}{\val} \\ \val \neq \VShrLoan{\_}{\_} \\ \nobotsmutloans{\val} \\ \lfresh{\loc}{\st} \\ \sharpshrborrowat{\fpath}{\val}{\st}{\loc}{\st'}}
    {\sharprvevals{\st}{\RvRefShr{\place}}{\VShrBorTy{\loc}{\tyof{\val}}}{\st'}}
\end{rulefig}

\begin{rulefig}{fig:llbc:symbolic:shr-borrow-at}{Placing a new shared loan for a shared borrow of the value $\val$ at $\fpath$, $\sharpshrborrowat{\fpath}{\val}{\st}{\loc}{\st'}$.}
\inferrule*[lab={\rulename{Shared-Borrow-Wrap}}]
    {\notinsideregionabs{\fpath} \\ \sharpstateupd{\fpath}{\st}{\VShrLoan{\loc}{\val}}{\st'}}
    {\sharpshrborrowat{\fpath}{\val}{\st}{\loc}{\st'}}
\and
\inferrule*[lab={\rulename{Shared-Borrow-Abs}}]
    {\sharpvaapplies{\InAbs{r}{i}}{\st}{\VShrLoan{\loc_{k}}{\val_{k}}}}
    {\sharpshrborrowat{\fplit{\InAbs{r}{i}}{\pathcomp{\dpath}{\dpunloan}}}{\val}{\st}{\loc}{\pushabs{\st}{\absliteral{\VShrBorTy{\loc_{k}}{\tyof{\val_{k}}},\ \VShrLoan{\loc}{\symof{\tyof{\val}}}}}}}
\end{rulefig}

\begin{definition}[Unit and other results]\label{def:llbc:symbolic:unit-results}
\[\unitstates{\results} \defeq [\,\st \mid (\RUnit, \st) \in \results\,] \qquad \nonunit{\results} \defeq [\,(r, \st) \in \results \mid r \neq \RUnit\,]\]
\end{definition}

\begin{rulefig}{fig:llbc:symbolic:stmt-evals}{Evaluating a statement to a list of results, $\sharpeval{\prog}{\st}{\stmt}{\results}$.}
\inferrule*[lab={\rulename{E-Return}}]
    {}
    {\sharpeval{\prog}{\st}{\SRet}{[(\RRet, \st)]}}
\and
\inferrule*[lab={\rulename{E-Panic}}]
    {}
    {\sharpeval{\prog}{\st}{\SPanic}{[(\RPanic, \st)]}}
\and
\inferrule*[lab={\rulename{E-Nil}}]
    {}
    {\sharpeval{\prog}{\st}{\SNil}{[(\RUnit, \st)]}}
\and
\inferrule*[lab={\rulename{E-Seq}}]
    {\sharpeval{\prog}{\st}{\stmt_{0}}{\results_{0}} \\ \unitstates{\results_{0}} = [\st_{1},\ \dots,\ \st_{n}] \\ \sharpeval{\prog}{\st_{i}}{\stmt_{1}}{\results_{i}}}
    {\sharpeval{\prog}{\st}{\SSeq{\stmt_{0}}{\stmt_{1}}}{\results_{1} \mathbin{+\!\!+} \cdots \mathbin{+\!\!+} \results_{n} \mathbin{+\!\!+} \nonunit{\results_{0}}}}
\and
\inferrule*[lab={\rulename{E-Assign}}]
    {\sharpassignevals{\st}{\place}{\rv}{\st'}}
    {\sharpeval{\prog}{\st}{\SAssign{\place}{\rv}}{[(\RUnit, \st')]}}
\and
\inferrule*[lab={\rulename{E-Reorg}}]
    {\sharpenvreorg{\st_{0}}{\st_{1}} \\ \sharpeval{\prog}{\st_{1}}{\stmt}{\results}}
    {\sharpeval{\prog}{\st_{0}}{\stmt}{\results}}
\and
\inferrule*[lab={\rulename{E-Call}}]
    {\sharpcallevals{\prog}{\st}{\place}{\fn}{\op}{\st'}}
    {\sharpeval{\prog}{\st}{\SCall{\place}{\fn}{\op}}{[(\RUnit, \st')]}}
\\
\inferrule*[lab={\rulename{E-Match-Symbolic}}]
    {\sharpreadat{\st}{\place}{\Imm}{\fpath}{\VSymbolicV{\TSum{\ty_{0}}{\ty_{1}}}} \\ \sharpstateupd{\fpath}{\st}{\VLeftTy{\symof{\ty_{0}}}{\ty_{1}}}{\st_{0}} \\ \sharpstateupd{\fpath}{\st}{\VRightTy{\ty_{0}}{\symof{\ty_{1}}}}{\st_{1}} \\\\ \sharpeval{\prog}{\st_{0}}{\stmt_{0}}{\results_{0}} \\ \sharpeval{\prog}{\st_{1}}{\stmt_{1}}{\results_{1}}}
    {\sharpeval{\prog}{\st}{\SMatch{\place}{\stmt_{0}}{\stmt_{1}}}{\results_{0} \mathbin{+\!\!+} \results_{1}}}
\and
\inferrule*[lab={\rulename{E-Match-Symbolic-Shared}}]
    {\sharpreadat{\st}{\place}{\Imm}{\fpath}{\VShrLoan{\loc}{\VSymbolicV{\TSum{\ty_{0}}{\ty_{1}}}}} \\ \sharpstateupd{\pathcomp{\dpunloan}{\fpath}}{\st}{\VLeftTy{\symof{\ty_{0}}}{\ty_{1}}}{\st_{0}} \\ \sharpstateupd{\pathcomp{\dpunloan}{\fpath}}{\st}{\VRightTy{\ty_{0}}{\symof{\ty_{1}}}}{\st_{1}} \\\\ \sharpeval{\prog}{\st_{0}}{\stmt_{0}}{\results_{0}} \\ \sharpeval{\prog}{\st_{1}}{\stmt_{1}}{\results_{1}}}
    {\sharpeval{\prog}{\st}{\SMatch{\place}{\stmt_{0}}{\stmt_{1}}}{\results_{0} \mathbin{+\!\!+} \results_{1}}}
\and
\inferrule*[lab={\rulename{E-Match-Left}}]
    {\sharpreads{\st}{\place}{\Imm}{w} \\ w = \VLeft{\val} \vee w = \VShrLoan{\loc}{\VLeft{\val}} \\ \sharpeval{\prog}{\st}{\stmt_{0}}{\results}}
    {\sharpeval{\prog}{\st}{\SMatch{\place}{\stmt_{0}}{\stmt_{1}}}{\results}}
\and
\inferrule*[lab={\rulename{E-Match-Right}}]
    {\sharpreads{\st}{\place}{\Imm}{w} \\ w = \VRight{\val} \vee w = \VShrLoan{\loc}{\VRight{\val}} \\ \sharpeval{\prog}{\st}{\stmt_{1}}{\results}}
    {\sharpeval{\prog}{\st}{\SMatch{\place}{\stmt_{0}}{\stmt_{1}}}{\results}}

\end{rulefig}

\subsubsection{Reorganisation}\label{app:llbc:symbolic:reorg}

\begin{rulefig}{fig:llbc:symbolic:end-elem}{Ending an element of a region abstraction, $\elemend{\val}{\val'}$.}
\inferrule*[lab={\rulename{End-Elem-Shared}}]
    {}
    {\elemend{\VShrBorTy{\loc}{\ty}}{\VShrBorTy{\loc}{\ty}}}
\and
\inferrule*[lab={\rulename{End-Elem-Mut}}]
    {\nolocs{\val}}
    {\elemend{\VMutBor{\loc}{\val}}{\VMutBor{\loc}{\symof{\tyof{\val}}}}}
\end{rulefig}

\begin{definition}[Ending a region abstraction]\label{def:llbc:symbolic:end-abstraction}
An abstraction ends element by element (\Cref{fig:llbc:symbolic:end-elem}).
\begin{mathpar}
\inferrule*[lab={\rulename{End-Abs}}]
    {\elemend{\val_{1}}{\val_{1}'} \\ \cdots \\ \elemend{\val_{n}}{\val_{n}'}}
    {\abstractionend{\absliteral{\val_{1},\ \dots,\ \val_{n}}}{[\val_{1}',\ \dots,\ \val_{n}']}}
\end{mathpar}
\end{definition}

\begin{rulefig}{fig:llbc:symbolic:give-back}{Giving the value $\val$ back in place of the loan $w$ at $\fpath$, $\sharpgiveback{\fpath}{w}{\st}{\val}{\st'}$.}
\inferrule*[lab={\rulename{Give-Back-In-Place}}]
    {\notinsideregionabs{\fpath} \\ \sharpfpreplace{\fpath}{\st}{w}{\val}{\st'}}
    {\sharpgiveback{\fpath}{w}{\st}{\val}{\st'}}
\and
\inferrule*[lab={\rulename{Give-Back-Abs}}]
    {\lindex{\vect{\abs}}{\nat_{0}} = \abs_{\mathit{hd}} \mathbin{+\!\!+} [w] \mathbin{+\!\!+} \abs_{\mathit{tl}} \\ \nat_{1} = \length{\abs_{\mathit{hd}}} \\ \nolocs{\val}}
    {\sharpgiveback{\InAbs{\nat_{0}}{\nat_{1}}}{w}{(\mathit{frame},\ \mathit{anons},\ \vect{\abs})}{\val}{(\mathit{frame},\ \mathit{anons},\ \lset{\vect{\abs}}{\nat_{0}}{\abs_{\mathit{hd}} \mathbin{+\!\!+} \abs_{\mathit{tl}}})}}
\end{rulefig}
\noindent A loan inside a value is replaced by the value given back; a loan that is a whole element of an abstraction takes back only a location-free value, and the element is dropped.

\begin{rulefig}{fig:llbc:symbolic:env-reorg}{State reorganisation, $\sharpenvreorg{\st}{\st'}$.}
\inferrule*[lab={\rulename{Reorg-End-SharedLoan}}]
    {\notcontainsshrbor{\loc}{\st} \\ \sharpgiveback{\fpath}{\VShrLoan{\loc}{\val}}{\st}{\val}{\st'}}
    {\sharpenvreorg{\st}{\st'}}
\and
\inferrule*[lab={\rulename{Reorg-End-SharedBorrow}}]
    {\dpunloan, \dpunbor, \abs \notin \fpath \\ \containsshrloan{\loc}{\st} \\ \sharpfpreplace{\fpath}{\st}{\VShrBorTy{\loc}{\ty}}{\VBotTy{\TShrRef{\ty}}}{\st'}}
    {\sharpenvreorg{\st}{\st'}}
\and
\inferrule*[lab={\rulename{Reorg-End-MutBorrow}}]
    {\dpunloan, \dpunbor, \abs \notin \fpath \\ \sharpfpreplace{\fpath}{\st}{\VMutBor{\loc}{\val}}{\VBotTy{\TMutRef{\tyof{\val}}}}{\st_{1}} \\\\ \noloans{\val} \\ \sharpvalty{\val}{\ty} \\ \sharpgiveback{\fpath_{\mathsf{own}}}{\VMutLoanTy{\loc}{\ty}}{\st_{1}}{\val}{\st_{2}}}
    {\sharpenvreorg{\st}{\st_{2}}}
\and
\inferrule*[lab={\rulename{Reorg-End-Abstraction}}]
    {\abstractionend{\abs}{\vect{\val}}}
    {\sharpenvreorg{(\mathit{frame},\ \mathit{anons},\ \vect{\abs}_{\mathit{hd}} \mathbin{+\!\!+} [\abs] \mathbin{+\!\!+} \vect{\abs}_{\mathit{tl}})}{(\mathit{frame},\ \vect{\val} \mathbin{+\!\!+} \mathit{anons},\ \vect{\abs}_{\mathit{hd}} \mathbin{+\!\!+} \vect{\abs}_{\mathit{tl}})}}
\end{rulefig}

\subsubsection{Function calls}\label{app:llbc:symbolic:calls}

\begin{rulefig}{fig:llbc:symbolic:proj-input}{Input projection, $\projinput{\val}{\abs}$.}
\inferrule*[lab={\rulename{Proj-Input-Ignore}}]
    {\reffree{\tyof{\val}}}
    {\projinput{\val}{[]}}
\and
\inferrule*[lab={\rulename{Proj-Input-Pair}}]
    {\notreffree{\tyof{\VPair{\val_{0}}{\val_{1}}}} \\ \projinput{\val_{0}}{\vect{\val}_{0}} \\ \projinput{\val_{1}}{\vect{\val}_{1}}}
    {\projinput{\VPair{\val_{0}}{\val_{1}}}{\vect{\val}_{0} \mathbin{+\!\!+} \vect{\val}_{1}}}
\and
\inferrule*[lab={\rulename{Proj-Input-MutBorrow}}]
    {\reffree{\tyof{\val}}}
    {\projinput{(\VMutBor{\loc}{\val})}{[\VMutBor{\loc}{\VIgnored{\tyof{\val}}}]}}
\and
\inferrule*[lab={\rulename{Proj-Input-SharedBorrow}}]
    {\reffree{\ty}}
    {\projinput{(\VShrBorTy{\loc}{\ty})}{[\VShrBorTy{\loc}{\ty}]}}
\end{rulefig}

\begin{rulefig}{fig:llbc:symbolic:proj-output}{Output projection, $\projoutput{\ty}{\val}{\abs}$.}
\inferrule*[lab={\rulename{Proj-Output-Symbolic}}]
    {\reffree{\ty}}
    {\projoutput{\ty}{(\symof{\ty})}{[]}}
\and
\inferrule*[lab={\rulename{Proj-Output-Pair}}]
    {\notreffree{\TPair{\ty_{0}}{\ty_{1}}} \\ \projoutput{\ty_{0}}{\val_{0}}{\abs_{0}} \\ \projoutput{\ty_{1}}{\val_{1}}{\abs_{1}} \\ \valabsdisjoint{\val_{0}}{\abs_{0}}{\val_{1}}{\abs_{1}}}
    {\projoutput{\TPair{\ty_{0}}{\ty_{1}}}{\VPair{\val_{0}}{\val_{1}}}{\abs_{0} \mathbin{+\!\!+} \abs_{1}}}
\and
\inferrule*[lab={\rulename{Proj-Output-Shared}}]
    {\reffree{\ty}}
    {\projoutput{(\TShrRef{\ty})}{(\VShrBorTy{\loc}{\ty})}{[\VShrLoan{\loc}{\symof{\ty}}]}}
\and
\inferrule*[lab={\rulename{Proj-Output-Mut}}]
    {\reffree{\ty}}
    {\projoutput{(\TMutRef{\ty})}{(\VMutBor{\loc}{\symof{\ty}})}{[\VMutLoanTy{\loc}{\ty}]}}
\end{rulefig}

\begin{definition}[Disjoint labels]\label{def:llbc:symbolic:val-abs-disjoint}
\[\valabsdisjoint{\val_{0}}{\abs_{0}}{\val_{1}}{\abs_{1}} \defeq \forall \loc.\ (\valcontainsloc{\loc}{\val_{0}} \vee \abscontainsloc{\loc}{\abs_{0}}) \Rightarrow \neg (\valcontainsloc{\loc}{\val_{1}} \vee \abscontainsloc{\loc}{\abs_{1}})\]
\end{definition}

\begin{definition}[Instantiating a signature]\label{def:llbc:symbolic:inst-sig}
\begin{mathpar}
\inferrule*[lab={\rulename{Inst-Sig}}]
    {\projinput{\val}{\abs_{\mathit{in}}} \\ \projoutput{\ty_{\mathit{out}}}{\val_{\mathit{out}}}{\abs_{\mathit{outs}}} \\\\ \forall \loc.\ (\valcontainsloc{\loc}{\val_{\mathit{out}}} \vee \abscontainsloc{\loc}{\abs_{\mathit{outs}}}) \Rightarrow \lfresh{\loc}{\st}}
    {\instsig{\st}{\val}{\ty_{\mathit{out}}}{\abs_{\mathit{in}} \mathbin{+\!\!+} \abs_{\mathit{outs}}}{\val_{\mathit{out}}}}
\end{mathpar}
\end{definition}

\begin{definition}[Calling a function]\label{def:llbc:symbolic:call-evals}
\begin{mathpar}
\inferrule*[lab={\rulename{Call}}]
    {\progfn{\prog}{\fn} = \fndecl{\fn}{\ty}{\ty_{\mathit{loc}}}{\ty_{\mathit{ret}}}{\stmt} \\ \sharpopevals{\st}{\op}{\val}{\st'} \\ \sharpvalty{\val}{\ty} \\\\ \instsig{\st}{\val}{\ty_{\mathit{ret}}}{\abs}{\val_{\mathit{out}}} \\ \sharpassignevals{\pushabs{\st'}{\abs}}{\place}{\val_{\mathit{out}}}{\st''}}
    {\sharpcallevals{\prog}{\st}{\place}{\fn}{\op}{\st''}}
\end{mathpar}
\end{definition}

\subsubsection{Refinement}\label{app:llbc:symbolic:refinement}

\begin{rulefig}{fig:llbc:symbolic:to-abs}{Moving the loans and borrows of a value into region abstractions, $\toabs{\val}{\vect{\abs}}$.}
\inferrule*[lab={\rulename{To-Abs-Empty}}]
    {\nolocs{\val}}
    {\toabs{\val}{[]}}
\and
\inferrule*[lab={\rulename{To-Abs-Left}}]
    {\toabs{\val}{\vect{\abs}}}
    {\toabs{\VLeft{\val}}{\vect{\abs}}}
\and
\inferrule*[lab={\rulename{To-Abs-Right}}]
    {\toabs{\val}{\vect{\abs}}}
    {\toabs{\VRight{\val}}{\vect{\abs}}}
\and
\inferrule*[lab={\rulename{To-Abs-Pair}}]
    {\toabs{\val_{0}}{\vect{\abs}_{0}} \\ \toabs{\val_{1}}{\vect{\abs}_{1}}}
    {\toabs{\VPair{\val_{0}}{\val_{1}}}{\vect{\abs}_{0} \mathbin{+\!\!+} \vect{\abs}_{1}}}
\and
\inferrule*[lab={\rulename{To-Abs-MutLoan}}]
    {}
    {\toabs{\VMutLoanTy{\loc}{\ty}}{[[\VMutLoanTy{\loc}{\ty}]]}}
\and
\inferrule*[lab={\rulename{To-Abs-SharedLoan}}]
    {\nolocs{\val}}
    {\toabs{\VShrLoan{\loc}{\val}}{[[\VShrLoan{\loc}{\val}]]}}
\and
\inferrule*[lab={\rulename{To-Abs-SharedBorrow}}]
    {}
    {\toabs{\VShrBorTy{\loc}{\ty}}{[[\VShrBorTy{\loc}{\ty}]]}}
\and
\inferrule*[lab={\rulename{To-Abs-MutBorrow}}]
    {\bot, \mathsf{borrow} \notin \val \\ \toabs{\val}{[\abs_{1},\ \dots,\ \abs_{n}]} \\\\ \abs = [\VMutBor{\loc}{\VIgnored{\tyof{\val}}}] \mathbin{+\!\!+} \abs_{1} \mathbin{+\!\!+} \cdots \mathbin{+\!\!+} \abs_{n}}
    {\toabs{\VMutBor{\loc}{\val}}{[\abs]}}
\end{rulefig}

\begin{rulefig}{fig:llbc:symbolic:merge-abs}{Merging two region abstractions, $\sharpmergeabs{\abs_{0}}{\abs_{1}}{\abs}$.}
\inferrule*[lab={\rulename{Merge-Abs-Union}}]
    {}
    {\sharpmergeabs{\abs_{0}}{\abs_{1}}{\abs_{0} \mathbin{+\!\!+} \abs_{1}}}
\and
\inferrule*[lab={\rulename{Merge-Abs-Mut}}]
    {\sharpmergeabs{\abs_{0}}{\abs_{1}}{\abs}}
    {\sharpmergeabs{(\VMutLoanTy{\loc}{\ty}) :: \abs_{0}}{(\VMutBor{\loc}{\VIgnored{\ty}}) :: \abs_{1}}{\abs}}
\and
\inferrule*[lab={\rulename{Merge-Abs-Shared}}]
    {\VShrLoan{\loc}{\val} \in \abs_{0} \\ \sharpmergeabs{\abs_{0}}{\abs_{1}}{\abs}}
    {\sharpmergeabs{\abs_{0}}{(\VShrBor{\loc}) :: \abs_{1}}{\abs}}
\end{rulefig}
\noindent A mutable loan and its borrow cancel out; a shared borrow whose loan is in the first abstraction is dropped.

\begin{rulefig}{fig:llbc:symbolic:le1}{One refinement step, $\lesharp{\st}{\st'}$: moving values and abstractions.}
\inferrule*[lab={\rulename{Le-To-Abs}}]
    {\toabs{\val}{\vect{\abs}}}
    {\lesharp{(\mathit{frame},\ \val :: \mathit{anons},\ \vect{\abs}')}{(\mathit{frame},\ \mathit{anons},\ \vect{\abs} \mathbin{+\!\!+} \vect{\abs}')}}
\and
\inferrule*[lab={\rulename{Le-To-Symbolic}}]
    {\nolocs{\val} \\ \bot, \_ \notin \val \\ \tyof{\val} \neq \TPair{\_}{\_} \\\\ \sharpfpreplace{\fpath}{\st}{\val}{\VSymbolicV{\tyof{\val}}}{\st'}}
    {\lesharp{\st}{\st'}}
\and
\inferrule*[lab={\rulename{Le-Move-Value}}]
    {\dpunloan, \dpunbor, \abs \notin \fpath \\ \nohasouterloan{\val} \\ \tyof{\val} \neq \TPair{\_}{\_} \\\\ \sharpfpreplace{\fpath}{\st}{\val}{\VBotTy{\tyof{\val}}}{\st'}}
    {\lesharp{\st}{\pushanon{\val}{\st'}}}
\and
\inferrule*[lab={\rulename{Le-Merge-Abs}}]
    {\sharpmergeabs{\abs_{0}}{\abs_{1}}{\abs}}
    {\lesharp{(\mathit{frame},\ \mathit{anons},\ \abs_{0} :: \abs_{1} :: \vect{\abs})}{(\mathit{frame},\ \mathit{anons},\ \abs :: \vect{\abs})}}
\and
\inferrule*[lab={\rulename{Le-Perm-Abs}}]
    {\Permutation{\vect{\abs}}{\vect{\abs}'}}
    {\lesharp{(\mathit{frame},\ \mathit{anons},\ \vect{\abs})}{(\mathit{frame},\ \mathit{anons},\ \vect{\abs}')}}
\and
\inferrule*[lab={\rulename{Le-Perm-In-Abs}}]
    {\Permutation{\abs}{\abs'}}
    {\lesharp{(\mathit{frame},\ \mathit{anons},\ \abs :: \vect{\abs})}{(\mathit{frame},\ \mathit{anons},\ \abs' :: \vect{\abs})}}
\and
\inferrule*[lab={\rulename{Le-Perm-Anons}}]
    {\Permutation{\mathit{anons}}{\mathit{anons}'}}
    {\lesharp{(\mathit{frame},\ \mathit{anons},\ \vect{\abs})}{(\mathit{frame},\ \mathit{anons}',\ \vect{\abs})}}
\end{rulefig}

\begin{rulefig}{fig:llbc:symbolic:le1-labels}{One refinement step, $\lesharp{\st}{\st'}$: labels and loans.}
\inferrule*[lab={\rulename{Le-Reborrow-MutBorrow}}]
    {\lfresh{\loc_{1}}{\st} \\ \sharpfpreplace{\fpath}{\st}{\VMutBor{\loc_{0}}{\val}}{\VMutBor{\loc_{1}}{\val}}{\st'}}
    {\lesharp{\st}{\pushanon{\VMutBor{\loc_{0}}{\VMutLoanTy{\loc_{1}}{\tyof{\val}}}}{\st'}}}
\and
\inferrule*[lab={\rulename{Le-End-SharedLoan-Abs}}]
    {\fpinsideregionabs{\fpath} \\ \sharpgiveback{\fpath}{\VShrLoan{\loc}{\val}}{\st}{\val}{\st'} \\ \lfresh{\loc}{\st'}}
    {\lesharp{\st}{\st'}}
\and
\inferrule*[lab={\rulename{Le-End-SharedLoan-Inner}}]
    {\notinsideregionabs{\fpath} \\ \unloanin{\fpath} \\ \sharpfpreplace{\fpath}{\st}{\VShrLoan{\loc}{\val}}{\val}{\st'} \\ \lfresh{\loc}{\st'}}
    {\lesharp{\st}{\st'}}
\and
\inferrule*[lab={\rulename{Le-Reborrow-SharedBorrow}}]
    {\lfresh{\loc_{1}}{\st_{0}} \\ \bot, \mathsf{loan}^{m}, \mathsf{borrow} \notin \val \\ \statecontains{\VShrLoan{\loc_{0}}{\val}}{\st_{0}} \\\\ \sharpfpreplace{\fpath_{i}}{\st_{i}}{\VShrBor{\loc_{0}}}{\VShrBor{\loc_{1}}}{\st_{i+1}}}
    {\lesharp{\st_{0}}{\pushabs{\st_{\nat}}{\absliteral{\VShrBor{\loc_{0}},\ \VShrLoan{\loc_{1}}{\stripouterloans{\val}}}}}}
\and
\inferrule*[lab={\rulename{Le-Abs-Drop-Covered-Borrow}}]
    {\VShrBor{\loc_{0}} \in \abs \\ \statecontains{\VShrLoan{\loc_{0}}{\val}}{(\mathit{frame},\ \mathit{anons},\ \abs :: \vect{\abs})} \\ \shrloaninside{\loc}{\val}}
    {\lesharp{(\mathit{frame},\ \mathit{anons},\ ((\VShrBor{\loc}) :: \abs) :: \vect{\abs})}{(\mathit{frame},\ \mathit{anons},\ \abs :: \vect{\abs})}}
\end{rulefig}

\noindent Refinement $\lesharpstar{\st}{\st'}$ is the reflexive transitive closure of $\lesharp{\cdot}{\cdot}$.

\subsubsection{Borrow checking}\label{app:llbc:symbolic:borrow-check}

A function is checked once, against its signature. Its body runs from an entry state built from its signature, and every state it returns in must refine an exit state built from the return type.

\begin{rulefig}{fig:llbc:symbolic:init-extern}{The borrows a caller lends through an argument of type $\ty$, $\initextern{\ty}{\abs}$.}
\inferrule*[lab={\rulename{Init-Extern-Symbolic}}]
    {\reffree{\ty}}
    {\initextern{\ty}{[]}}
\and
\inferrule*[lab={\rulename{Init-Extern-Pair}}]
    {\notreffree{\TPair{\ty_{0}}{\ty_{1}}} \\ \initextern{\ty_{0}}{\abs_{0}} \\ \initextern{\ty_{1}}{\abs_{1}} \\ \absdisjoint{\abs_{0}}{\abs_{1}}}
    {\initextern{\TPair{\ty_{0}}{\ty_{1}}}{\abs_{0} \mathbin{+\!\!+} \abs_{1}}}
\and
\inferrule*[lab={\rulename{Init-Extern-Mut}}]
    {\reffree{\ty}}
    {\initextern{(\TMutRef{\ty})}{[\VMutBor{\loc}{\VIgnored{\ty}}]}}
\and
\inferrule*[lab={\rulename{Init-Extern-Shared}}]
    {\reffree{\ty}}
    {\initextern{(\TShrRef{\ty})}{[\VShrBorTy{\loc}{\ty}]}}
\end{rulefig}

\begin{definition}[Disjoint abstractions]\label{def:llbc:symbolic:abs-disjoint}
\[\absdisjoint{\abs_{0}}{\abs_{1}} \defeq \forall \loc.\ \abscontainsloc{\loc}{\abs_{0}} \Rightarrow \loc \notin \abs_{1}\]
\end{definition}

\begin{definition}[Instantiating an argument or a result]\label{def:llbc:symbolic:inst-arg}
\begin{mathpar}
\inferrule*[lab={\rulename{Inst-Arg}}]
    {\projoutput{\ty}{\val}{\abs_{\mathit{out}}} \\ \forall \loc.\ \abscontainsloc{\loc}{\abs_{\mathit{ext}}} \Rightarrow \loc \notin \val, \abs_{\mathit{out}}}
    {\metafn{inst{\_}arg}\ \ty\ \abs_{\mathit{ext}}\ \val\ (\abs_{\mathit{ext}} \mathbin{+\!\!+} \abs_{\mathit{out}})}
\end{mathpar}
\end{definition}

\begin{definition}[Entry and exit states]\label{def:llbc:symbolic:fn-init-end-states}
\begin{mathpar}
\inferrule*[lab={\rulename{Fn-Init-End-States}}]
    {\initextern{\ty_{\mathit{in}}}{\mathit{ext}} \\ \metafn{inst{\_}arg}\ \ty_{\mathit{in}}\ \mathit{ext}\ \val_{\mathit{in}}\ \abs_{\mathit{in}} \\ \metafn{inst{\_}arg}\ \ty_{\mathit{ret}}\ \mathit{ext}\ \val_{\mathit{out}}\ \abs_{\mathit{out}} \\\\ \st_{\mathit{start}} = (\frameval{\botof{\ty_{\mathit{loc}}}}{\val_{\mathit{in}}}{\botof{\ty_{\mathit{ret}}}},\ [],\ [\abs_{\mathit{in}}]) \\\\ \st_{\mathit{end}} = (\frameval{\botof{\ty_{\mathit{loc}}}}{\botof{\ty_{\mathit{in}}}}{\val_{\mathit{out}}},\ [],\ [\abs_{\mathit{out}}])}
    {\fninitendstates{\fndecl{\fn}{\ty_{\mathit{in}}}{\ty_{\mathit{loc}}}{\ty_{\mathit{ret}}}{\stmt}}{\st_{\mathit{start}}}{\st_{\mathit{end}}}}
\end{mathpar}
\end{definition}

\begin{definition}[Borrow checking a function]\label{def:llbc:symbolic:borrow-checks}
\begin{mathpar}
\inferrule*[lab={\rulename{Borrow-Checks}}]
    {\fninitendstates{\mfndecl}{\st_{\mathit{start}}}{\st_{\mathit{end}}} \\ \sharpeval{\prog}{\st_{\mathit{start}}}{\fnbody{\mfndecl}}{\results} \\\\ \forall (\res, \st) \in \results.\ \res = \RPanic \vee (\res = \RRet \wedge \lesharpstar{\st}{\st_{\mathit{end}}})}
    {\borrowchecks{\prog}{\mfndecl}}
\end{mathpar}
\end{definition}
\noindent A body that panics passes; one that returns must end in a state that refines the exit state.

\subsection{Combined Semantics (\LLBCPlus)}
\label{app:llbc:combined}

We use \LLBCPlus in the soundness proof. Its judgements carry $\vdash^{+}$.

\subsubsection{Values and states}\label{app:llbc:combined:values}

Values and region abstractions are those of \LLBCsharp
(\Cref{fig:llbc:symbolic:values}). Frames are those of \LLBC
(\Cref{fig:llbc:concrete:values}).

\begin{deffig}{fig:llbc:combined:state}{\LLBCPlus states.}
\[\begin{grammar}
  \gprod{\st}{(\vect{\tyname{frame}},\ \vect{\val_{\mathit{anon}}},\ \vect{\abs})}{A stack of frames, anonymous values and region abstractions}
\end{grammar}\]
\end{deffig}

\subsubsection{Auxiliary semantic data types}\label{app:llbc:combined:aux-data}

Full paths and statement results are as in \LLBC (\Cref{fig:llbc:state-addr,fig:llbc:modes-results}), over the addresses below.

\begin{deffig}{fig:llbc:combined:state-addr}{Addresses of values in an \LLBCPlus state.}
\[\begin{grammar}
  \gprod{\valaddr}{\InFrames{\nat}{\var}}{Variable $x$ in $n$-th frame} \\
  \galt{\Anon{\nat}}{$n$-th anonymous variable} \\
  \galt{\InAbs{\nat_{0}}{\nat_{1}}}{$n_1$-th value of the $n_0$-th region abstraction}
\end{grammar}\]
\end{deffig}

\noindent Paths into a region abstraction, $\fpinsideregionabs{\fpath}$, are as in \LLBCsharp (\Cref{def:llbc:symbolic:fp-inside-region-abs}).

\subsubsection{Operations on values}\label{app:llbc:combined:value-ops}

The operations on values are those of \LLBCsharp (\Cref{app:llbc:symbolic:value-ops}), with $\vdash^{+}$ for $\vdash_\#$.

\begin{definition}[Relabelling a state]\label{def:llbc:combined:relabel-state}
$\relabel{\st}{\loc_{0}}{\loc_{1}}$ swaps the labels $\loc_{0}$ and $\loc_{1}$ in every value of $\st$: its frames, anonymous values and region abstractions (\Cref{def:llbc:symbolic:relabel-val}).
\end{definition}

\subsubsection{Operations on states}\label{app:llbc:combined:state-ops}

\begin{definition}[Adding an anonymous value or a region abstraction]\label{def:llbc:combined:push}
\[\pushanon{\val}{(\mathit{frames},\ \mathit{anons},\ \vect{\abs})} \defeq (\mathit{frames},\ \val :: \mathit{anons},\ \vect{\abs})\]
\[\pushabs{(\mathit{frames},\ \mathit{anons},\ \vect{\abs})}{\abs} \defeq (\mathit{frames},\ \mathit{anons},\ \abs :: \vect{\abs})\]
\end{definition}

\begin{definition}[Reading the value at an address]\label{def:llbc:combined:state-get}
\begin{mathpar}
\inferrule*[lab={\rulename{State-Get-Frames}}]
    {\lindex{\mathit{frames}}{\nat} = \mathit{frame}}
    {\plusvaapplies{\InFrames{\nat}{\var}}{(\mathit{frames},\ \mathit{anons},\ \vect{\abs})}{\frameget{\var}{\mathit{frame}}}}
\and
\inferrule*[lab={\rulename{State-Get-Anon}}]
    {\lindex{\mathit{anons}}{\nat} = \val}
    {\plusvaapplies{\Anon{\nat}}{(\mathit{frames},\ \mathit{anons},\ \vect{\abs})}{\val}}
\and
\inferrule*[lab={\rulename{State-Get-Abs}}]
    {\lindex{\vect{\abs}}{\nat_{0}} = \abs \\ \lindex{\abs}{\nat_{1}} = \val}
    {\plusvaapplies{\InAbs{\nat_{0}}{\nat_{1}}}{(\mathit{frames},\ \mathit{anons},\ \vect{\abs})}{\val}}
\end{mathpar}
\end{definition}

\begin{rulefig}{fig:llbc:combined:state-set}{Writing a value at an address of an \LLBCPlus state, $\plusstateupd{\valaddr}{\st}{\val}{\st'}$.}
\inferrule*[lab={\rulename{State-Set-Frames}}]
    {\lindex{\mathit{frames}}{\nat} = \mathit{frame}}
    {\plusstateupd{\InFrames{\nat}{\var}}{(\mathit{frames},\ \mathit{anons},\ \vect{\abs})}{\val}{(\lset{\mathit{frames}}{\nat}{\frameset{\var}{\val}{\mathit{frame}}},\ \mathit{anons},\ \vect{\abs})}}
\and
\inferrule*[lab={\rulename{State-Set-Anon}}]
    {}
    {\plusstateupd{\Anon{\nat}}{(\mathit{frames},\ \mathit{anons},\ \vect{\abs})}{\val}{(\mathit{frames},\ \lset{\mathit{anons}}{\nat}{\val},\ \vect{\abs})}}
\and
\inferrule*[lab={\rulename{State-Set-Abs}}]
    {\lindex{\vect{\abs}}{\nat_{0}} = \abs}
    {\plusstateupd{\InAbs{\nat_{0}}{\nat_{1}}}{(\mathit{frames},\ \mathit{anons},\ \vect{\abs})}{\val}{(\mathit{frames},\ \mathit{anons},\ \lset{\vect{\abs}}{\nat_{0}}{\lset{\abs}{\nat_{1}}{\val}})}}
\end{rulefig}

\noindent The following are the same in \LLBC and \LLBCsharp, and so in \LLBCPlus, with $\vdash^{+}$ for their turnstiles:
\begin{itemize}
  \item reading and writing along a full path, and replacing the value there (\Cref{def:llbc:fp-applies,def:llbc:state-updates});
  \item following a projection (\Cref{fig:llbc:place-address-aux});
  \item reading a place, $\plusreadat{\st}{\place}{\mode}{\fpath}{\val}$ and $\plusreads{\st}{\place}{\mode}{\val}$ (\Cref{def:llbc:read-at,def:llbc:read});
  \item reading and writing a place, $\plusreadwrite{\st}{\place}{\mode}{\val}{w}{\st'}$ (\Cref{def:llbc:read-write}).
\end{itemize}
\noindent The address of a place is as in \LLBC (\Cref{def:llbc:place-address}).

\subsubsection{Evaluation}\label{app:llbc:combined:eval}

\noindent The following are the same in \LLBC and \LLBCsharp, and so in \LLBCPlus, with $\vdash^{+}$ for their turnstiles:
\begin{itemize}
  \item evaluating an operand (\Cref{fig:llbc:op-evals});
  \item assignment (\Cref{fig:llbc:assign}).
\end{itemize}
\noindent Rvalue evaluation, including shared borrowing, follows \LLBCsharp
(\Cref{fig:llbc:symbolic:rv-evals,fig:llbc:symbolic:shr-borrow-at}).
Leaving a called function, $\exitfn{\place}{r}{r'}$, follows \LLBC
(\Cref{fig:llbc:exit-fn}).

\begin{definition}[Popping a frame]\label{def:llbc:combined:pop-frame}
\begin{mathpar}
\inferrule*[lab={\rulename{Pop-Frame}}]
    {\nohasouterloan{\val_{l}} \\ \nohasouterloan{\val_{i}} \\ \nobotsloans{\val} \\\\ \st' = (\mathit{frames},\ \val_{i} :: \val_{l} :: \mathit{anons},\ \vect{\abs})}
    {\popframe{(\frameval{\val_{l}}{\val_{i}}{\val} :: \mathit{frames},\ \mathit{anons},\ \vect{\abs})}{\val}{\st'}}
\end{mathpar}
\end{definition}

\begin{definition}[Unit and other results]\label{def:llbc:combined:unit-results}
\[\unitstates{\results} \defeq [\,\st \mid \StmtFin{\RUnit}{\st} \in \results\,] \qquad \nonunit{\results} \defeq [\,r \in \results \mid r \neq \StmtFin{\RUnit}{\_}\,]\]
\end{definition}

\begin{rulefig}{fig:llbc:combined:stmt-evals}{Evaluating a statement with fuel $\nat$ to a list of results, $\pluseval{\prog}{\st}{\stmt}{\nat}{\results}$.}
\inferrule*[lab={\rulename{E-Step-Zero}}]
    {}
    {\pluseval{\prog}{\st}{\stmt}{0}{[\NoFuel]}}
\and
\inferrule*[lab={\rulename{E-Return}}]
    {}
    {\pluseval{\prog}{\st}{\SRet}{\nat + 1}{[\StmtFin{\RRet}{\st}]}}
\and
\inferrule*[lab={\rulename{E-Panic}}]
    {}
    {\pluseval{\prog}{\st}{\SPanic}{\nat + 1}{[\StmtFin{\RPanic}{\st}]}}
\and
\inferrule*[lab={\rulename{E-Nil}}]
    {}
    {\pluseval{\prog}{\st}{\SNil}{\nat + 1}{[\StmtFin{\RUnit}{\st}]}}
\and
\inferrule*[lab={\rulename{E-Seq}}]
    {\pluseval{\prog}{\st}{\stmt_{0}}{\nat + 1}{\results_{0}} \\ \unitstates{\results_{0}} = [\st_{1},\ \dots,\ \st_{k}] \\ \pluseval{\prog}{\st_{i}}{\stmt_{1}}{\nat + 1}{\results_{i}}}
    {\pluseval{\prog}{\st}{\SSeq{\stmt_{0}}{\stmt_{1}}}{\nat + 1}{\results_{1} \mathbin{+\!\!+} \cdots \mathbin{+\!\!+} \results_{k} \mathbin{+\!\!+} \nonunit{\results_{0}}}}
\and
\inferrule*[lab={\rulename{E-Assign}}]
    {\plusassignevals{\st}{\place}{\rv}{\st'}}
    {\pluseval{\prog}{\st}{\SAssign{\place}{\rv}}{\nat + 1}{[\StmtFin{\RUnit}{\st'}]}}
\and
\inferrule*[lab={\rulename{E-Reorg}}]
    {\plusenvreorg{\st_{0}}{\st_{1}} \\ \pluseval{\prog}{\st_{1}}{\stmt}{\nat + 1}{\results}}
    {\pluseval{\prog}{\st_{0}}{\stmt}{\nat + 1}{\results}}
\and
\inferrule*[lab={\rulename{E-Call}}]
    {\progfn{\prog}{\fn} = \fndecl{\fn}{\ty_{\mathit{in}}}{\ty_{\mathit{loc}}}{\ty_{\mathit{ret}}}{\stmt} \\ \plusopevals{\st}{\op}{\val}{(\mathit{frames},\ \mathit{anons},\ \vect{\abs})} \\ \plusvalty{\val}{\ty_{\mathit{in}}} \\\\ \pluseval{\prog}{(\frameval{\botof{\ty_{\mathit{loc}}}}{\val}{\botof{\ty_{\mathit{ret}}}} :: \mathit{frames},\ \mathit{anons},\ \vect{\abs})}{\stmt}{\nat}{[r_{1},\ \dots,\ r_{k}]} \\ \exitfn{\place}{r_{i}}{r_{i}'}}
    {\pluseval{\prog}{\st}{\SCall{\place}{\fn}{\op}}{\nat + 1}{[r_{1}',\ \dots,\ r_{k}']}}
\end{rulefig}

\begin{rulefig}{fig:llbc:combined:stmt-evals-match}{Evaluating a match with fuel $\nat$, $\pluseval{\prog}{\st}{\SMatch{\place}{\stmt_{0}}{\stmt_{1}}}{\nat}{\results}$.}
\inferrule*[lab={\rulename{E-Match-Symbolic}}]
    {\plusreadat{\st}{\place}{\Imm}{\fpath}{\VSymbolicV{\TSum{\ty_{0}}{\ty_{1}}}} \\ \plusstateupd{\fpath}{\st}{\VLeftTy{\symof{\ty_{0}}}{\ty_{1}}}{\st_{0}} \\ \plusstateupd{\fpath}{\st}{\VRightTy{\ty_{0}}{\symof{\ty_{1}}}}{\st_{1}} \\\\ \pluseval{\prog}{\st_{0}}{\stmt_{0}}{\nat + 1}{\results_{0}} \\ \pluseval{\prog}{\st_{1}}{\stmt_{1}}{\nat + 1}{\results_{1}}}
    {\pluseval{\prog}{\st}{\SMatch{\place}{\stmt_{0}}{\stmt_{1}}}{\nat + 1}{\results_{0} \mathbin{+\!\!+} \results_{1}}}
\and
\inferrule*[lab={\rulename{E-Match-Symbolic-Shared}}]
    {\plusreadat{\st}{\place}{\Imm}{\fpath}{\VShrLoan{\loc}{\VSymbolicV{\TSum{\ty_{0}}{\ty_{1}}}}} \\ \plusstateupd{\pathcomp{\dpunloan}{\fpath}}{\st}{\VLeftTy{\symof{\ty_{0}}}{\ty_{1}}}{\st_{0}} \\ \plusstateupd{\pathcomp{\dpunloan}{\fpath}}{\st}{\VRightTy{\ty_{0}}{\symof{\ty_{1}}}}{\st_{1}} \\\\ \pluseval{\prog}{\st_{0}}{\stmt_{0}}{\nat + 1}{\results_{0}} \\ \pluseval{\prog}{\st_{1}}{\stmt_{1}}{\nat + 1}{\results_{1}}}
    {\pluseval{\prog}{\st}{\SMatch{\place}{\stmt_{0}}{\stmt_{1}}}{\nat + 1}{\results_{0} \mathbin{+\!\!+} \results_{1}}}
\and
\inferrule*[lab={\rulename{E-Match-Left}}]
    {\plusreads{\st}{\place}{\Imm}{w} \\ w = \VLeft{\val} \vee w = \VShrLoan{\loc}{\VLeft{\val}} \\ \pluseval{\prog}{\st}{\stmt_{0}}{\nat + 1}{\results}}
    {\pluseval{\prog}{\st}{\SMatch{\place}{\stmt_{0}}{\stmt_{1}}}{\nat + 1}{\results}}
\and
\inferrule*[lab={\rulename{E-Match-Right}}]
    {\plusreads{\st}{\place}{\Imm}{w} \\ w = \VRight{\val} \vee w = \VShrLoan{\loc}{\VRight{\val}} \\ \pluseval{\prog}{\st}{\stmt_{1}}{\nat + 1}{\results}}
    {\pluseval{\prog}{\st}{\SMatch{\place}{\stmt_{0}}{\stmt_{1}}}{\nat + 1}{\results}}
\end{rulefig}

\subsubsection{Reorganisation and refinement}\label{app:llbc:combined:refinement}

\noindent The following are as in \LLBCsharp, with $\vdash^{+}$ for $\vdash_\#$:
\begin{itemize}
  \item ending abstraction elements and abstractions, giving back, and reorganisation, $\plusenvreorg{\st}{\st'}$ (\Cref{fig:llbc:symbolic:end-elem,def:llbc:symbolic:end-abstraction,fig:llbc:symbolic:give-back,fig:llbc:symbolic:env-reorg});
  \item moving values into abstractions and merging abstractions (\Cref{fig:llbc:symbolic:to-abs,fig:llbc:symbolic:merge-abs}).
\end{itemize}

\begin{rulefig}{fig:llbc:combined:le1}{One \LLBCPlus refinement step, $\leplus{\st}{\st'}$: moving values and abstractions.}
\inferrule*[lab={\rulename{Le-To-Abs}}]
    {\toabs{\val}{\vect{\abs}}}
    {\leplus{(\mathit{frames},\ \val :: \mathit{anons},\ \vect{\abs}')}{(\mathit{frames},\ \mathit{anons},\ \vect{\abs} \mathbin{+\!\!+} \vect{\abs}')}}
\and
\inferrule*[lab={\rulename{Le-To-Symbolic}}]
    {\nolocs{\val} \\ \bot, \_ \notin \val \\ \tyof{\val} \neq \TPair{\_}{\_} \\\\ \plusfpreplace{\fpath}{\st}{\val}{\VSymbolicV{\tyof{\val}}}{\st'}}
    {\leplus{\st}{\st'}}
\and
\inferrule*[lab={\rulename{Le-Move-Value}}]
    {\dpunloan, \dpunbor, \abs \notin \fpath \\ \nohasouterloan{\val} \\ \tyof{\val} \neq \TPair{\_}{\_} \\\\ \plusfpreplace{\fpath}{\st}{\val}{\VBotTy{\tyof{\val}}}{\st'}}
    {\leplus{\st}{\pushanon{\val}{\st'}}}
\and
\inferrule*[lab={\rulename{Le-Merge-Abs}}]
    {\plusmergeabs{\abs_{0}}{\abs_{1}}{\abs}}
    {\leplus{(\mathit{frames},\ \mathit{anons},\ \abs_{0} :: \abs_{1} :: \vect{\abs})}{(\mathit{frames},\ \mathit{anons},\ \abs :: \vect{\abs})}}
\and
\inferrule*[lab={\rulename{Le-Empty-Abs}}]
    {}
    {\leplus{(\mathit{frames},\ \mathit{anons},\ \vect{\abs})}{(\mathit{frames},\ \mathit{anons},\ [] :: \vect{\abs})}}
\and
\inferrule*[lab={\rulename{Le-Perm-Abs}}]
    {\Permutation{\vect{\abs}}{\vect{\abs}'}}
    {\leplus{(\mathit{frames},\ \mathit{anons},\ \vect{\abs})}{(\mathit{frames},\ \mathit{anons},\ \vect{\abs}')}}
\and
\inferrule*[lab={\rulename{Le-Perm-In-Abs}}]
    {\Permutation{\abs}{\abs'}}
    {\leplus{(\mathit{frames},\ \mathit{anons},\ \abs :: \vect{\abs})}{(\mathit{frames},\ \mathit{anons},\ \abs' :: \vect{\abs})}}
\and
\inferrule*[lab={\rulename{Le-Perm-Anons}}]
    {\Permutation{\mathit{anons}}{\mathit{anons}'}}
    {\leplus{(\mathit{frames},\ \mathit{anons},\ \vect{\abs})}{(\mathit{frames},\ \mathit{anons}',\ \vect{\abs})}}
\and
\inferrule*[lab={\rulename{Le-Anon-Value}}]
    {}
    {\leplus{\st}{\pushanon{\botof{\ty}}{\st}}}
\end{rulefig}

\begin{rulefig}{fig:llbc:combined:le1-labels}{One \LLBCPlus refinement step, $\leplus{\st}{\st'}$: labels and loans.}
\inferrule*[lab={\rulename{Le-Reborrow-MutBorrow}}]
    {\lfresh{\loc_{1}}{\st} \\ \plusfpreplace{\fpath}{\st}{\VMutBor{\loc_{0}}{\val}}{\VMutBor{\loc_{1}}{\val}}{\st'}}
    {\leplus{\st}{\pushanon{\VMutBor{\loc_{0}}{\VMutLoanTy{\loc_{1}}{\tyof{\val}}}}{\st'}}}
\and
\inferrule*[lab={\rulename{Le-Reborrow-SharedBorrow}}]
    {\lfresh{\loc_{1}}{\st_{0}} \\ \bot, \mathsf{loan}^{m}, \mathsf{borrow} \notin \val \\ \statecontains{\VShrLoan{\loc_{0}}{\val}}{\st_{0}} \\\\ \plusfpreplace{\fpath_{i}}{\st_{i}}{\VShrBor{\loc_{0}}}{\VShrBor{\loc_{1}}}{\st_{i+1}}}
    {\leplus{\st_{0}}{\pushabs{\st_{\nat}}{\absliteral{\VShrBor{\loc_{0}},\ \VShrLoan{\loc_{1}}{\stripouterloans{\val}}}}}}
\and
\inferrule*[lab={\rulename{Le-Fresh-SharedBorrow}}]
    {\statecontains{\VShrLoan{\loc}{\val}}{\st}}
    {\leplus{\st}{\pushanon{\VShrBorTy{\loc}{\tyof{\val}}}{\st}}}
\and
\inferrule*[lab={\rulename{Le-End-SharedLoan-Abs}}]
    {\fpinsideregionabs{\fpath} \\ \plusgiveback{\fpath}{\VShrLoan{\loc}{\val}}{\st}{\val}{\st'} \\ \lfresh{\loc}{\st'}}
    {\leplus{\st}{\st'}}
\and
\inferrule*[lab={\rulename{Le-End-SharedLoan-Inner}}]
    {\notinsideregionabs{\fpath} \\ \unloanin{\fpath} \\ \plusfpreplace{\fpath}{\st}{\VShrLoan{\loc}{\val}}{\val}{\st'} \\ \lfresh{\loc}{\st'}}
    {\leplus{\st}{\st'}}
\and
\inferrule*[lab={\rulename{Le-Abs-Drop-Covered-Borrow}}]
    {\VShrBor{\loc_{0}} \in \abs \\ \statecontains{\VShrLoan{\loc_{0}}{\val}}{(\mathit{frames},\ \mathit{anons},\ \abs :: \vect{\abs})} \\ \shrloaninside{\loc}{\val}}
    {\leplus{(\mathit{frames},\ \mathit{anons},\ ((\VShrBor{\loc}) :: \abs) :: \vect{\abs})}{(\mathit{frames},\ \mathit{anons},\ \abs :: \vect{\abs})}}
\and
\inferrule*[lab={\rulename{Le-Rename-Shared}}]
    {\containsshrloan{\loc_{0}}{\st} \\ \lfresh{\loc_{1}}{\st}}
    {\leplus{\st}{\relabel{\st}{\loc_{0}}{\loc_{1}}}}
\and
\inferrule*[lab={\rulename{Le-Rename-Mut}}]
    {\containsmutloan{\loc_{0}}{\st} \\ \containsmutbor{\loc_{0}}{\st} \\ \lfresh{\loc_{1}}{\st}}
    {\leplus{\st}{\relabel{\st}{\loc_{0}}{\loc_{1}}}}
\end{rulefig}

\noindent Refinement $\leplusstar{\st}{\st'}$ is the reflexive transitive closure of $\leplus{\cdot}{\cdot}$. Beyond the \LLBCsharp rules, it may add an empty abstraction, a bottom or a fresh shared borrow of an existing loan, and rename a label.

\section{\LLBCView Syntax and Semantics}
\label{app:views}

\subsection{\LLBCView Syntax}\label{app:views:syntax}

\LLBCView extends \LLBC with views in reference types and borrow expressions.

\begin{deffig}{fig:views:syntax}{\LLBCView syntax.}
\[\begin{grammar}
  \gprod{k}{\Mut}{Mutable} \\
  \galt{\Shr}{Shared}
\end{grammar}\]

\[\begin{grammar}
  \gprod{\vp}{\varepsilon}{The whole value} \\
  \galt{0}{Its first field} \\
  \galt{1}{Its second field}
\end{grammar}\]

\[\begin{grammar}
  \gprod{\view}{\singleview{\vp}{k}}{Field $\vp$ with mode $k$} \\
  \galt{\doubleview{k_{0}}{k_{1}}}{Both fields, each with its own mode}
\end{grammar}\]

\[\begin{grammar}
  \gprod{\ty}{\TUnit}{} \\
  \galt{\TPair{\ty_{0}}{\ty_{1}}}{} \\
  \galt{\TSum{\ty_{0}}{\ty_{1}}}{} \\
  \galt{\TRef{\view}{\ty}}{A reference to $\ty$ with view $\view$}
\end{grammar}\]

\[\begin{grammar}
  \gprod{\op}{\OpMove{\place}}{} \\
  \galt{\OpUnit}{} \\
  \galt{\OpPair{\op_{0}}{\op_{1}}}{} \\
  \galt{\OpLeft{\op}{\ty}}{} \\
  \galt{\OpRight{\ty}{\op}}{}
\end{grammar}\]

\[\begin{grammar}
  \gprod{\rv}{\op}{} \\
  \galt{\RvBorrow{\view}{\place}}{A borrow of $\place$ with view $\view$}
\end{grammar}\]

\[\begin{grammar}
  \gprod{\stmt}{\SNil}{} \\
  \galt{\SSeq{\stmt_{0}}{\stmt_{1}}}{} \\
  \galt{\SAssign{\place}{\rv}}{} \\
  \galt{\SMatch{\place}{\stmt_{0}}{\stmt_{1}}}{} \\
  \galt{\SRet}{} \\
  \galt{\SPanic}{} \\
  \galt{\SCall{\place}{\fn}{\op}}{}
\end{grammar}\]

\[\begin{grammar}
  \gprod{\mfndecl}{\fndecl{f}{\tau_{\mathit{in}}}{\tau_{\mathit{loc}}}{\tau_{\mathit{ret}}}{s}}{}
\end{grammar}\]

\[\begin{grammar}
  \gabbr{\prog}{\vect{\fn \times \mfndecl}}{}
\end{grammar}\]
\end{deffig}

\subsection{Concrete Semantics (\LLBCView)}
\label{app:views:concrete}

\subsubsection{Values and states}\label{app:views:concrete:values}

Frames and states are as in \LLBC (\Cref{fig:llbc:concrete:values}).

\begin{deffig}{fig:views:concrete:values}{\LLBCView values.}
\[\begin{grammar}
  \gprod{\val}{\VUnit}{} \\
  \galt{\VPair{\val_{0}}{\val_{1}}}{} \\
  \galt{\VLeftTy{\val}{\ty}}{} \\
  \galt{\VRightTy{\ty}{\val}}{} \\
  \galt{\VBotTy{\ty}}{} \\
  \galt{\VMutLoanTy{\loc}{\ty}}{} \\
  \galt{\VShrLoan{\loc}{\val}}{} \\
  \galt{\VBorV{\singleborrow{\vp}{\entry}} : \TRef{\singleview{\vp}{\entrymode{\entry}}}{\ty}}{Field $\vp$} \\
  \galt{\VBorV{\doubleborrow{\entry_{0}}{\entry_{1}}} : \TRef{\doubleview{\entrymode{\entry_{0}}}{\entrymode{\entry_{1}}}}{\ty}}{Both fields}
\end{grammar}\]

\[\begin{grammar}
  \gprod{\entry}{\BMut{\loc}{\val}}{A mutable entry, holding the borrowed value} \\
  \galt{\BShared{\loc}{\ty}}{A shared entry} \\
  \galt{\BDead{k}}{A dead entry, keeping only its mode}
\end{grammar}\]
\end{deffig}

\subsubsection{Auxiliary semantic data types}\label{app:views:concrete:aux-data}

Modes, results, addresses and full paths are as in \LLBC
(\Cref{fig:llbc:modes-results,fig:llbc:state-addr}). Direct paths also follow
\LLBC (\Cref{fig:llbc:direct-path}), except that $\dpunbor$ carries a
viewpoint identifying the reference entry to enter.

\begin{deffig}{fig:views:concrete:paths}{Direct paths and entry addresses.}
\[\begin{grammar}
  \gprod{\seg}{\dpunloan}{} \\
  \galt{\dpproj{\nat}}{} \\
  \galt{\dpunbor_{\vp}}{Into the value of the mutable entry for field $\vp$}
\end{grammar}\]

\[\begin{grammar}
  \gabbr{\feaddr}{\fplit{\valaddr}{\dealit{\dpath}{\vp}}}{Entry address: the entry for field $\vp$ of the reference at $\fplit{\valaddr}{\dpath}$}
\end{grammar}\]
\end{deffig}

\subsubsection{Operations on types}\label{app:views:concrete:type-ops}

\begin{rulefig}{fig:views:projected-type}{Following a projection over a type, $\projty{\ty}{\proj}{\ty'}$.}
\inferrule*[lab={\rulename{PT-Nil}}]
    {}
    {\projty{\ty}{\dpnil}{\ty}}
\and
\inferrule*[lab={\rulename{PT-Deref}}]
    {\projty{\ty}{\proj}{\ty'}}
    {\projty{\TRef{\view}{\ty}}{\pathcomp{\proj}{\PDeref}}{\ty'}}
\and
\inferrule*[lab={\rulename{PT-Pair}}]
    {\projty{\ty_{i}}{\proj}{\ty'}}
    {\projty{\TPair{\ty_{0}}{\ty_{1}}}{\pathcomp{\proj}{\dpproj{i}}}{\ty'}}
\and
\inferrule*[lab={\rulename{PT-Sum}}]
    {\projty{\ty_{i}}{\proj}{\ty'}}
    {\projty{\TSum{\ty_{0}}{\ty_{1}}}{\pathcomp{\proj}{\dpproj{i}}}{\ty'}}
\end{rulefig}

\begin{rulefig}{fig:views:view-wf}{Well-formed reference types, $\viewapplicable{\ty}{\view}$.}
\inferrule*[lab={\rulename{WF-Whole}}]
    {}
    {\viewapplicable{\ty}{\singleview{\varepsilon}{k}}}
\and
\inferrule*[lab={\rulename{WF-Field-0}}]
    {}
    {\viewapplicable{\TPair{\ty_{0}}{\ty_{1}}}{\singleview{0}{k}}}
\and
\inferrule*[lab={\rulename{WF-Field-1}}]
    {}
    {\viewapplicable{\TPair{\ty_{0}}{\ty_{1}}}{\singleview{1}{k}}}
\and
\inferrule*[lab={\rulename{WF-Both}}]
    {}
    {\viewapplicable{\TPair{\ty_{0}}{\ty_{1}}}{\doubleview{k_{0}}{k_{1}}}}
\end{rulefig}

\subsubsection{Operations on values}\label{app:views:concrete:value-ops}

Bottoms of a type (\Cref{def:llbc:botof}) and outer loans (\Cref{def:llbc:outer-loan}) are as in \LLBC.

\begin{definition}[Entries]\label{def:views:entries}
\[\entrycontent{\BMut{\_}{\val}} = \val \qquad \entrycontent{\entry} = \VUnit \quad \text{otherwise}\]
\[\begin{array}{@{}r@{\;}l@{}}
  \viewof{\singleborrow{\vp}{\entry}} &= \singleview{\vp}{\entrymode{\entry}} \\
  \viewof{\doubleborrow{\entry_{0}}{\entry_{1}}} &= \doubleview{\entrymode{\entry_{0}}}{\entrymode{\entry_{1}}}
\end{array}\]
\end{definition}
\noindent $\entries$ stands for the entries of a reference, $\singleborrow{\vp}{\entry}$ or $\doubleborrow{\entry_{0}}{\entry_{1}}$.

\begin{definition}[Type of a value]\label{def:views:tyof}
\[\begin{array}{@{}r@{\;}l@{\qquad}r@{\;}l@{}}
  \tyof{\VUnit} &= \TUnit & \tyof{\VBotTy{\ty}} &= \ty \\
  \tyof{\VPair{\val_{0}}{\val_{1}}} &= \TPair{\tyof{\val_{0}}}{\tyof{\val_{1}}} & \tyof{\VMutLoanTy{\_}{\ty}} &= \ty \\
  \tyof{\VLeftTy{\val_{0}}{\ty}} &= \TSum{\tyof{\val_{0}}}{\ty} & \tyof{\VShrLoan{\_}{\val_{0}}} &= \tyof{\val_{0}} \\
  \tyof{\VRightTy{\ty}{\val_{0}}} &= \TSum{\ty}{\tyof{\val_{0}}} & \tyof{\VBorrow{\ty}{\entries}} &= \TRef{\viewof{\entries}}{\ty}
\end{array}\]
\end{definition}

\begin{rulefig}{fig:views:vp-applies}{The entry for field $\vp$, $\vpapplies{\vp}{\entries}{\entry}$, and replacing it, $\borrowupd{\vp}{\entries}{\entry'}{\entries'}$.}
\inferrule*[lab={\rulename{Entry-Single}}]
    {}
    {\vpapplies{\vp}{\singleborrow{\vp}{\entry}}{\entry}}
\and
\inferrule*[lab={\rulename{Entry-Double-0}}]
    {}
    {\vpapplies{0}{\doubleborrow{\entry_{0}}{\entry_{1}}}{\entry_{0}}}
\and
\inferrule*[lab={\rulename{Entry-Double-1}}]
    {}
    {\vpapplies{1}{\doubleborrow{\entry_{0}}{\entry_{1}}}{\entry_{1}}}
\\
\inferrule*[lab={\rulename{Borrow-Upd-Single}}]
    {}
    {\borrowupd{\vp}{\singleborrow{\vp}{\entry}}{\entry'}{\singleborrow{\vp}{\entry'}}}
\and
\inferrule*[lab={\rulename{Borrow-Upd-Double-0}}]
    {}
    {\borrowupd{0}{\doubleborrow{\entry_{0}}{\entry_{1}}}{\entry_{0}'}{\doubleborrow{\entry_{0}'}{\entry_{1}}}}
\and
\inferrule*[lab={\rulename{Borrow-Upd-Double-1}}]
    {}
    {\borrowupd{1}{\doubleborrow{\entry_{0}}{\entry_{1}}}{\entry_{1}'}{\doubleborrow{\entry_{0}}{\entry_{1}'}}}
\end{rulefig}

\begin{rulefig}{fig:views:dp-applies}{Following a direct path in a value, $\dpapplies{\dpath}{\val}{\val'}$.}
\inferrule*[lab={\rulename{DP-Nil}}]
    {}
    {\dpapplies{\dpnil}{\val}{\val}}
\and
\inferrule*[lab={\rulename{DP-Pair-0}}]
    {\dpapplies{\dpath}{\val_{0}}{\val}}
    {\dpapplies{\pathcomp{\dpath}{\dpproj{0}}}{\VPair{\val_{0}}{\val_{1}}}{\val}}
\and
\inferrule*[lab={\rulename{DP-Pair-1}}]
    {\dpapplies{\dpath}{\val_{1}}{\val}}
    {\dpapplies{\pathcomp{\dpath}{\dpproj{1}}}{\VPair{\val_{0}}{\val_{1}}}{\val}}
\and
\inferrule*[lab={\rulename{DP-Left}}]
    {\dpapplies{\dpath}{\val_{0}}{\val}}
    {\dpapplies{\pathcomp{\dpath}{\dpproj{0}}}{\VLeft{\val_{0}}}{\val}}
\and
\inferrule*[lab={\rulename{DP-Right}}]
    {\dpapplies{\dpath}{\val_{0}}{\val}}
    {\dpapplies{\pathcomp{\dpath}{\dpproj{1}}}{\VRight{\val_{0}}}{\val}}
\and
\inferrule*[lab={\rulename{DP-Unloan}}]
    {\dpapplies{\dpath}{\val_{0}}{\val}}
    {\dpapplies{\pathcomp{\dpath}{\dpunloan}}{\VShrLoan{\loc}{\val_{0}}}{\val}}
\and
\inferrule*[lab={\rulename{DP-Unbor}}]
    {\vpapplies{\vp}{\entries}{\BMut{\loc}{\val_{0}}} \\ \dpapplies{\dpath}{\val_{0}}{\val}}
    {\dpapplies{\pathcomp{\dpath}{\dpunbor_{\vp}}}{\VBorrow{\ty}{\entries}}{\val}}
\end{rulefig}

\begin{rulefig}{fig:views:val-updates}{Writing along a direct path in a value, $\valupd{\dpath}{\val}{w}{\val'}$.}
\inferrule*[lab={\rulename{V-Base}}]
    {\tyof{w} = \tyof{\val}}
    {\valupd{\dpnil}{\val}{w}{w}}
\and
\inferrule*[lab={\rulename{V-Sum-Left}}]
    {\valupd{\dpath}{\val}{w}{\val'}}
    {\valupd{\pathcomp{\dpath}{\dpproj{0}}}{\VLeftTy{\val}{\ty}}{w}{\VLeftTy{\val'}{\ty}}}
\and
\inferrule*[lab={\rulename{V-Sum-Right}}]
    {\valupd{\dpath}{\val}{w}{\val'}}
    {\valupd{\pathcomp{\dpath}{\dpproj{1}}}{\VRightTy{\ty}{\val}}{w}{\VRightTy{\ty}{\val'}}}
\and
\inferrule*[lab={\rulename{V-Pair-Left}}]
    {\valupd{\dpath}{\val_{0}}{w}{\val_{0}'}}
    {\valupd{\pathcomp{\dpath}{\dpproj{0}}}{\VPair{\val_{0}}{\val_{1}}}{w}{\VPair{\val_{0}'}{\val_{1}}}}
\and
\inferrule*[lab={\rulename{V-Pair-Right}}]
    {\valupd{\dpath}{\val_{1}}{w}{\val_{1}'}}
    {\valupd{\pathcomp{\dpath}{\dpproj{1}}}{\VPair{\val_{0}}{\val_{1}}}{w}{\VPair{\val_{0}}{\val_{1}'}}}
\and
\inferrule*[lab={\rulename{V-SharedUnloan}}]
    {\valupd{\dpath}{\val}{w}{\val'}}
    {\valupd{\pathcomp{\dpath}{\dpunloan}}{\VShrLoan{\loc}{\val}}{w}{\VShrLoan{\loc}{\val'}}}
\and
\inferrule*[lab={\rulename{V-Unbor}}]
    {\vpapplies{\vp}{\entries}{\BMut{\loc}{\val}} \\ \valupd{\dpath}{\val}{w}{\val'} \\ \borrowupd{\vp}{\entries}{\BMut{\loc}{\val'}}{\entries'}}
    {\valupd{\pathcomp{\dpath}{\dpunbor_{\vp}}}{\VBorrow{\ty}{\entries}}{w}{\VBorrow{\ty}{\entries'}}}
\end{rulefig}

\subsubsection{Operations on states}\label{app:views:concrete:state-ops}

\noindent The following are as in \LLBC (\Cref{app:llbc:concrete:state-ops}):
\begin{itemize}
  \item reading and writing an address, and along a full path, and replacing the value there (\Cref{def:llbc:state-get,fig:llbc:state-set,def:llbc:fp-applies,def:llbc:state-updates});
  \item the address of a place (\Cref{def:llbc:place-address});
  \item reading a place, and reading and writing it (\Cref{def:llbc:read-at,def:llbc:read,def:llbc:read-write}).
\end{itemize}

\begin{rulefig}{fig:views:place-address-aux}{Following a projection: from the value $\val$ at the full path $\fpath$, the projection $\proj$ in mode $\mode$ leads to the full path $\fpath'$, $\placeaddraux{\st}{\fpath}{\val}{\proj}{\mode}{\fpath'}$.}
\inferrule*[lab={\rulename{PA-Base}}]
    {}
    {\placeaddraux{\st}{\fpath_{\mathit{cur}}}{\val}{\dpnil}{\mode}{\fpath_{\mathit{cur}}}}
\and
\inferrule*[lab={\rulename{PA-Proj-Sum-Left}}]
    {\placeaddraux{\st}{\pathcomp{\dpproj{0}}{\fpath_{\mathit{cur}}}}{\val}{\proj}{\mode}{\fpath_{\mathit{final}}}}
    {\placeaddraux{\st}{\fpath_{\mathit{cur}}}{\VLeft{\val}}{\pathcomp{\proj}{\dpproj{0}}}{\mode}{\fpath_{\mathit{final}}}}
\and
\inferrule*[lab={\rulename{PA-Proj-Sum-Right}}]
    {\placeaddraux{\st}{\pathcomp{\dpproj{1}}{\fpath_{\mathit{cur}}}}{\val}{\proj}{\mode}{\fpath_{\mathit{final}}}}
    {\placeaddraux{\st}{\fpath_{\mathit{cur}}}{\VRight{\val}}{\pathcomp{\proj}{\dpproj{1}}}{\mode}{\fpath_{\mathit{final}}}}
\and
\inferrule*[lab={\rulename{PA-Proj-Pair-Left}}]
    {\placeaddraux{\st}{\pathcomp{\dpproj{0}}{\fpath_{\mathit{cur}}}}{\val_{0}}{\proj}{\mode}{\fpath_{\mathit{final}}}}
    {\placeaddraux{\st}{\fpath_{\mathit{cur}}}{\VPair{\val_{0}}{\val_{1}}}{\pathcomp{\proj}{\dpproj{0}}}{\mode}{\fpath_{\mathit{final}}}}
\and
\inferrule*[lab={\rulename{PA-Proj-Pair-Right}}]
    {\placeaddraux{\st}{\pathcomp{\dpproj{1}}{\fpath_{\mathit{cur}}}}{\val_{1}}{\proj}{\mode}{\fpath_{\mathit{final}}}}
    {\placeaddraux{\st}{\fpath_{\mathit{cur}}}{\VPair{\val_{0}}{\val_{1}}}{\pathcomp{\proj}{\dpproj{1}}}{\mode}{\fpath_{\mathit{final}}}}
\and
\inferrule*[lab={\rulename{PA-SharedLoan}}]
    {\proj \neq \dpnil \\ \placeaddraux{\st}{\pathcomp{\dpunloan}{\fpath_{\mathit{cur}}}}{\val}{\proj}{\Imm}{\fpath_{\mathit{final}}}}
    {\placeaddraux{\st}{\fpath_{\mathit{cur}}}{\VShrLoan{\loc}{\val}}{\proj}{\Imm}{\fpath_{\mathit{final}}}}
\and
\inferrule*[lab={\rulename{PA-Deref-MutBorrow}}]
    {\vpapplies{\vp}{\entries}{\BMut{\loc}{\val}} \\ \mode \neq \MMove \\ \placeaddraux{\st}{\pathcomp{\dpunbor_{\vp}}{\fpath_{\mathit{cur}}}}{\val}{\proj}{\mode}{\fpath_{\mathit{final}}}}
    {\placeaddraux{\st}{\fpath_{\mathit{cur}}}{\VBorrow{\ty}{\entries}}{\pathcomp{\pathcomp{\proj}{\vp}}{\PDeref}}{\mode}{\fpath_{\mathit{final}}}}
\and
\inferrule*[lab={\rulename{PA-Deref-SharedBorrow}}]
    {\vpapplies{\vp}{\entries}{\BShared{\loc}{\ty'}} \\ \fpapplies{\fpath_{\mathit{loan}}}{\st}{\VShrLoan{\loc}{\val}} \\ \placeaddraux{\st}{\fpath_{\mathit{loan}}}{\VShrLoan{\loc}{\val}}{\proj}{\Imm}{\fpath_{\mathit{final}}}}
    {\placeaddraux{\st}{\fpath_{\mathit{cur}}}{\VBorrow{\ty}{\entries}}{\pathcomp{\pathcomp{\proj}{\vp}}{\PDeref}}{\Imm}{\fpath_{\mathit{final}}}}
\end{rulefig}

\begin{definition}[Type of a place]\label{def:views:place-type}
\begin{mathpar}
\inferrule*[lab={\rulename{Place-Type}}]
    {\vaapplies{\InFrames{0}{\var}}{\st}{\val} \\ \projty{\tyof{\val}}{\proj}{\ty}}
    {\placety{\st}{\placelit{\proj}{\var}}{\ty}}
\end{mathpar}
\end{definition}

\subsubsection{Evaluation}\label{app:views:concrete:eval}

\noindent The following are as in \LLBC:
\begin{itemize}
  \item evaluating an operand (\Cref{fig:llbc:op-evals});
  \item assignment (\Cref{fig:llbc:assign});
  \item popping a frame and leaving a called function (\Cref{def:llbc:pop-frame,fig:llbc:exit-fn});
  \item evaluating a statement (\Cref{fig:llbc:stmt-evals}).
\end{itemize}

\begin{rulefig}{fig:views:borrow-at}{Borrowing the place $\place$ with mode $k$, giving the entry $\entry$, $\borrowat{\st}{\place}{k}{\entry}{\st'}$.}
\inferrule*[lab={\rulename{Mut-Borrow-At}}]
    {\readwrite{\st}{\place}{\Mut}{\val}{\VMutLoanTy{\loc}{\tyof{\val}}}{\st'} \\ \nobotsloans{\val} \\ \lfresh{\loc}{\st}}
    {\borrowat{\st}{\place}{\Mut}{\BMut{\loc}{\val}}{\st'}}
\and
\inferrule*[lab={\rulename{Shared-Borrow-At-Reuse}}]
    {\reads{\st}{\place}{\Imm}{\VShrLoan{\loc}{\val}} \\ \nobotsmutloans{\val}}
    {\borrowat{\st}{\place}{\Shr}{\BShared{\loc}{\tyof{\val}}}{\st}}
\and
\inferrule*[lab={\rulename{Shared-Borrow-At-Wrap}}]
    {\readwrite{\st}{\place}{\Imm}{\val}{\VShrLoan{\loc}{\val}}{\st'} \\ \val \neq \VShrLoan{\_}{\_} \\ \nobotsmutloans{\val} \\ \lfresh{\loc}{\st}}
    {\borrowat{\st}{\place}{\Shr}{\BShared{\loc}{\tyof{\val}}}{\st'}}
\end{rulefig}

\begin{rulefig}{fig:views:borrow-evals}{Borrowing the fields of $\place$ a view names, $\borrowevals{\st}{\place}{\view}{\entries}{\st'}$.}
\inferrule*[lab={\rulename{Borrow-Single}}]
    {\borrowat{\st}{(\pathcomp{\vp}{\place})}{k}{\entry}{\st'}}
    {\borrowevals{\st}{\place}{\singleview{\vp}{k}}{\singleborrow{\vp}{\entry}}{\st'}}
\and
\inferrule*[lab={\rulename{Borrow-Double}}]
    {\borrowat{\st_{0}}{(\pathcomp{\dpproj{0}}{\place})}{k_{0}}{\entry_{0}}{\st_{1}} \\\\ \borrowat{\pushanon{\entrycontent{\entry_{0}}}{\st_{1}}}{(\pathcomp{\dpproj{1}}{\place})}{k_{1}}{\entry_{1}}{\pushanon{\entrycontent{\entry_{0}}}{\st_{2}}}}
    {\borrowevals{\st_{0}}{\place}{\doubleview{k_{0}}{k_{1}}}{\doubleborrow{\entry_{0}}{\entry_{1}}}{\st_{2}}}
\end{rulefig}

\begin{rulefig}{fig:views:rv-evals}{Evaluating an rvalue, $\rvevals{\st}{\rv}{\val}{\st'}$.}
\inferrule*[lab={\rulename{E-Op}}]
    {\opevals{\st}{\op}{\val}{\st'}}
    {\rvevals{\st}{\op}{\val}{\st'}}
\and
\inferrule*[lab={\rulename{E-Borrow}}]
    {\placety{\st}{\place}{\ty} \\ \viewapplicable{\ty}{\view} \\ \borrowevals{\st}{\place}{\view}{\entries}{\st'}}
    {\rvevals{\st}{\RvBorrow{\view}{\place}}{\VBorrow{\ty}{\entries}}{\st'}}
\end{rulefig}
\noindent A view applies only to a type that has the fields it borrows, so no view accidentally enters a sum.

\begin{definition}[Replacing an entry]\label{def:views:entry-access}
\begin{mathpar}
\inferrule*[lab={\rulename{Entry-Replace}}]
    {\fpreplace{\valaddr}{\st}{\val}{\val'}{\st'} \\ \fpreplace{\dpath}{\val}{\VBorrow{\ty}{\entries}}{\VBorrow{\ty}{\entries'}}{\val'} \\ \fpreplace{\vp}{\entries}{\entry}{\entry'}{\entries'}}
    {\fpreplace{\fplit{\valaddr}{\dealit{\dpath}{\vp}}}{\st}{\entry}{\entry'}{\st'}}
\end{mathpar}
\end{definition}
\noindent The entry for field $\vp$ of the reference at $\fplit{\valaddr}{\dpath}$ is $\entry$, and becomes $\entry'$: the reference's entries $\entries$ become $\entries'$, its value $\val$ becomes $\val'$, and the state $\st$ becomes $\st'$. That reference is at $\feapath{\feaddr} = \fplit{\valaddr}{\dpath}$ for $\feaddr = \fplit{\valaddr}{\dealit{\dpath}{\vp}}$.

\begin{rulefig}{fig:views:env-reorg}{State reorganisation, $\envreorg{\st}{\st'}$.}
\inferrule*[lab={\rulename{Reorg-End-SharedLoan}}]
    {\notcontainsshrbor{\loc}{\st} \\ \fpreplace{\fpath}{\st}{\VShrLoan{\loc}{\val}}{\val}{\st'}}
    {\envreorg{\st}{\st'}}
\and
\inferrule*[lab={\rulename{Reorg-End-SharedEntry}}]
    {\notinsideborrowed{\feapath{\feaddr}} \\ \fpreplace{\feaddr}{\st}{\BShared{\loc}{\ty}}{\BDead{\Shr}}{\st'}}
    {\envreorg{\st}{\st'}}
\and
\inferrule*[lab={\rulename{Reorg-End-MutEntry}}]
    {\notinsideborrowed{\feapath{\feaddr}} \\ \fpreplace{\feaddr}{\st}{\BMut{\loc}{\val}}{\BDead{\Mut}}{\st_{1}} \\ \noloans{\val} \\\\ \fpreplace{\fpath}{\st_{1}}{\VMutLoan{\loc}}{\val}{\st_{2}}}
    {\envreorg{\st}{\st_{2}}}
\and
\inferrule*[lab={\rulename{Reorg-Drop-Dead}}]
    {\alldead{\entries} \\ \fpreplace{\fpath}{\st}{\VBorrow{\ty}{\entries}}{\botof{\TRef{\viewof{\entries}}{\ty}}}{\st'}}
    {\envreorg{\st}{\st'}}
\end{rulefig}
\noindent A reference ends one entry at a time; once every entry is dead, the reference becomes a bottom.

\subsection{Symbolic Semantics (\LLBCSharpView)}
\label{app:views:symbolic}

\LLBCSharpView extends \LLBCsharp with partial borrows, as \LLBCView extends \LLBC.

\subsubsection{Values and states}\label{app:views:symbolic:values}

Entries are as in \LLBCView (\Cref{fig:views:concrete:values}).
Region abstractions, frames and states are as in \LLBCsharp
(\Cref{fig:llbc:symbolic:values}).

\begin{deffig}{fig:views:symbolic:values}{\LLBCSharpView values.}
\[\begin{grammar}
  \gprod{\val}{\VUnit}{} \\
  \galt{\VSymbolicV{\ty}}{} \\
  \galt{\VPair{\val_{0}}{\val_{1}}}{} \\
  \galt{\VLeftTy{\val}{\ty}}{} \\
  \galt{\VRightTy{\ty}{\val}}{} \\
  \galt{\VBotTy{\ty}}{} \\
  \galt{\VMutLoanTy{\loc}{\ty}}{} \\
  \galt{\VShrLoan{\loc}{\val}}{} \\
  \galt{\VBorV{\singleborrow{\vp}{\entry}} : \TRef{\singleview{\vp}{\entrymode{\entry}}}{\ty}}{Field $\vp$} \\
  \galt{\VBorV{\doubleborrow{\entry_{0}}{\entry_{1}}} : \TRef{\doubleview{\entrymode{\entry_{0}}}{\entrymode{\entry_{1}}}}{\ty}}{Both fields} \\
  \galt{\VIgnored{\ty}}{}
\end{grammar}\]
\end{deffig}

\noindent A region abstraction holds loans and \emph{full-view references}, references whose single entry covers the whole referent, $\fullref{k}{\entry}{\ty}$. We write them as the borrows of \LLBCsharp:
\[\begin{array}{@{}r@{\;}l@{}}
  \FullMutBor{\loc}{\val} &\defeq \fullref{\Mut}{\BMut{\loc}{\val}}{\tyof{\val}} \\
  \FullShrBor{\loc}{\ty} &\defeq \fullref{\Shr}{\BShared{\loc}{\ty}}{\ty}
\end{array}\]

\subsubsection{Auxiliary semantic data types}\label{app:views:symbolic:aux-data}

Addresses, full paths and paths into a region abstraction are as in \LLBCsharp (\Cref{fig:llbc:symbolic:state-addr,def:llbc:symbolic:fp-inside-region-abs}); direct paths and entry addresses as in \LLBCView (\Cref{fig:views:concrete:paths}).

\subsubsection{Operations on types and values}\label{app:views:symbolic:value-ops}

The following are as in \LLBCView:
\begin{itemize}
  \item following a projection over a type, and well-formed reference types (\Cref{fig:views:projected-type,fig:views:view-wf});
  \item entries (\Cref{def:views:entries});
  \item the entry for a field, and replacing it (\Cref{fig:views:vp-applies});
  \item reading and writing along a direct path (\Cref{fig:views:dp-applies,fig:views:val-updates}).
\end{itemize}
Bottoms of a type and outer loans are as in \LLBC
(\Cref{def:llbc:botof,def:llbc:outer-loan}). The value typing judgement,
symbolic values of a type and stripping outer shared loans are as in
\LLBCsharp
(\Cref{def:llbc:symbolic:val-ty,def:llbc:symbolic:symof,def:llbc:symbolic:strip-outer-loans}).

\begin{definition}[Type of a value]\label{def:views:symbolic:tyof}
\[\begin{array}{@{}r@{\;}l@{\qquad}r@{\;}l@{}}
  \tyof{\VUnit} &= \TUnit & \tyof{\VBotTy{\ty}} &= \ty \\
  \tyof{\VPair{\val_{0}}{\val_{1}}} &= \TPair{\tyof{\val_{0}}}{\tyof{\val_{1}}} & \tyof{\VMutLoanTy{\_}{\ty}} &= \ty \\
  \tyof{\VLeftTy{\val_{0}}{\ty}} &= \TSum{\tyof{\val_{0}}}{\ty} & \tyof{\VShrLoan{\_}{\val_{0}}} &= \tyof{\val_{0}} \\
  \tyof{\VRightTy{\ty}{\val_{0}}} &= \TSum{\ty}{\tyof{\val_{0}}} & \tyof{\VBorrow{\ty}{\entries}} &= \TRef{\viewof{\entries}}{\ty} \\
  \tyof{\VSymbolicV{\ty}} &= \ty & \tyof{\VIgnored{\ty}} &= \ty
\end{array}\]
\end{definition}

\subsubsection{Operations on states}\label{app:views:symbolic:state-ops}

The following are as in \LLBCsharp (\Cref{app:llbc:symbolic:state-ops}):
\begin{itemize}
  \item adding an anonymous value or a region abstraction (\Cref{def:llbc:symbolic:push-anon,def:llbc:symbolic:push-abs});
  \item reading and writing an address, and along a full path (\Cref{def:llbc:symbolic:state-get,fig:llbc:symbolic:state-set});
  \item the address of a place, reading a place, and reading and writing it.
\end{itemize}
The following are as in \LLBCView, with $\vdash_\#$ for $\vdash$:
\begin{itemize}
  \item following a projection, $\sharpplaceaddraux{\st}{\fpath}{\val}{\proj}{\mode}{\fpath'}$ (\Cref{fig:views:place-address-aux});
  \item the type of a place (\Cref{def:views:place-type}), which reads the variable $\var$ of the single frame instead of $\InFrames{0}{\var}$ of the top frame;
  \item replacing an entry, $\sharpfpreplace{\feaddr}{\st}{\entry}{\entry'}{\st'}$ (\Cref{def:views:entry-access}).
\end{itemize}

\subsubsection{Evaluation}\label{app:views:symbolic:eval}

The following are as in \LLBCsharp:
\begin{itemize}
  \item evaluating an operand and assigning to a place (\Cref{fig:llbc:op-evals,fig:llbc:assign}, with $\vdash_\#$ for $\vdash$);
  \item evaluating a statement and calling a function (\Cref{fig:llbc:symbolic:stmt-evals,def:llbc:symbolic:call-evals}), over the function calls of \Cref{app:views:symbolic:calls}.
\end{itemize}
Rvalue evaluation and borrowing the fields selected by a view follow
\LLBCView (\Cref{fig:views:rv-evals,fig:views:borrow-evals}), with
$\vdash_\#$ for $\vdash$. They use the rules below to borrow an individual place.

\begin{rulefig}{fig:views:symbolic:borrow-at}{Borrowing the place $\place$ with mode $k$, giving the entry $\entry$, $\sharpborrowat{\st}{\place}{k}{\entry}{\st'}$.}
\inferrule*[lab={\rulename{Mut-Borrow-At}}]
    {\sharpreadwrite{\st}{\place}{\Mut}{\val}{\VMutLoanTy{\loc}{\tyof{\val}}}{\st'} \\ \nobotsloans{\val} \\ \lfresh{\loc}{\st}}
    {\sharpborrowat{\st}{\place}{\Mut}{\BMut{\loc}{\val}}{\st'}}
\and
\inferrule*[lab={\rulename{Shared-Borrow-At-Reuse}}]
    {\sharpreads{\st}{\place}{\Imm}{\VShrLoan{\loc}{\val}} \\ \nobotsmutloans{\val}}
    {\sharpborrowat{\st}{\place}{\Shr}{\BShared{\loc}{\tyof{\val}}}{\st}}
\and
\inferrule*[lab={\rulename{Shared-Borrow-At}}]
    {\sharpreadat{\st}{\place}{\Imm}{\fpath}{\val} \\ \val \neq \VShrLoan{\_}{\_} \\ \nobotsmutloans{\val} \\ \lfresh{\loc}{\st} \\ \sharpshrborrowat{\fpath}{\val}{\st}{\loc}{\st'}}
    {\sharpborrowat{\st}{\place}{\Shr}{\BShared{\loc}{\tyof{\val}}}{\st'}}
\end{rulefig}

\begin{rulefig}{fig:views:symbolic:shr-borrow-at}{Placing a new shared loan for a shared borrow of the value $\val$ at $\fpath$, $\sharpshrborrowat{\fpath}{\val}{\st}{\loc}{\st'}$.}
\inferrule*[lab={\rulename{Shared-Borrow-Wrap}}]
    {\notinsideregionabs{\fpath} \\ \sharpstateupd{\fpath}{\st}{\VShrLoan{\loc}{\val}}{\st'}}
    {\sharpshrborrowat{\fpath}{\val}{\st}{\loc}{\st'}}
\and
\inferrule*[lab={\rulename{Shared-Borrow-Abs}}]
    {\sharpvaapplies{\InAbs{r}{i}}{\st}{\VShrLoan{\loc_{k}}{\val_{k}}} \\\\ \abs = [\FullShrBor{\loc_{k}}{\tyof{\val_{k}}},\ \VShrLoan{\loc}{\symof{\tyof{\val}}}]}
    {\sharpshrborrowat{\fplit{\InAbs{r}{i}}{\pathcomp{\dpath}{\dpunloan}}}{\val}{\st}{\loc}{\pushabs{\st}{\abs}}}
\end{rulefig}

\subsubsection{Reorganisation}\label{app:views:symbolic:reorg}

Ending an element of a region abstraction, ending a region abstraction and giving a value back are as in \LLBCsharp (\Cref{fig:llbc:symbolic:end-elem,def:llbc:symbolic:end-abstraction,fig:llbc:symbolic:give-back}): a region abstraction holds only full-view references, written as the borrows of \LLBCsharp, so these rules stay the same.

\begin{rulefig}{fig:views:symbolic:env-reorg}{State reorganisation, $\sharpenvreorg{\st}{\st'}$.}
\inferrule*[lab={\rulename{Reorg-End-SharedLoan}}]
    {\notcontainsshrbor{\loc}{\st} \\ \sharpgiveback{\fpath}{\VShrLoan{\loc}{\val}}{\st}{\val}{\st'}}
    {\sharpenvreorg{\st}{\st'}}
\and
\inferrule*[lab={\rulename{Reorg-End-SharedEntry}}]
    {\dpunloan, \dpunbor, \abs \notin \feapath{\feaddr} \\ \containsshrloan{\loc}{\st} \\ \sharpfpreplace{\feaddr}{\st}{\BShared{\loc}{\ty}}{\BDead{\Shr}}{\st'}}
    {\sharpenvreorg{\st}{\st'}}
\and
\inferrule*[lab={\rulename{Reorg-End-MutEntry}}]
    {\dpunloan, \dpunbor, \abs \notin \feapath{\feaddr} \\ \sharpfpreplace{\feaddr}{\st}{\BMut{\loc}{\val}}{\BDead{\Mut}}{\st_{1}} \\\\ \noloans{\val} \\ \sharpgiveback{\fpath}{\VMutLoanTy{\loc}{\ty}}{\st_{1}}{\val}{\st_{2}}}
    {\sharpenvreorg{\st}{\st_{2}}}
\and
\inferrule*[lab={\rulename{Reorg-Drop-Dead}}]
    {\alldead{\entries} \\ \sharpfpreplace{\fpath}{\st}{\VBorrow{\ty}{\entries}}{\botof{\TRef{\viewof{\entries}}{\ty}}}{\st'}}
    {\sharpenvreorg{\st}{\st'}}
\and
\inferrule*[lab={\rulename{Reorg-End-Abstraction}}]
    {\abstractionend{\abs}{\vect{\val}}}
    {\sharpenvreorg{(\mathit{frame},\ \mathit{anons},\ \vect{\abs}_{\mathit{hd}} \mathbin{+\!\!+} [\abs] \mathbin{+\!\!+} \vect{\abs}_{\mathit{tl}})}{(\mathit{frame},\ \vect{\val} \mathbin{+\!\!+} \mathit{anons},\ \vect{\abs}_{\mathit{hd}} \mathbin{+\!\!+} \vect{\abs}_{\mathit{tl}})}}
\end{rulefig}

\subsubsection{Function calls}\label{app:views:symbolic:calls}

Disjointness of labels and signature instantiation are as in \LLBCsharp
(\Cref{def:llbc:symbolic:val-abs-disjoint,def:llbc:symbolic:inst-sig}),
using the projections below. Each entry of a reference argument enters the
abstraction as a separate full-view reference. We omit a reference's type
where no premise uses it.

\begin{rulefig}{fig:views:symbolic:proj-input}{Input projection, $\projinput{\val}{\abs}$, and of one entry, $\projinputentry{\entry}{\val}$.}
\inferrule*[lab={\rulename{Proj-Input-Ignore}}]
    {\reffree{\tyof{\val}}}
    {\projinput{\val}{[]}}
\and
\inferrule*[lab={\rulename{Proj-Input-Pair}}]
    {\notreffree{\tyof{\VPair{\val_{0}}{\val_{1}}}} \\ \projinput{\val_{0}}{\abs_{0}} \\ \projinput{\val_{1}}{\abs_{1}}}
    {\projinput{\VPair{\val_{0}}{\val_{1}}}{\abs_{0} \mathbin{+\!\!+} \abs_{1}}}
\\
\inferrule*[lab={\rulename{Proj-Input-Single}}]
    {\projinputentry{\entry}{\val}}
    {\projinput{(\VBorV{\singleborrow{\vp}{\entry}})}{[\val]}}
\and
\inferrule*[lab={\rulename{Proj-Input-Double}}]
    {\projinputentry{\entry_{0}}{\val_{0}} \\ \projinputentry{\entry_{1}}{\val_{1}}}
    {\projinput{(\VBorV{\doubleborrow{\entry_{0}}{\entry_{1}}})}{[\val_{0},\ \val_{1}]}}
\\
\inferrule*[lab={\rulename{Proj-Input-Mut}}]
    {\reffree{\tyof{\val}}}
    {\projinputentry{\BMut{\loc}{\val}}{\FullMutBor{\loc}{\VIgnored{\tyof{\val}}}}}
\and
\inferrule*[lab={\rulename{Proj-Input-Shared}}]
    {\reffree{\ty}}
    {\projinputentry{\BShared{\loc}{\ty}}{\FullShrBor{\loc}{\ty}}}
\end{rulefig}

\begin{rulefig}{fig:views:symbolic:proj-output}{Output projection, $\projoutput{\ty}{\val}{\abs}$, and of one entry, $\projoutputentry{k}{\ty}{\entry}{\val}$.}
\inferrule*[lab={\rulename{Proj-Output-Symbolic}}]
    {\reffree{\ty}}
    {\projoutput{\ty}{(\symof{\ty})}{[]}}
\and
\inferrule*[lab={\rulename{Proj-Output-Pair}}]
    {\notreffree{\TPair{\ty_{0}}{\ty_{1}}} \\ \projoutput{\ty_{0}}{\val_{0}}{\abs_{0}} \\ \projoutput{\ty_{1}}{\val_{1}}{\abs_{1}} \\ \valabsdisjoint{\val_{0}}{\abs_{0}}{\val_{1}}{\abs_{1}}}
    {\projoutput{\TPair{\ty_{0}}{\ty_{1}}}{\VPair{\val_{0}}{\val_{1}}}{\abs_{0} \mathbin{+\!\!+} \abs_{1}}}
\\
\inferrule*[lab={\rulename{Proj-Output-Whole}}]
    {\projoutputentry{k}{\ty}{\entry}{\val}}
    {\projoutput{(\TRef{\singleview{\varepsilon}{k}}{\ty})}{(\VBorV{\singleborrow{\varepsilon}{\entry}})}{[\val]}}
\and
\inferrule*[lab={\rulename{Proj-Output-Field}}]
    {\projoutputentry{k}{\ty_{i}}{\entry}{\val}}
    {\projoutput{(\TRef{\singleview{i}{k}}{\TPair{\ty_{0}}{\ty_{1}}})}{(\VBorV{\singleborrow{i}{\entry}})}{[\val]}}
\and
\inferrule*[lab={\rulename{Proj-Output-Both}}]
    {\projoutputentry{k_{0}}{\ty_{0}}{\entry_{0}}{\val_{0}} \\ \projoutputentry{k_{1}}{\ty_{1}}{\entry_{1}}{\val_{1}} \\ \absdisjoint{[\val_{0}]}{[\val_{1}]}}
    {\projoutput{(\TRef{\doubleview{k_{0}}{k_{1}}}{\TPair{\ty_{0}}{\ty_{1}}})}{(\VBorV{\doubleborrow{\entry_{0}}{\entry_{1}}})}{[\val_{0},\ \val_{1}]}}
\\
\inferrule*[lab={\rulename{Proj-Output-Mut}}]
    {\reffree{\ty}}
    {\projoutputentry{\Mut}{\ty}{\BMut{\loc}{\symof{\ty}}}{\VMutLoanTy{\loc}{\ty}}}
\and
\inferrule*[lab={\rulename{Proj-Output-Shared}}]
    {\reffree{\ty}}
    {\projoutputentry{\Shr}{\ty}{\BShared{\loc}{\ty}}{\VShrLoan{\loc}{\symof{\ty}}}}
\end{rulefig}

\subsubsection{Refinement}\label{app:views:symbolic:refinement}

\begin{rulefig}{fig:views:symbolic:to-abs}{Moving the loans and references of a value into region abstractions, $\toabs{\val}{\vect{\abs}}$, and of one entry, $\entrytoabsarrow{\entry}{\vect{\abs}}$.}
\inferrule*[lab={\rulename{To-Abs-Empty}}]
    {\nolocs{\val}}
    {\toabs{\val}{[]}}
\and
\inferrule*[lab={\rulename{To-Abs-Left}}]
    {\toabs{\val}{\vect{\abs}}}
    {\toabs{\VLeft{\val}}{\vect{\abs}}}
\and
\inferrule*[lab={\rulename{To-Abs-Right}}]
    {\toabs{\val}{\vect{\abs}}}
    {\toabs{\VRight{\val}}{\vect{\abs}}}
\and
\inferrule*[lab={\rulename{To-Abs-Pair}}]
    {\toabs{\val_{0}}{\vect{\abs}_{0}} \\ \toabs{\val_{1}}{\vect{\abs}_{1}}}
    {\toabs{\VPair{\val_{0}}{\val_{1}}}{\vect{\abs}_{0} \mathbin{+\!\!+} \vect{\abs}_{1}}}
\and
\inferrule*[lab={\rulename{To-Abs-MutLoan}}]
    {}
    {\toabs{\VMutLoanTy{\loc}{\ty}}{[[\VMutLoanTy{\loc}{\ty}]]}}
\and
\inferrule*[lab={\rulename{To-Abs-SharedLoan}}]
    {\nolocs{\val}}
    {\toabs{\VShrLoan{\loc}{\val}}{[[\VShrLoan{\loc}{\val}]]}}
\\
\inferrule*[lab={\rulename{To-Abs-Single}}]
    {\entrytoabsarrow{\entry}{\vect{\abs}}}
    {\toabs{\VBorV{\singleborrow{\vp}{\entry}}}{\vect{\abs}}}
\and
\inferrule*[lab={\rulename{To-Abs-Double}}]
    {\entrytoabsarrow{\entry_{0}}{\vect{\abs}_{0}} \\ \entrytoabsarrow{\entry_{1}}{\vect{\abs}_{1}}}
    {\toabs{\VBorV{\doubleborrow{\entry_{0}}{\entry_{1}}}}{\vect{\abs}_{0} \mathbin{+\!\!+} \vect{\abs}_{1}}}
\\
\inferrule*[lab={\rulename{Entry-To-Abs-Shared}}]
    {}
    {\entrytoabsarrow{\BShared{\loc}{\ty}}{[[\FullShrBor{\loc}{\ty}]]}}
\and
\inferrule*[lab={\rulename{Entry-To-Abs-Dead}}]
    {}
    {\entrytoabsarrow{\BDead{k}}{[]}}
\and
\inferrule*[lab={\rulename{Entry-To-Abs-Mut}}]
    {\bot, \mathsf{borrow} \notin \val \\ \toabs{\val}{[\abs_{1},\ \dots,\ \abs_{n}]} \\\\ \abs = [\FullMutBor{\loc}{\VIgnored{\tyof{\val}}}] \mathbin{+\!\!+} \abs_{1} \mathbin{+\!\!+} \cdots \mathbin{+\!\!+} \abs_{n}}
    {\entrytoabsarrow{\BMut{\loc}{\val}}{[\abs]}}
\end{rulefig}
\noindent A shared or mutable entry becomes an abstraction holding a full-view reference; a dead entry borrows nothing.

\noindent Merging two region abstractions is as in \LLBCsharp (\Cref{fig:llbc:symbolic:merge-abs}): a region abstraction holds only full-view references, written as the borrows of \LLBCsharp, so these rules stay the same.

\begin{rulefig}{fig:views:symbolic:le1}{One refinement step, $\lesharp{\st}{\st'}$: moving values and abstractions. Identical to \LLBCsharp (\Cref{fig:llbc:symbolic:le1}).}
\inferrule*[lab={\rulename{Le-To-Abs}}]
    {\toabs{\val}{\vect{\abs}}}
    {\lesharp{(\mathit{frame},\ \val :: \mathit{anons},\ \vect{\abs}')}{(\mathit{frame},\ \mathit{anons},\ \vect{\abs} \mathbin{+\!\!+} \vect{\abs}')}}
\and
\inferrule*[lab={\rulename{Le-To-Symbolic}}]
    {\nolocs{\val} \\ \bot, \_ \notin \val \\ \tyof{\val} \neq \TPair{\_}{\_} \\\\ \sharpfpreplace{\fpath}{\st}{\val}{\VSymbolicV{\tyof{\val}}}{\st'}}
    {\lesharp{\st}{\st'}}
\and
\inferrule*[lab={\rulename{Le-Move-Value}}]
    {\dpunloan, \dpunbor, \abs \notin \fpath \\ \nohasouterloan{\val} \\ \tyof{\val} \neq \TPair{\_}{\_} \\\\ \sharpfpreplace{\fpath}{\st}{\val}{\VBotTy{\tyof{\val}}}{\st'}}
    {\lesharp{\st}{\pushanon{\val}{\st'}}}
\and
\inferrule*[lab={\rulename{Le-Merge-Abs}}]
    {\sharpmergeabs{\abs_{0}}{\abs_{1}}{\abs}}
    {\lesharp{(\mathit{frame},\ \mathit{anons},\ \abs_{0} :: \abs_{1} :: \vect{\abs})}{(\mathit{frame},\ \mathit{anons},\ \abs :: \vect{\abs})}}
\and
\inferrule*[lab={\rulename{Le-Perm-Abs}}]
    {\Permutation{\vect{\abs}}{\vect{\abs}'}}
    {\lesharp{(\mathit{frame},\ \mathit{anons},\ \vect{\abs})}{(\mathit{frame},\ \mathit{anons},\ \vect{\abs}')}}
\and
\inferrule*[lab={\rulename{Le-Perm-In-Abs}}]
    {\Permutation{\abs}{\abs'}}
    {\lesharp{(\mathit{frame},\ \mathit{anons},\ \abs :: \vect{\abs})}{(\mathit{frame},\ \mathit{anons},\ \abs' :: \vect{\abs})}}
\and
\inferrule*[lab={\rulename{Le-Perm-Anons}}]
    {\Permutation{\mathit{anons}}{\mathit{anons}'}}
    {\lesharp{(\mathit{frame},\ \mathit{anons},\ \vect{\abs})}{(\mathit{frame},\ \mathit{anons}',\ \vect{\abs})}}
\end{rulefig}

\begin{rulefig}{fig:views:symbolic:le1-labels}{One refinement step, $\lesharp{\st}{\st'}$: labels and loans.}
\inferrule*[lab={\rulename{Le-Reborrow-MutBorrow}}]
    {\lfresh{\loc_{1}}{\st} \\ \sharpfpreplace{\feaddr}{\st}{\BMut{\loc_{0}}{\val}}{\BMut{\loc_{1}}{\val}}{\st'}}
    {\lesharp{\st}{\pushanon{\FullMutBor{\loc_{0}}{\VMutLoanTy{\loc_{1}}{\tyof{\val}}}}{\st'}}}
\and
\inferrule*[lab={\rulename{Le-End-SharedLoan-Abs}}]
    {\fpinsideregionabs{\fpath} \\ \sharpgiveback{\fpath}{\VShrLoan{\loc}{\val}}{\st}{\val}{\st'} \\ \lfresh{\loc}{\st'}}
    {\lesharp{\st}{\st'}}
\and
\inferrule*[lab={\rulename{Le-End-SharedLoan-Inner}}]
    {\notinsideregionabs{\fpath} \\ \unloanin{\fpath} \\ \sharpfpreplace{\fpath}{\st}{\VShrLoan{\loc}{\val}}{\val}{\st'} \\ \lfresh{\loc}{\st'}}
    {\lesharp{\st}{\st'}}
\and
\inferrule*[lab={\rulename{Le-Reborrow-SharedBorrow}}]
    {\lfresh{\loc_{1}}{\st_{0}} \\ \bot, \mathsf{loan}^{m}, \mathsf{borrow} \notin \val \\ \statecontains{\VShrLoan{\loc_{0}}{\val}}{\st_{0}} \\\\ \sharpfpreplace{\feaddr_{i}}{\st_{i}}{\BShared{\loc_{0}}{\tyof{\val}}}{\BShared{\loc_{1}}{\tyof{\val}}}{\st_{i+1}}}
    {\lesharp{\st_{0}}{\pushabs{\st_{\nat}}{\absliteral{\FullShrBor{\loc_{0}}{\tyof{\val}},\ \VShrLoan{\loc_{1}}{\stripouterloans{\val}}}}}}
\and
\inferrule*[lab={\rulename{Le-Abs-Drop-Covered-Borrow}}]
    {\FullShrBor{\loc_{0}}{\ty_{0}} \in \abs \\ \statecontains{\VShrLoan{\loc_{0}}{\val}}{(\mathit{frame},\ \mathit{anons},\ \abs :: \vect{\abs})} \\ \shrloaninside{\loc}{\val}}
    {\lesharp{(\mathit{frame},\ \mathit{anons},\ (\FullShrBor{\loc}{\ty} :: \abs) :: \vect{\abs})}{(\mathit{frame},\ \mathit{anons},\ \abs :: \vect{\abs})}}
\end{rulefig}
\noindent A mutable reborrow names its entry by its address $\feaddr$; a shared reborrow relabels $\nat$ shared entries of the label $\loc_{0}$, one address $\feaddr_{i}$ each. Refinement $\lesharpstar{\st}{\st'}$ is the reflexive transitive closure of $\lesharp{\cdot}{\cdot}$.

\subsubsection{Borrow checking}\label{app:views:symbolic:borrow-check}

Argument and result instantiation, entry and exit states, and function
borrow checking are as in \LLBCsharp
(\Cref{def:llbc:symbolic:inst-arg,def:llbc:symbolic:fn-init-end-states,def:llbc:symbolic:borrow-checks}).
They use the following rules for the borrows supplied by the caller.

\begin{rulefig}{fig:views:symbolic:init-extern}{The references a caller lends through an argument of type $\ty$, $\initextern{\ty}{\abs}$, and one entry of mode $k$ over a field of type $\ty$, $\initexternentry{k}{\ty}{\entry}$.}
\inferrule*[lab={\rulename{Init-Extern-Symbolic}}]
    {\reffree{\ty}}
    {\initextern{\ty}{[]}}
\and
\inferrule*[lab={\rulename{Init-Extern-Pair}}]
    {\notreffree{\TPair{\ty_{0}}{\ty_{1}}} \\ \initextern{\ty_{0}}{\abs_{0}} \\ \initextern{\ty_{1}}{\abs_{1}} \\ \absdisjoint{\abs_{0}}{\abs_{1}}}
    {\initextern{\TPair{\ty_{0}}{\ty_{1}}}{\abs_{0} \mathbin{+\!\!+} \abs_{1}}}
\\
\inferrule*[lab={\rulename{Init-Extern-Whole}}]
    {\initexternentry{k}{\ty}{\entry}}
    {\initextern{(\TRef{\singleview{\varepsilon}{k}}{\ty})}{[\fullref{k}{\entry}{\ty}]}}
\and
\inferrule*[lab={\rulename{Init-Extern-Field}}]
    {\initexternentry{k}{\ty_{i}}{\entry}}
    {\initextern{(\TRef{\singleview{i}{k}}{\TPair{\ty_{0}}{\ty_{1}}})}{[\fullref{k}{\entry}{\ty_{i}}]}}
\and
\inferrule*[lab={\rulename{Init-Extern-Both}}]
    {\initexternentry{k_{0}}{\ty_{0}}{\entry_{0}} \\ \initexternentry{k_{1}}{\ty_{1}}{\entry_{1}} \\ \absdisjoint{[\fullref{k_{0}}{\entry_{0}}{\ty_{0}}]}{[\fullref{k_{1}}{\entry_{1}}{\ty_{1}}]}}
    {\initextern{(\TRef{\doubleview{k_{0}}{k_{1}}}{\TPair{\ty_{0}}{\ty_{1}}})}{[\fullref{k_{0}}{\entry_{0}}{\ty_{0}},\ \fullref{k_{1}}{\entry_{1}}{\ty_{1}}]}}
\\
\inferrule*[lab={\rulename{Init-Extern-Mut}}]
    {\reffree{\ty}}
    {\initexternentry{\Mut}{\ty}{\BMut{\loc}{\VIgnored{\ty}}}}
\and
\inferrule*[lab={\rulename{Init-Extern-Shared}}]
    {\reffree{\ty}}
    {\initexternentry{\Shr}{\ty}{\BShared{\loc}{\ty}}}
\end{rulefig}
\noindent For each field the argument's view borrows, the abstraction of the entry state holds a full-view reference that stands for the caller's borrow of that field.

\subsection{Combined Semantics (\LLBCPlusView)}
\label{app:views:combined}

\LLBCPlusView extends \LLBCPlus with partial borrows, as \LLBCSharpView extends \LLBCsharp. Its judgements carry $\vdash^{+}$.

\subsubsection{Values and states}\label{app:views:combined:values}

Values, full-view references and region abstractions are those of \LLBCSharpView (\Cref{fig:views:symbolic:values}); states are as in \LLBCPlus (\Cref{fig:llbc:combined:state}).

\subsubsection{Auxiliary semantic data types}\label{app:views:combined:aux-data}

Addresses, full paths and statement results are as in \LLBCPlus (\Cref{fig:llbc:combined:state-addr,fig:llbc:modes-results}); direct paths and entry addresses as in \LLBCView (\Cref{fig:views:concrete:paths}).

\subsubsection{Operations on types and values}\label{app:views:combined:value-ops}

The operations on types and values are those of \LLBCSharpView (\Cref{app:views:symbolic:value-ops}).

\begin{definition}[Relabelling]\label{def:views:combined:relabel}
$\relabel{\val}{\loc_{0}}{\loc_{1}}$ swaps the labels $\loc_{0}$ and $\loc_{1}$ everywhere in $\val$, the labels of its loans and of its entries included. Relabelling a state is as in \LLBCPlus (\Cref{def:llbc:combined:relabel-state}).
\end{definition}

\subsubsection{Operations on states}\label{app:views:combined:state-ops}

The following are as in \LLBCPlus (\Cref{app:llbc:combined:state-ops}): adding an anonymous value or a region abstraction, reading and writing an address, along a full path and at a place, and the address of a place.
The following are as in \LLBCView, with $\vdash^{+}$ for $\vdash$:
\begin{itemize}
  \item following a projection (\Cref{fig:views:place-address-aux});
  \item the type of a place (\Cref{def:views:place-type});
  \item replacing an entry, $\plusfpreplace{\feaddr}{\st}{\entry}{\entry'}{\st'}$ (\Cref{def:views:entry-access}).
\end{itemize}

\subsubsection{Evaluation}\label{app:views:combined:eval}

The following are as in \LLBCSharpView, with $\vdash^{+}$ for $\vdash_\#$: evaluating an rvalue and borrowing the fields a view names (\Cref{fig:views:rv-evals,fig:views:borrow-evals}), borrowing one place (\Cref{fig:views:symbolic:borrow-at}) and placing a new shared loan (\Cref{fig:views:symbolic:shr-borrow-at}).
Evaluating an operand, assignment, popping a frame, leaving a called function and evaluating a statement are as in \LLBCPlus (\Cref{app:llbc:combined:eval}).

\subsubsection{Reorganisation and refinement}\label{app:views:combined:refinement}

\noindent The following are as in \LLBCSharpView, with $\vdash^{+}$ for $\vdash_\#$:
\begin{itemize}
  \item reorganisation, $\plusenvreorg{\st}{\st'}$ (\Cref{fig:views:symbolic:env-reorg}), with ending abstraction elements and abstractions and giving back;
  \item moving values into abstractions and merging abstractions (\Cref{fig:views:symbolic:to-abs,fig:llbc:symbolic:merge-abs}).
\end{itemize}

\begin{rulefig}{fig:views:combined:le1}{One \LLBCPlusView refinement step, $\leplus{\st}{\st'}$: moving values and abstractions.}
\inferrule*[lab={\rulename{Le-To-Abs}}]
    {\toabs{\val}{\vect{\abs}}}
    {\leplus{(\mathit{frames},\ \val :: \mathit{anons},\ \vect{\abs}')}{(\mathit{frames},\ \mathit{anons},\ \vect{\abs} \mathbin{+\!\!+} \vect{\abs}')}}
\and
\inferrule*[lab={\rulename{Le-To-Symbolic}}]
    {\nolocs{\val} \\ \bot, \_ \notin \val \\ \tyof{\val} \neq \TPair{\_}{\_} \\\\ \plusfpreplace{\fpath}{\st}{\val}{\VSymbolicV{\tyof{\val}}}{\st'}}
    {\leplus{\st}{\st'}}
\and
\inferrule*[lab={\rulename{Le-Move-Value}}]
    {\dpunloan, \dpunbor, \abs \notin \fpath \\ \nohasouterloan{\val} \\ \tyof{\val} \neq \TPair{\_}{\_} \\\\ \plusfpreplace{\fpath}{\st}{\val}{\VBotTy{\tyof{\val}}}{\st'}}
    {\leplus{\st}{\pushanon{\val}{\st'}}}
\and
\inferrule*[lab={\rulename{Le-Move-Entry}}]
    {\dpunloan, \dpunbor, \abs \notin \feapath{\feaddr} \\ \plusfpreplace{\feaddr}{\st}{\BMut{\loc}{\val}}{\BDead{\Mut}}{\st'}}
    {\leplus{\st}{\pushanon{\FullMutBor{\loc}{\val}}{\st'}}}
\and
\inferrule*[lab={\rulename{Le-Merge-Abs}}]
    {\plusmergeabs{\abs_{0}}{\abs_{1}}{\abs}}
    {\leplus{(\mathit{frames},\ \mathit{anons},\ \abs_{0} :: \abs_{1} :: \vect{\abs})}{(\mathit{frames},\ \mathit{anons},\ \abs :: \vect{\abs})}}
\and
\inferrule*[lab={\rulename{Le-Empty-Abs}}]
    {}
    {\leplus{(\mathit{frames},\ \mathit{anons},\ \vect{\abs})}{(\mathit{frames},\ \mathit{anons},\ [] :: \vect{\abs})}}
\and
\inferrule*[lab={\rulename{Le-Perm-Abs}}]
    {\Permutation{\vect{\abs}}{\vect{\abs}'}}
    {\leplus{(\mathit{frames},\ \mathit{anons},\ \vect{\abs})}{(\mathit{frames},\ \mathit{anons},\ \vect{\abs}')}}
\and
\inferrule*[lab={\rulename{Le-Perm-In-Abs}}]
    {\Permutation{\abs}{\abs'}}
    {\leplus{(\mathit{frames},\ \mathit{anons},\ \abs :: \vect{\abs})}{(\mathit{frames},\ \mathit{anons},\ \abs' :: \vect{\abs})}}
\and
\inferrule*[lab={\rulename{Le-Perm-Anons}}]
    {\Permutation{\mathit{anons}}{\mathit{anons}'}}
    {\leplus{(\mathit{frames},\ \mathit{anons},\ \vect{\abs})}{(\mathit{frames},\ \mathit{anons}',\ \vect{\abs})}}
\and
\inferrule*[lab={\rulename{Le-Anon-Value}}]
    {}
    {\leplus{\st}{\pushanon{\botof{\ty}}{\st}}}
\and
\inferrule*[lab={\rulename{Le-Anon-Dead}}]
    {\alldead{\entries}}
    {\leplus{\st}{\pushanon{\VBorrow{\ty}{\entries}}{\st}}}
\end{rulefig}

\begin{rulefig}{fig:views:combined:le1-labels}{One \LLBCPlusView refinement step, $\leplus{\st}{\st'}$: labels and loans.}
\inferrule*[lab={\rulename{Le-Reborrow-MutBorrow}}]
    {\lfresh{\loc_{1}}{\st} \\ \plusfpreplace{\feaddr}{\st}{\BMut{\loc_{0}}{\val}}{\BMut{\loc_{1}}{\val}}{\st'}}
    {\leplus{\st}{\pushanon{\FullMutBor{\loc_{0}}{\VMutLoanTy{\loc_{1}}{\tyof{\val}}}}{\st'}}}
\and
\inferrule*[lab={\rulename{Le-Reborrow-SharedBorrow}}]
    {\lfresh{\loc_{1}}{\st_{0}} \\ \bot, \mathsf{loan}^{m}, \mathsf{borrow} \notin \val \\ \statecontains{\VShrLoan{\loc_{0}}{\val}}{\st_{0}} \\\\ \plusfpreplace{\feaddr_{i}}{\st_{i}}{\BShared{\loc_{0}}{\tyof{\val}}}{\BShared{\loc_{1}}{\tyof{\val}}}{\st_{i+1}}}
    {\leplus{\st_{0}}{\pushabs{\st_{\nat}}{\absliteral{\FullShrBor{\loc_{0}}{\tyof{\val}},\ \VShrLoan{\loc_{1}}{\stripouterloans{\val}}}}}}
\and
\inferrule*[lab={\rulename{Le-Fresh-SharedBorrow}}]
    {\statecontains{\VShrLoan{\loc}{\val}}{\st}}
    {\leplus{\st}{\pushanon{\FullShrBor{\loc}{\tyof{\val}}}{\st}}}
\and
\inferrule*[lab={\rulename{Le-End-SharedLoan-Abs}}]
    {\fpinsideregionabs{\fpath} \\ \plusgiveback{\fpath}{\VShrLoan{\loc}{\val}}{\st}{\val}{\st'} \\ \lfresh{\loc}{\st'}}
    {\leplus{\st}{\st'}}
\and
\inferrule*[lab={\rulename{Le-End-SharedLoan-Inner}}]
    {\notinsideregionabs{\fpath} \\ \unloanin{\fpath} \\ \plusfpreplace{\fpath}{\st}{\VShrLoan{\loc}{\val}}{\val}{\st'} \\ \lfresh{\loc}{\st'}}
    {\leplus{\st}{\st'}}
\and
\inferrule*[lab={\rulename{Le-Abs-Drop-Covered-Borrow}}]
    {\FullShrBor{\loc_{0}}{\ty_{0}} \in \abs \\ \statecontains{\VShrLoan{\loc_{0}}{\val}}{(\mathit{frames},\ \mathit{anons},\ \abs :: \vect{\abs})} \\ \shrloaninside{\loc}{\val}}
    {\leplus{(\mathit{frames},\ \mathit{anons},\ (\FullShrBor{\loc}{\ty} :: \abs) :: \vect{\abs})}{(\mathit{frames},\ \mathit{anons},\ \abs :: \vect{\abs})}}
\and
\inferrule*[lab={\rulename{Le-Rename-Shared}}]
    {\containsshrloan{\loc_{0}}{\st} \\ \lfresh{\loc_{1}}{\st}}
    {\leplus{\st}{\relabel{\st}{\loc_{0}}{\loc_{1}}}}
\and
\inferrule*[lab={\rulename{Le-Rename-Mut}}]
    {\containsmutloan{\loc_{0}}{\st} \\ \containsmutbor{\loc_{0}}{\st} \\ \lfresh{\loc_{1}}{\st}}
    {\leplus{\st}{\relabel{\st}{\loc_{0}}{\loc_{1}}}}
\end{rulefig}

\noindent Refinement $\leplusstar{\st}{\st'}$ is the reflexive transitive closure of $\leplus{\cdot}{\cdot}$. Beyond the \LLBCSharpView rules, it may add an empty abstraction, a bottom, a reference whose entries are all dead or a fresh full-view shared reference of an existing loan, move a mutable entry out to an anonymous full-view reference, leaving the entry dead, and rename a label.

\section{Correspondence with the Original Formalization}\label{app:thesis}

We relate the rules of \LLBC, \LLBCsharp and \LLBCPlus to the original
formalization in Ho's thesis~\cite{ho2024thesis}. Figure and page numbers in
the thesis column refer to that thesis, using its printed page numbers.

\subsection{Concrete semantics (\LLBC)}\label{app:thesis:llbc}
\paragraph{Judgements.}
{\small
\begin{longtable}{@{}>{\raggedright\arraybackslash}p{0.28\linewidth}>{\raggedright\arraybackslash}p{0.29\linewidth}>{\raggedright\arraybackslash}p{0.35\linewidth}@{}}
\toprule
Ours & Thesis & Note \\
\midrule
\endhead
\texttt{place\_address}, \texttt{place\_address\_aux}, \texttt{read}, \texttt{read\_at}, \texttt{read\_write} (\Cref{def:llbc:place-address,fig:llbc:place-address-aux,def:llbc:read,def:llbc:read-at,def:llbc:read-write}) & read (Fig.~9.3, p.~161); read along a path (Fig.~9.3, p.~161); write (Fig.~9.4, p.~162); write along a path (Fig.~9.4, p.~162) & Ours computes an address, then reads or writes at it; the thesis follows the path while reading or writing. \\
\midrule
\texttt{fp\_applies}, \texttt{dp\_applies}, \texttt{state\_get\_va}, \texttt{state\_set\_va}, \texttt{state\_updates}, \texttt{val\_updates} (\Cref{def:llbc:fp-applies,fig:llbc:dp-applies,def:llbc:state-get,fig:llbc:state-set,def:llbc:state-updates,fig:llbc:val-updates}) & $\Omega[.]$, $V[.]$ (\S9.5, p.~170) & The thesis writes these with its hole notation; ours defines them. \\
\midrule
\texttt{push\_anon} (\Cref{def:llbc:push-anon}) & $\Omega, \_ \to v$ (Fig.~9.2, p.~160) & We store anonymous values in a list. \\
\midrule
\texttt{op\_evals}, \texttt{rv\_evals} (\Cref{fig:llbc:op-evals,fig:llbc:rv-evals}) & rvalue evaluation (Fig.~9.5, p.~164) & The thesis has one judgement for operands and rvalues; ours has two. \\
\midrule
--- & copy (Fig.~9.6, p.~165) & Not in the paper's calculus. \\
\midrule
\texttt{stmt\_evals}, \texttt{assign\_evals}, \texttt{assign\_val\_evals}, \texttt{exit\_fn}, \texttt{pop\_frame} (\Cref{fig:llbc:stmt-evals,fig:llbc:assign,fig:llbc:exit-fn,def:llbc:pop-frame}) & statement evaluation (Fig.~9.7, p.~166; Fig.~9.8, p.~168; Fig.~11.6, p.~200); push\_stack (Fig.~9.8, p.~168); pop\_stack (Fig.~9.8, p.~168) & \texttt{push\_stack} is inline in our E-Call, \texttt{pop\_stack} is \texttt{exit\_fn} and \texttt{pop\_frame}; \texttt{call\_evals} (LLBC\#) and \texttt{assign\_evals}, \texttt{assign\_val\_evals} hold the premises of E-Call and E-Assign. \\
\midrule
\texttt{env\_reorg} (\Cref{fig:llbc:env-reorg}) & reorganisation (Fig.~9.10, p.~169) & \texttt{give\_back} and \texttt{end\_elem} are inline in the thesis's Reorg-End-* rules. \\
\midrule
\texttt{ty\_of} (\Cref{def:llbc:tyof}) & $v : \tau$ (Fig.~9.4, p.~162) & The thesis gives no rules for it; ours defines it. \\
\midrule
\texttt{has\_outer\_loan} (\Cref{def:llbc:outer-loan}) & no outer loans in $v$ (\S9.4.2, p.~166) & Defined in the thesis's text (\S{}9.4.2, p. 166); ours as two rules. \\
\midrule
\texttt{bot\_of} (\Cref{def:llbc:botof}) & $\bot$ (Fig.~9.2, p.~160) & Ours is a tree: no bottom of pair type. \\
\bottomrule
\end{longtable}
}
\paragraph{Statements.}
We compare the step-indexed statement rules. A $^\dagger$ marks rules whose
step-indexed forms appear explicitly in Fig.~11.6 (p.~200); the step index
is implicit in the other thesis rules.
{\small
\begin{longtable}{@{}>{\raggedright\arraybackslash}p{0.28\linewidth}>{\raggedright\arraybackslash}p{0.29\linewidth}>{\raggedright\arraybackslash}p{0.35\linewidth}@{}}
\toprule
Ours & Thesis & Note \\
\midrule
\endhead
\rulename{E-Step-Zero} (\Cref{fig:llbc:stmt-evals}) & \rulename{E-Step-Zero}$^\dagger$ (Fig.~11.6, p.~200) &  \\
\rulename{E-Return} (\Cref{fig:llbc:stmt-evals}) & \rulename{E-Return}$^\dagger$ (Fig.~9.7, p.~166) &  \\
\rulename{E-Panic} (\Cref{fig:llbc:stmt-evals}) & \rulename{E-Panic} (Fig.~9.7, p.~166) &  \\
\rulename{E-Reorg} (\Cref{fig:llbc:stmt-evals}) & \rulename{E-Reorg} (Fig.~9.7, p.~166) &  \\
\rulename{E-Seq-Unit} (\Cref{fig:llbc:stmt-evals}) & \rulename{E-Seq-Unit}$^\dagger$ (Fig.~9.7, p.~166) &  \\
\rulename{E-Seq-Propagate} (\Cref{fig:llbc:stmt-evals}) & \rulename{E-Seq-Propagate} (Fig.~9.7, p.~166) &  \\
\rulename{E-Match-Left}, \rulename{E-Match-Right} (\Cref{fig:llbc:stmt-evals}) & \rulename{E-Match} (Fig.~9.7, p.~166) & The thesis has one rule. \\
\rulename{E-Assign}, \rulename{Assign-Evals}, \rulename{Assign-Val-Evals} (\Cref{fig:llbc:stmt-evals,fig:llbc:assign}) & \rulename{E-Assign} (Fig.~9.7, p.~166) &  \\
\rulename{E-Call}, \rulename{Exit-No-Fuel}, \rulename{Exit-Panic}, \rulename{Exit-Ret}, \rulename{Pop-Frame} (\Cref{fig:llbc:stmt-evals,fig:llbc:exit-fn,def:llbc:pop-frame}) & \rulename{E-Call}$^\dagger$, \rulename{E-PushStack}, \rulename{E-PopStack} (Fig.~9.8, p.~168) & The Exit-* rules are the three cases of E-Step-Call (Fig. 11.6). \\
\rulename{E-Nil} (\Cref{fig:llbc:stmt-evals}) & --- & No thesis rule, though its grammar has the empty statement (Fig. 9.1). \\
--- & \rulename{E-Break}, \rulename{E-Continue} (Fig.~9.7, p.~166); \rulename{E-Loop-Break-Inner}, \rulename{E-Loop-Break-Outer}, \rulename{E-Loop-Continue-Inner}$^\dagger$, \rulename{E-Loop-Continue-Outer}, \rulename{E-Loop-Panic}, \rulename{E-Loop-Return} (Fig.~9.9, p.~168) & Loops: not in the paper's calculus \\
--- & \rulename{E-IfThenElse-T}, \rulename{E-IfThenElse-F} (Fig.~9.7, p.~166) & If-then-else: not in the paper's calculus \\
--- & \rulename{E-Box-New}, \rulename{E-Box-Free} (Fig.~9.7, p.~166) & Boxes: not in the paper's calculus \\
\bottomrule
\end{longtable}
}
\paragraph{Rvalues.}
{\small
\begin{longtable}{@{}>{\raggedright\arraybackslash}p{0.28\linewidth}>{\raggedright\arraybackslash}p{0.29\linewidth}>{\raggedright\arraybackslash}p{0.35\linewidth}@{}}
\toprule
Ours & Thesis & Note \\
\midrule
\endhead
\rulename{E-Op} (\Cref{fig:llbc:rv-evals}) & --- & Implicit in the thesis: an operand is an rvalue. \\
\rulename{E-Move} (\Cref{fig:llbc:op-evals}) & \rulename{E-Move} (Fig.~9.5, p.~164) &  \\
\rulename{E-Pair} (\Cref{fig:llbc:op-evals}) & \rulename{E-Pair} (Fig.~9.5, p.~164) &  \\
\rulename{E-Sum-Left} (\Cref{fig:llbc:op-evals}) & \rulename{E-Sum-Left} (Fig.~9.5, p.~164) &  \\
\rulename{E-Sum-Right} (\Cref{fig:llbc:op-evals}) & \rulename{E-Sum-Right} (Fig.~9.5, p.~164) &  \\
\rulename{E-Unit} (\Cref{fig:llbc:op-evals}) & --- & No thesis rule, though its grammar has the unit operand (Fig. 9.1). \\
\rulename{E-MutBorrow} (\Cref{fig:llbc:rv-evals}) & \rulename{E-MutBorrow} (Fig.~9.5, p.~164) &  \\
\rulename{E-SharedBorrow-Reuse}, \rulename{E-SharedBorrow-Wrap} (\Cref{fig:llbc:rv-evals}) & \rulename{E-SharedBorrow} (Fig.~9.5, p.~164) & The thesis has one rule; ours splits it by cases. \\
--- & \rulename{E-ReservedBorrow} (Fig.~9.5, p.~164) & Reserved (two-phase) borrows: not in the paper's calculus \\
--- & \rulename{E-Copy} (Fig.~9.5, p.~164) & Copies: not in the paper's calculus \\
\midrule
--- & \rulename{Copy-SharedBorrow}, \rulename{Copy-SharedLoan}, \rulename{Copy-Scalar}, \rulename{Copy-Sum}, \rulename{Copy-Pair} (Fig.~9.6, p.~165) & Copies: not in the paper's calculus \\
\bottomrule
\end{longtable}
}
\paragraph{Places.}
{\small
\begin{longtable}{@{}>{\raggedright\arraybackslash}p{0.28\linewidth}>{\raggedright\arraybackslash}p{0.29\linewidth}>{\raggedright\arraybackslash}p{0.35\linewidth}@{}}
\toprule
Ours & Thesis & Note \\
\midrule
\endhead
\rulename{Place-Address}, \rulename{Read-At}, \rulename{Read}, \rulename{Read-Write} (\Cref{def:llbc:place-address,def:llbc:read-at,def:llbc:read,def:llbc:read-write}) & \rulename{Read} (Fig.~9.3, p.~161); \rulename{Write} (Fig.~9.4, p.~162) & Read: Place-Address, Read-At, Read. Write: Place-Address, Read-At, Read-Write. \\
\rulename{PA-Base} (\Cref{fig:llbc:place-address-aux}) & \rulename{R-Base} (Fig.~9.3, p.~161); \rulename{W-Base} (Fig.~9.4, p.~162) &  \\
\rulename{PA-Proj-Sum-Left} (\Cref{fig:llbc:place-address-aux}) & \rulename{R-ProjSum-Left} (Fig.~9.3, p.~161); \rulename{W-ProjSum-Left} (Fig.~9.4, p.~162) &  \\
\rulename{PA-Proj-Sum-Right} (\Cref{fig:llbc:place-address-aux}) & \rulename{R-ProjSum-Right} (Fig.~9.3, p.~161); \rulename{W-ProjSum-Right} (Fig.~9.4, p.~162) &  \\
\rulename{PA-Proj-Pair-Left} (\Cref{fig:llbc:place-address-aux}) & \rulename{R-ProjPair-Left} (Fig.~9.3, p.~161); \rulename{W-ProjPair-Left} (Fig.~9.4, p.~162) &  \\
\rulename{PA-Proj-Pair-Right} (\Cref{fig:llbc:place-address-aux}) & \rulename{R-ProjPair-Right} (Fig.~9.3, p.~161); \rulename{W-ProjPair-Right} (Fig.~9.4, p.~162) &  \\
\rulename{PA-SharedLoan} (\Cref{fig:llbc:place-address-aux}) & \rulename{R-SharedLoan} (Fig.~9.3, p.~161); \rulename{W-SharedLoan} (Fig.~9.4, p.~162) &  \\
\rulename{PA-Deref-MutBorrow} (\Cref{fig:llbc:place-address-aux}) & \rulename{R-Deref-MutBorrow} (Fig.~9.3, p.~161); \rulename{W-Deref-MutBorrow} (Fig.~9.4, p.~162) &  \\
\rulename{PA-Deref-SharedBorrow} (\Cref{fig:llbc:place-address-aux}) & \rulename{R-Deref-SharedBorrow} (Fig.~9.3, p.~161); \rulename{W-Deref-SharedBorrow} (Fig.~9.4, p.~162) &  \\
--- & \rulename{R-Deref-Box} (Fig.~9.3, p.~161); \rulename{W-Deref-Box} (Fig.~9.4, p.~162) & Boxes: not in the paper's calculus \\
\bottomrule
\end{longtable}
}
\paragraph{Reorganisation.}
{\small
\begin{longtable}{@{}>{\raggedright\arraybackslash}p{0.28\linewidth}>{\raggedright\arraybackslash}p{0.29\linewidth}>{\raggedright\arraybackslash}p{0.35\linewidth}@{}}
\toprule
Ours & Thesis & Note \\
\midrule
\endhead
\rulename{Reorg-End-MutBorrow} (\Cref{fig:llbc:env-reorg}) & \rulename{Reorg-End-MutBorrow} (Fig.~9.10, p.~169) &  \\
\rulename{Reorg-End-SharedBorrow} (\Cref{fig:llbc:env-reorg}) & \rulename{Reorg-End-SharedReservedBorrow} (Fig.~9.10, p.~169) & The thesis's rule also ends reserved borrows. \\
\rulename{Reorg-End-SharedLoan} (\Cref{fig:llbc:env-reorg}) & \rulename{Reorg-End-SharedLoan} (Fig.~9.10, p.~169) &  \\
--- & \rulename{Reorg-None}, \rulename{Reorg-Seq} (Fig.~9.10, p.~169) & Ours: one reorganisation step per E-Reorg. \\
--- & \rulename{Reorg-Activate-Reserved} (Fig.~9.10, p.~169) & Reserved (two-phase) borrows: not in the paper's calculus \\
\bottomrule
\end{longtable}
}
\paragraph{States and values.}
{\small
\begin{longtable}{@{}>{\raggedright\arraybackslash}p{0.28\linewidth}>{\raggedright\arraybackslash}p{0.29\linewidth}>{\raggedright\arraybackslash}p{0.35\linewidth}@{}}
\toprule
Ours & Thesis & Note \\
\midrule
\endhead
\rulename{FP-Applies}, \rulename{DP-Nil}, \rulename{DP-Pair-0}, \rulename{DP-Pair-1}, \rulename{DP-Left}, \rulename{DP-Right}, \rulename{DP-Unloan}, \rulename{DP-Unbor}, \rulename{State-Get-Frames}, \rulename{State-Get-Anon}, \rulename{State-Set-Anon}, \rulename{State-Set-Frames}, \rulename{State-Updates}, \rulename{V-Base}, \rulename{V-Sum-Left}, \rulename{V-Sum-Right}, \rulename{V-Pair-Left}, \rulename{V-Pair-Right}, \rulename{V-SharedUnloan}, \rulename{V-Unbor} (\Cref{def:llbc:fp-applies,fig:llbc:dp-applies,def:llbc:state-get,fig:llbc:state-set,def:llbc:state-updates,fig:llbc:val-updates}) & $\Omega[.]$, $V[.]$ (\S9.5, p.~170) & The thesis writes these with its hole notation; ours defines them. \\
\midrule
\texttt{push\_anon} (\Cref{def:llbc:push-anon}) & $\Omega, \_ \to v$ (Fig.~9.2, p.~160) & We store anonymous values in a list. \\
\midrule
\texttt{ty\_of} (\Cref{def:llbc:tyof}) & $v : \tau$ (Fig.~9.4, p.~162) & The thesis gives no rules for it; ours defines it. \\
\midrule
\rulename{Outer-Loan-Mut}, \rulename{Outer-Loan-Shared} (\Cref{def:llbc:outer-loan}) & no outer loans in $v$ (\S9.4.2, p.~166) & Defined in the thesis's text (\S{}9.4.2, p. 166); ours as two rules. \\
\midrule
\texttt{bot\_of} (\Cref{def:llbc:botof}) & $\bot$ (Fig.~9.2, p.~160) & Ours is a tree: no bottom of pair type. \\
\bottomrule
\end{longtable}
}

\subsection{Symbolic semantics (\LLBCsharp)}\label{app:thesis:llbc-sharp}
\paragraph{Judgements.}
{\small
\begin{longtable}{@{}>{\raggedright\arraybackslash}p{0.28\linewidth}>{\raggedright\arraybackslash}p{0.29\linewidth}>{\raggedright\arraybackslash}p{0.35\linewidth}@{}}
\toprule
Ours & Thesis & Note \\
\midrule
\endhead
\texttt{place\_address}, \texttt{place\_address\_aux}, \texttt{read}, \texttt{read\_at}, \texttt{read\_write} (\Cref{def:llbc:place-address,fig:llbc:place-address-aux,def:llbc:read,def:llbc:read-at,def:llbc:read-write}) & read (Fig.~9.3, p.~161); read along a path (Fig.~9.3, p.~161); write (Fig.~9.4, p.~162); write along a path (Fig.~9.4, p.~162) & Ours computes an address, then reads or writes at it; the thesis follows the path while reading or writing. \\
\midrule
\texttt{fp\_applies}, \texttt{dp\_applies}, \texttt{state\_get\_va}, \texttt{state\_set\_va}, \texttt{state\_updates}, \texttt{val\_updates}, \texttt{get\_abs}, \texttt{fp\_inside\_region\_abs} (\Cref{def:llbc:fp-applies,fig:llbc:dp-applies,def:llbc:symbolic:state-get,fig:llbc:symbolic:state-set,def:llbc:state-updates,fig:llbc:val-updates,def:llbc:symbolic:fp-inside-region-abs}) & $\Omega[.]$, $V[.]$ (\S9.5, p.~170) & The thesis writes these with its hole notation; ours defines them. \\
\midrule
\texttt{push\_anon} (\Cref{def:llbc:symbolic:push-anon}) & $\Omega, \_ \to v$ (Fig.~9.2, p.~160) & We store anonymous values in a list. \\
\midrule
\texttt{push\_abs} (\Cref{def:llbc:symbolic:push-abs}) & $\Omega, A$ (Fig.~10.1, p.~178) &  \\
\midrule
\texttt{op\_evals}, \texttt{rv\_evals}, \texttt{shr\_borrow\_at} (\Cref{fig:llbc:op-evals,fig:llbc:symbolic:rv-evals,fig:llbc:symbolic:shr-borrow-at}) & rvalue evaluation (Fig.~9.5, p.~164) & The thesis has one judgement for operands and rvalues; ours has two. \\
\midrule
--- & copy (Fig.~9.6, p.~165) & Not in the paper's calculus. \\
\midrule
\texttt{stmt\_evals}, \texttt{stmt\_evals\_lifted}, \texttt{unit\_states}, \texttt{non\_unit}, \texttt{call\_evals}, \texttt{assign\_evals}, \texttt{assign\_val\_evals} (\Cref{fig:llbc:symbolic:stmt-evals,def:llbc:symbolic:unit-results,def:llbc:symbolic:call-evals,fig:llbc:assign}) & statement evaluation (Fig.~9.7, p.~166; Fig.~10.2, p.~179; Fig.~10.3, p.~181) & \texttt{push\_stack} is inline in our E-Call, \texttt{pop\_stack} is \texttt{exit\_fn} and \texttt{pop\_frame}; \texttt{call\_evals} (LLBC\#) and \texttt{assign\_evals}, \texttt{assign\_val\_evals} hold the premises of E-Call and E-Assign. \\
\midrule
\texttt{env\_reorg}, \texttt{give\_back}, \texttt{end\_elem}, \texttt{end\_abstraction} (\Cref{fig:llbc:symbolic:env-reorg,fig:llbc:symbolic:give-back,fig:llbc:symbolic:end-elem,def:llbc:symbolic:end-abstraction}) & reorganisation (Fig.~9.10, p.~169; Fig.~10.2, p.~179) & \texttt{give\_back} and \texttt{end\_elem} are inline in the thesis's Reorg-End-* rules. \\
\midrule
\texttt{inst\_sig} (\Cref{def:llbc:symbolic:inst-sig}) & inst\_sig (Fig.~10.3, p.~181) & Ours has one region; the thesis a vector of regions $\vec{\rho}$. \\
\midrule
\texttt{proj\_input} (\Cref{fig:llbc:symbolic:proj-input}) & proj\_input (Fig.~10.4, p.~182) & Ours has one region; the thesis a vector of regions $\vec{\rho}$. \\
\midrule
\texttt{proj\_output}, \texttt{val\_abs\_disjoint} (\Cref{fig:llbc:symbolic:proj-output,def:llbc:symbolic:val-abs-disjoint}) & proj\_output (Fig.~10.5, p.~182) & Ours has one region; the thesis a vector of regions $\vec{\rho}$. \\
\midrule
\texttt{le\_sharp}, \texttt{le1}, \texttt{abs\_covers}, \texttt{replace\_values}, \texttt{replace\_values\_at} (\Cref{app:llbc:symbolic:refinement,fig:llbc:symbolic:le1,fig:llbc:symbolic:le1-labels}) & refinement (Fig.~11.8, p.~205; Fig.~11.9, p.~206) & \texttt{le\_sharp} is the closure of le1. \\
\midrule
\texttt{to\_abs} (\Cref{fig:llbc:symbolic:to-abs}) & to-abs (Fig.~11.10, p.~206) &  \\
\midrule
\texttt{merge\_abs} (\Cref{fig:llbc:symbolic:merge-abs}) & merge (Fig.~11.11, p.~207) &  \\
\midrule
\texttt{borrow\_checks}, \texttt{fn\_res\_refines\_state} (\Cref{def:llbc:symbolic:borrow-checks}) & borrow\_checks (Fig.~11.12, p.~208) &  \\
\midrule
\texttt{fn\_init\_end\_states}, \texttt{inst\_arg} (\Cref{def:llbc:symbolic:fn-init-end-states,def:llbc:symbolic:inst-arg}) & init (Fig.~11.13, p.~209); final (Fig.~11.13, p.~209) & Ours has one region; the thesis a vector of regions $\vec{\rho}$. \texttt{inst\_arg} is what init and final share. \\
\midrule
\texttt{init\_extern}, \texttt{abs\_disjoint} (\Cref{fig:llbc:symbolic:init-extern,def:llbc:symbolic:abs-disjoint}) & init\_extern (Fig.~11.14, p.~209) & Ours has one region; the thesis a vector of regions $\vec{\rho}$. \\
\midrule
\texttt{ty\_of}, \texttt{val\_ty} (\Cref{def:llbc:symbolic:tyof,def:llbc:symbolic:val-ty}) & $v : \tau$ (Fig.~9.4, p.~162) & The thesis gives no rules for it; ours defines it. \\
\midrule
\texttt{has\_outer\_loan} (\Cref{def:llbc:outer-loan}) & no outer loans in $v$ (\S9.4.2, p.~166) & Defined in the thesis's text (\S{}9.4.2, p. 166); ours as two rules. \\
\midrule
\texttt{bot\_of} (\Cref{def:llbc:botof}) & $\bot$ (Fig.~9.2, p.~160) & Ours is a tree: no bottom of pair type. \\
\midrule
\texttt{sym\_of} (\Cref{def:llbc:symbolic:symof}) & $\sigma : \tau$ (Fig.~10.1, p.~178) & Ours is a tree: no symbolic value of pair type. \\
\midrule
\texttt{strip\_outer\_loans} (\Cref{def:llbc:symbolic:strip-outer-loans}) & --- & Ours only; Le-Reborrow-SharedBorrow uses it. \\
\bottomrule
\end{longtable}
}
\paragraph{Statements.}
{\small
\begin{longtable}{@{}>{\raggedright\arraybackslash}p{0.28\linewidth}>{\raggedright\arraybackslash}p{0.29\linewidth}>{\raggedright\arraybackslash}p{0.35\linewidth}@{}}
\toprule
Ours & Thesis & Note \\
\midrule
\endhead
\rulename{E-Return} (\Cref{fig:llbc:symbolic:stmt-evals}) & \rulename{E-Return} (Fig.~9.7, p.~166) &  \\
\rulename{E-Panic} (\Cref{fig:llbc:symbolic:stmt-evals}) & \rulename{E-Panic} (Fig.~9.7, p.~166) &  \\
\rulename{E-Reorg} (\Cref{fig:llbc:symbolic:stmt-evals}) & \rulename{E-Reorg} (Fig.~9.7, p.~166) &  \\
\rulename{E-Seq}, \texttt{unit\_states}, \texttt{non\_unit} (\Cref{fig:llbc:symbolic:stmt-evals,def:llbc:symbolic:unit-results}) & \rulename{E-Seq-Symbolic} (Fig.~10.2, p.~179) &  \\
\rulename{E-Match-Left}, \rulename{E-Match-Right} (\Cref{fig:llbc:symbolic:stmt-evals}) & \rulename{E-Match} (Fig.~9.7, p.~166) & The thesis has one rule. \\
\rulename{E-Match-Symbolic}, \rulename{E-Match-Symbolic-Shared} (\Cref{fig:llbc:symbolic:stmt-evals}) & \rulename{E-Match-Symbolic} (Fig.~10.2, p.~179) & The thesis has one rule. \\
\rulename{E-Assign}, \rulename{Assign-Evals}, \rulename{Assign-Val-Evals} (\Cref{fig:llbc:symbolic:stmt-evals,fig:llbc:assign}) & \rulename{E-Assign} (Fig.~9.7, p.~166) &  \\
\rulename{E-Call}, \rulename{Call} (\Cref{fig:llbc:symbolic:stmt-evals,def:llbc:symbolic:call-evals}) & \rulename{E-Call-Symbolic} (Fig.~10.3, p.~181) &  \\
\rulename{E-Nil} (\Cref{fig:llbc:symbolic:stmt-evals}) & --- & No thesis rule, though its grammar has the empty statement (Fig. 9.1). \\
--- & \rulename{E-Break}, \rulename{E-Continue} (Fig.~9.7, p.~166); \rulename{E-Loop-Break-Inner}, \rulename{E-Loop-Break-Outer}, \rulename{E-Loop-Continue-Inner}, \rulename{E-Loop-Continue-Outer}, \rulename{E-Loop-Panic}, \rulename{E-Loop-Return} (Fig.~9.9, p.~168) & Loops: not in the paper's calculus \\
--- & \rulename{E-Box-Free} (Fig.~9.7, p.~166) & Boxes: not in the paper's calculus \\
--- & \rulename{E-IfThenElse-Symbolic} (Fig.~10.2, p.~179) & If-then-else: not in the paper's calculus \\
\bottomrule
\end{longtable}
}
\paragraph{Rvalues.}
{\small
\begin{longtable}{@{}>{\raggedright\arraybackslash}p{0.28\linewidth}>{\raggedright\arraybackslash}p{0.29\linewidth}>{\raggedright\arraybackslash}p{0.35\linewidth}@{}}
\toprule
Ours & Thesis & Note \\
\midrule
\endhead
\rulename{E-Op} (\Cref{fig:llbc:symbolic:rv-evals}) & --- & Implicit in the thesis: an operand is an rvalue. \\
\rulename{E-Move} (\Cref{fig:llbc:op-evals}) & \rulename{E-Move} (Fig.~9.5, p.~164) &  \\
\rulename{E-Pair} (\Cref{fig:llbc:op-evals}) & \rulename{E-Pair} (Fig.~9.5, p.~164) &  \\
\rulename{E-Sum-Left} (\Cref{fig:llbc:op-evals}) & \rulename{E-Sum-Left} (Fig.~9.5, p.~164) &  \\
\rulename{E-Sum-Right} (\Cref{fig:llbc:op-evals}) & \rulename{E-Sum-Right} (Fig.~9.5, p.~164) &  \\
\rulename{E-Unit} (\Cref{fig:llbc:op-evals}) & --- & No thesis rule, though its grammar has the unit operand (Fig. 9.1). \\
\rulename{E-MutBorrow} (\Cref{fig:llbc:symbolic:rv-evals}) & \rulename{E-MutBorrow} (Fig.~9.5, p.~164) &  \\
\rulename{E-SharedBorrow-Reuse}, \rulename{E-SharedBorrow}, \rulename{Shared-Borrow-Wrap}, \rulename{Shared-Borrow-Abs} (\Cref{fig:llbc:symbolic:rv-evals,fig:llbc:symbolic:shr-borrow-at}) & \rulename{E-SharedBorrow} (Fig.~9.5, p.~164) & The thesis has one rule; ours splits it by where the value is: already shared, elsewhere, or under a shared loan held by a region abstraction (a new abstraction). \\
--- & \rulename{E-Copy} (Fig.~9.5, p.~164) & Copies: not in the paper's calculus \\
\midrule
--- & \rulename{Copy-SharedBorrow}, \rulename{Copy-SharedLoan}, \rulename{Copy-Scalar}, \rulename{Copy-Sum}, \rulename{Copy-Pair} (Fig.~9.6, p.~165); \rulename{Copy-Symbolic} (Fig.~10.2, p.~179) & Copies: not in the paper's calculus \\
\bottomrule
\end{longtable}
}
\paragraph{Places.}
{\small
\begin{longtable}{@{}>{\raggedright\arraybackslash}p{0.28\linewidth}>{\raggedright\arraybackslash}p{0.29\linewidth}>{\raggedright\arraybackslash}p{0.35\linewidth}@{}}
\toprule
Ours & Thesis & Note \\
\midrule
\endhead
\rulename{Place-Address}, \rulename{Read-At}, \rulename{Read}, \rulename{Read-Write} (\Cref{def:llbc:place-address,def:llbc:read-at,def:llbc:read,def:llbc:read-write}) & \rulename{Read} (Fig.~9.3, p.~161); \rulename{Write} (Fig.~9.4, p.~162) & Read: Place-Address, Read-At, Read. Write: Place-Address, Read-At, Read-Write. \\
\rulename{PA-Base} (\Cref{fig:llbc:place-address-aux}) & \rulename{R-Base} (Fig.~9.3, p.~161); \rulename{W-Base} (Fig.~9.4, p.~162) &  \\
\rulename{PA-Proj-Sum-Left} (\Cref{fig:llbc:place-address-aux}) & \rulename{R-ProjSum-Left} (Fig.~9.3, p.~161); \rulename{W-ProjSum-Left} (Fig.~9.4, p.~162) &  \\
\rulename{PA-Proj-Sum-Right} (\Cref{fig:llbc:place-address-aux}) & \rulename{R-ProjSum-Right} (Fig.~9.3, p.~161); \rulename{W-ProjSum-Right} (Fig.~9.4, p.~162) &  \\
\rulename{PA-Proj-Pair-Left} (\Cref{fig:llbc:place-address-aux}) & \rulename{R-ProjPair-Left} (Fig.~9.3, p.~161); \rulename{W-ProjPair-Left} (Fig.~9.4, p.~162) &  \\
\rulename{PA-Proj-Pair-Right} (\Cref{fig:llbc:place-address-aux}) & \rulename{R-ProjPair-Right} (Fig.~9.3, p.~161); \rulename{W-ProjPair-Right} (Fig.~9.4, p.~162) &  \\
\rulename{PA-SharedLoan} (\Cref{fig:llbc:place-address-aux}) & \rulename{R-SharedLoan} (Fig.~9.3, p.~161); \rulename{W-SharedLoan} (Fig.~9.4, p.~162) &  \\
\rulename{PA-Deref-MutBorrow} (\Cref{fig:llbc:place-address-aux}) & \rulename{R-Deref-MutBorrow} (Fig.~9.3, p.~161); \rulename{W-Deref-MutBorrow} (Fig.~9.4, p.~162) &  \\
\rulename{PA-Deref-SharedBorrow} (\Cref{fig:llbc:place-address-aux}) & \rulename{R-Deref-SharedBorrow} (Fig.~9.3, p.~161); \rulename{W-Deref-SharedBorrow} (Fig.~9.4, p.~162) &  \\
--- & \rulename{R-Deref-Box} (Fig.~9.3, p.~161); \rulename{W-Deref-Box} (Fig.~9.4, p.~162) & Boxes: not in the paper's calculus \\
\bottomrule
\end{longtable}
}
\paragraph{Reorganisation.}
{\small
\begin{longtable}{@{}>{\raggedright\arraybackslash}p{0.28\linewidth}>{\raggedright\arraybackslash}p{0.29\linewidth}>{\raggedright\arraybackslash}p{0.35\linewidth}@{}}
\toprule
Ours & Thesis & Note \\
\midrule
\endhead
\rulename{Reorg-End-MutBorrow} (\Cref{fig:llbc:symbolic:env-reorg}) & \rulename{Reorg-End-MutBorrow\#} (Fig.~10.2, p.~179) &  \\
\rulename{Reorg-End-SharedBorrow} (\Cref{fig:llbc:symbolic:env-reorg}) & \rulename{Reorg-End-SharedReservedBorrow\#} (Fig.~10.2, p.~179) & The thesis's rule also ends reserved borrows. \\
\rulename{Reorg-End-SharedLoan} (\Cref{fig:llbc:symbolic:env-reorg}) & \rulename{Reorg-End-SharedLoan} (Fig.~9.10, p.~169) &  \\
\rulename{Give-Back-In-Place}, \rulename{Give-Back-Abs} (\Cref{fig:llbc:symbolic:give-back}) & --- & The thesis writes the value in place wherever the loan is; ours drops a loan held by an abstraction, with its value, which must hold no location. \\
\rulename{Reorg-End-Abstraction}, \rulename{End-Abs}, \rulename{End-Elem-Shared}, \rulename{End-Elem-Mut} (\Cref{fig:llbc:symbolic:env-reorg,def:llbc:symbolic:end-abstraction,fig:llbc:symbolic:end-elem}) & \rulename{Reorg-End-Abstraction} (Fig.~10.2, p.~179) & The thesis has one rule, which also drops elements holding no loan or borrow. \\
--- & \rulename{Reorg-None}, \rulename{Reorg-Seq} (Fig.~9.10, p.~169) & Ours: one reorganisation step per E-Reorg. \\
--- & \rulename{Reorg-SymbolicPair} (Fig.~10.2, p.~179) & Ours has no symbolic value of pair type. \\
--- & \rulename{Reorg-SymbolicBox} (Fig.~10.2, p.~179) & Boxes: not in the paper's calculus \\
\bottomrule
\end{longtable}
}
\paragraph{Calls.}
{\small
\begin{longtable}{@{}>{\raggedright\arraybackslash}p{0.28\linewidth}>{\raggedright\arraybackslash}p{0.29\linewidth}>{\raggedright\arraybackslash}p{0.35\linewidth}@{}}
\toprule
Ours & Thesis & Note \\
\midrule
\endhead
\rulename{Inst-Sig} (\Cref{def:llbc:symbolic:inst-sig}) & \rulename{InstSig} (Fig.~10.3, p.~181) &  \\
\midrule
\rulename{Proj-Input-Ignore} (\Cref{fig:llbc:symbolic:proj-input}) & \rulename{ProjInput-Ignore} (Fig.~10.4, p.~182) &  \\
\rulename{Proj-Input-Pair} (\Cref{fig:llbc:symbolic:proj-input}) & \rulename{ProjInput-Pair} (Fig.~10.4, p.~182) &  \\
\rulename{Proj-Input-MutBorrow} (\Cref{fig:llbc:symbolic:proj-input}) & \rulename{ProjInput-Mut} (Fig.~10.4, p.~182) &  \\
\rulename{Proj-Input-SharedBorrow} (\Cref{fig:llbc:symbolic:proj-input}) & \rulename{ProjInput-Shared} (Fig.~10.4, p.~182) &  \\
\midrule
\rulename{Proj-Output-Symbolic} (\Cref{fig:llbc:symbolic:proj-output}) & \rulename{ProjOutput-Symbolic} (Fig.~10.5, p.~182) &  \\
\rulename{Proj-Output-Pair}, \texttt{val\_abs\_disjoint} (\Cref{fig:llbc:symbolic:proj-output,def:llbc:symbolic:val-abs-disjoint}) & \rulename{ProjOutput-Pair} (Fig.~10.5, p.~182) & We require disjoint labels for the two components. The thesis requires fresh labels. \\
\rulename{Proj-Output-Shared} (\Cref{fig:llbc:symbolic:proj-output}) & \rulename{ProjOutput-Shared} (Fig.~10.5, p.~182) &  \\
\rulename{Proj-Output-Mut} (\Cref{fig:llbc:symbolic:proj-output}) & \rulename{ProjOutput-Mut} (Fig.~10.5, p.~182) &  \\
\bottomrule
\end{longtable}
}
\paragraph{Refinement.}
{\small
\begin{longtable}{@{}>{\raggedright\arraybackslash}p{0.28\linewidth}>{\raggedright\arraybackslash}p{0.29\linewidth}>{\raggedright\arraybackslash}p{0.35\linewidth}@{}}
\toprule
Ours & Thesis & Note \\
\midrule
\endhead
\texttt{le\_sharp} (\Cref{app:llbc:symbolic:refinement}) & --- & The thesis takes the closure in its text (\S{}11.3.2, p. 204). \\
\rulename{Le-To-Symbolic} (\Cref{fig:llbc:symbolic:le1}) & \rulename{Le-ToSymbolic} (Fig.~11.8, p.~205) & Ours: not for a value of pair type. \\
\rulename{Le-To-Abs} (\Cref{fig:llbc:symbolic:le1}) & \rulename{Le-ToAbs} (Fig.~11.8, p.~205) & Ours: a shared loan only with a location-free content. \\
\rulename{Le-Move-Value} (\Cref{fig:llbc:symbolic:le1}) & \rulename{Le-MoveValue} (Fig.~11.8, p.~205) & Ours: the path passes no mutable borrow, shared loan or region abstraction, and not for a value of pair type. \\
\rulename{Le-Merge-Abs} (\Cref{fig:llbc:symbolic:le1}) & \rulename{Le-MergeAbs} (Fig.~11.8, p.~205) &  \\
\rulename{Le-Reborrow-MutBorrow} (\Cref{fig:llbc:symbolic:le1-labels}) & \rulename{Le-Reborrow-MutBorrow} (Fig.~11.8, p.~205) &  \\
\rulename{Le-End-SharedLoan-Abs} (\Cref{fig:llbc:symbolic:le1-labels}) & \rulename{Le-Abs-End-SharedLoan} (Fig.~11.8, p.~205) & Ours: the label absent from the result, and the loan element dropped with its content. \\
\rulename{Le-End-SharedLoan-Inner} (\Cref{fig:llbc:symbolic:le1-labels}) & --- & Ours only: a loan under a shared loan, outside region abstractions. \\
\rulename{Le-Reborrow-SharedBorrow} (\Cref{fig:llbc:symbolic:le1-labels}) & \rulename{Le-Reborrow-SharedBorrow} (Fig.~11.8, p.~205) & We create the loan over \texttt{strip\_outer\_loans} v. The thesis uses $\sigma$. \\
--- & \rulename{Le-AnonValue} (Fig.~11.8, p.~205) & Not in our LLBC\#. \\
--- & \rulename{Le-Fresh-SharedBorrow} (Fig.~11.8, p.~205) & Not in our LLBC\#. \\
\rulename{Le-Abs-Drop-Covered-Borrow} (\Cref{fig:llbc:symbolic:le1-labels}) & --- &  \\
\rulename{Le-Perm-Abs}, \rulename{Le-Perm-In-Abs}, \rulename{Le-Perm-Anons} (\Cref{fig:llbc:symbolic:le1}) & --- & The thesis's states and abstractions are unordered; ours are lists. \\
--- & \rulename{Le-ClearAbs} (Fig.~11.8, p.~205) & Not in ours. \\
--- & \rulename{Le-Abs-ClearValue} (Fig.~11.8, p.~205) & Not in ours. \\
--- & \rulename{Le-Abs-End-DupSharedBorrow} (Fig.~11.8, p.~205) & Not in ours. \\
--- & \rulename{Le-Fresh-MutLoan}, \rulename{Le-Fresh-SharedLoan} (Fig.~11.8, p.~205) & Not in ours; the thesis uses them at loop joins. \\
--- & \rulename{Le-Abs-DeconstructPair}, \rulename{Le-Abs-DeconstructSum} (Fig.~11.8, p.~205) & Not in ours. \\
--- & \rulename{Le-Reborrow-SharedLoan} (Fig.~11.8, p.~205) & Not in ours. \\
\midrule
\rulename{To-Abs-Empty} (\Cref{fig:llbc:symbolic:to-abs}) & \rulename{ToAbs-Empty} (Fig.~11.10, p.~206) &  \\
\rulename{To-Abs-Left}, \rulename{To-Abs-Right} (\Cref{fig:llbc:symbolic:to-abs}) & \rulename{ToAbs-Sum} (Fig.~11.10, p.~206) &  \\
\rulename{To-Abs-MutLoan} (\Cref{fig:llbc:symbolic:to-abs}) & \rulename{ToAbs-MutLoan} (Fig.~11.10, p.~206) &  \\
\rulename{To-Abs-SharedLoan} (\Cref{fig:llbc:symbolic:to-abs}) & \rulename{ToAbs-SharedLoan} (Fig.~11.10, p.~206) &  \\
\rulename{To-Abs-Pair} (\Cref{fig:llbc:symbolic:to-abs}) & \rulename{ToAbs-Pair} (Fig.~11.10, p.~206) &  \\
\rulename{To-Abs-SharedBorrow} (\Cref{fig:llbc:symbolic:to-abs}) & \rulename{ToAbs-SharedBorrow} (Fig.~11.10, p.~206) &  \\
\rulename{To-Abs-MutBorrow} (\Cref{fig:llbc:symbolic:to-abs}) & \rulename{ToAbs-MutBorrow} (Fig.~11.10, p.~206) &  \\
--- & \rulename{ToAbs-Box} (Fig.~11.10, p.~206) & Boxes: not in the paper's calculus \\
\midrule
\rulename{Merge-Abs-Union} (\Cref{fig:llbc:symbolic:merge-abs}) & \rulename{MergeAbs-Union} (Fig.~11.11, p.~207) &  \\
\rulename{Merge-Abs-Mut} (\Cref{fig:llbc:symbolic:merge-abs}) & \rulename{MergeAbs-Mut} (Fig.~11.11, p.~207) &  \\
\rulename{Merge-Abs-Shared} (\Cref{fig:llbc:symbolic:merge-abs}) & \rulename{MergeAbs-Shared} (Fig.~11.11, p.~207) &  \\
\bottomrule
\end{longtable}
}
\paragraph{Borrow checking.}
{\small
\begin{longtable}{@{}>{\raggedright\arraybackslash}p{0.28\linewidth}>{\raggedright\arraybackslash}p{0.29\linewidth}>{\raggedright\arraybackslash}p{0.35\linewidth}@{}}
\toprule
Ours & Thesis & Note \\
\midrule
\endhead
\texttt{borrow\_checks}, \texttt{fn\_res\_refines\_state} (\Cref{def:llbc:symbolic:borrow-checks}) & \rulename{borrow\_checks} (Fig.~11.12, p.~208) &  \\
\midrule
\rulename{Fn-Init-End-States}, \rulename{Inst-Arg} (\Cref{def:llbc:symbolic:fn-init-end-states,def:llbc:symbolic:inst-arg}) & \rulename{Init}, \rulename{Final} (Fig.~11.13, p.~209) &  \\
\midrule
\rulename{Init-Extern-Symbolic} (\Cref{fig:llbc:symbolic:init-extern}) & \rulename{InitExtern-Symbolic} (Fig.~11.14, p.~209) &  \\
\rulename{Init-Extern-Pair}, \texttt{abs\_disjoint} (\Cref{fig:llbc:symbolic:init-extern,def:llbc:symbolic:abs-disjoint}) & \rulename{InitExtern-Pair} (Fig.~11.14, p.~209) & We require disjoint labels for the two components. The thesis requires fresh labels. \\
\rulename{Init-Extern-Shared} (\Cref{fig:llbc:symbolic:init-extern}) & \rulename{InitExtern-Shared} (Fig.~11.14, p.~209) &  \\
\rulename{Init-Extern-Mut} (\Cref{fig:llbc:symbolic:init-extern}) & \rulename{InitExtern-Mut} (Fig.~11.14, p.~209) &  \\
\bottomrule
\end{longtable}
}
\paragraph{States and values.}
{\small
\begin{longtable}{@{}>{\raggedright\arraybackslash}p{0.28\linewidth}>{\raggedright\arraybackslash}p{0.29\linewidth}>{\raggedright\arraybackslash}p{0.35\linewidth}@{}}
\toprule
Ours & Thesis & Note \\
\midrule
\endhead
\rulename{FP-Applies}, \rulename{DP-Nil}, \rulename{DP-Pair-0}, \rulename{DP-Pair-1}, \rulename{DP-Left}, \rulename{DP-Right}, \rulename{DP-Unloan}, \rulename{DP-Unbor}, \texttt{state\_get\_va}, \rulename{State-Set-Anon}, \rulename{State-Set-Var}, \rulename{State-Set-Abs}, \rulename{State-Updates}, \rulename{V-Base}, \rulename{V-Sum-Left}, \rulename{V-Sum-Right}, \rulename{V-Pair-Left}, \rulename{V-Pair-Right}, \rulename{V-SharedUnloan}, \rulename{V-Unbor}, \texttt{get\_abs}, \rulename{In-Abs} (\Cref{def:llbc:fp-applies,fig:llbc:dp-applies,def:llbc:symbolic:state-get,fig:llbc:symbolic:state-set,def:llbc:state-updates,fig:llbc:val-updates,def:llbc:symbolic:fp-inside-region-abs}) & $\Omega[.]$, $V[.]$ (\S9.5, p.~170) & The thesis writes these with its hole notation; ours defines them. \\
\midrule
\texttt{push\_anon} (\Cref{def:llbc:symbolic:push-anon}) & $\Omega, \_ \to v$ (Fig.~9.2, p.~160) & We store anonymous values in a list. \\
\midrule
\texttt{push\_abs} (\Cref{def:llbc:symbolic:push-abs}) & $\Omega, A$ (Fig.~10.1, p.~178) &  \\
\midrule
\texttt{ty\_of}, \texttt{val\_ty} (\Cref{def:llbc:symbolic:tyof,def:llbc:symbolic:val-ty}) & $v : \tau$ (Fig.~9.4, p.~162) & The thesis gives no rules for it; ours defines it. \\
\midrule
\rulename{Outer-Loan-Mut}, \rulename{Outer-Loan-Shared} (\Cref{def:llbc:outer-loan}) & no outer loans in $v$ (\S9.4.2, p.~166) & Defined in the thesis's text (\S{}9.4.2, p. 166); ours as two rules. \\
\midrule
\texttt{bot\_of} (\Cref{def:llbc:botof}) & $\bot$ (Fig.~9.2, p.~160) & Ours is a tree: no bottom of pair type. \\
\midrule
\texttt{sym\_of} (\Cref{def:llbc:symbolic:symof}) & $\sigma : \tau$ (Fig.~10.1, p.~178) & Ours is a tree: no symbolic value of pair type. \\
\midrule
\texttt{strip\_outer\_loans} (\Cref{def:llbc:symbolic:strip-outer-loans}) & --- & Ours only; Le-Reborrow-SharedBorrow uses it. \\
\bottomrule
\end{longtable}
}

\subsection{Combined semantics (\LLBCPlus)}\label{app:thesis:llbc-plus}
\paragraph{Judgements.}
{\small
\begin{longtable}{@{}>{\raggedright\arraybackslash}p{0.28\linewidth}>{\raggedright\arraybackslash}p{0.29\linewidth}>{\raggedright\arraybackslash}p{0.35\linewidth}@{}}
\toprule
Ours & Thesis & Note \\
\midrule
\endhead
\texttt{place\_address}, \texttt{place\_address\_aux}, \texttt{read}, \texttt{read\_at}, \texttt{read\_write} (\Cref{def:llbc:place-address,fig:llbc:place-address-aux,def:llbc:read,def:llbc:read-at,def:llbc:read-write}) & read (Fig.~9.3, p.~161); read along a path (Fig.~9.3, p.~161); write (Fig.~9.4, p.~162); write along a path (Fig.~9.4, p.~162) & Ours computes an address, then reads or writes at it; the thesis follows the path while reading or writing. \\
\midrule
\texttt{fp\_applies}, \texttt{dp\_applies}, \texttt{state\_get\_va}, \texttt{state\_set\_va}, \texttt{state\_updates}, \texttt{val\_updates}, \texttt{get\_abs}, \texttt{fp\_inside\_region\_abs} (\Cref{def:llbc:fp-applies,fig:llbc:dp-applies,def:llbc:combined:state-get,fig:llbc:combined:state-set,def:llbc:state-updates,fig:llbc:val-updates,def:llbc:symbolic:state-get,def:llbc:symbolic:fp-inside-region-abs}) & $\Omega[.]$, $V[.]$ (\S9.5, p.~170) & The thesis writes these with its hole notation; ours defines them. \\
\midrule
\texttt{push\_anon} (\Cref{def:llbc:combined:push}) & $\Omega, \_ \to v$ (Fig.~9.2, p.~160) & We store anonymous values in a list. \\
\midrule
\texttt{push\_abs} (\Cref{def:llbc:combined:push}) & $\Omega, A$ (Fig.~10.1, p.~178) &  \\
\midrule
\texttt{op\_evals}, \texttt{rv\_evals}, \texttt{shr\_borrow\_at} (\Cref{fig:llbc:op-evals,fig:llbc:symbolic:rv-evals,fig:llbc:symbolic:shr-borrow-at}) & rvalue evaluation (Fig.~9.5, p.~164) & The thesis has one judgement for operands and rvalues; ours has two. \\
\midrule
--- & copy (Fig.~9.6, p.~165) & Not in the paper's calculus. \\
\midrule
\texttt{stmt\_evals}, \texttt{stmt\_evals\_lifted}, \texttt{unit\_states}, \texttt{non\_unit}, \texttt{assign\_evals}, \texttt{assign\_val\_evals}, \texttt{exit\_fn}, \texttt{exit\_fn\_lifted}, \texttt{pop\_frame} (\Cref{fig:llbc:combined:stmt-evals,def:llbc:combined:unit-results,fig:llbc:assign,fig:llbc:exit-fn,def:llbc:combined:pop-frame}) & statement evaluation (Fig.~9.7, p.~166; Fig.~9.8, p.~168; Fig.~10.2, p.~179; Fig.~11.6, p.~200); push\_stack (Fig.~9.8, p.~168); pop\_stack (Fig.~9.8, p.~168) & \texttt{push\_stack} is inline in our E-Call, \texttt{pop\_stack} is \texttt{exit\_fn} and \texttt{pop\_frame}; \texttt{call\_evals} (LLBC\#) and \texttt{assign\_evals}, \texttt{assign\_val\_evals} hold the premises of E-Call and E-Assign. \\
\midrule
\texttt{env\_reorg}, \texttt{give\_back}, \texttt{end\_elem}, \texttt{end\_abstraction} (\Cref{fig:llbc:symbolic:env-reorg,fig:llbc:symbolic:give-back,fig:llbc:symbolic:end-elem,def:llbc:symbolic:end-abstraction}) & reorganisation (Fig.~9.10, p.~169; Fig.~10.2, p.~179) & \texttt{give\_back} and \texttt{end\_elem} are inline in the thesis's Reorg-End-* rules. \\
\midrule
\texttt{le\_sharp}, \texttt{le1}, \texttt{abs\_covers}, \texttt{replace\_values}, \texttt{replace\_values\_at}, \texttt{relabel\_state}, \texttt{relabel\_val}, \texttt{ren}, \texttt{map\_state}, \texttt{map\_frame} (\Cref{app:llbc:combined:refinement,fig:llbc:combined:le1,fig:llbc:combined:le1-labels,def:llbc:combined:relabel-state,def:llbc:symbolic:relabel-val,def:llbc:relabel}) & refinement (Fig.~11.8, p.~205; Fig.~11.4, p.~193; Fig.~11.9, p.~206) & \texttt{le\_sharp} is the closure of le1; the relabelling is for Le-Rename. \\
\midrule
\texttt{to\_abs} (\Cref{fig:llbc:symbolic:to-abs}) & to-abs (Fig.~11.10, p.~206) &  \\
\midrule
\texttt{merge\_abs} (\Cref{fig:llbc:symbolic:merge-abs}) & merge (Fig.~11.11, p.~207) &  \\
\midrule
\texttt{ty\_of}, \texttt{val\_ty} (\Cref{def:llbc:symbolic:tyof,def:llbc:symbolic:val-ty}) & $v : \tau$ (Fig.~9.4, p.~162) & The thesis gives no rules for it; ours defines it. \\
\midrule
\texttt{has\_outer\_loan} (\Cref{def:llbc:outer-loan}) & no outer loans in $v$ (\S9.4.2, p.~166) & Defined in the thesis's text (\S{}9.4.2, p. 166); ours as two rules. \\
\midrule
\texttt{bot\_of} (\Cref{def:llbc:botof}) & $\bot$ (Fig.~9.2, p.~160) & Ours is a tree: no bottom of pair type. \\
\midrule
\texttt{sym\_of} (\Cref{def:llbc:symbolic:symof}) & $\sigma : \tau$ (Fig.~10.1, p.~178) & Ours is a tree: no symbolic value of pair type. \\
\midrule
\texttt{strip\_outer\_loans} (\Cref{def:llbc:symbolic:strip-outer-loans}) & --- & Ours only; Le-Reborrow-SharedBorrow uses it. \\
\bottomrule
\end{longtable}
}
\paragraph{Statements.}
We compare the step-indexed statement rules. A $^\dagger$ marks rules whose
step-indexed forms appear explicitly in Fig.~11.6 (p.~200); the step index
is implicit in the other thesis rules.
{\small
\begin{longtable}{@{}>{\raggedright\arraybackslash}p{0.28\linewidth}>{\raggedright\arraybackslash}p{0.29\linewidth}>{\raggedright\arraybackslash}p{0.35\linewidth}@{}}
\toprule
Ours & Thesis & Note \\
\midrule
\endhead
\rulename{E-Step-Zero} (\Cref{fig:llbc:combined:stmt-evals}) & \rulename{E-Step-Zero}$^\dagger$ (Fig.~11.6, p.~200) &  \\
\rulename{E-Return} (\Cref{fig:llbc:combined:stmt-evals}) & \rulename{E-Return}$^\dagger$ (Fig.~9.7, p.~166) &  \\
\rulename{E-Panic} (\Cref{fig:llbc:combined:stmt-evals}) & \rulename{E-Panic} (Fig.~9.7, p.~166) &  \\
\rulename{E-Reorg} (\Cref{fig:llbc:combined:stmt-evals}) & \rulename{E-Reorg} (Fig.~9.7, p.~166) &  \\
\rulename{E-Seq}, \texttt{unit\_states}, \texttt{non\_unit} (\Cref{fig:llbc:combined:stmt-evals,def:llbc:combined:unit-results}) & \rulename{E-Seq-Unit}$^\dagger$, \rulename{E-Seq-Propagate} (Fig.~9.7, p.~166); \rulename{E-Seq-Symbolic} (Fig.~10.2, p.~179) & Ours: one rule over lists of results. \\
\rulename{E-Match-Left}, \rulename{E-Match-Right} (\Cref{fig:llbc:combined:stmt-evals-match}) & \rulename{E-Match} (Fig.~9.7, p.~166) & The thesis has one rule. \\
\rulename{E-Match-Symbolic}, \rulename{E-Match-Symbolic-Shared} (\Cref{fig:llbc:combined:stmt-evals-match}) & \rulename{E-Match-Symbolic} (Fig.~10.2, p.~179) & The thesis has one rule. \\
\rulename{E-Assign}, \rulename{Assign-Evals}, \rulename{Assign-Val-Evals} (\Cref{fig:llbc:combined:stmt-evals,fig:llbc:assign}) & \rulename{E-Assign} (Fig.~9.7, p.~166) &  \\
\rulename{E-Call}, \rulename{Exit-No-Fuel}, \rulename{Exit-Panic}, \rulename{Exit-Ret}, \texttt{exit\_fn\_lifted}, \rulename{Pop-Frame} (\Cref{fig:llbc:combined:stmt-evals,fig:llbc:exit-fn,def:llbc:combined:pop-frame}) & \rulename{E-Call}$^\dagger$, \rulename{E-PushStack}, \rulename{E-PopStack} (Fig.~9.8, p.~168) & The Exit-* rules are the three cases of E-Step-Call (Fig. 11.6). \\
\rulename{E-Nil} (\Cref{fig:llbc:combined:stmt-evals}) & --- & No thesis rule, though its grammar has the empty statement (Fig. 9.1). \\
--- & \rulename{E-Break}, \rulename{E-Continue} (Fig.~9.7, p.~166); \rulename{E-Loop-Break-Inner}, \rulename{E-Loop-Break-Outer}, \rulename{E-Loop-Continue-Inner}$^\dagger$, \rulename{E-Loop-Continue-Outer}, \rulename{E-Loop-Panic}, \rulename{E-Loop-Return} (Fig.~9.9, p.~168) & Loops: not in the paper's calculus \\
--- & \rulename{E-IfThenElse-T}, \rulename{E-IfThenElse-F} (Fig.~9.7, p.~166); \rulename{E-IfThenElse-Symbolic} (Fig.~10.2, p.~179) & If-then-else: not in the paper's calculus \\
--- & \rulename{E-Box-New}, \rulename{E-Box-Free} (Fig.~9.7, p.~166) & Boxes: not in the paper's calculus \\
\bottomrule
\end{longtable}
}
\paragraph{Rvalues.}
{\small
\begin{longtable}{@{}>{\raggedright\arraybackslash}p{0.28\linewidth}>{\raggedright\arraybackslash}p{0.29\linewidth}>{\raggedright\arraybackslash}p{0.35\linewidth}@{}}
\toprule
Ours & Thesis & Note \\
\midrule
\endhead
\rulename{E-Op} (\Cref{fig:llbc:symbolic:rv-evals}) & --- & Implicit in the thesis: an operand is an rvalue. \\
\rulename{E-Move} (\Cref{fig:llbc:op-evals}) & \rulename{E-Move} (Fig.~9.5, p.~164) &  \\
\rulename{E-Pair} (\Cref{fig:llbc:op-evals}) & \rulename{E-Pair} (Fig.~9.5, p.~164) &  \\
\rulename{E-Sum-Left} (\Cref{fig:llbc:op-evals}) & \rulename{E-Sum-Left} (Fig.~9.5, p.~164) &  \\
\rulename{E-Sum-Right} (\Cref{fig:llbc:op-evals}) & \rulename{E-Sum-Right} (Fig.~9.5, p.~164) &  \\
\rulename{E-Unit} (\Cref{fig:llbc:op-evals}) & --- & No thesis rule, though its grammar has the unit operand (Fig. 9.1). \\
\rulename{E-MutBorrow} (\Cref{fig:llbc:symbolic:rv-evals}) & \rulename{E-MutBorrow} (Fig.~9.5, p.~164) &  \\
\rulename{E-SharedBorrow-Reuse}, \rulename{E-SharedBorrow}, \rulename{Shared-Borrow-Wrap}, \rulename{Shared-Borrow-Abs} (\Cref{fig:llbc:symbolic:rv-evals,fig:llbc:symbolic:shr-borrow-at}) & \rulename{E-SharedBorrow} (Fig.~9.5, p.~164) & The thesis has one rule; ours splits it by where the value is: already shared, elsewhere, or under a shared loan held by a region abstraction (a new abstraction). \\
--- & \rulename{E-ReservedBorrow} (Fig.~9.5, p.~164) & Reserved (two-phase) borrows: not in the paper's calculus \\
--- & \rulename{E-Copy} (Fig.~9.5, p.~164) & Copies: not in the paper's calculus \\
\midrule
--- & \rulename{Copy-SharedBorrow}, \rulename{Copy-SharedLoan}, \rulename{Copy-Scalar}, \rulename{Copy-Sum}, \rulename{Copy-Pair} (Fig.~9.6, p.~165); \rulename{Copy-Symbolic} (Fig.~10.2, p.~179) & Copies: not in the paper's calculus \\
\bottomrule
\end{longtable}
}
\paragraph{Places.}
{\small
\begin{longtable}{@{}>{\raggedright\arraybackslash}p{0.28\linewidth}>{\raggedright\arraybackslash}p{0.29\linewidth}>{\raggedright\arraybackslash}p{0.35\linewidth}@{}}
\toprule
Ours & Thesis & Note \\
\midrule
\endhead
\rulename{Place-Address}, \rulename{Read-At}, \rulename{Read}, \rulename{Read-Write} (\Cref{def:llbc:place-address,def:llbc:read-at,def:llbc:read,def:llbc:read-write}) & \rulename{Read} (Fig.~9.3, p.~161); \rulename{Write} (Fig.~9.4, p.~162) & Read: Place-Address, Read-At, Read. Write: Place-Address, Read-At, Read-Write. \\
\rulename{PA-Base} (\Cref{fig:llbc:place-address-aux}) & \rulename{R-Base} (Fig.~9.3, p.~161); \rulename{W-Base} (Fig.~9.4, p.~162) &  \\
\rulename{PA-Proj-Sum-Left} (\Cref{fig:llbc:place-address-aux}) & \rulename{R-ProjSum-Left} (Fig.~9.3, p.~161); \rulename{W-ProjSum-Left} (Fig.~9.4, p.~162) &  \\
\rulename{PA-Proj-Sum-Right} (\Cref{fig:llbc:place-address-aux}) & \rulename{R-ProjSum-Right} (Fig.~9.3, p.~161); \rulename{W-ProjSum-Right} (Fig.~9.4, p.~162) &  \\
\rulename{PA-Proj-Pair-Left} (\Cref{fig:llbc:place-address-aux}) & \rulename{R-ProjPair-Left} (Fig.~9.3, p.~161); \rulename{W-ProjPair-Left} (Fig.~9.4, p.~162) &  \\
\rulename{PA-Proj-Pair-Right} (\Cref{fig:llbc:place-address-aux}) & \rulename{R-ProjPair-Right} (Fig.~9.3, p.~161); \rulename{W-ProjPair-Right} (Fig.~9.4, p.~162) &  \\
\rulename{PA-SharedLoan} (\Cref{fig:llbc:place-address-aux}) & \rulename{R-SharedLoan} (Fig.~9.3, p.~161); \rulename{W-SharedLoan} (Fig.~9.4, p.~162) &  \\
\rulename{PA-Deref-MutBorrow} (\Cref{fig:llbc:place-address-aux}) & \rulename{R-Deref-MutBorrow} (Fig.~9.3, p.~161); \rulename{W-Deref-MutBorrow} (Fig.~9.4, p.~162) &  \\
\rulename{PA-Deref-SharedBorrow} (\Cref{fig:llbc:place-address-aux}) & \rulename{R-Deref-SharedBorrow} (Fig.~9.3, p.~161); \rulename{W-Deref-SharedBorrow} (Fig.~9.4, p.~162) &  \\
--- & \rulename{R-Deref-Box} (Fig.~9.3, p.~161); \rulename{W-Deref-Box} (Fig.~9.4, p.~162) & Boxes: not in the paper's calculus \\
\bottomrule
\end{longtable}
}
\paragraph{Reorganisation.}
{\small
\begin{longtable}{@{}>{\raggedright\arraybackslash}p{0.28\linewidth}>{\raggedright\arraybackslash}p{0.29\linewidth}>{\raggedright\arraybackslash}p{0.35\linewidth}@{}}
\toprule
Ours & Thesis & Note \\
\midrule
\endhead
\rulename{Reorg-End-MutBorrow} (\Cref{fig:llbc:symbolic:env-reorg}) & \rulename{Reorg-End-MutBorrow} (Fig.~9.10, p.~169); \rulename{Reorg-End-MutBorrow\#} (Fig.~10.2, p.~179) & Ours is the LLBC\# rule. \\
\rulename{Reorg-End-SharedBorrow} (\Cref{fig:llbc:symbolic:env-reorg}) & \rulename{Reorg-End-SharedReservedBorrow} (Fig.~9.10, p.~169); \rulename{Reorg-End-SharedReservedBorrow\#} (Fig.~10.2, p.~179) & Ours is the LLBC\# rule; the thesis's rules also end reserved borrows. \\
\rulename{Reorg-End-SharedLoan} (\Cref{fig:llbc:symbolic:env-reorg}) & \rulename{Reorg-End-SharedLoan} (Fig.~9.10, p.~169) &  \\
\rulename{Give-Back-In-Place}, \rulename{Give-Back-Abs} (\Cref{fig:llbc:symbolic:give-back}) & --- & The thesis writes the value in place wherever the loan is; ours drops a loan held by an abstraction, with its value, which must hold no location. \\
\rulename{Reorg-End-Abstraction}, \rulename{End-Abs}, \rulename{End-Elem-Shared}, \rulename{End-Elem-Mut} (\Cref{fig:llbc:symbolic:env-reorg,def:llbc:symbolic:end-abstraction,fig:llbc:symbolic:end-elem}) & \rulename{Reorg-End-Abstraction} (Fig.~10.2, p.~179) & The thesis has one rule, which also drops elements holding no loan or borrow. \\
--- & \rulename{Reorg-None}, \rulename{Reorg-Seq} (Fig.~9.10, p.~169) & Ours: one reorganisation step per E-Reorg. \\
--- & \rulename{Reorg-SymbolicPair} (Fig.~10.2, p.~179) & Ours has no symbolic value of pair type. \\
--- & \rulename{Reorg-Activate-Reserved} (Fig.~9.10, p.~169) & Reserved (two-phase) borrows: not in the paper's calculus \\
--- & \rulename{Reorg-SymbolicBox} (Fig.~10.2, p.~179) & Boxes: not in the paper's calculus \\
\bottomrule
\end{longtable}
}
\paragraph{Refinement.}
{\small
\begin{longtable}{@{}>{\raggedright\arraybackslash}p{0.28\linewidth}>{\raggedright\arraybackslash}p{0.29\linewidth}>{\raggedright\arraybackslash}p{0.35\linewidth}@{}}
\toprule
Ours & Thesis & Note \\
\midrule
\endhead
\texttt{le\_sharp} (\Cref{app:llbc:combined:refinement}) & --- & The thesis takes the closure in its text (\S{}11.3.2, p. 204). \\
\rulename{Le-To-Symbolic} (\Cref{fig:llbc:combined:le1}) & \rulename{Le-ToSymbolic} (Fig.~11.8, p.~205) & Ours: not for a value of pair type. \\
\rulename{Le-To-Abs} (\Cref{fig:llbc:combined:le1}) & \rulename{Le-ToAbs} (Fig.~11.8, p.~205) & Ours: a shared loan only with a location-free content. \\
\rulename{Le-Move-Value} (\Cref{fig:llbc:combined:le1}) & \rulename{Le-MoveValue} (Fig.~11.8, p.~205) & Ours: the path passes no mutable borrow, shared loan or region abstraction, and not for a value of pair type. \\
\rulename{Le-Merge-Abs} (\Cref{fig:llbc:combined:le1}) & \rulename{Le-MergeAbs} (Fig.~11.8, p.~205) &  \\
\rulename{Le-Reborrow-MutBorrow} (\Cref{fig:llbc:combined:le1-labels}) & \rulename{Le-Reborrow-MutBorrow} (Fig.~11.8, p.~205) &  \\
\rulename{Le-End-SharedLoan-Abs} (\Cref{fig:llbc:combined:le1-labels}) & \rulename{Le-Abs-End-SharedLoan} (Fig.~11.8, p.~205) & Ours: the label absent from the result, and the loan element dropped with its content. \\
\rulename{Le-End-SharedLoan-Inner} (\Cref{fig:llbc:combined:le1-labels}) & --- & Ours only: a loan under a shared loan, outside region abstractions. \\
\rulename{Le-Reborrow-SharedBorrow} (\Cref{fig:llbc:combined:le1-labels}) & \rulename{Le-Reborrow-SharedBorrow} (Fig.~11.8, p.~205) & We create the loan over \texttt{strip\_outer\_loans} v. The thesis uses $\sigma$. \\
\rulename{Le-Anon-Value} (\Cref{fig:llbc:combined:le1}) & \rulename{Le-AnonValue} (Fig.~11.8, p.~205) & Ours has the thesis's conclusion only. \\
\rulename{Le-Fresh-SharedBorrow} (\Cref{fig:llbc:combined:le1-labels}) & \rulename{Le-Fresh-SharedBorrow} (Fig.~11.8, p.~205) &  \\
\rulename{Le-Rename-Shared}, \rulename{Le-Rename-Mut}, \texttt{relabel\_state}, \texttt{relabel\_val}, \texttt{ren}, \texttt{map\_state}, \texttt{map\_frame} (\Cref{fig:llbc:combined:le1-labels,def:llbc:combined:relabel-state,def:llbc:symbolic:relabel-val,def:llbc:relabel}) & \rulename{Le-Subst} (Fig.~11.4, p.~193) & HLPL+ only in the thesis; ours only for a label whose loan is present. \\
\rulename{Le-Abs-Drop-Covered-Borrow} (\Cref{fig:llbc:combined:le1-labels}) & --- &  \\
\rulename{Le-Perm-Abs}, \rulename{Le-Perm-In-Abs}, \rulename{Le-Perm-Anons} (\Cref{fig:llbc:combined:le1}) & --- & The thesis's states and abstractions are unordered; ours are lists. \\
\rulename{Le-Empty-Abs} (\Cref{fig:llbc:combined:le1}) & --- & The converse of the thesis's Le-ClearAbs. \\
--- & \rulename{Le-ClearAbs} (Fig.~11.8, p.~205) & Not in ours. \\
--- & \rulename{Le-Abs-ClearValue} (Fig.~11.8, p.~205) & Not in ours. \\
--- & \rulename{Le-Abs-End-DupSharedBorrow} (Fig.~11.8, p.~205) & Not in ours. \\
--- & \rulename{Le-Fresh-MutLoan}, \rulename{Le-Fresh-SharedLoan} (Fig.~11.8, p.~205) & Not in ours; the thesis uses them at loop joins. \\
--- & \rulename{Le-Abs-DeconstructPair}, \rulename{Le-Abs-DeconstructSum} (Fig.~11.8, p.~205) & Not in ours. \\
--- & \rulename{Le-Reborrow-SharedLoan} (Fig.~11.8, p.~205) & Not in ours. \\
\midrule
\rulename{To-Abs-Empty} (\Cref{fig:llbc:symbolic:to-abs}) & \rulename{ToAbs-Empty} (Fig.~11.10, p.~206) &  \\
\rulename{To-Abs-Left}, \rulename{To-Abs-Right} (\Cref{fig:llbc:symbolic:to-abs}) & \rulename{ToAbs-Sum} (Fig.~11.10, p.~206) &  \\
\rulename{To-Abs-MutLoan} (\Cref{fig:llbc:symbolic:to-abs}) & \rulename{ToAbs-MutLoan} (Fig.~11.10, p.~206) &  \\
\rulename{To-Abs-SharedLoan} (\Cref{fig:llbc:symbolic:to-abs}) & \rulename{ToAbs-SharedLoan} (Fig.~11.10, p.~206) &  \\
\rulename{To-Abs-Pair} (\Cref{fig:llbc:symbolic:to-abs}) & \rulename{ToAbs-Pair} (Fig.~11.10, p.~206) &  \\
\rulename{To-Abs-SharedBorrow} (\Cref{fig:llbc:symbolic:to-abs}) & \rulename{ToAbs-SharedBorrow} (Fig.~11.10, p.~206) &  \\
\rulename{To-Abs-MutBorrow} (\Cref{fig:llbc:symbolic:to-abs}) & \rulename{ToAbs-MutBorrow} (Fig.~11.10, p.~206) &  \\
--- & \rulename{ToAbs-Box} (Fig.~11.10, p.~206) & Boxes: not in the paper's calculus \\
\midrule
\rulename{Merge-Abs-Union} (\Cref{fig:llbc:symbolic:merge-abs}) & \rulename{MergeAbs-Union} (Fig.~11.11, p.~207) &  \\
\rulename{Merge-Abs-Mut} (\Cref{fig:llbc:symbolic:merge-abs}) & \rulename{MergeAbs-Mut} (Fig.~11.11, p.~207) &  \\
\rulename{Merge-Abs-Shared} (\Cref{fig:llbc:symbolic:merge-abs}) & \rulename{MergeAbs-Shared} (Fig.~11.11, p.~207) &  \\
\bottomrule
\end{longtable}
}
\paragraph{States and values.}
{\small
\begin{longtable}{@{}>{\raggedright\arraybackslash}p{0.28\linewidth}>{\raggedright\arraybackslash}p{0.29\linewidth}>{\raggedright\arraybackslash}p{0.35\linewidth}@{}}
\toprule
Ours & Thesis & Note \\
\midrule
\endhead
\rulename{FP-Applies}, \rulename{DP-Nil}, \rulename{DP-Pair-0}, \rulename{DP-Pair-1}, \rulename{DP-Left}, \rulename{DP-Right}, \rulename{DP-Unloan}, \rulename{DP-Unbor}, \rulename{State-Get-Frames}, \rulename{State-Get-Anon}, \rulename{State-Get-Abs}, \rulename{State-Set-Anon}, \rulename{State-Set-Frames}, \rulename{State-Set-Abs}, \rulename{State-Updates}, \rulename{V-Base}, \rulename{V-Sum-Left}, \rulename{V-Sum-Right}, \rulename{V-Pair-Left}, \rulename{V-Pair-Right}, \rulename{V-SharedUnloan}, \rulename{V-Unbor}, \texttt{get\_abs}, \rulename{In-Abs} (\Cref{def:llbc:fp-applies,fig:llbc:dp-applies,def:llbc:combined:state-get,fig:llbc:combined:state-set,def:llbc:state-updates,fig:llbc:val-updates,def:llbc:symbolic:state-get,def:llbc:symbolic:fp-inside-region-abs}) & $\Omega[.]$, $V[.]$ (\S9.5, p.~170) & The thesis writes these with its hole notation; ours defines them. \\
\midrule
\texttt{push\_anon} (\Cref{def:llbc:combined:push}) & $\Omega, \_ \to v$ (Fig.~9.2, p.~160) & We store anonymous values in a list. \\
\midrule
\texttt{push\_abs} (\Cref{def:llbc:combined:push}) & $\Omega, A$ (Fig.~10.1, p.~178) &  \\
\midrule
\texttt{ty\_of}, \texttt{val\_ty} (\Cref{def:llbc:symbolic:tyof,def:llbc:symbolic:val-ty}) & $v : \tau$ (Fig.~9.4, p.~162) & The thesis gives no rules for it; ours defines it. \\
\midrule
\rulename{Outer-Loan-Mut}, \rulename{Outer-Loan-Shared} (\Cref{def:llbc:outer-loan}) & no outer loans in $v$ (\S9.4.2, p.~166) & Defined in the thesis's text (\S{}9.4.2, p. 166); ours as two rules. \\
\midrule
\texttt{bot\_of} (\Cref{def:llbc:botof}) & $\bot$ (Fig.~9.2, p.~160) & Ours is a tree: no bottom of pair type. \\
\midrule
\texttt{sym\_of} (\Cref{def:llbc:symbolic:symof}) & $\sigma : \tau$ (Fig.~10.1, p.~178) & Ours is a tree: no symbolic value of pair type. \\
\midrule
\texttt{strip\_outer\_loans} (\Cref{def:llbc:symbolic:strip-outer-loans}) & --- & Ours only; Le-Reborrow-SharedBorrow uses it. \\
\bottomrule
\end{longtable}
}

\section{Annotation Study}
\label{app:annotation-counts}

We give the full results of \S\ref{sec:eval-prevalence} and explain which
methods enter the comparison. The study covers 70 workspaces containing
86 of the top 100 ranked crates.

We use one configuration per workspace, targeting x86 with default features enabled.
Table~\ref{tab:annotation-features} lists the additional features.
Workspaces absent from the table use
only their default features. We check the original sources with the baseline
\code{rustc} and the annotated sources with our modified \code{rustc}.
Both builds use the same selected library packages, features, and dependency
lockfile.

\begin{table}[htbp]
\centering
\caption{Additional features enabled for the annotation study.}
\label{tab:annotation-features}
\small
\begin{tabular}{@{}lp{0.73\linewidth}@{}}
\toprule
\textbf{Workspace} & \textbf{Additional features} \\
\midrule
\codetext{syn} & \codetext{full} \\
\codetext{hashbrown} & \codetext{rayon} \\
\codetext{serde\_json} & \codetext{float\_roundtrip} \\
\codetext{cc} & \codetext{parallel} \\
\codetext{clap} & \codetext{clap\_builder/wrap\_help}, \codetext{clap\_complete/unstable-dynamic} \\
\codetext{rustix} & \codetext{all-apis} \\
\codetext{linux-raw-sys} & \codetext{net} \\
\codetext{tokio} & \codetext{tokio-util/full}, \codetext{tokio/full} \\
\codetext{time} & \codetext{time-macros/formatting}, \codetext{time-macros/parsing}, \codetext{time/formatting}, \codetext{time/parsing} \\
\codetext{hyper} & \codetext{full} \\
\codetext{futures-core} & \codetext{futures-util/compat} \\
\codetext{generic-array} & \codetext{alloc} \\
\codetext{tracing-core} & \codetext{tracing-mock/tracing-subscriber}, \codetext{tracing-subscriber/env-filter} \\
\bottomrule
\end{tabular}
\end{table}

A successful build does not necessarily check every method in its sources,
since conditional compilation can leave some methods inactive. We exclude
these methods, including code for other architectures and inactive
alternative implementations. We identify
compiled methods from \code{rustc}'s high-level intermediate representation
(HIR), after conditional compilation and macro expansion. A method must be compiled in both builds to enter the comparison.

Table~\ref{tab:annotation-full} gives the results for each workspace.
\emph{Recorded} counts the source method declarations that our analysis
identifies, including inactive declarations. Each declaration is counted
once, even if a macro expands it several times. Generated methods whose
source declarations our analysis cannot recover are absent from this count.
\emph{Included} counts the methods left after the exclusions described below
and corresponds to \emph{Methods} in Table~\ref{tab:prevalence}.

\emph{Annotated} counts the included methods that receive a view selecting
a strict subset of fields and pass \code{cargo check --lib}. Its percentage
is taken over \emph{Included}. The receiver columns divide annotated methods
by their original receiver type. \emph{Avg. excl.} is the mean percentage of
struct fields omitted by each annotated method's view. Dashes denote
undefined rates or averages.

\begingroup
\small
\setlength{\tabcolsep}{3pt}
\begin{longtable}{@{}lrrrrrr@{}}
\caption{Full method counts and annotation results.}\label{tab:annotation-full}\\
\toprule
\textbf{Workspace} & \textbf{Recorded} & \textbf{Included} & \textbf{Annotated} &
\multicolumn{2}{c}{\textbf{Receiver kind}} & \textbf{Avg. excl.} \\
\cmidrule(lr){5-6}
& & & & \codetext{\&self} & \codetext{\&mut self} & \\
\midrule
\endfirsthead
\toprule
\textbf{Workspace} & \textbf{Recorded} & \textbf{Included} & \textbf{Annotated} &
\multicolumn{2}{c}{\textbf{Receiver kind}} & \textbf{Avg. excl.} \\
\cmidrule(lr){5-6}
& & & & \codetext{\&self} & \codetext{\&mut self} & \\
\midrule
\endhead
\midrule
\multicolumn{7}{r}{Continued on the next page.}\\
\endfoot
\bottomrule
\endlastfoot
\codetext{syn} & 1,312 & 103 & 18 (17.5\%) & 16 & 2 & 54.1\% \\
\codetext{hashbrown} & 683 & 224 & 77 (34.4\%) & 43 & 34 & 57.1\% \\
\codetext{getrandom} & 12 & 3 & 0 (0.0\%) & 0 & 0 & \textemdash \\
\codetext{rand\_core} & 24 & 9 & 2 (22.2\%) & 1 & 1 & 50.0\% \\
\codetext{rand} & 115 & 13 & 1 (7.7\%) & 1 & 0 & 33.3\% \\
\codetext{libc} & 200 & 0 & 0 (\textemdash) & 0 & 0 & \textemdash \\
\codetext{proc-macro2} & 215 & 61 & 11 (18.0\%) & 8 & 3 & 62.1\% \\
\codetext{quote} & 64 & 2 & 0 (0.0\%) & 0 & 0 & \textemdash \\
\codetext{base64} & 40 & 9 & 5 (55.6\%) & 0 & 5 & 51.1\% \\
\codetext{indexmap} & 627 & 287 & 10 (3.5\%) & 5 & 5 & 48.3\% \\
\codetext{itertools} & 347 & 45 & 16 (35.6\%) & 7 & 9 & 54.0\% \\
\codetext{regex-syntax} & 2,456 & 1,350 & 342 (25.3\%) & 270 & 72 & 62.9\% \\
\codetext{thiserror-impl} & 47 & 22 & 18 (81.8\%) & 17 & 1 & 64.3\% \\
\codetext{serde} & 772 & 66 & 8 (12.1\%) & 3 & 5 & 66.8\% \\
\codetext{rand\_chacha} & 143 & 14 & 9 (64.3\%) & 0 & 9 & 50.0\% \\
\codetext{memchr} & 332 & 143 & 45 (31.5\%) & 42 & 3 & 52.2\% \\
\codetext{unicode-ident} & 0 & 0 & 0 (\textemdash) & 0 & 0 & \textemdash \\
\codetext{itoa} & 3 & 1 & 0 (0.0\%) & 0 & 0 & \textemdash \\
\codetext{autocfg} & 30 & 27 & 3 (11.1\%) & 3 & 0 & 41.7\% \\
\codetext{heck} & 19 & 0 & 0 (\textemdash) & 0 & 0 & \textemdash \\
\codetext{serde\_json} & 651 & 82 & 33 (40.2\%) & 8 & 25 & 59.4\% \\
\codetext{once\_cell} & 77 & 41 & 6 (14.6\%) & 5 & 1 & 50.0\% \\
\codetext{log} & 164 & 13 & 0 (0.0\%) & 0 & 0 & \textemdash \\
\codetext{cc} & 219 & 78 & 52 (66.7\%) & 48 & 4 & 81.2\% \\
\codetext{socket2} & 267 & 115 & 2 (1.7\%) & 1 & 1 & 50.0\% \\
\codetext{smallvec} & 110 & 32 & 2 (6.2\%) & 1 & 1 & 50.0\% \\
\codetext{clap} & 1,103 & 466 & 172 (36.9\%) & 125 & 47 & 71.0\% \\
\codetext{ryu} & 10 & 2 & 0 (0.0\%) & 0 & 0 & \textemdash \\
\codetext{aho-corasick} & 473 & 213 & 89 (41.8\%) & 61 & 28 & 63.7\% \\
\codetext{parking\_lot\_core} & 362 & 54 & 2 (3.7\%) & 1 & 1 & 37.5\% \\
\codetext{rustix} & 468 & 96 & 16 (16.7\%) & 11 & 5 & 56.1\% \\
\codetext{strsim} & 8 & 7 & 3 (42.9\%) & 2 & 1 & 50.0\% \\
\codetext{linux-raw-sys} & 2,908 & 94 & 86 (91.5\%) & 43 & 43 & 89.8\% \\
\codetext{num-traits} & 31 & 0 & 0 (\textemdash) & 0 & 0 & \textemdash \\
\codetext{version\_check} & 18 & 8 & 0 (0.0\%) & 0 & 0 & \textemdash \\
\codetext{bytes} & 194 & 44 & 9 (20.5\%) & 5 & 4 & 63.0\% \\
\codetext{digest} & 296 & 13 & 0 (0.0\%) & 0 & 0 & \textemdash \\
\codetext{mio} & 394 & 18 & 1 (5.6\%) & 1 & 0 & 50.0\% \\
\codetext{semver} & 43 & 9 & 2 (22.2\%) & 2 & 0 & 35.0\% \\
\codetext{either} & 123 & 0 & 0 (\textemdash) & 0 & 0 & \textemdash \\
\codetext{http} & 370 & 121 & 16 (13.2\%) & 4 & 12 & 48.1\% \\
\codetext{block-buffer} & 261 & 50 & 2 (4.0\%) & 0 & 2 & 33.3\% \\
\codetext{idna} & 206 & 87 & 24 (27.6\%) & 9 & 15 & 68.3\% \\
\codetext{anyhow} & 76 & 10 & 0 (0.0\%) & 0 & 0 & \textemdash \\
\codetext{tokio} & 3,350 & 1,343 & 366 (27.3\%) & 260 & 106 & 57.5\% \\
\codetext{scopeguard} & 4 & 0 & 0 (\textemdash) & 0 & 0 & \textemdash \\
\codetext{time} & 999 & 56 & 15 (26.8\%) & 11 & 4 & 46.7\% \\
\codetext{miniz\_oxide} & 82 & 66 & 26 (39.4\%) & 7 & 19 & 49.9\% \\
\codetext{hyper} & 524 & 213 & 138 (64.8\%) & 43 & 95 & 60.5\% \\
\codetext{rustls} & 1,227 & 385 & 182 (47.3\%) & 108 & 74 & 61.4\% \\
\codetext{ppv-lite86} & 213 & 17 & 6 (35.3\%) & 1 & 5 & 69.4\% \\
\codetext{slab} & 64 & 28 & 0 (0.0\%) & 0 & 0 & \textemdash \\
\codetext{futures-core} & 1,112 & 205 & 44 (21.5\%) & 38 & 6 & 55.0\% \\
\codetext{fastrand} & 26 & 14 & 0 (0.0\%) & 0 & 0 & \textemdash \\
\codetext{sha2} & 161 & 25 & 22 (88.0\%) & 0 & 22 & 53.3\% \\
\codetext{toml} & 1,462 & 321 & 57 (17.8\%) & 13 & 44 & 60.7\% \\
\codetext{generic-array} & 110 & 12 & 1 (8.3\%) & 0 & 1 & 33.3\% \\
\codetext{typenum} & 274 & 0 & 0 (\textemdash) & 0 & 0 & \textemdash \\
\codetext{http-body} & 93 & 20 & 0 (0.0\%) & 0 & 0 & \textemdash \\
\codetext{tracing-core} & 1,357 & 284 & 79 (27.8\%) & 72 & 7 & 56.6\% \\
\codetext{zerocopy} & 230 & 22 & 12 (54.5\%) & 11 & 1 & 46.5\% \\
\codetext{byteorder} & 0 & 0 & 0 (\textemdash) & 0 & 0 & \textemdash \\
\codetext{winnow} & 393 & 18 & 1 (5.6\%) & 1 & 0 & 50.0\% \\
\codetext{h2} & 724 & 516 & 110 (21.3\%) & 39 & 71 & 62.7\% \\
\codetext{unicode-width} & 16 & 0 & 0 (\textemdash) & 0 & 0 & \textemdash \\
\codetext{errno} & 3 & 0 & 0 (\textemdash) & 0 & 0 & \textemdash \\
\codetext{fnv} & 2 & 0 & 0 (\textemdash) & 0 & 0 & \textemdash \\
\codetext{chrono} & 646 & 165 & 40 (24.2\%) & 36 & 4 & 80.0\% \\
\codetext{tempfile} & 82 & 16 & 0 (0.0\%) & 0 & 0 & \textemdash \\*
\codetext{ahash} & 109 & 9 & 3 (33.3\%) & 2 & 1 & 50.0\% \\*
\midrule
Total & 29,536 & 7,767 & 2,184 (28.1\%) & 1,385 & 799 & 62.2\% \\
\end{longtable}
\endgroup

Table~\ref{tab:annotation-exclusions} gives the reasons for exclusion.
\emph{Traits} covers methods constrained by trait signatures.
\emph{By-value} covers receivers that take ownership of \code{self}, and
\emph{Custom} covers receivers written with an explicit type,
as in \code{self: ...}.
\emph{Non-struct} covers enum and union receivers.
\emph{Whole-self} covers uses of \code{self} as a whole, such as passing
it to another function, that our analysis does not narrow. \emph{Tooling} covers inactive methods and limitations of
our analysis, rewriting, or compiler.

These exclusions can also apply to callers. When our analysis resolves a
call on \code{self} to an excluded method, it propagates the exclusion to
the caller. When several reasons apply, we count the method under the first
applicable category from left to right in the table. Each excluded method
therefore appears in exactly one category, and the category counts sum to
\emph{Recorded} minus \emph{Included}. Methods that access every field
remain in \emph{Included} but unannotated.

Public methods may have views that name public fields, and private methods
may have views that name private fields. However, a public method's view
cannot expose private fields. Methods that cannot be annotated for this
reason remain included in the comparison but unannotated.

\begingroup
\small
\setlength{\tabcolsep}{2pt}
\begin{longtable}{@{}lrrrrrrr@{}}
\caption{Method-level exclusions from the annotation comparison.}\label{tab:annotation-exclusions}\\
\toprule
\textbf{Workspace} & \textbf{Excluded} & \textbf{Traits} & \textbf{By-value} &
\textbf{Custom} & \textbf{Non-struct} & \textbf{Whole-self} & \textbf{Tooling} \\
\midrule
\endfirsthead
\toprule
\textbf{Workspace} & \textbf{Excluded} & \textbf{Traits} & \textbf{By-value} &
\textbf{Custom} & \textbf{Non-struct} & \textbf{Whole-self} & \textbf{Tooling} \\
\midrule
\endhead
\midrule
\multicolumn{8}{r}{Continued on the next page.}\\
\endfoot
\bottomrule
\endlastfoot
\codetext{syn} & 1,209 & 1,022 & 34 & 0 & 26 & 8 & 119 \\
\codetext{hashbrown} & 459 & 284 & 126 & 0 & 4 & 28 & 17 \\
\codetext{getrandom} & 9 & 5 & 1 & 0 & 0 & 1 & 2 \\
\codetext{rand\_core} & 15 & 15 & 0 & 0 & 0 & 0 & 0 \\
\codetext{rand} & 102 & 96 & 1 & 0 & 4 & 1 & 0 \\
\codetext{libc} & 200 & 96 & 0 & 0 & 0 & 33 & 71 \\
\codetext{proc-macro2} & 154 & 72 & 22 & 0 & 26 & 6 & 28 \\
\codetext{quote} & 62 & 59 & 2 & 0 & 0 & 0 & 1 \\
\codetext{base64} & 31 & 25 & 6 & 0 & 0 & 0 & 0 \\
\codetext{indexmap} & 340 & 203 & 82 & 6 & 3 & 14 & 32 \\
\codetext{itertools} & 302 & 236 & 46 & 0 & 18 & 0 & 2 \\
\codetext{regex-syntax} & 1,106 & 515 & 182 & 0 & 52 & 225 & 132 \\
\codetext{thiserror-impl} & 25 & 22 & 0 & 0 & 2 & 1 & 0 \\
\codetext{serde} & 706 & 649 & 25 & 0 & 8 & 1 & 23 \\
\codetext{rand\_chacha} & 129 & 96 & 0 & 0 & 1 & 30 & 2 \\
\codetext{memchr} & 189 & 99 & 9 & 0 & 2 & 20 & 59 \\
\codetext{unicode-ident} & 0 & 0 & 0 & 0 & 0 & 0 & 0 \\
\codetext{itoa} & 2 & 2 & 0 & 0 & 0 & 0 & 0 \\
\codetext{autocfg} & 3 & 3 & 0 & 0 & 0 & 0 & 0 \\
\codetext{heck} & 19 & 19 & 0 & 0 & 0 & 0 & 0 \\
\codetext{serde\_json} & 569 & 486 & 19 & 0 & 29 & 5 & 30 \\
\codetext{once\_cell} & 36 & 20 & 5 & 0 & 0 & 2 & 9 \\
\codetext{log} & 151 & 108 & 1 & 0 & 17 & 12 & 13 \\
\codetext{cc} & 141 & 22 & 1 & 0 & 12 & 81 & 25 \\
\codetext{socket2} & 152 & 28 & 23 & 0 & 0 & 2 & 99 \\
\codetext{smallvec} & 78 & 51 & 7 & 0 & 8 & 7 & 5 \\
\codetext{clap} & 637 & 264 & 252 & 0 & 19 & 33 & 69 \\
\codetext{ryu} & 8 & 8 & 0 & 0 & 0 & 0 & 0 \\
\codetext{aho-corasick} & 260 & 184 & 12 & 0 & 6 & 41 & 17 \\
\codetext{parking\_lot\_core} & 308 & 182 & 9 & 29 & 0 & 65 & 23 \\
\codetext{rustix} & 372 & 245 & 55 & 0 & 1 & 15 & 56 \\
\codetext{strsim} & 1 & 1 & 0 & 0 & 0 & 0 & 0 \\
\codetext{linux-raw-sys} & 2,814 & 314 & 0 & 0 & 0 & 692 & 1,808 \\
\codetext{num-traits} & 31 & 31 & 0 & 0 & 0 & 0 & 0 \\
\codetext{version\_check} & 10 & 4 & 0 & 0 & 0 & 6 & 0 \\
\codetext{bytes} & 150 & 124 & 8 & 0 & 0 & 10 & 8 \\
\codetext{digest} & 283 & 263 & 4 & 0 & 0 & 1 & 15 \\
\codetext{mio} & 376 & 147 & 9 & 0 & 2 & 9 & 209 \\
\codetext{semver} & 34 & 31 & 0 & 0 & 0 & 3 & 0 \\
\codetext{either} & 123 & 70 & 43 & 2 & 8 & 0 & 0 \\
\codetext{http} & 249 & 178 & 39 & 0 & 9 & 19 & 4 \\
\codetext{block-buffer} & 211 & 136 & 60 & 0 & 0 & 7 & 8 \\
\codetext{idna} & 119 & 63 & 25 & 0 & 7 & 19 & 5 \\
\codetext{anyhow} & 66 & 46 & 18 & 0 & 0 & 2 & 0 \\
\codetext{tokio} & 2,007 & 1,079 & 230 & 72 & 49 & 218 & 359 \\
\codetext{scopeguard} & 4 & 4 & 0 & 0 & 0 & 0 & 0 \\
\codetext{time} & 943 & 481 & 426 & 0 & 21 & 7 & 8 \\
\codetext{miniz\_oxide} & 16 & 7 & 3 & 0 & 2 & 3 & 1 \\
\codetext{hyper} & 311 & 170 & 28 & 8 & 15 & 65 & 25 \\
\codetext{rustls} & 842 & 623 & 108 & 0 & 69 & 20 & 22 \\
\codetext{ppv-lite86} & 196 & 172 & 16 & 0 & 5 & 3 & 0 \\
\codetext{slab} & 36 & 32 & 2 & 0 & 0 & 2 & 0 \\
\codetext{futures-core} & 907 & 788 & 37 & 50 & 7 & 20 & 5 \\
\codetext{fastrand} & 12 & 2 & 0 & 0 & 0 & 0 & 10 \\
\codetext{sha2} & 136 & 124 & 8 & 0 & 4 & 0 & 0 \\
\codetext{toml} & 1,141 & 928 & 59 & 0 & 113 & 11 & 30 \\
\codetext{generic-array} & 98 & 81 & 7 & 2 & 0 & 8 & 0 \\
\codetext{typenum} & 274 & 274 & 0 & 0 & 0 & 0 & 0 \\
\codetext{http-body} & 73 & 62 & 8 & 3 & 0 & 0 & 0 \\
\codetext{tracing-core} & 1,073 & 781 & 185 & 8 & 5 & 26 & 68 \\
\codetext{zerocopy} & 208 & 82 & 103 & 0 & 16 & 6 & 1 \\
\codetext{byteorder} & 0 & 0 & 0 & 0 & 0 & 0 & 0 \\
\codetext{winnow} & 375 & 349 & 23 & 0 & 3 & 0 & 0 \\
\codetext{h2} & 208 & 85 & 22 & 0 & 21 & 38 & 42 \\
\codetext{unicode-width} & 16 & 4 & 12 & 0 & 0 & 0 & 0 \\
\codetext{errno} & 3 & 3 & 0 & 0 & 0 & 0 & 0 \\
\codetext{fnv} & 2 & 2 & 0 & 0 & 0 & 0 & 0 \\
\codetext{chrono} & 481 & 329 & 64 & 0 & 16 & 40 & 32 \\
\codetext{tempfile} & 66 & 42 & 16 & 0 & 0 & 8 & 0 \\*
\codetext{ahash} & 100 & 92 & 2 & 0 & 0 & 1 & 5 \\*
\midrule
Total & 21,769 & 13,120 & 2,485 & 180 & 610 & 1,875 & 3,499 \\
\end{longtable}
\endgroup

\FloatBarrier
\section{Machine-Code Comparison}
\label{app:runtime-comparison}

We check whether adding views changes the generated machine code. For each workspace, we build the original sources with the baseline
\code{rustc} and the annotated sources with our modified \code{rustc}.
We also build the original sources with our modified \code{rustc} to identify
differences that occur without annotations. All three builds use the same
selected library packages, features, and dependency lockfile. We use
optimization level 3 and one codegen unit, with debug information, link-time
optimization, and incremental compilation disabled.

We compare the machine code byte for byte in the executable sections of
native object files extracted from library archives. For procedural macros
and dynamic libraries, we compare the executable sections of the emitted
shared objects. Regex's C API also produces static and shared libraries that
bundle dependency and standard-library code. We exclude those duplicate
outputs and compare its Rust library archive.

To match sections between builds, we ignore compiler-generated hashes in
section names only when each section still has a unique match. We still compare the original machine-code bytes. Ten workspaces require a further comparison
of disassembled instructions after normalizing symbol names, addresses, and
corresponding reference \code{TypeId} constants. These differences also occur
when compiling the original sources with the two compilers. When the original and annotated sources
are compiled with our modified \code{rustc}, their machine code matches byte
for byte without this further normalization.

The comparison covers code emitted for the selected library configurations.
It excludes inactive conditional code and generic instantiations generated by
downstream users. Data and relocations are also excluded, so matching machine
code does not establish equality of complete linked binaries. We do not
measure execution time.

Table~\ref{tab:runtime-full} gives the comparison for each workspace.
\begin{table}[htbp]
\caption{Machine-code comparison for the 70 workspaces. Yes denotes a byte-for-byte match. Yes\textsuperscript{*} denotes a match after the normalization described above. A dash means that no machine code was emitted. Read down the left block, then the right.}
\label{tab:runtime-full}
\centering
\small
\begin{tabular}{@{}lr@{\qquad}lr@{}}
\toprule
\textbf{Workspace} & \textbf{Match} & \textbf{Workspace} & \textbf{Match} \\
\midrule
\codetext{syn} & Yes & \codetext{bytes} & Yes \\
\codetext{hashbrown} & Yes & \codetext{digest} & Yes \\
\codetext{getrandom} & Yes & \codetext{mio} & Yes \\
\codetext{rand\_core} & --- & \codetext{semver} & Yes \\
\codetext{rand} & Yes & \codetext{either} & --- \\
\codetext{libc} & Yes & \codetext{http} & Yes \\
\codetext{proc-macro2} & Yes & \codetext{block-buffer} & Yes\textsuperscript{*} \\
\codetext{quote} & Yes & \codetext{idna} & Yes \\
\codetext{base64} & Yes & \codetext{anyhow} & Yes \\
\codetext{indexmap} & Yes & \codetext{tokio} & Yes\textsuperscript{*} \\
\codetext{itertools} & Yes & \codetext{scopeguard} & --- \\
\codetext{regex-syntax} & Yes\textsuperscript{*} & \codetext{time} & Yes\textsuperscript{*} \\
\codetext{thiserror-impl} & Yes\textsuperscript{*} & \codetext{miniz\_oxide} & Yes \\
\codetext{serde} & Yes\textsuperscript{*} & \codetext{hyper} & Yes \\
\codetext{rand\_chacha} & Yes & \codetext{rustls} & Yes \\
\codetext{memchr} & Yes & \codetext{ppv-lite86} & Yes \\
\codetext{unicode-ident} & --- & \codetext{slab} & Yes \\
\codetext{itoa} & Yes & \codetext{futures-core} & Yes\textsuperscript{*} \\
\codetext{autocfg} & Yes & \codetext{fastrand} & Yes \\
\codetext{heck} & Yes & \codetext{sha2} & Yes \\
\codetext{serde\_json} & Yes & \codetext{toml} & Yes \\
\codetext{once\_cell} & Yes & \codetext{generic-array} & Yes \\
\codetext{log} & Yes & \codetext{typenum} & Yes \\
\codetext{cc} & Yes & \codetext{http-body} & Yes \\
\codetext{socket2} & Yes & \codetext{tracing-core} & Yes\textsuperscript{*} \\
\codetext{smallvec} & Yes & \codetext{zerocopy} & Yes\textsuperscript{*} \\
\codetext{clap} & Yes\textsuperscript{*} & \codetext{byteorder} & Yes \\
\codetext{ryu} & Yes & \codetext{winnow} & Yes \\
\codetext{aho-corasick} & Yes & \codetext{h2} & Yes \\
\codetext{parking\_lot\_core} & Yes & \codetext{unicode-width} & Yes \\
\codetext{rustix} & Yes & \codetext{errno} & Yes \\
\codetext{strsim} & Yes & \codetext{fnv} & --- \\
\codetext{linux-raw-sys} & --- & \codetext{chrono} & Yes \\
\codetext{num-traits} & Yes & \codetext{tempfile} & Yes \\
\codetext{version\_check} & Yes & \codetext{ahash} & Yes \\
\midrule
\multicolumn{4}{c}{Total 70 workspaces: 54 exact matches, 10 normalized matches, 6 without machine code.} \\
\bottomrule
\end{tabular}
\end{table}

\FloatBarrier

\section{Compile-Time Overhead}
\label{app:compilation-overhead}

We give the full compilation-time results from \S\ref{sec:eval-perf}.
We measure the wall time of clean \codetext{cargo check} runs, using
\codetext{cargo check --lib} for the annotated workspaces.
All measurements were taken on a single-socket Intel Xeon w9-3495X
workstation (56 cores, 112 threads, up to 4.8\,GHz, 105\,MiB L3 cache,
256\,GB DDR5-4800 ECC memory) running Ubuntu 24.04 LTS (Linux 6.8.0).

For the unannotated crates, we use 30 timed runs when baseline compilation
takes under $1\,\text{sec}$, 20 when it takes at least $1$ but under
$5\,\text{sec}$, 15 when it takes at least $5$ but under $15\,\text{sec}$,
and 12 for $15\,\text{sec}$ or longer.
We compute each crate's or workspace's compilation time as a 20\% trimmed
mean over its timed runs. Let $t_i^{\mathit{baseline}}$ and
$t_i^{\mathit{modified}}$ be these means for crate or workspace $i$ under the two
configurations. We report overall overhead as the time-weighted mean,
\[
  100 \times
  \frac{\sum_i (t_i^{\mathit{modified}} - t_i^{\mathit{baseline}})}
       {\sum_i t_i^{\mathit{baseline}}}\%.
\]
Table~\ref{tab:perf} groups the unannotated crates by baseline compilation
time and reports the overhead within each group.

\begin{table}[htbp]
  \centering
  \caption{Compilation overhead for the 487 unannotated crates.
    Groups use baseline \codetext{cargo check} wall time.
    \emph{Weighted mean} gives the time-weighted overhead for each group.
    \emph{Min} and \emph{Max} give the smallest and largest per-crate overheads.}
  \label{tab:perf}
  \begin{tabular}{lrrrr}
    \toprule
    \textbf{Compile-time group} & \textbf{Crates} &
    \textbf{Weighted mean} & \textbf{Min} & \textbf{Max} \\
    \midrule
    $<1\,\text{sec}$           & 222 & $+0.07\%$ & $-3.02\%$ & $+3.05\%$ \\
    $1\text{--}5\,\text{sec}$  & 192 & $+0.17\%$ & $-2.43\%$ & $+1.69\%$ \\
    $5\text{--}15\,\text{sec}$ &  68 & $+0.41\%$ & $-1.57\%$ & $+1.62\%$ \\
    $\geq 15\,\text{sec}$          &   5 & $+0.43\%$ & $-0.14\%$ & $+1.18\%$ \\
    \midrule
    Overall                  & 487 & $+0.29\%$ & $-3.02\%$ & $+3.05\%$ \\
    \bottomrule
  \end{tabular}
\end{table}

Tables~\ref{tab:annotated-times} and~\ref{tab:annotated-overheads} report
the compilation times and overheads for all 70 workspaces in the
annotated-code experiment. We compare their original sources under the
baseline \code{rustc} with their annotated sources under our modified
\code{rustc}.

\begin{longtable}{@{}lrrrrrr@{}}
\caption{Compilation times for the 70 workspaces.
\emph{Baseline} uses the original sources with baseline \code{rustc}, and
\emph{With view types} uses annotated sources with our modified \code{rustc}.
Times measure clean \codetext{cargo check --lib} runs and are given in sec.
\emph{Mean} is the 20\% trimmed mean. \emph{Min} and \emph{Max} are
 the fastest and slowest individual runs.}
\label{tab:annotated-times}\\
\toprule
& \multicolumn{3}{c}{Baseline (sec)} & \multicolumn{3}{c}{With view types (sec)} \\
\cmidrule(lr){2-4}\cmidrule(lr){5-7}
Workspace & Mean & Min & Max & Mean & Min & Max \\
\midrule
\endfirsthead
\multicolumn{7}{c}{Table~\thetable\ continued}\\
\toprule
& \multicolumn{3}{c}{Baseline (sec)} & \multicolumn{3}{c}{With view types (sec)} \\
\cmidrule(lr){2-4}\cmidrule(lr){5-7}
Workspace & Mean & Min & Max & Mean & Min & Max \\
\midrule
\endhead
\midrule
\multicolumn{7}{r}{Continued on next page}\\
\endfoot
\bottomrule
\endlastfoot
\code{syn} & 2.687 & 2.657 & 2.728 & 2.686 & 2.661 & 2.717 \\
\code{hashbrown} & 1.313 & 1.301 & 1.335 & 1.314 & 1.293 & 1.325 \\
\code{getrandom} & 1.513 & 1.487 & 1.546 & 1.512 & 1.493 & 1.540 \\
\code{rand\_core} & 0.229 & 0.219 & 0.242 & 0.228 & 0.212 & 0.240 \\
\code{rand} & 2.274 & 2.246 & 2.350 & 2.277 & 2.250 & 2.322 \\
\code{libc} & 1.732 & 1.706 & 1.769 & 1.726 & 1.699 & 1.750 \\
\code{proc-macro2} & 0.856 & 0.841 & 0.871 & 0.855 & 0.840 & 0.867 \\
\code{quote} & 0.927 & 0.912 & 0.953 & 0.928 & 0.911 & 0.944 \\
\code{base64} & 0.441 & 0.426 & 0.455 & 0.443 & 0.429 & 0.454 \\
\code{indexmap} & 1.729 & 1.711 & 1.747 & 1.732 & 1.708 & 1.759 \\
\code{itertools} & 2.406 & 2.389 & 2.427 & 2.408 & 2.385 & 2.435 \\
\code{regex-syntax} & 5.966 & 5.924 & 6.037 & 5.962 & 5.913 & 6.052 \\
\code{thiserror-impl} & 4.243 & 4.207 & 4.285 & 4.236 & 4.210 & 4.268 \\
\code{serde} & 4.017 & 3.972 & 4.058 & 4.013 & 3.986 & 4.057 \\
\code{rand\_chacha} & 2.649 & 2.624 & 2.686 & 2.647 & 2.620 & 2.676 \\
\code{memchr} & 0.693 & 0.674 & 0.706 & 0.692 & 0.679 & 0.711 \\
\code{unicode-ident} & 0.189 & 0.176 & 0.203 & 0.189 & 0.178 & 0.199 \\
\code{itoa} & 0.252 & 0.241 & 0.266 & 0.251 & 0.235 & 0.262 \\
\code{autocfg} & 0.256 & 0.244 & 0.274 & 0.255 & 0.245 & 0.267 \\
\code{heck} & 0.183 & 0.173 & 0.201 & 0.182 & 0.173 & 0.198 \\
\code{serde\_json} & 3.686 & 3.639 & 3.733 & 3.689 & 3.644 & 3.728 \\
\code{once\_cell} & 0.296 & 0.285 & 0.307 & 0.295 & 0.280 & 0.311 \\
\code{log} & 0.244 & 0.232 & 0.256 & 0.244 & 0.233 & 0.255 \\
\code{cc} & 0.966 & 0.951 & 0.989 & 0.969 & 0.949 & 0.997 \\
\code{socket2} & 1.856 & 1.831 & 1.894 & 1.857 & 1.826 & 1.876 \\
\code{smallvec} & 0.333 & 0.323 & 0.345 & 0.332 & 0.317 & 0.347 \\
\code{clap} & 4.075 & 4.031 & 4.113 & 4.071 & 4.051 & 4.124 \\
\code{ryu} & 0.330 & 0.313 & 0.348 & 0.329 & 0.315 & 0.343 \\
\code{aho-corasick} & 1.722 & 1.693 & 1.754 & 1.725 & 1.703 & 1.753 \\
\code{parking\_lot\_core} & 1.997 & 1.974 & 2.024 & 2.003 & 1.972 & 2.030 \\
\code{rustix} & 1.007 & 0.989 & 1.023 & 1.003 & 0.990 & 1.018 \\
\code{strsim} & 0.251 & 0.242 & 0.268 & 0.254 & 0.239 & 0.264 \\
\code{linux-raw-sys} & 0.538 & 0.518 & 0.553 & 0.537 & 0.524 & 0.548 \\
\code{num-traits} & 1.515 & 1.499 & 1.535 & 1.519 & 1.491 & 1.542 \\
\code{version\_check} & 0.235 & 0.220 & 0.250 & 0.235 & 0.224 & 0.252 \\
\code{bytes} & 0.834 & 0.824 & 0.848 & 0.833 & 0.819 & 0.848 \\
\code{digest} & 4.858 & 4.822 & 4.912 & 4.862 & 4.831 & 4.908 \\
\code{mio} & 1.695 & 1.661 & 1.721 & 1.702 & 1.680 & 1.738 \\
\code{semver} & 0.546 & 0.530 & 0.567 & 0.547 & 0.532 & 0.567 \\
\code{either} & 0.343 & 0.329 & 0.354 & 0.341 & 0.327 & 0.354 \\
\code{http} & 1.660 & 1.638 & 1.686 & 1.666 & 1.642 & 1.677 \\
\code{block-buffer} & 4.089 & 4.052 & 4.128 & 4.105 & 4.061 & 4.154 \\
\code{idna} & 11.187 & 11.124 & 11.242 & 11.175 & 11.123 & 11.239 \\
\code{anyhow} & 0.623 & 0.601 & 0.641 & 0.624 & 0.606 & 0.645 \\
\code{tokio} & 3.366 & 3.341 & 3.398 & 3.364 & 3.338 & 3.388 \\
\code{scopeguard} & 0.151 & 0.139 & 0.163 & 0.150 & 0.140 & 0.161 \\
\code{time} & 2.354 & 2.325 & 2.384 & 2.349 & 2.336 & 2.378 \\
\code{miniz\_oxide} & 2.028 & 1.988 & 2.053 & 2.024 & 2.006 & 2.054 \\
\code{hyper} & 1.847 & 1.818 & 1.886 & 1.851 & 1.822 & 1.866 \\
\code{rustls} & 13.822 & 13.741 & 13.927 & 13.795 & 13.664 & 13.874 \\
\code{ppv-lite86} & 5.794 & 5.739 & 5.893 & 5.808 & 5.748 & 5.840 \\
\code{slab} & 0.257 & 0.244 & 0.275 & 0.255 & 0.243 & 0.265 \\
\code{futures-core} & 8.991 & 8.936 & 9.053 & 8.972 & 8.914 & 9.017 \\
\code{fastrand} & 0.314 & 0.305 & 0.329 & 0.315 & 0.307 & 0.330 \\
\code{sha2} & 5.876 & 5.849 & 5.915 & 5.874 & 5.841 & 5.910 \\
\code{toml} & 3.963 & 3.929 & 4.002 & 3.975 & 3.941 & 4.026 \\
\code{generic-array} & 1.781 & 1.744 & 2.267 & 1.785 & 1.741 & 2.285 \\
\code{typenum} & 1.110 & 1.007 & 1.311 & 1.097 & 1.024 & 1.334 \\
\code{http-body} & 1.923 & 1.894 & 1.944 & 1.920 & 1.902 & 1.946 \\
\code{tracing-core} & 8.297 & 8.247 & 8.344 & 8.292 & 8.241 & 8.355 \\
\code{zerocopy} & 6.147 & 6.090 & 6.218 & 6.172 & 6.117 & 6.313 \\
\code{byteorder} & 0.372 & 0.352 & 0.386 & 0.373 & 0.361 & 0.382 \\
\code{winnow} & 1.662 & 1.642 & 1.686 & 1.663 & 1.640 & 1.695 \\
\code{h2} & 6.598 & 6.544 & 6.677 & 6.597 & 6.568 & 6.647 \\
\code{unicode-width} & 0.297 & 0.281 & 0.312 & 0.294 & 0.280 & 0.309 \\
\code{errno} & 1.634 & 1.610 & 1.739 & 1.632 & 1.605 & 1.660 \\
\code{fnv} & 0.131 & 0.119 & 0.146 & 0.132 & 0.118 & 0.143 \\
\code{chrono} & 2.520 & 2.474 & 2.559 & 2.526 & 2.499 & 2.565 \\
\code{tempfile} & 2.584 & 2.558 & 2.616 & 2.590 & 2.553 & 2.616 \\
\code{ahash} & 5.224 & 5.185 & 5.264 & 5.229 & 5.187 & 5.292 \\
\end{longtable}

\begin{longtable}{@{}lrrr@{}}
\caption{Compilation overhead for the same workspaces.
\emph{Overhead} compares the two trimmed means.
\emph{Min} and \emph{Max} give the smallest and largest overheads across
paired baseline and annotated runs. The \emph{Overall} row gives the
time-weighted overhead and the smallest and largest per-workspace overheads.}
\label{tab:annotated-overheads}\\
\toprule
Workspace & Overhead (\%) & Min (\%) & Max (\%) \\
\midrule
\endfirsthead
\multicolumn{4}{c}{Table~\thetable\ continued}\\
\toprule
Workspace & Overhead (\%) & Min (\%) & Max (\%) \\
\midrule
\endhead
\midrule
\multicolumn{4}{r}{Continued on next page}\\
\endfoot
\bottomrule
\endlastfoot
\code{syn} & $-0.01$ & $-1.37$ & $+0.78$ \\
\code{hashbrown} & $+0.07$ & $-2.27$ & $+1.80$ \\
\code{getrandom} & $-0.08$ & $-2.51$ & $+2.86$ \\
\code{rand\_core} & $-0.56$ & $-9.77$ & $+9.20$ \\
\code{rand} & $+0.12$ & $-3.54$ & $+3.13$ \\
\code{libc} & $-0.35$ & $-2.51$ & $+1.88$ \\
\code{proc-macro2} & $-0.13$ & $-1.83$ & $+3.06$ \\
\code{quote} & $+0.08$ & $-2.01$ & $+1.96$ \\
\code{base64} & $+0.44$ & $-4.31$ & $+4.33$ \\
\code{indexmap} & $+0.19$ & $-1.62$ & $+1.22$ \\
\code{itertools} & $+0.11$ & $-1.01$ & $+1.05$ \\
\code{regex-syntax} & $-0.06$ & $-1.34$ & $+2.16$ \\
\code{thiserror-impl} & $-0.16$ & $-1.64$ & $+0.95$ \\
\code{serde} & $-0.08$ & $-1.24$ & $+2.10$ \\
\code{rand\_chacha} & $-0.05$ & $-1.09$ & $+1.40$ \\
\code{memchr} & $-0.05$ & $-2.76$ & $+2.86$ \\
\code{unicode-ident} & $-0.35$ & $-12.63$ & $+13.52$ \\
\code{itoa} & $-0.55$ & $-10.20$ & $+7.81$ \\
\code{autocfg} & $-0.36$ & $-8.92$ & $+9.28$ \\
\code{heck} & $-0.90$ & $-10.16$ & $+13.22$ \\
\code{serde\_json} & $+0.07$ & $-1.15$ & $+1.22$ \\
\code{once\_cell} & $-0.43$ & $-7.20$ & $+7.80$ \\
\code{log} & $-0.02$ & $-8.15$ & $+7.58$ \\
\code{cc} & $+0.29$ & $-2.61$ & $+2.22$ \\
\code{socket2} & $+0.04$ & $-3.23$ & $+1.86$ \\
\code{smallvec} & $-0.28$ & $-5.88$ & $+5.11$ \\
\code{clap} & $-0.10$ & $-1.24$ & $+1.51$ \\
\code{ryu} & $-0.30$ & $-8.12$ & $+7.31$ \\
\code{aho-corasick} & $+0.14$ & $-1.05$ & $+2.20$ \\
\code{parking\_lot\_core} & $+0.30$ & $-1.37$ & $+2.48$ \\
\code{rustix} & $-0.37$ & $-2.11$ & $+1.60$ \\
\code{strsim} & $+1.14$ & $-8.50$ & $+8.31$ \\
\code{linux-raw-sys} & $-0.20$ & $-3.89$ & $+4.13$ \\
\code{num-traits} & $+0.27$ & $-2.21$ & $+2.03$ \\
\code{version\_check} & $-0.11$ & $-9.12$ & $+8.13$ \\
\code{bytes} & $-0.14$ & $-2.47$ & $+2.13$ \\
\code{digest} & $+0.08$ & $-1.05$ & $+1.79$ \\
\code{mio} & $+0.39$ & $-1.28$ & $+2.74$ \\
\code{semver} & $+0.03$ & $-4.12$ & $+4.22$ \\
\code{either} & $-0.55$ & $-6.41$ & $+5.80$ \\
\code{http} & $+0.34$ & $-1.59$ & $+1.46$ \\
\code{block-buffer} & $+0.38$ & $-1.17$ & $+1.83$ \\
\code{idna} & $-0.11$ & $-0.70$ & $+0.68$ \\
\code{anyhow} & $+0.10$ & $-4.35$ & $+5.37$ \\
\code{tokio} & $-0.05$ & $-0.95$ & $+1.42$ \\
\code{scopeguard} & $-0.70$ & $-14.22$ & $+15.29$ \\
\code{time} & $-0.23$ & $-1.19$ & $+1.83$ \\
\code{miniz\_oxide} & $-0.19$ & $-1.95$ & $+2.36$ \\
\code{hyper} & $+0.23$ & $-1.88$ & $+1.80$ \\
\code{rustls} & $-0.19$ & $-1.12$ & $+0.31$ \\
\code{ppv-lite86} & $+0.24$ & $-1.67$ & $+1.63$ \\
\code{slab} & $-0.79$ & $-7.45$ & $+7.90$ \\
\code{futures-core} & $-0.21$ & $-1.00$ & $+0.78$ \\
\code{fastrand} & $+0.10$ & $-5.81$ & $+8.22$ \\
\code{sha2} & $-0.03$ & $-0.79$ & $+0.77$ \\
\code{toml} & $+0.28$ & $-0.34$ & $+1.63$ \\
\code{generic-array} & $+0.23$ & $-1.83$ & $+7.03$ \\
\code{typenum} & $-1.09$ & $-16.88$ & $+4.00$ \\
\code{http-body} & $-0.12$ & $-1.57$ & $+1.87$ \\
\code{tracing-core} & $-0.06$ & $-0.94$ & $+0.76$ \\
\code{zerocopy} & $+0.40$ & $-0.89$ & $+1.82$ \\
\code{byteorder} & $+0.25$ & $-4.60$ & $+6.11$ \\
\code{winnow} & $+0.06$ & $-1.18$ & $+1.82$ \\
\code{h2} & $-0.00$ & $-1.00$ & $+1.22$ \\
\code{unicode-width} & $-1.02$ & $-9.51$ & $+9.01$ \\
\code{errno} & $-0.13$ & $-5.28$ & $+1.24$ \\
\code{fnv} & $+1.14$ & $-18.57$ & $+19.19$ \\
\code{chrono} & $+0.23$ & $-1.66$ & $+2.34$ \\
\code{tempfile} & $+0.21$ & $-1.08$ & $+1.54$ \\
\code{ahash} & $+0.11$ & $-0.81$ & $+1.11$ \\
\midrule
Overall & $+0.00$ & $-1.09$ & $+1.14$ \\
\end{longtable}

\section{Workaround Study}
\label{app:workaround-study}

We examine the borrowing conflicts behind the workarounds collected in
\S\ref{sec:eval-cases}. We explain which conflicts views resolve, show
representative patches, and distinguish cases beyond field-level precision
from cases we could not evaluate or express with the prototype.

\subsection{Selecting and Checking Cases}

View types address the loss of field-level precision across a call.
A reference may borrow fields that the called operation does not need,
or borrow mutably a field that it only reads. A view makes those
requirements explicit. The form of a workaround alone does not tell us
whether views can replace it: copying data, for example, can avoid either
an unnecessarily broad borrow or an operation that would invalidate the
original reference. We therefore examine the conflict exposed by removing
each workaround.

We searched GitHub for comments attributing workarounds to borrow checking
and inspected the results in a fixed order, grouped by repository popularity,
until the study's time limit. We also examined cases gathered earlier from
developer discussions and previous investigations.
Table~\ref{tab:workaround-outcomes} keeps these collections separate.
A \emph{site} is one recorded workaround case, which may span several source
locations. These collections document the kinds of conflicts encountered,
not their prevalence in Rust programs.

Some collected comments do not identify a workaround that we can remove.
We exclude those entries. The selection also sets aside two cases consisting
only of an explicit \code{drop} or a scope block: these control when a borrow
ends, rather than restructure code to retain it across an operation. A copy
of the Rust compiler falls outside our Cargo build setup. Other cases cannot be evaluated because dependencies
are unavailable or incompatible, or because no selected target compiles the
relevant source. We record these as build or source exclusions, without
concluding that views would fail to help.

For a case we can build, we reconstruct the code without its workaround.
We then check whether that code fails to compile and whether views resolve
the conflict. Four reconstructions already compile with stock Rust, so they
provide no rejected program for views to improve. For five earlier cases,
the available evidence does not identify a sufficiently clear reconstruction.
For example, reversing h2's struct split would require choosing among several
possible redesigns. We do not treat these inconclusive cases as failures of
view types.

\begin{table}[htbp]
\centering
\caption{Disposition of the collected workaround cases. The rows below
``Classified sites'' partition those sites. The cases with unresolved conflicts
are explained in \S\ref{app:workaround-limits}.}
\label{tab:workaround-outcomes}
\small
\begin{tabular}{@{}lrr@{}}
\toprule
 & Search & Earlier collection \\
\midrule
Recorded sites & 40 & 33 \\
Initial exclusions & 8 & 2 \\
Build or source exclusions & 14 & 7 \\
Not processed before the time limit & 1 & 0 \\
\midrule
Classified sites & 17 & 24 \\
\addlinespace
Resolved with views alone & 5 & 6 \\
Resolved with views and supporting changes & 2 & 2 \\
Requires more than field-level precision & 5 & 8 \\
Not expressed through the studied interfaces & 2 & 2 \\
Already accepted without views & 3 & 1 \\
No sufficiently justified reconstruction & 0 & 5 \\
\bottomrule
\end{tabular}
\end{table}

We additionally checked Helix using the developer report discussed in
\S\ref{sec:eval-cases}. With this follow-up, views remove workarounds at
16 sites across 12 repositories, represented by 15 final patch sets.
Two Sway sites share one patch and test run. Replica's single record
combines restoring a socket field to its struct with replacing separate
field arguments and inlined lookups by a method.

\FloatBarrier
\subsection{Removing Workarounds with Views}

The main text develops the Moonfire and Helix cases. Here we show three
other changes, each starting from the operation that the code needs to
perform. The excerpts come from the final patches: \code{-} marks removed
lines and \code{+} marks added lines. We omit unrelated context and shorten
indentation.

\paragraph{Using borrowed text directly.}
In grok-build, updating suggestions requires reading the prompt text while
mutating the prompt's \code{suggestions} field. The \code{text} getter takes
\code{&self}, so its result keeps the whole prompt borrowed. The original
code copies the text into a separate string before updating suggestions.
Restricting the getter to \code{textarea} leaves \code{suggestions} available
for mutation and removes the copy:
\Needspace{4\baselineskip}
\begin{lstlisting}[style=diff,linebackgroundcolor={\CodeDiffLines{1,6,10}{2,11}}]
-pub fn text(&self) -> &str {
+pub fn text(&{textarea} self) -> &str {
     self.textarea.text()
 }
...
-let text = self.prompt.text().to_owned();
 let action = self
     .prompt
     .suggestions
-    .text_changed(&text, slash_active, has_inline_ghost)?;
+    .text_changed(self.prompt.text(), slash_active, has_inline_ghost)?;
\end{lstlisting}

\paragraph{Updating a field in place.}
Fish records the export-generation numbers of environment scopes in
\code{export_array_generations}. It fills this vector using the iterator from \code{iter_generations}. That method reads
\code{globals} and \code{locals}, but its \code{&self} receiver keeps the
whole environment borrowed while the iterator is used. The original code
therefore moves the destination vector into a local variable, fills it,
and puts it back. A view on the iterator method permits filling the vector
where it is stored:
\Needspace{4\baselineskip}
\begin{lstlisting}[style=diff,linebackgroundcolor={\CodeDiffLines{1,4,5,6,7}{2,8,9}}]
-fn iter_generations(&self) -> impl Iterator<Item = u64> {
+fn iter_generations(&{globals, locals} self) -> impl Iterator<Item = u64> {
...
-let mut generations = std::mem::take(&mut self.export_array_generations);
-generations.clear();
-generations.extend(self.iter_generations());
-self.export_array_generations = generations;
+self.export_array_generations.clear();
+self.export_array_generations.extend(self.iter_generations());
\end{lstlisting}

The Wasmtime WAST runner uses a more elaborate version of this workaround.
Its \code{get_export} method can return a mutable store reference that must
be used alongside \code{async_runtime}. The original code takes the runtime
out of its struct and introduces a guard whose destructor puts it back.
A receiver view selecting \code{core_linker}, \code{core_store}, and
\code{current} leaves \code{async_runtime} outside the borrow. The rewrite
keeps the runtime in place and deletes the guard and its destructor.

\paragraph{Retaining a reference across recursive calls.}
Sway looks up a pass descriptor in \code{passes}, recursively processes its
dependencies, and then uses the descriptor to run the pass. These recursive
helpers are free functions taking \code{&mut PassManager}. That parameter
prevents the caller from retaining the descriptor reference across their
calls, so the original code repeats the lookup afterward. Views let the helpers read \code{passes}
while modifying \code{analyses}, making the repeated lookup unnecessary:
\Needspace{4\baselineskip}
\begin{lstlisting}[style=diff,linebackgroundcolor={\CodeDiffLines{2,5,6}{3}}]
 fn run_module_pass(
-    pm: &mut PassManager,
+    pm: &mut {passes, mut analyses} PassManager,
...
-// Get the pass again to satisfy the borrow checker.
-let pass_t = pm.passes.get(pass).expect("Unregistered pass");
 let ScopedPass::ModulePass(mp) = pass_t.runner.clone() else {
\end{lstlisting}

We give \code{run_function_pass} the same parameter view.

\paragraph{Keeping field access inside methods.}
Apicula's getters return references into a stored database. The original
code passes that database as a separate argument, tying the returned
reference to that argument rather than to the whole viewer. This lets the
caller update fields such as \code{object_mats} while reading a model from
the database. With views, \code{cur_model} can instead borrow
\code{db} and \code{model_id} through its receiver and read \code{self.db}
directly. The extra database argument disappears, and the caller can still
update its matrices. These getters were already methods.

RhoJVM's rewrite consolidates six field arguments of two verification
helpers into a reference to \code{Env}. The reconstructed code looks up a
method in \code{env.methods} and passes it to a new descriptor-loading
method on \code{env}. An ordinary \code{&mut self} receiver conflicts with
the reference into \code{methods}. A view lets the loader borrow the fields
it updates while the caller retains the reference into \code{methods}. This is a proposed
method refactoring, not evidence that an upstream method had previously
been replaced by free functions. Some other calls still pass fields separately.

\paragraph{Changes to interfaces.}
Four rewrites also need changes to interfaces.
Three encounter visibility restrictions. A public method cannot name private
fields in its view. In Wasmtime's Cranelift helper,
we use \code{self.insts.len()} instead of \code{self.num_insts()}. The latter
is a public accessor and cannot carry a view naming the private
\code{insts} field. In its component type builder,
we restrict \code{resources_mut} from \code{pub} to \code{pub(super)}.
Binius instead retains its public entry point and puts the view on a private
recursive helper. That helper can read \code{oracles} while updating
\code{suffixes}, allowing iteration over polynomial identifiers without
first collecting them into a temporary vector.

Varisat's checker already uses the \code{partial_ref} library to express
partial borrowing. We replace that encoding in the selected component with
view-typed parameters, a stored view reference, and ordinary field access.
The prototype does not support copying the shared view reference used there,
so this rewrite also needs an explicit reborrow helper. Its final patch
adds 22 parameter views and one stored-reference view across six files,
with call-site changes in two further files.

\subsection{Conflicts That Views Do Not Resolve}
\label{app:workaround-limits}

Knowing which fields an operation accesses does not settle every borrowing
conflict. The distinction is visible in two nearby diamond-types cases.
Both reacquire a mutable leaf reference after calling a helper. In the first,
\code{create_new_root_node} reads \code{total_len} and updates \code{height},
\code{root}, and \code{nodes}, leaving \code{leaves} untouched. A view naming
those four fields lets the caller retain \code{old_leaf} across the call,
so the second lookup disappears. In the other case,
\code{split_child_of_node} can update the parent field of that same leaf.
The helper needs access that overlaps the retained reference, so narrowing
its receiver to the fields it uses does not resolve the conflict.

\paragraph{Separation within a collection.}
A Cranelift case copies a sibling node out of a pool, updates it, and puts
it back. Removing the copy requires borrowing
\code{pool[self.node[level]]} and \code{pool[rhs_node]} together.
The required separation is between two indices of the same collection,
not two struct fields. Three diamond-types cases similarly rely on tree
invariants to distinguish nodes within one collection. Views do not express
those index relationships.

\paragraph{Overlapping accesses and lifetimes.}
Other diamond-types reconstructions retain a reference into a vector while
a helper can grow that vector and reallocate its storage. Even an exact
field contract still permits that invalidating operation. A Graphite case
instead needs a returned callback to outlive the borrow of \code{self}
that it captures. Selecting fewer fields cannot extend that lifetime.

Overlapping accesses are not necessarily unsafe, either. In Hyper and wgpu,
a map entry stays borrowed for a later insertion while the program first
reads the map. The desired distinction concerns when the accesses occur,
rather than which fields they use. These cases need reasoning beyond field
selection.

\paragraph{Field separation hidden by an interface.}
Four cases encounter a different obstacle: the interface used by the code
does not expose the desired separation. Ruffle retains a field reference while passing \code{self} to a helper
that expects \code{&mut dyn RenderBackend}. The trait-object parameter cannot state which fields of the concrete
renderer it needs. Zola needs to separate fields of a \code{Library} stored
behind an \code{Arc}. View paths cannot follow that pointer to name those
fields.

Two diamond-types cases need to read a vector's length while keeping
an element mutably borrowed. The distinction between vector metadata and
element storage is hidden by the library interface. These cases require
additional interface support or a different rewrite. They do not establish
that every possible use of views in these programs would fail.

\subsection{Revisions and Validation}

Table~\ref{tab:workaround-revisions} identifies the upstream revisions for
the successful cases. The displayed patch excerpts include both removal
of the workaround and the view annotations.
``Views alone'' in Table~\ref{tab:workaround-outcomes} describes the step
from the reconstructed program to the compiling version, not the size of
the whole patch.

\begin{table}[htbp]
\centering
\caption{Upstream revisions for the successful workaround reversals.
Revision links identify the source repositories.}
\label{tab:workaround-revisions}
\small
\begin{tabular}{@{}ll@{}}
\toprule
\textbf{Project} & \textbf{Revision} \\
\midrule
Sway & \href{https://github.com/FuelLabs/sway/tree/7dafb1c7ae52ae8247e41e8e7b36a6f444b40b71}{\texttt{7dafb1c7ae}} \\
fish & \href{https://github.com/fish-shell/fish-shell/tree/76a3afcab93501f0f1b26f66a761c8d6dc0211fe}{\texttt{76a3afcab9}} \\
Wasmtime & \href{https://github.com/bytecodealliance/wasmtime/tree/e2f9ca6be1b06b5c5c9e78834d10b8132cea0c80}{\texttt{e2f9ca6be1}} \\
grok-build & \href{https://github.com/xai-org/grok-build/tree/2bdd1d6a6369de0e8c68132ea4539e9abd9e14a8}{\texttt{2bdd1d6a63}} \\
Apicula & \href{https://github.com/scurest/apicula/tree/3d4e91e14045392a49c89e86dab8cb936225588c}{\texttt{3d4e91e140}} \\
diamond-types & \href{https://github.com/josephg/diamond-types/tree/00f722d6ebdc9f3aec94d4406d9856b385939eae}{\texttt{00f722d6eb}} \\
Replica & \href{https://github.com/josephg/replica/tree/73ec5c52ce03068784bc11b701deb7a05e6c5e0a}{\texttt{73ec5c52ce}} \\
RhoJVM & \href{https://github.com/MinusGix/RhoJVM-old/tree/de5ae7103cb7e40ffa1b6bc24e34d7a975e44f15}{\texttt{de5ae7103c}} \\
moonfire-nvr & \href{https://github.com/scottlamb/moonfire-nvr/tree/2985214d87823d8555fecf591d651e10e2bd265a}{\texttt{2985214d87}} \\
Binius & \href{https://github.com/IrreducibleOSS/binius/tree/1b09443bbb08b13e0116be81ab09742589aed267}{\texttt{1b09443bbb}} \\
Varisat & \href{https://github.com/jix/varisat/tree/33e876937c5d22305664e9ae5601484b25cec23f}{\texttt{33e876937c}} \\
Helix & \href{https://github.com/helix-editor/helix/tree/ffb54b4eac3403f145679b9be3691d24e13a9599}{\texttt{ffb54b4eac}} \\
\bottomrule
\end{tabular}
\end{table}

For each successful case, the original code compiles under the selected
configuration. Removing the workaround produces borrowing errors under both
stock and modified \codetext{rustc}. Adding views, with the supporting
changes described above where needed, produces a version that compiles.
The recorded runs compare a prototype based on \codetext{rustc} 1.93.0
against \codetext{nightly-2025-11-25}. Repeating the checks for Apicula and
the diamond-types storage case with the updated prototype gives the same
acceptance and rejection results.

We also ran tests where available. Apicula passes three tests on each
version, and diamond-types passes 135 library tests with 20 ignored on each
in the repeated runs. The updated prototype lacks \codetext{rustdoc}, so
we could not run its diamond-types doctests. The Wasmtime WAST package runs
no tests. Moonfire's test executable fails to link because the system
SQLite library is missing. Fish, grok-build, and Binius have matching test
failures on the original and modified versions.

For Helix, both compilers accept the original macro-based editor and reject
the method-based reconstruction with 14 borrowing errors. Adding four views
allows the full \code{hx} executable to build and run its
\codetext{--version} command. We did not test interactive editing.
All variants use the same lockfile update from Ropey 1.3.0 to 1.6.1,
because the older version fails a layout assertion with the selected stock
compiler. The grammar submodules retain their upstream-pinned revisions.

\end{document}